\documentclass[]{interact}

\usepackage[numbers,sort&compress]{natbib}% 
\bibpunct[, ]{[}{]}{,}{n}{,}{,}% 
\makeatletter% 
\def\NAT@def@citea{\def\@citea{\NAT@separator}}% No spaces between citations using natbib
\makeatother% 

\usepackage[latin1]{inputenc}
\usepackage{mathrsfs} 
\usepackage{siunitx}
\usepackage{braket}
\usepackage{calc}
\usepackage{longtable}
\usepackage{tabularx}
\usepackage{dsfont}
\usepackage{color}
\usepackage{amssymb}
\usepackage{placeins}
\usepackage{marvosym}
\usepackage{xcolor,colortbl}
\newcommand{\TikzBilderNeuErzeugen}[1]{}\newcommand{\TikzBildEinfuegen}[2]{\centering\includegraphics[]{fig#1.pdf}#2}
\usepackage{varioref}
\usepackage{hyperref}
\usepackage[capitalise]{cleveref}
\crefname{paragraph}{Paragraph}{Paragraphs}
\Crefname{paragraph}{Paragraph}{Paragraphs}
\usepackage{ifthen}
\usepackage{seqsplit}
\usepackage{textcomp}%Fuer texttilde
\newcommand{\TextTilde}{\raisebox{0.5ex}{\texttildelow}}

\newcommand{\subsubsubsection}[1]{\paragraph{#1}}%
\allowdisplaybreaks

\newcommand{\Eins}{\mathds{1}}%
\newcommand{\ii}{\mathrm{i}}%
\newcommand{\dif}{\mathrm{d}}%
\newcommand{\tdif}[2]{\frac{\dif#1}{\dif#2}}%
\newcommand{\pdif}[2]{\frac{\partial#1}{\partial#2}}%
\newcommand{\Nabla}{\vec{\nabla}}%
\newcommand{\fdif}{\operatorname{\delta}}%
\newcommand{\Fdif}[2]{\frac{\fdif\!#1}{\fdif\!#2}}%
\newcommand{\Rea}{\operatorname{Re}}%
\newcommand{\abs}[1]{\lvert#1\rvert}%
\newcommand{\norm}[1]{\lVert#1\rVert}%
\newcommand{\Tr}{\operatorname{Tr}}%
\newcommand{\ZT}[1]{\textquotedblleft#1\textquotedblright}%

\newcolumntype{L}[1]{>{\raggedright\arraybackslash}p{#1}}% 
\newcolumntype{C}[1]{>{\centering\arraybackslash}p{#1}}%
\newcolumntype{R}[1]{>{\raggedleft\arraybackslash}p{#1}}% 
\newcolumntype{Y}{>{\centering\arraybackslash}X}%    
\newcolumntype{Z}{>{\raggedright\arraybackslash}X}%  

\newlength{\myl}%
\newcommand{\SUM}[2]{{\setlength{\myl}{\widthof{$\displaystyle\sum_{#1}^{#2}$}*\real{0.5}-\widthof{$\displaystyle\sum$}*\real{0.5}}\sum_{#1}^{#2}\;\hspace{-\the\myl}}}% Summen in abgesetzten Gleichungen
\newcommand{\INT}[3]{\settowidth{\myl}{$\displaystyle\int_{#1}^{#2}$}{\int_{#1}^{#2}\;\;\;\hspace{-\the\myl}\dif #3}\,}% Integrale in abgesetzten Gleichungen
\newcommand{\TINT}[3]{\settowidth{\myl}{$\int_{#1}^{#2}$}{\int_{#1}^{#2}\!\ifthenelse{\equal{#1#2}{}}{}{\;\;\;\;\hspace{-\the\myl}}\dif #3}\,}%
\newcommand{\EINT}[3]{\settowidth{\myl}{$\int_{#1}^{#2}$}{\int_{#1}^{#2}\;\;\;\,\hspace{-\the\myl}\dif #3}\,}% Integrale in Exponenten

\newcommand{\uu}{\vec{u}}%
\newcommand{\ffX}[3]{f_{#3}^{(#1\mathrm{D})}}%
\newcommand{\TT}[3]{\mathrm{T}_{#3}^{(#1\mathrm{D})}}%
\newcommand{\LS}{L_{\mathrm{Sp}}}% Schr\"odinger picture Liouvillian
\newcommand{\SmoluchowskiOperator}{L_{\mathrm{S}}}%
\newcommand{\FPOperator}{L_{\mathrm{FP}}}%
\newcommand{\KramersOperator}{L_{\mathrm{K}}}%
\newcommand{\ZustandssummeGaussianChain}{Z_{\mathrm{gc}}}%
\newcommand{\NumberOfParticleTypes}{N_{\mathrm{pt}}}% in chain
\newcommand{\KanonischeZustandssumme}{Z}%
\newcommand{\ProjectionOperator}{\mathcal{P}}%
\newcommand{\NumberOfSpecies}{N_{\mathrm{s}}}% in mixture
\newcommand{\rt}{(\vec{r},t)}%
\newcommand{\memory}{\underline{\mathcal{M}}}%
\newcommand{\randomf}{\mathfrak{z}}%
\newcommand{\pftkernel}{\mathcal{M}_{4}}%
\newcommand{\covariance}{\sigma_{\mathrm{c}}}%
\newcommand{\WeightedDensity}{n}%
\newcommand{\WeightFunction}{w}%
\newcommand{\SubfunktionalFMT}{g}%
\newcommand{\SpeedSound}{c_{\mathrm{s}}}%
\newcommand{\ShearViscosity}{\eta_{\mathrm{s}}}%
\newcommand{\AttenuationLength}{\ell}%
\newcommand{\SmallAmplitude}{\mathfrak{e}}%
\newcommand{\organized}{\mathfrak{v}}%
\newcommand{\frequenz}{\underline{\Omega}}%
\newcommand{\RObstacle}{R_{\mathrm{o}}}%
\newcommand{\winkel}{\varsigma}%
\newcommand{\MobilityGradientDynamics}{M_{\mathrm{g}}}%
\newcommand{\MobilityCahnHilliard}{M_{\mathrm{ch}}}%
\newcommand{\FieldCahnHilliard}{\varphi_{\mathrm{ch}}}%
\newcommand{\ConstantCahnHilliard}{\varphi_{\mathrm{ch},0}}%
\newcommand{\CPCahnHilliard}{\kappa_{\mathrm{ch}}}%
\newcommand{\FEDCahnHilliard}{f_{\mathrm{ch}}}%
\newcommand{\DissipativeMatrix}{\mathcal{M}}%
\newcommand{\InjectionRate}{q_{\mathrm{in}}}%
\newcommand{\DissipativeContribution}{\mathfrak{D}}%
\newcommand{\PSDistributionRVT}{\Psi}%
\newcommand{\MicroscopicPhaseSpaceDistribution}{\widehat{\Psi}}%
\newcommand{\SmoluchowskiDistribution}{\widehat{\Psi}}%
\newcommand{\coupling}{\mathcal{J}}%
\newcommand{\EntropyProduction}{\Phi_{\mathfrak{s}}}%
\newcommand{\DifferenceOperator}{\mathcal{Z}}%
\newcommand{\FpfcII}{F}%
\newcommand{\IntegrationConstantJ}{J_{\mathrm{int}}}%
\newcommand{\Reducedfexc}{g_{\mathrm{fmt}}}%
\newcommand{\QuelltermMCT}{\underline{R}}%
\newcommand{\Stromoperator}[1]{\hat{#1}}%

\newcommand{\creflastconjunction}{, and~}

\begin{document}

\articletype{Review Article}

\title{Classical dynamical density functional theory: from fundamentals to applications}

\author{\name{Michael te Vrugt\textsuperscript{a}, Hartmut L\"owen\textsuperscript{b} and Raphael Wittkowski\textsuperscript{a}*\thanks{*Corresponding author: raphael.wittkowski@uni-muenster.de}}
\affil{\textsuperscript{a}Institut f\"ur Theoretische Physik, Center for Soft Nanoscience, Westf\"alische Wilhelms-Universit\"at M\"unster, D-48149 M\"unster, Germany;\\\textsuperscript{b}Institut f\"ur Theoretische Physik II: Weiche Materie, Heinrich-Heine-Universit\"at D\"usseldorf, D-40225 D\"usseldorf, Germany}}

\maketitle

\begin{abstract}	
Classical dynamical density functional theory (DDFT) is one of the cornerstones of modern statistical mechanics. It is an extension of the highly successful method of classical density functional theory (DFT) to nonequilibrium systems. Originally developed for the treatment of simple and complex fluids, DDFT is now applied in fields as diverse as hydrodynamics, materials science, chemistry, biology, and plasma physics. In this review, we give a broad overview over classical DDFT. We explain its theoretical foundations and the ways in which it can be derived. The relations between the different forms of deterministic and stochastic DDFT as well as between DDFT and related theories, such as quantum-mechanical time-dependent DFT, mode coupling theory, and phase field crystal models, are clarified. Moreover, we discuss the wide spectrum of extensions of DDFT, which covers methods with additional order parameters (like extended DDFT), exact approaches (like power functional theory), and systems with more complex dynamics (like active matter). Finally, the large variety of applications, ranging from fluid mechanics and polymer physics to solidification, pattern formation, biophysics, and electrochemistry, is presented.
\end{abstract}

\begin{pacscode}
05.20.Jj; %Statistical mechanics of classical fluids
05.70.Ln; %Nonequilibrium and irreversible thermodynamics
47.10.-g; %General theory in fluid dynamics
66.10.Cb; %Diffusion and thermal diffusion
68.15.+e; %Liquid thin films
81.30.Fb; %Solidification
64.70.Pf; %Glass transitions
82.35.Jk; %Copolymers, phase transitions, structure
64.60.-i; %General studies of phase transitions
05.65.+b; %Self-organized systems
82.70.Dd; %Colloids
71.15.Mb; %Density functional theory, local density approximation, gradient and other corrections
05.30.-d; %Quantum statistical mechanics
05.90.+m %Other topics in statistical physics, thermodynamics, and nonlinear dynamical systems
87.10.Ed; %Ordinary differential equations (ODE), partial differential equations (PDE), integrodifferential models
\end{pacscode}


\maketitle
\tableofcontents

\section*{List of abbreviations}
\begin{tabularx}{\linewidth}{@{}lX@{}}
0D & zero spatial dimensions\\
1D & one spatial dimension \\
2D & two spatial dimensions \\
3D & three spatial dimensions \\
ABP & active Brownian particle\\
BBGKY & Bogoliubov-Born-Green-Kirkwood-Yvon\\
BD & Brownian dynamics \\
DDFT & dynamical density functional theory \\
DFT & density functional theory \\
DMFT & dynamic mean-field theory\\
EDDFT & extended dynamical density functional theory\\
ENE & ergodic-to-nonergodic\\
EPD & external potential dynamics\\
FMT & fundamental measure theory\\
FTD & functional thermodynamics\\
GCM & Gaussian core model\\
GDA & Generalized diffusion approach\\
MC & Monte Carlo\\
MCT & mode coupling theory\\
MSR & Martin-Siggia-Rose\\
NE-SCGLE & nonequilibrium self-consistent generalized Langevin equations\\
OPD & order-preserving dynamics\\
PCD & particle-conserving dynamics\\
PFC & phase field crystal \\
PFT & power functional theory\\
PNP & Poisson-Nernst-Planck\\
RDDFT & reaction-diffusion density functional theory\\
TDDFT & time-dependent density functional theory\\
TR & time reversal\\
\end{tabularx}

\section{Introduction}
Variational methods have been used in statistical physics for a long time and with tremendous success. Density functional theory (DFT) for quantum systems \cite{HohenbergK1964,KohnS1965,Mermin1965}, for which the Nobel Prize in chemistry 1998 was awarded \cite{Feldman2000}, is one of the central methods of many-electron theory. It allows for the description of a many-particle system based on a variational principle in which the one-body density is the only variable. The same idea has also been applied to classical fluids \cite{EbnerSS1976,SaamE1977,YangFG1976}. Here, a grand-canonical free energy functional is minimized by the equilibrium one-body density $\rho(\vec{r})$ with position $\vec{r}$. Classical DFT has found many applications in soft matter physics.

Dynamical extensions of classical DFT, known as \ZT{dynamical density functional theory} (DDFT) or \ZT{time-dependent density functional theory}, were first suggested on a phenomenological basis \cite{Evans1979,DieterichFM1990,Fraaije1993,Munakata1989} and later derived from the microscopic equations of motion of the individual particles \cite{MarconiT1999,MarconiT2000,ArcherE2004,Yoshimori2005,EspanolL2009}. They describe the time evolution of the one-body density $\rho\rt$ with time $t$ using a continuity equation in which the current is proportional to the gradient of the functional derivative of the free energy that is zero in the equilibrium case corresponding to static DFT. The resulting equation of motion thus describes the relaxation towards the equilibrium state described by DFT. It can, however, also be applied to  systems that do not approach equilibrium, such as driven \cite{DzubiellaL2003,PennaDT2003,RexLL2005} or active \cite{WensinkL2008,WittkowskiL2011,MenzelSHL2016,WittmannB2016} soft matter. In general, DDFT describes not only the equilibrium configuration, but also the dynamical behavior out of equilibrium, in excellent agreement with Brownian dynamics simulations. 

Unlike its static counterpart DFT, DDFT (as presented in Ref.\ \cite{MarconiT1999}) is not an exact theory. It is based on the \textit{adiabatic approximation}, which is the assumption that the pair correlation of the nonequilibrium system is identical to that of an equilibrium system with the same one-body density. While this approximation works well for many systems, \textit{superadiabatic forces}, which include memory effects, have been found to be important in some cases  \cite{BraderS2013,SchindlerS2016}. These can be dealt with using formally exact methods such as power functional theory  \cite{SchmidtB2013,BraderS2015,Schmidt2015,Schmidt2018,delasHerasRS2019,BraderS2014,KrinningerSB2016} or projection operator methods \cite{Nakajima1958,Mori1965,Zwanzig1960,Yoshimori2005,EspanolL2009,teVrugtW2019,teVrugtW2019d}. Moreover, the practical application of DDFT requires additional approximations. This involves, in particular, the choice of the free energy functional, which is not known exactly and is mostly assumed to have a grand-canonical rather than a canonical form \cite{delasHerasBFS2016,SchindlerWB2019}. The limitations of DDFT are discussed in \cref{limit}.

The early forms of DDFT were extended in a vast number of directions, making the theory applicable to systems with nonuniform temperature \cite{WittkowskiLB2012}, hydrodynamic interactions \cite{RexL2008}, or superadiabatic forces \cite{SchmidtB2013} and to particles with inertia \cite{MarconiT2006}, nonspherical shape \cite{RexWL2007}, or self-propulsion \cite{WensinkL2008}. It has also found a significant number of applications, ranging from \ZT{typical} ones, such as the derivation of phase field crystal (PFC) models \cite{vanTeeffelenBVL2009}, phase separation \cite{ArcherE2004}, and solidification \cite{KahlL2009}, to more exotic ones, such as cancer growth \cite{ChauviereLC2012} or quantum hydrodynamics \cite{DiawM2017}. DDFT is now used in many areas of physics, as well as in related subjects such as biology \cite{FangSS2005,AlSaediHAW2018,teVrugtBW2020}, chemistry \cite{YeNTZM2019,LiuL2020,BabelEL2018}, materials science \cite{EmmerichVWS2014,KnollLHKSZM2004,LudwigsBVRMK2003}, engineering \cite{AltevogtEFMvV1999}, mathematics \cite{FehrmanG2019,KonarovskyiLvR2019b}, and philosophy \cite{teVrugt2020}. This diversity arises, since the problem of finding a useful and accurate description of the collective dynamics in many-particle systems is of importance in a large number of areas in and beyond physics. In this review, we present this broad range of applications, together with a detailed discussion of the theoretical foundations and the various extensions of DDFT.

Our review has the form of a reference work, such that even if it is not read from the beginning to the end, all important information can be found quickly by consulting the section one is interested in. Thus, some references are discussed more than once (e.g., a DDFT for active particles with hydrodynamic interactions \cite{MenzelSHL2016} is relevant for \cref{hydro,active}), and many sections contain cross-references or longer lists of articles. Of course, the article can also be read entirely, which gives a complete overview over variants and applications of DDFT.

This review is structured as follows: In \cref{fromstatictodynamic}, we give a historical overview and a brief introduction to static DFT. We explain the standard forms of DDFT, including their derivation, in \cref{traditional}. Extensions of DDFT are presented in \cref{extensions}. In \cref{exact}, formally exact approaches that generalize DDFT are discussed. We explain theories related to DDFT in \cref{theories}. Analytical and numerical methods used to analyze a DDFT are the topic of \cref{methods}. Applications are reviewed in \cref{applications}. An outlook to possible future work is given in \cref{outlook}. We conclude in \cref{conclusion}.

\section{\label{fromstatictodynamic}From static to dynamical density functional theory}
In this section, we first give an overview over the historical development of DDFT in \cref{history}. Afterwards, we provide an introduction to static DFT in \cref{static}, which includes several concepts that are important also for the dynamical case, such as the free energy functional and the direct correlation function. In \cref{dynamicdft}, we explain how DFT can be extended to dynamical situations. 

\subsection{\label{history}Historical overview}
Pioneering work on the theory of Brownian motion (discovered by \citet{Brown1828}) was done by \citet{Einstein1905c,Einstein1906a}, Smoluchowski \cite{Smoluchowski1906,Smoluchowski1916}, \citet{Langevin1908}, \citet{Fokker1913,Fokker1918}, and \citet{Planck1917} (although, as pointed out in Ref.\ \cite{Chavanis2019}, some basics were already developed by Lord \citet{Rayleigh1891}). The discussion of the inhomogeneous liquid-gas interface by \citet{vanderWaals1894} is often considered the first density functional study \cite{Loewen2002}. Further early contributions were made by \citet{Percus1962}, \citet{LebowitzP1963b}, and \citet{StillingerB1962}.

Important foundational work on the statistical mechanics of classical systems, in particular the theory of ensembles, was done by \citet{Gibbs1902}. DFT, however, has its roots mainly in quantum mechanics. Early functional treatments of electron systems where developed by \citet{Thomas1927} and \citet{Fermi1928}. Modern quantum DFT is based on the famous results by Hohenberg, Kohn, and Sham \cite{HohenbergK1964,KohnS1965}, which allow to describe many-electron systems using a variational principle. An extension to systems at nonzero temperature was derived by \citet{Mermin1965}. Moreover, DFT is applied to atomic nuclei (see Ref.\ \cite{Jones2015} for a short overview over such approaches). \citet{RungeG1984} obtained a dynamical extension of quantum DFT, known as time-dependent density functional theory (TDDFT).

The results from Hohenberg and Kohn were applied to classical fluids by Ebner, Saam, and Stroud \cite{EbnerSS1976,SaamE1977}. A transformation between density and one-body potential was, for classical systems, proposed by \citet{YangFG1976} (influenced by work from \citet{DeDominicisM1964}). Another early classical DFT is the generalized van der Waals theory \cite{NordholmH1980,JohnsonN1981}, further contributions were made by Haymet and Oxtoby \cite{HaymetO1981,OxtobyH1982}. Brief historical overviews over DFT can be found in Refs.\ \cite{Loewen2002,Jones2015,EvansORK2016,vanTeeffelenBVL2009,BurghardtB2006,Chavanis2019}.

Dynamical theories in statistical mechanics that provide a connection between microscopic particle dynamics and macroscopic physics date back to the famous Boltzmann equation \cite{Boltzmann1872} (see Ref.\ \cite{BrownUM2009} for a historical overview). The Cahn-Hilliard equation \cite{Cahn1965,CahnH1958} is  an early equation of the gradient dynamics type (see \cref{net}). An extension of the DFT principle towards dynamic equations has been suggested by \citet{Evans1979} for spinodal decomposition. This work introduced what is now known as the governing equation of deterministic DDFT (\cref{trddft}). Methods of this type were also applied to spinodal decomposition in Refs.\ \cite{EvansTdG1979,Abraham1976}. \citet{KirkpatrickW1987} employed a dynamic DFT in the context of the glass transition (see also Refs.\ \cite{KirkpatrickN1986,KirkpatrickN1986b}). Munakata \cite{Munakata1989,Munakata1990,Munakata1990b,Munakata1990c,Munakata1994,Munakata2003} derived a dynamical extension of DFT as the overdamped limit of this result for application to supercooled liquids (see \cref{phenomenological,kawasaki}), thereby extending earlier work on this topic \cite{Munakata1977}. \citet{DieterichFM1990} presented a phenomenological derivation of DDFT. Nonlinear diffusion equations for freezing were justified by the argument that they contain DFT as a static limit \cite{Bagchi1987}. A stochastic dynamic equation for the one-body density of a system of Brownian particles in the form of DDFT was derived by \citet{Dean1996}. \citet{ReinelD1996} developed a generalization of classical DFT to time-dependent situations for lattice gases referred to as \ZT{time-dependent density functional theory}. 

Early approaches to DDFT had in mind a variety of specific applications. They included solvation dynamics and dipolar molecules. A Smoluchowski-Vlasov equation for orientable particles was obtained by \citet{CalefW1983}, whose results were used in the study of orientational relaxation by Chandra and Bagchi \cite{BagchiC1988,ChandraB1988,ChandraB1989,ChandraB1989b,ChandraB1990}. These methods constituted a first DDFT with orientational degrees of freedom (see \cref{nonspherical}). Moreover, DDFT had a very close relation to mode coupling theory (MCT) at its early stages. \citet{KirkpatrickW1987} discussed the relation of a phenomenological DDFT to MCT. A description of glassy dynamics based on a density functional Hamiltonian was developed by \citet{KirkpatrickT1989}. \citet{Kawasaki1994,Kawasaki1998} developed a stochastic DDFT in the form of a Fokker-Planck equation for the probability distribution functional $P[\rho]$, in particular for glassy systems as an alternative to standard MCT. This method was also formulated in path integral form \cite{KawasakiM1997}. Chandra and Bagchi combined DDFT and MCT in theoretical studies of electrolytes \cite{ChandraB1999,ChandraB2000,ChandraB2000b,ChandraB2000c} (see \cref{electro}). An overview over orientational relaxation dynamics is given in Ref.\ \cite{BagchiC1991} and Ref.\ \cite{Bagchi2015} contains historical remarks. \citet{Yoshimori2004} has reviewed early forms of DDFT. Another important line of development concerns polymers. A \ZT{dynamic density functional theory} for polymer dynamics was proposed by Fraaije and coworkers \cite{Fraaije1993,FraaijevVMPEHAGW1997} (see also Refs.\ \cite{Fraaije1994,MauritsvVF1997,HasegawaD1997,HuininkBvDS2000}). \citet{KawasakiK1993} applied a DDFT-type model to polymer blends and binary fluids. 

Deterministic DDFT was first microscopically derived by \citet{MarconiT1999,MarconiT2000} from \textit{Langevin equations}. The theory was later rederived from a \textit{Smoluchowski equation} by \citet{ArcherE2004} and from the \textit{Mori-Zwanzig projection operator method} by \citet{Yoshimori2005} and \citet{EspanolL2009}.

Standard DDFT has since then been developed in various directions. An existence proof was given by \citet{ChanF2005}. \citet{Archer2006} derived a DDFT for atomic fluids, being together with \citet{MarconiT2006} among the first ones to incorporate inertia. \citet{RauscherDKP2007} generalized DDFT to flowing solvents. Hydrodynamic interactions, not present in the original formalism, where included by \citet{RoyallDSvB2007} and by \citet{RexL2008}. Further important developments are the treatment of nonspherical particles \cite{RexWL2007,WittkowskiL2011}, active particles \cite{WensinkL2008}, nonisothermal systems \cite{LopezM2007,WittkowskiLB2012}, and superadiabatic forces \cite{SchmidtB2013}.
\begin{NoHyper}%
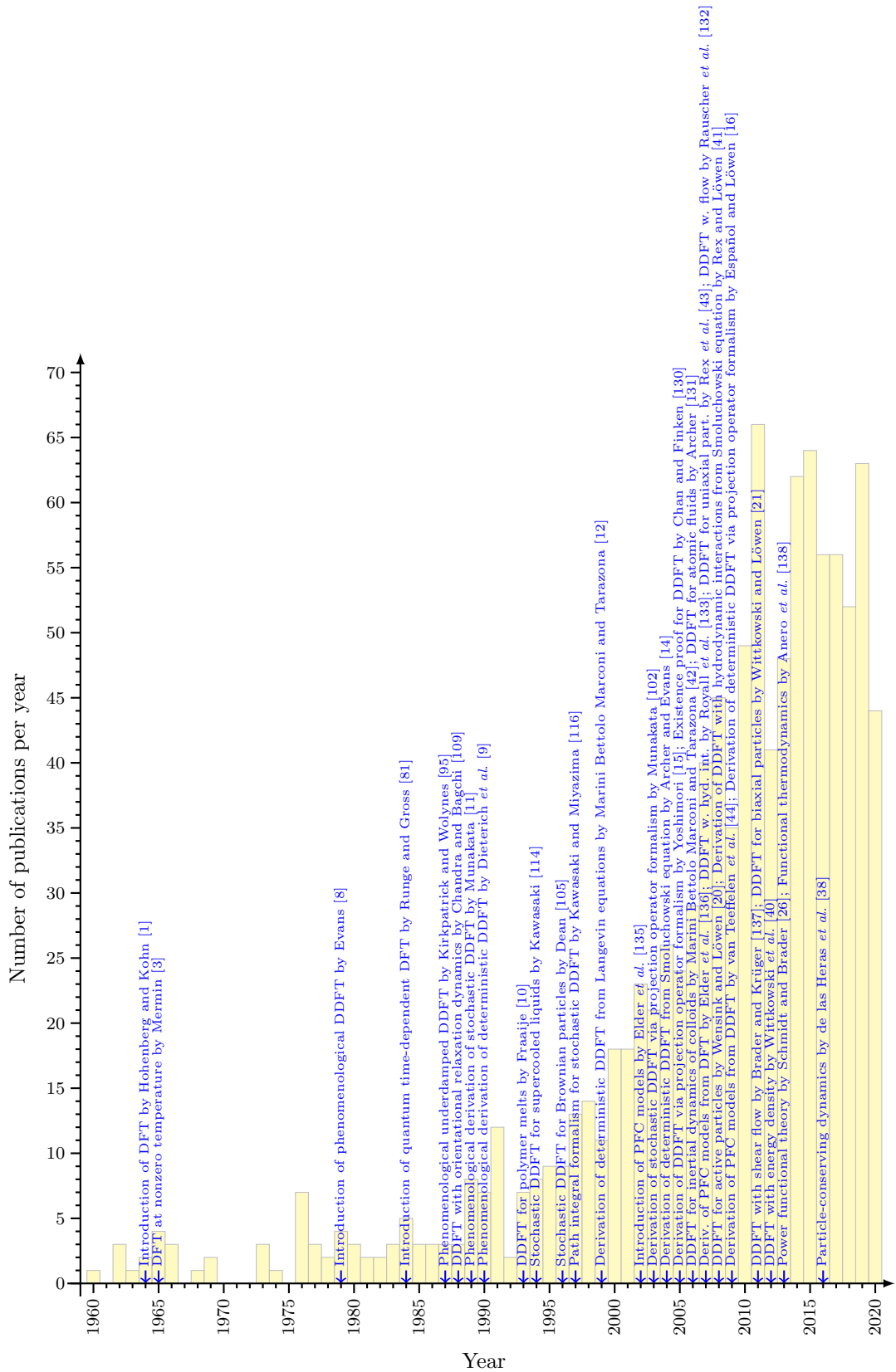
\begin{figure}[htb]
\TikzBilderNeuErzeugen{\tikzstyle{arrow} = [thick,->,>=latex]
\tikzstyle{strichI} = [thick]
\tikzstyle{strichII} = [thick,blue,<-]
\newcommand{\ErstesJahr}{1959}%
\newcommand{\LetztesJahr}{2021}%
\newcount\differenz
\differenz=\LetztesJahr
\advance\differenz by -\ErstesJahr
\newdimen\xl
\xl=\textwidth
\advance\xl by -40pt
\divide\xl by \differenz
\newcommand{\EintragI}[3]{\draw[strichII](#1*\xl,0) -- node[rotate=90,anchor=west,xshift=0.1*\xl]{\color{blue}\fontsize{6}{6}\selectfont #2~by \citet{#3}} (#1*\xl,1.0*\xl);}%
\newcommand{\EintragII}[5]{\draw[strichII](#1*\xl,0) -- node[rotate=90,anchor=west,xshift=0.1*\xl]{\color{blue}\fontsize{6}{6}\selectfont #2~by \citet{#3}; #4~by \citet{#5}} (#1*\xl,1.0*\xl);}%
\newcommand{\EintragIII}[7]{\draw[strichII](#1*\xl,0) -- node[rotate=90,anchor=west,xshift=0.1*\xl]{\color{blue}\fontsize{6}{6}\selectfont #2~by \citet{#3}; #4~by \citet{#5}; #6~by \citet{#7}} (#1*\xl,1.0*\xl);}%
\newcommand{\EintragIV}[9]{\draw[strichII](#1*\xl,0) -- node[rotate=90,anchor=west,xshift=0.1*\xl]{\color{blue}\fontsize{6}{6}\selectfont #2~by \citet{#3}; #4~by \citet{#5}; #6~by \citet{#7}; #8~by \citet{#9}} (#1*\xl,1.0*\xl);}%
\newcommand{\JahrI}[1]{\draw[strichI](#1*\xl,-0.75*\xl) -- node[rotate=90,anchor=east,xshift=-0.1*\xl](N#1){\scriptsize #1} (#1*\xl,0);}%
\newcommand{\JahrII}[1]{\draw[strichI](#1*\xl,-0.75*\xl) -- node[rotate=90,anchor=east,xshift=-0.1*\xl](N#1){\scriptsize #1} (#1*\xl,0);}%
\newcommand{\JahrIII}[1]{\draw[strichI](#1*\xl,-0.375*\xl) -- node[rotate=90,anchor=east,xshift=-0.1*\xl](N#1){} (#1*\xl,0);}%
\newdimen\xll
\xll=\xl
\divide\xll by 2
\newdimen\yI
\yI=\xl
\newcommand{\HistogrammBalken}[2]{\draw[fill=yellow!30,draw=gray!50] (#1*\xl-\xll,0) rectangle (#1*\xl+\xll,#2*\yI);}%
\newcommand{\AnzahlI}[1]{\draw[strichI](\ErstesJahr*\xl-0.75*\xl,#1*\yI) -- node[anchor=east,xshift=-0.1*\xl](N#1){\scriptsize #1} (\ErstesJahr*\xl,#1*\yI);}%
\newcommand{\AnzahlII}[1]{\draw[strichI](\ErstesJahr*\xl-0.75*\xl,#1*\yI) -- node[anchor=east,xshift=-0.1*\xl](N#1){\scriptsize #1} (\ErstesJahr*\xl,#1*\yI);}%
\newcommand{\AnzahlIII}[1]{\draw[strichI](\ErstesJahr*\xl-0.375*\xl,#1*\yI) -- node[anchor=east,xshift=-0.1*\xl](N#1){} (\ErstesJahr*\xl,#1*\yI);}%
\centering
\begin{tikzpicture}%
%
%\HistogrammBalken{1828}{1}\HistogrammBalken{1872}{1}\HistogrammBalken{1891}{1}\HistogrammBalken{1894}{1}
%\HistogrammBalken{1902}{1}\HistogrammBalken{1905}{1}\HistogrammBalken{1906}{2}\HistogrammBalken{1908}{1}
%\HistogrammBalken{1913}{1}\HistogrammBalken{1916}{1}\HistogrammBalken{1917}{1}\HistogrammBalken{1918}{1}
%\HistogrammBalken{1927}{2}\HistogrammBalken{1928}{1}
%\HistogrammBalken{1931}{2}\HistogrammBalken{1932}{1}
%\HistogrammBalken{1946}{2}\HistogrammBalken{1952}{1}\HistogrammBalken{1953}{1}\HistogrammBalken{1954}{1}\HistogrammBalken{1958}{2}
\HistogrammBalken{1960}{1}\HistogrammBalken{1962}{3}\HistogrammBalken{1963}{1}\HistogrammBalken{1964}{2}
\HistogrammBalken{1965}{4}\HistogrammBalken{1966}{3}\HistogrammBalken{1968}{1}\HistogrammBalken{1969}{2}
\HistogrammBalken{1973}{3}\HistogrammBalken{1974}{1}
\HistogrammBalken{1976}{7}\HistogrammBalken{1977}{3}\HistogrammBalken{1978}{2}\HistogrammBalken{1979}{4}
\HistogrammBalken{1980}{3}\HistogrammBalken{1981}{2}\HistogrammBalken{1982}{2}\HistogrammBalken{1983}{3}\HistogrammBalken{1984}{5}
\HistogrammBalken{1985}{3}\HistogrammBalken{1986}{3}\HistogrammBalken{1987}{3}\HistogrammBalken{1988}{3}\HistogrammBalken{1989}{8}
\HistogrammBalken{1990}{7}\HistogrammBalken{1991}{12}\HistogrammBalken{1992}{2}\HistogrammBalken{1993}{7}\HistogrammBalken{1994}{8}
\HistogrammBalken{1995}{9}\HistogrammBalken{1996}{9}\HistogrammBalken{1997}{13}\HistogrammBalken{1998}{14}\HistogrammBalken{1999}{11}
\HistogrammBalken{2000}{18}\HistogrammBalken{2001}{18}\HistogrammBalken{2002}{23}\HistogrammBalken{2003}{21}\HistogrammBalken{2004}{26}
\HistogrammBalken{2005}{24}\HistogrammBalken{2006}{29}\HistogrammBalken{2007}{40}\HistogrammBalken{2008}{45}\HistogrammBalken{2009}{35}
\HistogrammBalken{2010}{49}\HistogrammBalken{2011}{66}\HistogrammBalken{2012}{41}\HistogrammBalken{2013}{48}\HistogrammBalken{2014}{62}
\HistogrammBalken{2015}{64}\HistogrammBalken{2016}{56}\HistogrammBalken{2017}{56}\HistogrammBalken{2018}{52}\HistogrammBalken{2019}{63}
\HistogrammBalken{2020}{44}
\JahrI{1960}\JahrIII{1961}\JahrIII{1962}\JahrIII{1963}\JahrIII{1964}
\JahrII{1965}\JahrIII{1966}\JahrIII{1967}\JahrIII{1968}\JahrIII{1969}
\JahrI{1970}\JahrIII{1971}\JahrIII{1972}\JahrIII{1973}\JahrIII{1974}
\JahrII{1975}\JahrIII{1976}\JahrIII{1977}\JahrIII{1978}\JahrIII{1979}
\JahrI{1980}\JahrIII{1981}\JahrIII{1982}\JahrIII{1983}\JahrIII{1984}
\JahrII{1985}\JahrIII{1986}\JahrIII{1987}\JahrIII{1988}\JahrIII{1989}
\JahrI{1990}\JahrIII{1991}\JahrIII{1992}\JahrIII{1993}\JahrIII{1994}
\JahrII{1995}\JahrIII{1996}\JahrIII{1997}\JahrIII{1998}\JahrIII{1999}
\JahrI{2000}\JahrIII{2001}\JahrIII{2002}\JahrIII{2003}\JahrIII{2004}
\JahrII{2005}\JahrIII{2006}\JahrIII{2007}\JahrIII{2008}\JahrIII{2009}
\JahrI{2010}\JahrIII{2011}\JahrIII{2012}\JahrIII{2013}\JahrIII{2014}
\JahrII{2015}\JahrIII{2016}\JahrIII{2017}\JahrIII{2018}\JahrIII{2019}
\JahrI{2020}
\AnzahlI{0}\AnzahlIII{1}\AnzahlIII{2}\AnzahlIII{3}\AnzahlIII{4}
\AnzahlII{5}\AnzahlIII{6}\AnzahlIII{7}\AnzahlIII{8}\AnzahlIII{9}
\AnzahlI{10}\AnzahlIII{11}\AnzahlIII{12}\AnzahlIII{13}\AnzahlIII{14}
\AnzahlII{15}\AnzahlIII{16}\AnzahlIII{17}\AnzahlIII{18}\AnzahlIII{19}
\AnzahlI{20}\AnzahlIII{21}\AnzahlIII{22}\AnzahlIII{23}\AnzahlIII{24}
\AnzahlII{25}\AnzahlIII{26}\AnzahlIII{27}\AnzahlIII{28}\AnzahlIII{29}
\AnzahlI{30}\AnzahlIII{31}\AnzahlIII{32}\AnzahlIII{33}\AnzahlIII{34}
\AnzahlII{35}\AnzahlIII{36}\AnzahlIII{37}\AnzahlIII{38}\AnzahlIII{39}
\AnzahlI{40}\AnzahlIII{41}\AnzahlIII{42}\AnzahlIII{43}\AnzahlIII{44}
\AnzahlII{45}\AnzahlIII{46}\AnzahlIII{47}\AnzahlIII{48}\AnzahlIII{49}
\AnzahlI{50}\AnzahlIII{51}\AnzahlIII{52}\AnzahlIII{53}\AnzahlIII{54}
\AnzahlII{55}\AnzahlIII{56}\AnzahlIII{57}\AnzahlIII{58}\AnzahlIII{59}
\AnzahlI{60}\AnzahlIII{61}\AnzahlIII{62}\AnzahlIII{63}\AnzahlIII{64}
\AnzahlII{65}\AnzahlIII{66}\AnzahlIII{67}\AnzahlIII{68}\AnzahlIII{69}
\AnzahlI{70}
\EintragI{1964}{Introduction of DFT}{HohenbergK1964}
\EintragI{1965}{DFT at nonzero temperature}{Mermin1965}
\EintragI{1979}{Introduction of phenomenological DDFT}{Evans1979}
\EintragI{1984}{Introduction of quantum time-dependent DFT}{RungeG1984}
\EintragI{1987}{Phenomenological underdamped DDFT}{KirkpatrickW1987}
\EintragI{1988}{DDFT with orientational relaxation dynamics}{ChandraB1988}
\EintragI{1989}{Phenomenological derivation of stochastic DDFT}{Munakata1989}
\EintragI{1990}{Phenomenological derivation of deterministic DDFT}{DieterichFM1990}
\EintragI{1993}{DDFT for polymer melts}{Fraaije1993}
\EintragI{1994}{Stochastic DDFT for supercooled liquids}{Kawasaki1994}
\EintragI{1996}{Stochastic DDFT for Brownian particles}{Dean1996}
\EintragI{1997}{Path integral formalism for stochastic DDFT}{KawasakiM1997}
\EintragI{1999}{Derivation of deterministic DDFT from Langevin equations}{MarconiT1999}
\EintragI{2002}{Introduction of PFC models}{ElderKHG2002}
\EintragI{2003}{Derivation of stochastic DDFT via projection operator formalism}{Munakata2003}
\EintragI{2004}{Derivation of deterministic DDFT from Smoluchowski equation}{ArcherE2004}
\EintragII{2005}{Derivation of DDFT via projection operator formalism}{Yoshimori2005}%
{Existence proof for DDFT}{ChanF2005}
\EintragII{2006}{DDFT for inertial dynamics of colloids}{MarconiT2006}%
{DDFT for atomic fluids}{Archer2006}
\EintragIV{2007}{Deriv.\ of PFC models from DFT}{ElderPBSG2007}%
{DDFT w.\ hyd.\ int.}{RoyallDSvB2007}%
{DDFT for uniaxial part.}{RexWL2007}%
{DDFT w.\ flow}{RauscherDKP2007}
\EintragII{2008}{DDFT for active particles}{WensinkL2008}%
{Derivation of DDFT with hydrodynamic interactions from Smoluchowski equation}{RexL2008}
\EintragII{2009}{Derivation of PFC models from DDFT}{vanTeeffelenBVL2009}%
{Derivation of deterministic DDFT via projection operator formalism}{EspanolL2009}
\EintragII{2011}{DDFT with shear flow}{BraderK2011}%
{DDFT for biaxial particles}{WittkowskiL2011}%
\EintragI{2012}{DDFT with energy density}{WittkowskiLB2012}
\EintragII{2013}{Power functional theory}{SchmidtB2013}%
{Functional thermodynamics}{AneroET2013}
\EintragI{2016}{Particle-conserving dynamics}{delasHerasBFS2016}
\draw[arrow](\ErstesJahr*\xl-\xl,0) -- node[anchor=north]{} (\LetztesJahr*\xl+\xll,0);
\draw[arrow](\ErstesJahr*\xl,-\xl) -- node[anchor=south,rotate=90]{} (\ErstesJahr*\xl,71.5*\xl);
\draw node[anchor=north,font=\small]at(N1990.west){Year};
\draw node[anchor=south,rotate=90,font=\small]at(N35.west){Number of publications per year};
\end{tikzpicture}}%
\TikzBildEinfuegen{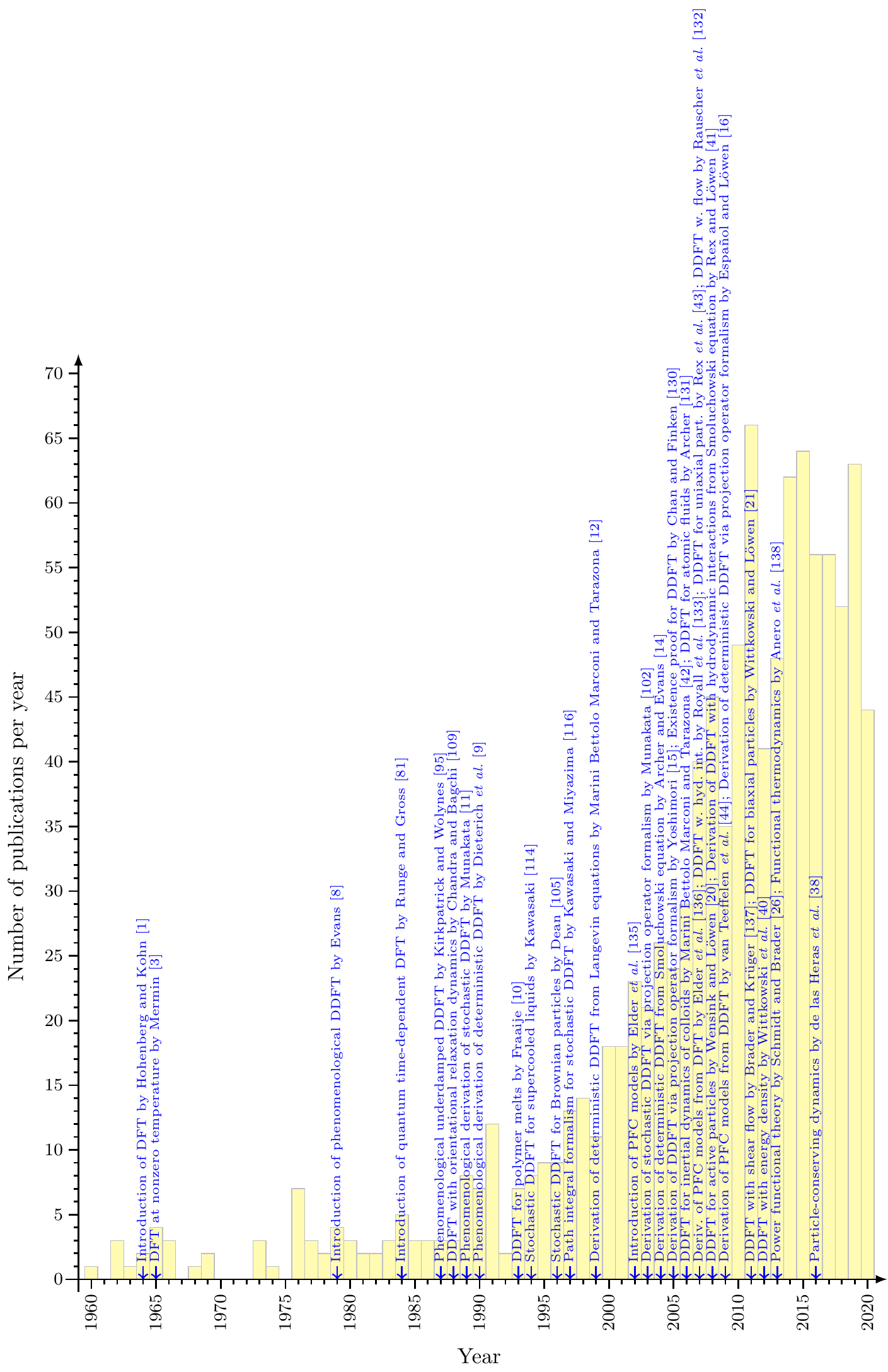}{\nocite{HohenbergK1964,Mermin1965,Evans1979,RungeG1984,KirkpatrickW1987,ChandraB1988,Munakata1989,DieterichFM1990,Fraaije1993,Kawasaki1994,Dean1996,KawasakiM1997,MarconiT1999,ElderKHG2002,Munakata2003,ArcherE2004,Yoshimori2005,ChanF2005,MarconiT2006,Archer2006,ElderPBSG2007,RoyallDSvB2007,RexWL2007,RauscherDKP2007,ElderPBSG2007,RoyallDSvB2007,RexWL2007,RauscherDKP2007,WensinkL2008,RexL2008,vanTeeffelenBVL2009,EspanolL2009,BraderK2011,WittkowskiL2011,WittkowskiLB2012,SchmidtB2013,AneroET2013,delasHerasBFS2016}}%
\caption{\label{fig:Zeitpfeil}Important steps in the development of DDFT and publications considered in this review over the course of time.}%
\end{figure}%
\end{NoHyper}%
\FloatBarrier\noindent%
A closely related approach are phase field crystal models \cite{EmmerichEtAl2012} (see \cref{phasefield}), proposed by \citet{ElderKHG2002} and later derived from DDFT by \citet{vanTeeffelenBVL2009}. 

An overview over the historical development of DDFT is given in \vref{fig:Zeitpfeil}.

\subsection{\label{static}Static density functional theory}
\subsubsection{\label{variation}Variational principle}
Classical static density functional theory (DFT) is a statistical-mechanical theory for the equilibrium state of a classical many-body system. We present it following Ref.\ \cite{Loewen1994a}. Microscopically, a system of $N$ particles is described by the phase-space distribution function $\MicroscopicPhaseSpaceDistribution(\{\vec{r}_i,\vec{p}_i\})$ that depends on the positions $\{\vec{r}_i\}$ and momenta $\{\vec{p}_i\}$ of the individual particles and is normalized as $\Tr(\MicroscopicPhaseSpaceDistribution)=1$ with the trace $\Tr$. We introduce a grand-canonical free energy functional $\overline{\Omega}(T,\mu,[\MicroscopicPhaseSpaceDistribution])$ that depends on the temperature $T$ and the chemical potential $\mu$, which are treated as fixed parameters, and that is moreover a functional of the distribution $\MicroscopicPhaseSpaceDistribution$. In the grand-canonical case, the trace of an arbitrary function $Y$ on phase space is (for systems in 3D) given by
\begin{equation}
\Tr(Y) = \sum_{N=0}^{\infty}\frac{1}{N!h^{3N}}\INT{}{}{^3 r_1}\INT{}{}{^3 p_1}\dotsb\INT{}{}{^3 r_N}\INT{}{}{^3 p_N}Y,
\end{equation}
where $h$ is the Planck constant. Denoting the $N$-particle Hamiltonian by $H_N$, the grand-canonical free energy functional is given by 
\begin{equation}
\overline{\Omega}(T,\mu,[\MicroscopicPhaseSpaceDistribution])=\Tr(\MicroscopicPhaseSpaceDistribution(H_N - \mu N + k_B T \ln(\MicroscopicPhaseSpaceDistribution))),
\label{grandfunctional}
\end{equation}
where $k_B$ is the Boltzmann constant. When evaluated at the equilibrium distribution
\begin{equation}
\MicroscopicPhaseSpaceDistribution_{\mathrm{eq}} = \frac{1}{\Xi}e^{-\beta(H_N - \mu N)},
\label{psi0}
\end{equation}
with the thermodynamic beta $\beta = 1/(k_B T)$ and the grand-canonical partition function
\begin{equation}
\Xi = \Tr(\exp(-\beta(H_N - \mu N))),    
\end{equation}
the functional \eqref{grandfunctional} is minimized and equal to the actual equilibrium grand-canonical free energy. If we fix the particle interactions, $\MicroscopicPhaseSpaceDistribution_{\mathrm{eq}}$ is completely determined by the external potential $U_1(\vec{r})$. 

Since it is defined on a $6N$-dimensional phase space, the function $\MicroscopicPhaseSpaceDistribution$ can be very complicated. DFT makes it possible to work with the one-body density
\begin{equation}
\rho(\vec{r}) =  \bigg\langle \sum_{i=1}^{N}\delta(\vec{r}-\vec{r}_i) \bigg\rangle
\end{equation}
instead, where $\braket{\cdot}$ denotes an ensemble average and $\delta(\vec{r})$ is the Dirac delta distribution. The one-body density gives the probability of finding a particle at position $\vec{r}$. We denote the one-body distribution at equilibrium by $\rho_{\mathrm{eq}}(\vec{r})$. \citet{Mermin1965} showed that the external potential $U_1(\vec{r})$ entering the equation \eqref{psi0} for $\MicroscopicPhaseSpaceDistribution_{\mathrm{eq}}$ is uniquely determined once $\rho_{\mathrm{eq}}(\vec{r})$ is known, such that $\MicroscopicPhaseSpaceDistribution_{\mathrm{eq}}$ is a functional of $\rho_{\mathrm{eq}}$. Moreover, the mapping $\rho \to U_1$ can be shown to exist under very general conditions \cite{ChayesC1984}. Hence, we can introduce a well-defined functional (\ZT{intrinsic free energy})
\begin{equation}
F_\mathrm{in}(T,[\rho])=\Tr(\MicroscopicPhaseSpaceDistribution_{\mathrm{eq}}[\rho](H_{\mathrm{kin}} + H_{\mathrm{int}} + k_B T \ln(\MicroscopicPhaseSpaceDistribution_{\mathrm{eq}}[\rho]))),
\label{fzero}
\end{equation}
with the kinetic part $H_{\mathrm{kin}}$ and the interaction part $H_{\mathrm{int}}$ of the Hamiltonian, that is independent of the external potential. A further functional is given by
\begin{equation}
\Omega(T,\mu,[\rho]) = F_\mathrm{in}(T,[\rho]) + \INT{}{}{^3r}\rho(\vec{r})U_1(\vec{r}) - \mu\INT{}{}{^3r}\rho(\vec{r}).
\end{equation}
This functional is minimal and equal to the equilibrium grand-canonical free energy if evaluated at the equilibrium density $\rho_{\mathrm{eq}}$. These considerations lead to the central \textit{variational principle of DFT}
\begin{equation}
\frac{\delta \Omega(T,\mu,[\rho])}{\delta \rho(\vec{r})}\bigg|_{\rho = \rho_{\mathrm{eq}}} = 0.
\label{dft}
\end{equation}
$\Omega(T,\mu,[\rho_{\mathrm{eq}}])$ then gives the actual grand-canonical free energy of the system \cite{Loewen2002}.

For finding the equilibrium configuration, one needs to determine the functional $\Omega(T,\mu,[\rho])$, which in most cases cannot be done exactly. Through a Legendre transformation it can be related to a Helmholtz free energy functional $F(T,[\rho])$ in the form \cite{EmmerichEtAl2012}
\begin{equation}
\Omega(T,\mu,[\rho]) = F(T,[\rho]) - \mu\INT{}{}{^3r}\rho(\vec{r}).
\end{equation}
We can split the free energy into three parts:
\begin{equation}
F(T,[\rho])= F_{\mathrm{id}}(T,[\rho]) + F_{\mathrm{exc}}(T,[\rho]) + F_{\mathrm{ext}}([\rho]).
\label{freeenergy}
\end{equation}
The first term
\begin{equation}
F_{\mathrm{id}}(T,[\rho]) = k_B T \INT{}{}{^3r}\rho(\vec{r})(\ln(\Lambda^3\rho(\vec{r}))-1),
\label{idealfreeenergy}
\end{equation}
with the thermal de Broglie wavelength $\Lambda$, is the exact expression for the free energy of an ideal gas. An external potential $U_1(\vec{r})$ is taken into account through the last term
\begin{equation}
F_{\mathrm{ext}}([\rho]) = \INT{}{}{^3r}\rho(\vec{r})U_1(\vec{r}).
\end{equation}
Finally, the term $F_{\mathrm{exc}}(T,[\rho])$, known as \textit{excess free energy}, gives the contribution from the particle interactions. In general, it cannot be calculated exactly and needs to be approximated. Approximation methods (which are also required in DDFT) are discussed in \cref{fmt}. Note that many authors refer to the intrinsic free energy functional $F_\mathrm{in}$ given by \cref{fzero} as \ZT{free energy} and use the symbol $F$ for the intrinsic free energy. This is simply a different convention. It is, however, important to clarify whether $F$ denotes (in our terminology) the intrinsic or the full free energy, since this affects the form of the DDFT equation (see \cref{dynamicdft}). In this review, we employ the convention \eqref{freeenergy}. From now on, we will drop the parametric dependence on $T$ in our notation. 

Reviews of DFT (some of which include comments on DDFT) can be found in Refs.\ \cite{Loewen1994a,Evans1992,Lutsko2010,Ram2014,Wu2006,WuL2007,Evans2009,TarazonaCM2008}. Books discussing DFT include Refs.\ \cite{HansenMD2009,RowlinsonW2013}.

\subsubsection{\label{fmt}Approximations for the free energy functional}
For practical applications of DFT, one requires an expression for the free energy \cite{Loewen2003}. More precisely, since the ideal gas contribution $F_{\mathrm{id}}$ and the external potential contribution $F_{\mathrm{ext}}$ are known analytically, one has to find a good approximation for the excess free energy $F_{\mathrm{exc}}$. Since the same problem arises in DDFT, which is in most cases based on a free energy functional from DFT, we here explain various ways of constructing an approximate excess free energy.

A useful starting point are functional Taylor expansions of the free energy around a homogeneous reference density $\rho_0$, which we present following Ref.\ \cite{EmmerichEtAl2012}. They take the form
\begin{equation}
F_{\mathrm{exc}}[\rho] = F_{\mathrm{exc}}^{(0)}(\rho_0) - \sum_{n=1}^{\infty}\frac{k_B T}{n!} \INT{}{}{^3 r_1}\dotsb \INT{}{}{^3 r_n}c^{(n)}(\vec{r}_1, \dotsc, \vec{r}_n)\prod_{i=1}^{n}\Delta\rho(\vec{r}_i)
\label{functionaltaylorexpansion}
\end{equation}
with an irrelevant constant $F_{\mathrm{exc}}^{(0)}(\rho_0)$, the $n$-th-order direct correlation function
\begin{equation}
c^{(n)}(\vec{r}_1, \dotsc, \vec{r}_n) = - \frac{1}{k_B T}\frac{\delta^{n} F[\rho]}{\delta\rho(\vec{r}_1)\dotsb\delta\rho(\vec{r}_n)}\bigg|_{\rho=\rho_0},
\end{equation}
and the reduced density $\Delta\rho(\vec{r}) = \rho(\vec{r})- \rho_0$. Truncating the expansion \eqref{functionaltaylorexpansion} at second order gives the Ramakrishnan-Yussouff approximation \cite{RamakrishnanY1979}
\begin{equation}
F_{\mathrm{exc}}[\rho] = - \frac{1}{2}k_B T\INT{}{}{^3r_1}\INT{}{}{^3r_2}c^{(2)}(\vec{r}_1 - \vec{r}_2)\Delta\rho(\vec{r}_1)\Delta\rho(\vec{r}_2),
\label{ramak}
\end{equation}
which allows to predict the freezing transition. The correlation function can be obtained, e.g., from the random phase approximation
\begin{equation}
c^{(2)}(\vec{r}_1 - \vec{r}_2) = - \frac{U_2(\vec{r}_1 - \vec{r}_2)}{k_B T}  
\label{rpa}
\end{equation}
or the virial expression
\begin{equation}
c^{(2)}(\vec{r}_1 - \vec{r}_2) = \exp\!\bigg(\! - \frac{U_2(\vec{r}_1 - \vec{r}_2)}{k_B T}\bigg)-1. 
\end{equation}

Which approximation gives good results depends on the system under consideration. A very simple form of $F_{\mathrm{exc}}$ that is particularly useful in the case of soft interactions and high densities is the \textit{mean-field approximation} \cite{Loewen2010b,LangLWL2000,ArcherE2001,Archer2005c,HansenMD2009,LouisBH2000}
\begin{equation}
F_{\mathrm{exc}}=\frac{1}{2}\INT{}{}{^3r}\INT{}{}{^3r'}U_2(\vec{r}-\vec{r}')\rho(\vec{r})\rho(\vec{r}').
\label{meanfieldapproximation}%
\end{equation}
Since the second functional derivative of $F_{\mathrm{exc}}$ gives the direct pair-correlation function, the mean-field approximation generates the random phase approximation \eqref{rpa}. On the other hand, the form \eqref{meanfieldapproximation} can be obtained by inserting the random phase approximation \eqref{rpa} into the Ramakrishnan-Yussouff approximation \eqref{ramak}. (This also requires the replacement $\Delta \rho \to \rho$, which is possible in DDFT on this level of approximation, since the difference arising from this replacement vanishes in the DDFT equation \eqref{trddft} due to a combination of functional derivatives and gradient operator.) The mean-field approximation has been used, e.g., in DDFT models for ultrasoft particles \cite{DzubiellaL2003}, active particles \cite{WittmannB2016}, and social interactions \cite{teVrugtBW2020}.

For particles with hard-core interactions, such as hard spheres, \textit{fundamental measure theory} (FMT), which has been introduced by \citet{Rosenfeld1989}, is a very successful approach. We present it here following Ref.\ \cite{StopperMRH2015}. A review can be found in Ref.\ \cite{Roth2010}. Note that FMT can also be combined with the mean-field approximation \eqref{meanfieldapproximation}: For example, when describing particles interacting through a combination of hard-sphere repulsion and attractive interaction, one can approximate $F_{\mathrm{exc}}$ as a sum of a term for the hard-sphere interaction (obtained from FMT) and a mean-field term for the attractive interaction \cite{ArcherE2004}.

The starting point is the fact that the excess free energy for a dilute hard-sphere mixture (with $\NumberOfSpecies$ species) can be written as
\begin{equation}
F_{\mathrm{exc}} = k_B T\INT{}{}{^3r}(\WeightedDensity_0(\vec{r})\WeightedDensity_3(\vec{r}) + \WeightedDensity_1(\vec{r})\WeightedDensity_2(\vec{r}) - \vec{\WeightedDensity}_1(\vec{r})\cdot\vec{\WeightedDensity}_2(\vec{r}))
\label{fexcdilute}
\end{equation}
with the weighted densities
\begin{equation}
\WeightedDensity_\alpha(\vec{r}) = \sum_{i=1}^{\NumberOfSpecies}\INT{}{}{^3r'}\rho_i(\vec{r}')\WeightFunction_\alpha^i(\vec{r}-\vec{r}')
\end{equation}
and the weight functions
\begin{align}
\WeightFunction^i_3(\vec{r}) &= \Theta(R_i - \norm{\vec{r}}),\\
\WeightFunction_2^i(\vec{r})&= \delta(R_i - \norm{\vec{r}}),\\
\WeightFunction_1^i(\vec{r})&=\frac{\WeightFunction_2^i}{4\pi R_i},\\
\WeightFunction_0^i(\vec{r})&=\frac{\WeightFunction_2^i}{4\pi R_i^2},\\
\vec{\WeightFunction}_2^i(\vec{r})&=\frac{\vec{r}}{r}\WeightFunction_2^i,\\
\vec{\WeightFunction}_1^i(\vec{r})&=\frac{\vec{\WeightFunction}_2^i}{4\pi R_i},
\end{align}
where $\Theta(\cdot)$ denotes the Heaviside step function, $R_i$ is the radius of the spheres of the $i$-th species, and $\norm{\cdot}$ is the Euclidean norm. 
The form \eqref{fexcdilute} is extended to larger densities by the ansatz
\begin{equation}
F_{\mathrm{exc}} = k_B T\INT{}{}{^3r}\Reducedfexc(\{\WeightedDensity_\alpha(\vec{r})\})
\end{equation}
with the function $\Reducedfexc$ that has to recover the low-density limit. Various forms of $\Reducedfexc$ are used in the literature. \citet{Rosenfeld1989} proposed the form
\begin{equation}
\begin{split}
\Reducedfexc = -\WeightedDensity_0\ln(1-\WeightedDensity_3) + \frac{\WeightedDensity_1\WeightedDensity_2 - \vec{\WeightedDensity}_1\cdot\vec{\WeightedDensity}_2}{1-\WeightedDensity_3} + \frac{\WeightedDensity_2^3 - 3\WeightedDensity_2\vec{\WeightedDensity}_2\cdot\vec{\WeightedDensity}_2}{24\pi(1-\WeightedDensity_3)^2}.
\label{rosenfeldfmt}
\end{split}    
\end{equation}
Numerical aspects of this expression are discussed in Ref.\ \cite{NoldGYSK2017}. This form, however, cannot describe hard-sphere crystals \cite{StopperMRH2015}. A problem occurs due to a divergence in the last term on the right-hand side of \cref{rosenfeldfmt}. This can be avoided using the $q_3$ correction \cite{RosenfeldSLT1996,RosenfeldSLT1997}, in which the last term of \cref{rosenfeldfmt} is modified as
\begin{equation}
\frac{1}{24\pi(1-\WeightedDensity_3)^2}\bigg(\WeightedDensity_2 - \frac{\vec{\WeightedDensity}_2 \cdot \vec{\WeightedDensity}_2}{\WeightedDensity_2}\bigg)^3.
\end{equation}
An alternative proposed by \citet{Tarazona2000} is to introduce the tensorial weight function (here in the notation of Refs.\ \cite{SchmidtLBE2000,Roth2010})
\begin{equation}
\WeightFunction^i_{m_2}(\vec{r}) = \bigg(\frac{\vec{r}\otimes\vec{r}}{r^2} - \frac{1}{3}\Eins \bigg)\WeightFunction^i_2(\vec{r})
\end{equation}
with the dyadic product $\otimes$ and the 3D identity matrix $\Eins$. From $\WeightFunction^i_{m_2}$ one obtains a tensorial weighted density $\WeightedDensity_{m_2}$,
which allows to rewrite the third term on the right-hand side of \cref{rosenfeldfmt} as \cite{Roth2010}
\begin{equation}
\frac{\WeightedDensity_2^3 - 3\WeightedDensity_2\vec{\WeightedDensity}_2\cdot\vec{\WeightedDensity}_2}{24\pi(1-\WeightedDensity_3)^2} + \frac{9(\vec{\WeightedDensity}_2 \cdot \WeightedDensity_{m_2}\vec{\WeightedDensity}_2 - \frac{1}{2} \Tr(\WeightedDensity^3_{m_2}))}{24\pi(1-\WeightedDensity_3)^2}, 
\end{equation}
where $\Tr$ denotes the trace of a matrix (and not the trace on phase space as in the rest of this article). Using the $q_3$ correction or the tensorial modification, a description of hard-sphere crystals in FMT is possible  \cite{StopperMRH2015}. Additional modifications of FMT have also been proposed, such as the White Bear FMT \cite{RothELK2002}. A state-of-the-art form is White Bear mark II \cite{HansenGoosR2006}, given by \cite{StopperMRH2015}
\begin{equation}
\begin{split}
\Reducedfexc &= -\WeightedDensity_0\ln(1-\WeightedDensity_3) + \frac{\WeightedDensity_1\WeightedDensity_2 - \vec{\WeightedDensity}_1\cdot\vec{\WeightedDensity}_2}{1-\WeightedDensity_3}\bigg(1+\frac{1}{3}\SubfunktionalFMT_2(\WeightedDensity_3)\bigg)\\
&\quad\:\! + \frac{\WeightedDensity_2^3 - 3\WeightedDensity_2\vec{\WeightedDensity}_2\cdot\vec{\WeightedDensity}_2}{24\pi(1-\WeightedDensity_3)^2}\bigg(1-\frac{1}{3}\SubfunktionalFMT_3(\WeightedDensity_3)\bigg)
\end{split}
\end{equation}
with the functionals
\begin{align}
\SubfunktionalFMT_2(\WeightedDensity_3)&=\frac{1}{\WeightedDensity_3}(2\WeightedDensity_3 - \WeightedDensity_3^2 + 2(1-\WeightedDensity_3)\ln(1-\WeightedDensity_3)),\\
\SubfunktionalFMT_3(\WeightedDensity_3)&=\frac{1}{\WeightedDensity_3^2}(2\WeightedDensity_3 - 3\WeightedDensity_3^2 + 2 \WeightedDensity_3^3+ 2(1-\WeightedDensity_3)^2\ln(1-\WeightedDensity_3)).
\end{align}
When combined with a tensorial modification, White Bear mark II gives accurate results in the description of hard-sphere crystals \cite{OettelGHLRS2010}. FMT can be extended to general nonspherical convex particle shapes \cite{HansenGoosM2009,HansenGoosM2010,WittmannMM2015,WittmannSSL2017,MarechalGHL2011,WittmannMM2016}, which allows to apply it to orientational dynamics \cite{HaertelL2010,HaertelBL2010}. An FMT-based DDFT for hard polyhedra was developed by \citet{MarechalL2013}. Moreover, lattice FMT \cite{LafuenteC2002,LafuenteC2004,OettelKDESH2016} can be used in lattice DDFT models of hard rods \cite{KlopotekHGDSSO2017}. Applications of FMT to DDFT can be found, e.g., in Refs.\ \cite{AerovK2014,BabelEL2018,BiervR2007,delasHerasBFS2016,JiangCJW2014,GaoX2018,GoddardHO2020,GonzalezMV2017,HaertelL2010,HaertelBL2010,HowardNP2017,HouDZTM2017,KelkarFC2014,KelkarFC2015,LianZLW2016,LiuGHLH2017,LiuWDZM2020,NeuhausHMSL2014,NeuhausSL2013,NeuhausSL2013b,NiuLLH2019,OettelDBNS2012,RexL2008,RexL2009,RothRA2009,RoyallDSvB2007,StopperMRH2015,StopperR2018,StopperRH2016,StopperTDR2018,WittmannMMB2017,WittmannB2016,YatsyshinSK2012,YeNTZM2016,YeTZDM2016,YeNTZM2018,ZhaoWZDTM2014}.

\subsection{\label{dynamicdft}Dynamical density functional theory}
Up to now, we have only considered equilibrium situations, in which the density $\rho(\vec{r})$ is constant with respect to time and minimizes the grand-canonical free energy. In this case, the density distribution can be determined from the variational principle \eqref{dft}. The central idea of DDFT is to extend this principle to nonequilibrium situations. If the equilibrium state is the one in which the free energy is minimized, the dynamic equation will have a form which ensures that the free energy decreases over time. This allows to relate the rate of change $\partial_t\rho\rt$ of the particle number density to the variation of the free energy. Moreover, since $\rho\rt$ is a conserved quantity, its time evolution will have the form of a continuity equation, i.e., $\partial_t\rho\rt$ is proportional to the gradient of a flux. These considerations lead to the central equation of deterministic \textit{dynamical density functional theory}
\begin{equation}
\pdif{}{t}\rho\rt=\Gamma\vec{\nabla}\cdot\bigg(\rho\rt\vec{\nabla}\frac{\delta F[\rho]}{\delta\rho\rt}\bigg)
\label{trddft}%
\end{equation}
with the mobility $\Gamma$. Note that the theory is formally based on the free energy $F$, in contrast to DFT which uses the grand-canonical free energy $\Omega$. However, this makes no significant difference in most practical applications, and one typically uses free energy functionals from DFT, which are grand-canonical (see \cref{particleconserving} for exceptions). If $F$ denotes the intrinsic free energy (see \cref{static} for a discussion of this convention), a term $\Gamma \Nabla \cdot (\rho\Nabla U_1)$ has to be added to the right-hand side of \cref{trddft}. The general result \eqref{trddft} can be derived in a variety of ways (see \cref{derivation} for an overview). Some authors add a noise term to \cref{trddft}, this corresponds to the stochastic DDFT equation
\begin{equation}
\pdif{}{t}\rho\rt=\Gamma\Nabla\cdot\bigg(\rho\rt\Nabla\frac{\delta F[\rho]}{\delta\rho\rt}\bigg) + \Nabla\cdot(\sqrt{2\Gamma k_B T\rho\rt}\vec{\eta}\rt),
\label{trddfts}
\end{equation}
where $\vec{\eta}\rt$ is a multiplicative noise. Deterministic DDFT describes the ensemble-averaged one-body density, whereas stochastic DDFT is concerned either with the exact microscopic or with a coarse-grained density \cite{ArcherR2004}. The relation of deterministic and stochastic approaches is discussed in \cref{noise}. 

At the heart of deterministic DDFT is the assumption that the equal-time two-point correlation function of the nonequilibrium system is identical to that of the equilibrium system with the same one-body density \cite{MarconiT1999}. This \textit{adiabatic approximation}, which is strictly valid only if the system density evolves infinitely slowly (quasi-statically), allows to express the interaction term in the dynamic equation in terms of the functional derivative of the excess free energy. For obtaining the time evolution of $\rho\rt$ from \cref{trddft}, one needs to determine the form of $F[\rho]$, which, in general, cannot be done exactly. In many situations, one uses the equilibrium density functional, making the approximation that it can still be applied close to equilibrium. This approximation can break down in the case of, e.g., driven or active systems, which require the construction of a different functional. Note that the \textit{adiabatic approximation} made in the derivation of stochastic DDFT from underdamped dynamics (see \cref{phenomenological}) is the weaker assumption that the density is a slow variable compared to the momentum. In this review, we denote by \textit{adiabatic approximation} the assumption of equilibrium pair correlations unless stated otherwise. Stochastic DDFTs describe a coarse-grained rather than an ensemble-averaged density, and can thus differ in their degree of coarse graining. If they are set up as an exact (as compared to Brownian dynamics) theory for the microscopic density operator, they have no adiabatic approximation. Similar approximations can, however, be involved in further coarse graining \cite{ArcherR2004}.

The dynamical principle of DDFT can be expressed in four ways:
\begin{enumerate}
\item As a (deterministic or stochastic) differential equation for the one-body density $\rho\rt$ in the form \eqref{trddft} or \eqref{trddfts}.
\item As a variational principle based on a dissipation functional (see \cref{net}).
\item As a stochastic Fokker-Planck equation for the probability distribution $P[\rho]$ (see \cref{kawasaki}).
\item As an action functional in the path integral formalism (see \cref{renormalization}).
\end{enumerate}
The formulation as a differential equation is by far the most common one, although others can have advantages in specific contexts (such as the dissipation functional for calculating entropy production \cite{WittkowskiLB2013} or the action functional for renormalization \cite{KimKJvW2014}). The last two ways are associated with stochastic DDFT.

Introductions to DDFT can be found in Refs.\ \cite{Loewen2010b,Loewen2010W,Loewen2017,Lutsko2010}. 

\section{\label{traditional}Standard DDFT}
\subsection{Overview}
In the standard form of DDFT, the equation of motion takes the form \eqref{trddft} for the deterministic and \eqref{trddfts} for the stochastic case. This equation describes the dynamics of the one-body density $\rho\rt$ of a fluid. Standard DDFT assumes that the particles interact through a two-body potential $U_2(\vec{r}_1,\vec{r}_2)=U_2(\vec{r}_1 - \vec{r}_2)$ and have no degrees of freedom other than their center-of-mass position. ($N$-body interactions can also be incorporated \cite{ArcherE2004}, this is discussed in \cref{derivationsmolu}.) In practice, the pair potential will generally be assumed to have the form $U_2(\norm{\vec{r}_1 - \vec{r}_2})$, such that standard DDFT describes spherical particles. (Strictly speaking, this assumption is made for the interaction potential and not for the physical particles. For example, the effective interaction between polymers is often assumed to have a Gaussian form \cite{LouisBH2000,ArcherE2001}. The same form has been used for a DDFT model of humans \cite{teVrugtBW2020}. A Gaussian interaction potential has spherical symmetry, but this does not mean that polymers or humans are spheres.) Hydrodynamic interactions and effects of inertia are neglected and the system is assumed to be isothermal. In this section, we will cover deterministic (see \cref{marconi}) and stochastic (see \cref{kawasaki}) DDFT for simple and colloidal fluids and DDFT for polymer melts (see \cref{pd}) as these are the three main conceptually distinct lines of development. The deterministic form \eqref{trddft} is the most widely used one, such that we will, for simplicity, often write \ZT{(standard) DDFT} to refer only to this theory if there is no danger of confusion.

Equation \eqref{trddft} describes, for time-independent external fields, a diffusive relaxation to an equilibrium state. In the classification by \citet{HohenbergH1977}, it is a \textit{model B} equation, i.e., it describes a conserved order parameter in the overdamped limit without hydrodynamics \cite{MullerdP2013}. The fact that \cref{trddft,trddfts} are overdamped can have two origins. First, it is possible that the underlying dynamics is itself overdamped, which is the case if it is given by overdamped Langevin equations (as in the derivations by \citet{MarconiT1999} and \citet{Dean1996}). In this case, the underlying particles do (within the theory) not have momentum degrees of freedom, such that a momentum density cannot even be defined. Second, it is possible that the underlying dynamics is underdamped (e.g., Hamiltonian) such that a full theory would also contain the momentum density. In this case, one makes the further approximation that the momentum density relaxes quickly compared to the number density on the level of the mesoscopic theory (as done by \citet{Munakata1989}).

The main difference of \cref{trddft} to the standard diffusion equation
\begin{equation}
\pdif{}{t}\rho\rt= D \vec{\nabla}^2\rho\rt
\label{diffusionequation}
\end{equation}
with the diffusion constant $D=\Gamma k_B T$, to which \cref{trddft} reduces for noninteracting particles (ideal gas free energy), is that interactions of the particles with each other and with external fields can also be incorporated \cite{Rauscher2015}. 

\subsection{Variants}
\subsubsection{\label{marconi}Deterministic DDFT for simple and colloidal fluids}
The deterministic form of DDFT was derived microscopically for colloidal fluids by \citet{MarconiT1999} as well as \citet{ArcherE2004} and for simple fluids by \citet{Archer2006}. These theories and their derivations are discussed in detail in \cref{langevin}, \cref{derivationsmolu}, and \cref{inertia}, respectively. 

\subsubsection{\label{kawasaki}Stochastic DDFT for simple and colloidal fluids}
Stochastic DDFT was developed by \citet{Kawasaki1994,Kawasaki1995,Kawasaki1998}, \citet{Munakata1989,Munakata1990,Munakata1994}, \citet{KirkpatrickW1987}, and \citet{Dean1996}, and is thus historically older than deterministic DDFT. The diversity of its roots can make this form difficult to access, since there are approaches that are equivalent even though they look very different (such as the Fokker-Planck and the path-integral formalism by Kawasaki), but also approaches that are not equivalent despite looking identical (such as the stochastic DDFT equations by Kawasaki and Dean).

A pioneer of stochastic DDFT was Munakata \cite{Munakata1989}, whose derivation is discussed in \cref{phenomenological}. This approach, which starts from an equation presented by \citet{KirkpatrickW1987} and from which earlier theories for supercooled liquids \cite{Munakata1977} can be recovered, allows to study density fluctuations in liquids \cite{Munakata1990,Munakata1990b,Munakata1995,Munakata1996}, polymer conformation \cite{TakahashiM1997,Munakata2001}, and shear viscosity \cite{ArakiM1995,YoshidaHM1996}. Within this framework, various H-theorems can be proven \cite{Munakata1994}.

Kawasaki's DDFT, first derived in Ref.\ \cite{Kawasaki1994}, is a stochastic theory for the coarse-grained probability distribution functional $P([\rho],t)$ and applicable to both simple and complex fluids. This method has been developed as an alternative to and extension of MCT for the description of supercooled liquids and the glass transition \cite{FuchizakiK1998,FuchizakiK1999,FuchizakiK1999b,KawasakiF1998,KawasakiK1996,Fuchizaki2000}. It can be solved by mapping it onto a kinetic lattice gas model \cite{FuchizakiK1998,FuchizakiK1998b,FuchizakiK1999}. A further discussion can be found in Ref.\ \cite{Das2011}. We present the derivation following Ref.\ \cite{FuchizakiK2002}: In the case of a colloidal fluid, one starts from the Smoluchowski equation
\begin{equation}
\pdif{}{t}\SmoluchowskiDistribution(\{\vec{r}_i\},t) = \SmoluchowskiOperator(\{\vec{r}_i\})\SmoluchowskiDistribution(\{\vec{r}_i\},t)
\label{smoluchk}
\end{equation}
for the many-particle distribution function $\SmoluchowskiDistribution$ with the Smoluchowski operator
\begin{equation}
\SmoluchowskiOperator(\{\vec{r}_i\}) = \sum_{i=1}^{N} D \vec{\nabla}_{\vec{r}_i}\cdot\bigg(\vec{\nabla}_{\vec{r}_i}+\frac{1}{k_B T}\vec{\nabla}_{\vec{r}_i} U_N(\{\vec{r}_i\})\bigg)
\end{equation}
with the total potential energy $U_N(\{\vec{r}_i\})$.

The system is then coarse-grained by splitting it into cells of volume $v_\rho$. To overcome the problem that the interaction $U_N$ depends on distances smaller than the coarse-graining size, one first makes $v_\rho$ infinitesimal, which allows to rewrite the interaction term in \cref{smoluchk} using a bilinear functional $U(\rho)$. It is then assumed that, after the coarse graining, one can express $U$ as a bilinear in the coarse-grained density $\rho$. Expanding around the homogeneous liquid state $\rho = \rho_0$ up to second order in $\Delta\rho=\rho(\vec{r})-\rho_0$ and using that Fourier transforming and averaging the second-order term gives the static structure factor and therefore the direct pair-correlation function $c^{(2)}(r)$, one finds \cite{FuchizakiK2002}
\begin{equation}
\pdif{}{t}P([\rho],t)=\FPOperator[\rho]P([\rho],t) 
\label{eq:kawasaki}
\end{equation}
with the Fokker-Planck operator 
\begin{equation}
\FPOperator[\rho] = -\frac{D}{k_B T}\INT{}{}{^3r}\frac{\delta}{\delta \rho\rt}\vec{\nabla}\cdot\bigg(\rho\rt\vec{\nabla}\bigg(k_B T \frac{\delta}{\delta \rho\rt} + \frac{\delta F[\rho]}{\delta \rho\rt}\bigg)\!\bigg)
\end{equation}
and the coarse-grained free energy 
\begin{equation}
\begin{split}
F[\rho] &= k_B T \INT{}{}{^3r}\rho\rt \ln\!\bigg(\frac{\rho\rt}{\rho_0}-1\bigg) \\
&\quad\:\!-\frac{1}{2}k_B T\INT{}{}{^3r}\INT{}{}{^3r'}c^{(2)}(\norm{\vec{r}-\vec{r}'})\Delta\rho\rt \Delta\rho(\vec{r}',t).
\end{split}
\end{equation}
Note that \cref{eq:kawasaki} has a stationary solution of the form $\exp(-\beta F[\rho])$ \cite{Munakata1989}. Equation \eqref{eq:kawasaki} is the Fokker-Planck equation corresponding to the stochastic dynamic equation
\begin{equation}
\pdif{}{t}\rho\rt=\Gamma\Nabla\cdot\bigg(\rho\rt\Nabla\frac{\delta F[\rho]}{\delta\rho\rt}\bigg) + \Nabla\cdot(\sqrt{2\Gamma k_B T\rho\rt}\vec{\eta}\rt)
\label{dk2}
\end{equation}
with the Gaussian noise $\vec{\eta}\rt$ satisfying
\begin{align}
\braket{\eta_i\rt}&=0,\label{eq:langevinetaI}\\
\braket{\eta_i\rt\eta_j(\vec{r}',t)}&=\delta_{ij}\delta(\vec{r}-\vec{r}')\delta(t-t').\label{eq:langevinetaII}
\end{align}
Since \cref{dk2} is simply the Langevin equation corresponding to the Fokker-Planck equation \eqref{eq:kawasaki} \cite{ArcherR2004}, the density field $\rho\rt$ has the same interpretation (spatially coarse-grained density profile) in both equations. One can, of course, make the coarse-graining cells very small, such that they are comparable to the interparticle distance, in which case the density field would be a very \ZT{spiky} and locally smoothened function for which noise is very important \cite{Kawasaki1994}.

A coarse-grained description can also be derived for an atomic fluid. Here, the microscopic description is given by the Liouville rather than by the Smoluchowski equation. As a starting point, the method of fluctuating hydrodynamics is used, which provides a coarse-grained Fokker-Planck equation for a functional $P([\rho,\vec{g}],t)$ that also depends on the momentum density $\vec{g}\rt$. Using the projection operator method \cite{Mori1965}, the momentum density is then eliminated, giving a closed equation of motion for $P([\rho],t)$ that, despite describing a different physical system, also has the form \eqref{eq:kawasaki}. The only difference is that $D/(k_B T)$ is replaced by $\tau/m$, where $\tau$ is the time scale of momentum relaxation that determines the time scale on which the coarse-grained description is valid \cite{Kawasaki1994,FuchizakiK2002} and $m$ is the particle mass. A detailed discussion of the derivation of stochastic DDFT can also be found in Ref.\ \cite{Kawasaki1998}. Reviews are given by Refs.\ \cite{Kawasaki2009,FuchizakiK2002}.

A closely related approach is the (generalized) stochastic Smoluchowski equation \cite{Chavanis2015,Chavanis2019} (see also Ref.\ \cite{Chavanis2008})
\begin{equation}
\pdif{}{t}\rho\rt = \Nabla\cdot\bigg(\Gamma g(\rho)\Nabla\frac{\delta F[\rho]}{\delta \rho\rt}\bigg) + \Nabla \cdot(\sqrt{2\Gamma k_B T g(\rho)}\vec{\eta}\rt) 
\label{generalizedsmoluchowski}
\end{equation}
with the function $g(\rho)$. This approach (which also exists in a deterministic form) is based on the idea of modeling the dynamics of the density using, instead of $\Gamma\rho$ and $D$, the more general expressions $\Gamma g(\rho)$  and $D\widehat{g}(\rho)$ with functions $g(\rho)$ and $\widehat{g}(\rho)$ that take into account microscopic constraints (the standard case is recovered by setting $g(\rho)=\rho$ and $\widehat{g}(\rho)=1$). For the same purpose, the standard Boltzmann entropy is replaced by a generalized entropy, which also leads to a modified free energy. A generalized Einstein relation, derived from the requirement that an equilibrium state of the generalized Smoluchowski equation is a thermodynamic equilibrium state as described by the generalized free energy, shows that the second derivative of the local density of the generalized entropy is proportional to the ratio $\widehat{g}(\rho)/g(\rho)$. This allows to eliminate $\widehat{g}(\rho)$ and to write the dynamics in the form \eqref{generalizedsmoluchowski}. Whether a noise term is required depends on whether $\rho$ describes the coarse-grained or the ensemble-averaged density (see \cref{noise}) \cite{Chavanis2019}.

Another form of stochastic DDFT was developed by \citet{Dean1996}. It was derived for a system of $N$ overdamped particles in a white-noise heat bath interacting via a pair potential $U_2(\vec{r})$. Using stochastic calculus, he showed that the density operator
\begin{equation}
\hat{\rho}\rt = \sum_{i=1}^{N}\delta(\vec{r}_i(t) - \vec{r})
\label{hatrho}%
\end{equation}
with the positions $\{\vec{r}_i\}$ of the individual particles obeys the Langevin equation
\begin{equation}
\begin{split}
\pdif{}{t}\hat{\rho}\rt &= \vec{\nabla}\cdot(\sqrt{2\Gamma k_B T\hat{\rho}\rt}\vec{\eta}\rt) \\
&\quad\:\!+ \Gamma\vec{\nabla}\cdot\bigg(\hat{\rho}\rt\INT{}{}{^3r'}\hat{\rho}(\vec{r}',t)\vec{\nabla}U_2(\vec{r}-\vec{r}')\bigg) \\
&\quad\:\!+ D\vec{\nabla}^2\hat{\rho}\rt.
\label{dk1}%
\end{split}
\end{equation}
With the free energy 
\begin{equation}
F[\hat{\rho}] = \frac{1}{2}\INT{}{}{^3r}\INT{}{}{^3r'}\hat{\rho}\rt U_2(\vec{r}-\vec{r}')\hat{\rho}(\vec{r}',t) + k_BT\INT{}{}{^3r}\hat{\rho}\rt(\ln(\Lambda^3\hat{\rho}\rt)-1), 
\label{deanfreeenergy}
\end{equation}
\cref{dk1} becomes
\begin{equation}
\pdif{}{t}\hat{\rho}\rt=\Gamma \Nabla\cdot\bigg(\hat{\rho}\rt\vec{\nabla}\frac{\delta F[\hat{\rho}]}{\delta\hat{\rho}\rt}\bigg) +\Nabla\cdot(\sqrt{2\Gamma k_B T\hat{\rho}\rt}\vec{\eta}\rt),
\label{dk}%
\end{equation}
making the relation to DDFT more obvious. (In Dean's original article \cite{Dean1996}, the last term of \cref{deanfreeenergy} is simply written with $\hat{\rho}\ln(\hat{\rho})$ instead of $\hat{\rho}(\ln(\Lambda^3\hat{\rho})-1)$, which makes no difference for \cref{dk}.) Mathematically, the notation in Eqs.\ \eqref{dk1}-\eqref{dk}, which involves square roots and logarithms of Dirac delta distributions, should be understood as formal \cite{DonevVE2014}. An extension to orientational degrees of freedom is also possible \cite{CugliandoloDLvW2015}. A notable aspect in the derivation of \cref{dk} is the treatment of the noise: The Langevin equations \eqref{eq:langevinglgnherleitung} (see below), which are the starting point for the derivation of \cref{dk}, contain an \textit{additive} noise term $\vec{\chi}_i(t)$ for each particle $i$. This leads to a noise term in the dynamic equation for $\hat{\rho}$ in which all $\vec{\chi}_i(t)$ appear. Then, this term is rewritten in such a way that it contains the \textit{multiplicative} noise field $\vec{\eta}\rt$ instead, giving the closed equation of motion \eqref{dk}. 

Dean's theory is a formally exact microscopic balance equation for the density field $\hat{\rho}\rt$ defined as a sum over Dirac delta distributions, in contrast to both deterministic DDFT and to \cref{dk2}, which provide coarse-grained dynamic equations. In consequence, \cref{dk,dk2} should, despite the fact that they look identical, not be confused \cite{GuptaDB2011,Das2011}. Equation \eqref{dk} can also be written as a Fokker-Planck equation, which takes the form of \cref{eq:kawasaki} (although, again, the meaning is different). Often, the name \textit{Dean-Kawasaki equation} is used for both \cref{dk} and \cref{dk2} (although, given that they are not equivalent, this name is not ideal \cite{Das2011}). The Dean-Kawasaki equation (in one or both of its forms) is discussed in more detail in Refs.\ \cite{FrusawaH2000,DasY2013,DeanLMP2016,DelfauOLBH2016,Frusawa2019,Frusawa2020,KonarovskyiLvR2019}. Moreover, Dean's results are of importance in active matter physics \cite{TjhungNC2018,SolonCT2015,BarreCMP2015,TailleurC2008,CatesT2015,SolonSWKKCT2015,FilyM2012,GelimsonG2015}.

For mathematical discussions of the well-posedness of stochastic differential equations of this type, see Refs.\ \cite{FehrmanG2019,KonarovskyiLvR2019b}. The stress tensor of stochastic DDFT is discussed in Ref.\ \cite{KrugerSDRD2018}. Extensions to underdamped dynamics that include the momentum density are possible \cite{NakamuraY2009,Lutsko2012,PerezMadridRR2002}. A third way of setting up stochastic DDFT in addition to using a Langevin or a Fokker-Planck equation is the path integral formalism \cite{KawasakiM1997,KimK2007,KimK2008,BidhoodiD2015} (see \cref{renormalization} for details).

\subsubsection{\label{pd}DDFT for polymer melts}
DDFT also plays an important role in the study of polymers. The theory was first presented by \citet{Fraaije1993} (see also Ref.\ \cite{Fraaije1994}) and then improved significantly by \citet{FraaijevVMPEHAGW1997}. Earlier work on models of this type was done by \citet{Binder1983} and \citet{KawasakiS1987,KawasakiS1988,KawasakiS1989}. Although there are similarities to the DDFTs presented in \cref{marconi,kawasaki}, differences arise from the fact that polymers are chains of connected beads rather than simple particles. This form of DDFT is also known as \ZT{dynamic mean-field density functional method} and typically introduced as generalized time-dependent Ginzburg-Landau theory \cite{FraaijevVMPEHAGW1997}. Historically, the methods from \cref{marconi,kawasaki} and the method presented in this section have had a relatively independent development.

Consider a mixture of $N$ chains of length $n_{\mathrm{c}}$, which consist of $\NumberOfParticleTypes$ particle types indexed by an integer $i$. For each particle type, one introduces a density field $\rho_i\rt$. The general evolution equation, obtained by \citet{KawasakiS1987}, reads
\begin{equation}
\pdif{}{t}\rho_i\rt = -\sum_{j=1}^{\NumberOfParticleTypes}\INT{}{}{^3r'}M_{ij}(\vec{r},\vec{r}',t)\mu_j(\vec{r}',t) + \eta_i\rt
\end{equation}
with the kinetic coefficient matrix $M_{ij}$, the chemical potential
\begin{equation}
\mu_i\rt = \frac{\delta F[\{\rho_k\}]}{\delta \rho_i\rt},
\end{equation}
and the noise $\eta_i$ (which we will drop from now on). The microscopic construction of polymer DDFT is discussed in Ref.\ \cite{ManthaQS2020}. 
Various approximations for the collective dynamics can be used \cite{MauritsF1997}. The simplest one, employed by Fraaije \cite{Fraaije1993,FraaijevVMPEHAGW1997}, is the \textit{local coupling}
\begin{equation}
\pdif{}{t}\rho_i\rt = \frac{D_{\mathrm{lo}}}{k_B T}\Nabla\cdot(\rho_i\rt\Nabla\mu_i\rt)
\label{loco}
\end{equation}
with the diffusion constant $D_{\mathrm{lo}}$. Here, the close relation to the standard DDFT equation \eqref{trddft} is obvious. The introduction of an additional pressure functional allows to account for incompressibility \cite{FraaijevVMPEHAGW1997}, which gives for a binary mixture
\begin{align}
\pdif{}{t}\rho_a\rt &= M_b\nu_m\Nabla\cdot\big(\rho_a\rt\rho_b\rt\Nabla(\mu_a\rt - \mu_b\rt)\big),\label{locoa}\\
\pdif{}{t}\rho_b\rt &= M_b\nu_m\Nabla\cdot\big(\rho_a\rt\rho_b\rt\Nabla(\mu_b\rt - \mu_a\rt)\big)\label{locob}
\end{align}
with the bead mobility parameter $M_b$ and the constant molecular volume $\nu_m$. More complex approaches are the Rouse dynamics
\begin{equation}
\pdif{}{t}\rho_i\rt = \frac{D_{\mathrm{ro}}}{k_B T}\sum_{j=1}^{\NumberOfParticleTypes}\Nabla_{\vec{r}}\cdot\INT{}{}{^3r'}P_{ij}(\vec{r},\vec{r}',t)\vec{\nabla}_{\vec{r}'}\mu_j(\vec{r}',t)
\label{rouse}%
\end{equation}
with the diffusion constant $D_{\mathrm{ro}}$ and the reptation dynamics for continuous chains parametrized by a real index $s$ \cite{KawasakiS1988}
\begin{equation}
\pdif{}{t}\rho_s\rt = -\frac{D_{\mathrm{c}}}{k_B T}\INT{}{}{^3r'}\INT{0}{n_{\mathrm{c}}}{s'}\bigg(\pdif{^2}{s\partial s'}P_{ss'}(\vec{r},\vec{r}',t)\bigg)\mu_{s'}(\vec{r}',t) 
\end{equation}
with the diffusion constant $D_{\mathrm{c}}$. The two-body correlators are defined as \cite{MauritsF1997}
\begin{align}
P_{ss'}(\vec{r},\vec{r}',t) &= N \langle\delta(\vec{r}-\vec{r}_s)\delta(\vec{r}'-\vec{r}_{s'})\rangle,\\ 
P_{ij}(\vec{r},\vec{r}',t) &= \sum_{s=1}^{n_\mathrm{c}}\sum_{s'=1}^{n_\mathrm{c}}\delta^{\mathrm{K}}_{is}\delta^{\mathrm{K}}_{js'}P_{ss'}(\vec{r},\vec{r}',t),
\end{align}
where $\vec{r}_s$ is the position of bead $s$ and $\delta^{\mathrm{K}}_{ij}$ is a Kronecker function that is $1$ if bead $j$ is of type $i$ and $0$ otherwise. (This means that $\delta^{\mathrm{K}}_{ij}$ is \textit{not} a standard Kronecker delta. For example, if we have two particle types which we call 1 and 2, $\delta^{\mathrm{K}}_{11}$ would be zero if the first bead is not of type 1.) Note that in the literature, time-dependence is sometimes dropped, i.e., one writes, e.g., $\partial_t \rho(\vec{r})$ instead of $\partial_t \rho\rt$ even though $\rho$ is time-dependent. 

The statistical theory is introduced following Ref.\ \cite{FraaijevVMPEHAGW1997}. Microscopically, the system is described using a distribution function $\SmoluchowskiDistribution(\vec{r}_{11},\dotsc,\vec{r}_{N n_{\mathrm{c}}})$ with $\vec{r}_{ls}$ being the position of bead $s$ from chain $l$, i.e., we now also indicate the chain number. An average of the microscopic density operator is given by 
\begin{equation}
\rho_i[\SmoluchowskiDistribution](\vec{r})=\sum_{l=1}^{N}\sum_{s=1}^{n_{\mathrm{c}}} \delta^{\mathrm{K}}_{is} \Tr(\SmoluchowskiDistribution\delta(\vec{r}-\vec{r}_{ls})),
\label{density}
\end{equation}
where the trace denotes an integral over $\mathbb{R}^{N n_{\mathrm{c}}}$. For an observed density $\rho_i(\vec{r})$, the requirement $\rho_i(\vec{r})=\rho_i[\SmoluchowskiDistribution](\vec{r})$ gives an equivalence class of microscopic distribution functions on which a free energy functional is defined. It is now assumed that (a) correlations between chains can be accounted for in the free energy by a mean-field approximation (while the chains are treated exactly) and that (b) the distribution function $\SmoluchowskiDistribution$ is chosen in such a way that the free energy is minimized. $\SmoluchowskiDistribution$ is thus independent of the history of the system and only constrained by the density. This allows to obtain a bijective relation between density fields and external potentials. Denoting the external potential for species $i$ by $U_i(\vec{r})$, the constrained minimization based on DFT gives the intrinsic free energy functional
\begin{equation}
F_{\mathrm{in}}[\rho]=-k_B T\ln(\ZustandssummeGaussianChain) + k_B T\ln(N!) - \sum_{i=1}^{\NumberOfParticleTypes}\INT{}{}{^3r} \rho_i(\vec{r}) U_i(\vec{r}) + F_{\mathrm{ni}}[\rho]
\end{equation}
with the partition function of a single Gaussian chain $\ZustandssummeGaussianChain$ and the nonideal free energy $F_{\mathrm{ni}}$. 

DDFT for polymer melts is reviewed in Refs.\ \cite{SevinkF2008b,SevinkZF2003}. Extensions are discussed in \cref{expolymer} and applications in \cref{polymer}.

\subsection{\label{derivation}Derivation}
\begin{figure*}[htb]
\centering
\TikzBilderNeuErzeugen{\tikzstyle{begriffsfeldI} = [rectangle, rounded corners, minimum width=5mm, minimum height=5mm, text width=20mm, text centered, draw=black, fill=green!10]
\tikzstyle{begriffsfeldII} = [rectangle, rounded corners, minimum width=5mm, minimum height=5mm, text width=34mm, text centered, draw=black, fill=yellow!30]%
\tikzstyle{begriffsfeldIII} = [rectangle, rounded corners, minimum width=5mm, minimum height=5mm, text width=20mm, text centered, draw=black, fill=yellow!30]%
\tikzstyle{arrow} = [thick,->,>=latex]
\begin{tikzpicture}[node distance=15mm and 0mm]%
\coordinate[](P1)at(0pt,0pt);
\coordinate[](P2)at(\textwidth,0pt);
\draw[arrow](P1) -- node[midway,above,text centered](N1){increasing length and time scales} (P2);
\draw[draw=none](P1)--(0pt,-3pt);
\node(K3)[begriffsfeldII,anchor=south west,xshift=10mm,yshift=10mm]at(P1){Smoluchowski equation};
\node(K2)[begriffsfeldII, above=of K3.north west,anchor=south west]{Langevin equations};
\node(K1)[begriffsfeldII, above=of K2.north west,anchor=south west,xshift=-10mm]{Hamiltonian equations};
\node(K0)[begriffsfeldI,right=of K2.east,anchor=west,xshift=40pt]{Dynamical density functional theory};
\node(K4)[begriffsfeldIII,anchor=east,align=center]at(P2|-K0.east){Phenome-\\nological\\level};
\draw[implies-implies, double equal sign distance, thick](K3.north) to node[text width=10mm,anchor=west]{statistically\newline equivalent} (K2.south); 
\draw[arrow](K1.south) --++(0,-0.75) --++(1,0) node[text width=15mm,anchor=west]{coarse graining} -- (K2.north);
\draw[arrow](K1.east) -| node[pos=0,anchor=south west,xshift=0mm]{projection operator formalism} (K0);
\draw[arrow](K2) -- node[anchor=south]{} (K0);
\draw[arrow](K3) -| node[pos=0.25, anchor=north]{} (K0);
\draw[arrow](K4) -- node[text width=30mm,midway,above,text centered]{phenomenological\newline considerations} (K0);
\end{tikzpicture}}%
\TikzBildEinfuegen{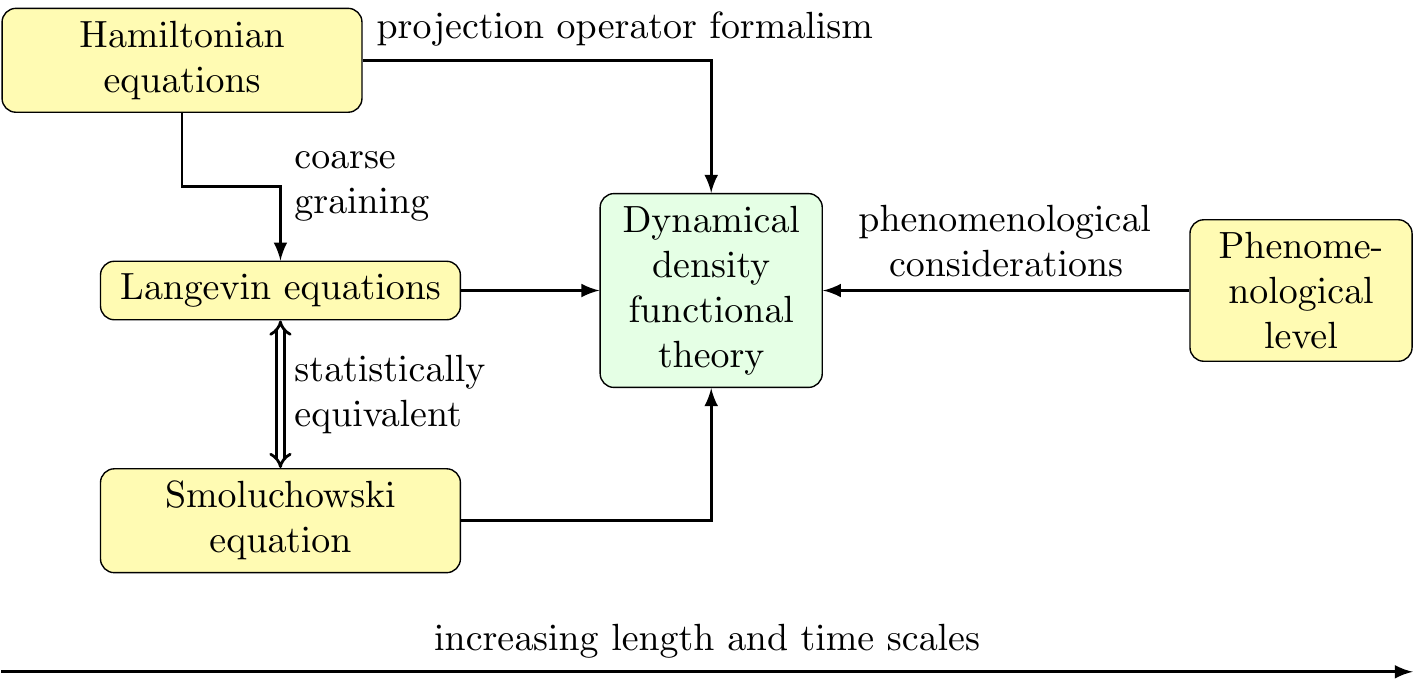}{}%
\caption{\label{fig:derivation}Different ways of deriving DDFT.}%
\end{figure*}
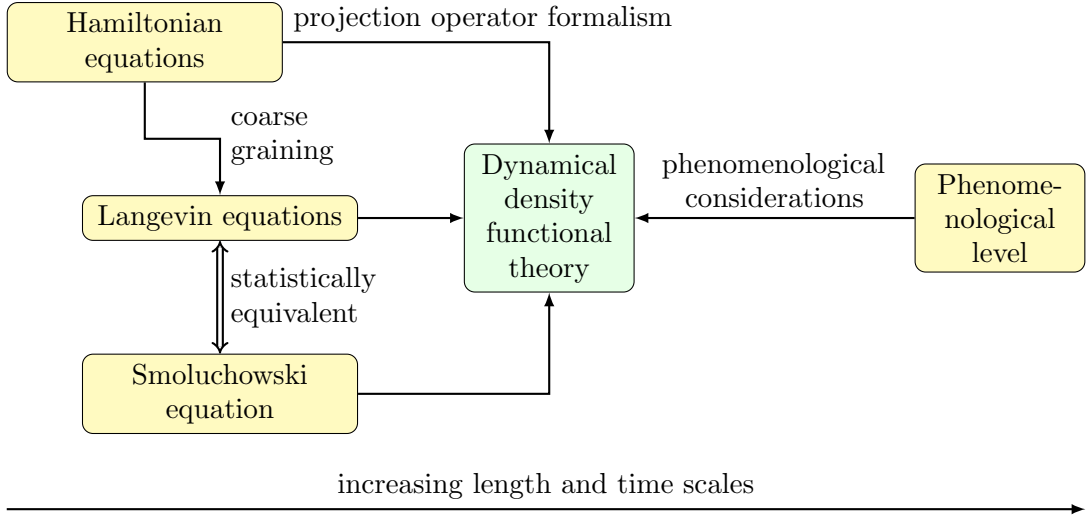
In this subsection, we discuss how DDFT can be derived from the microscopic particle dynamics or from phenomenological considerations. While we focus on the deterministic form of DDFT, we also include the result by \citet{Munakata1989}. The four main ways are a phenomenological derivation \cite{Munakata1989,DieterichFM1990}, a derivation from Langevin equations \cite{MarconiT1999}, a derivation from a Smoluchowski equation \cite{ArcherE2004}, and a derivation from the projection operator formalism \cite{EspanolL2009,Yoshimori2005}. They are visualized in \vref{fig:derivation}. Langevin equations and Smoluchowski equation are statistically equivalent descriptions of the overdamped motion of Brownian particles, from which the dynamics of the one-body density is obtained using ensemble averages and an adiabatic approximation. The derivation starting from the projection operator formalism, in contrast, bases on the deterministic Hamiltonian dynamics of the individual particles and introduces the assumption of overdamped motion at a later stage. By coarse graining, the Langevin equations can be derived from the Hamiltonian equations. Note that projection operators have also been used to derive stochastic DDFT (see \cref{kawasaki}) from stochastic descriptions in the form of fluctuating hydrodynamics \cite{Kawasaki1994} or the Smoluchowski equation \cite{EspanolV2002}.

A detailed conceptual understanding of the different types of derivations is of high importance especially for the development of further extensions of DDFT. These are often derived along the same lines as standard DDFT. For example, \citet{RexL2008} obtained a DDFT for particles with hydrodynamic interactions (see \cref{hydro}) from the underlying Smoluchowski equation by generalizing the derivation by \citet{ArcherE2004} (see \cref{derivationsmolu}) and \citet{WittkowskiLB2012} used the projection operator formalism to derive a DDFT for nonisothermal mixtures (see \cref{eddft}) based on the results by \citet{EspanolL2009} (see \cref{derivationmz}). When extending an existing derivation, it is important to understand where in this derivation the essential approximations are made, which in DDFT concerns the \textit{adiabatic approximation}. This is a closure relation whose precise effect can vary between different derivations. In the simplest scenario, a derivation starts from already overdamped Brownian dynamics \cite{MarconiT1999,ArcherE2004} (see \cref{langevin,derivationsmolu}). The adiabatic approximation then allows to express the unknown pair-correlation function in terms of the one-body density by assuming that the pair-correlation has the form it would have in an equilibrium system with the same one-body density. Alternatively, one can take as a microscopic starting point a system of underdamped particles, governed by Hamilton's equations or underdamped Langevin equations. In this case, the adiabatic approximation can appear in two forms: Some authors, such as \citet{Munakata1989} (see \cref{phenomenological}) or \citet{EspanolL2009} (see \cref{derivationmz}), employ an adiabatic approximation of a slightly different form to eliminate momentum degrees of freedom in order to arrive at an overdamped diffusive equation of motion. Other authors, such as \citet{Archer2005} (see \cref{inertia}), use the standard adiabatic approximation to evaluate an unknown correlation function also in this case, which leads to a DDFT equation that involves inertia. Finally, the adiabatic dynamics can depend on whether or not the conservation of particle order is taken into account \cite{WittmannLB2020}.

\subsubsection{\label{phenomenological}Via phenomenological considerations}
The fastest way of deriving the DDFT equation is to use phenomenological considerations. Among the first to derive DDFT was Munakata \cite{Munakata1989}. The starting point are the hydrodynamic equations \cite{KirkpatrickW1987}
\begin{align}
\pdif{}{t}\rho\rt &= -\frac{1}{m}\Nabla\cdot\vec{g}\rt\label{pdg0},\\
\pdif{}{t}\vec{g}\rt &=  - \rho\rt\Nabla\frac{\delta F}{\delta \rho\rt} - \INT{}{}{^3r'}\INT{0}{t}{t'}\DissipativeMatrix(\vec{r},\vec{r}',t-t')\vec{g}(\vec{r}',t') 
+ \vec{\mathfrak{f}}\rt\label{pdg1}
\end{align}
with the particle mass $m$, momentum density $\vec{g}$, dissipative matrix $\DissipativeMatrix$, and random force $\vec{\mathfrak{f}}$. When one assumes that
\begin{equation}
\DissipativeMatrix(\vec{r},\vec{r}',t-t') = 2\DissipativeMatrix_0 \Eins \delta(\vec{r}-\vec{r}')\delta(t-t')
\end{equation}
with a constant $\DissipativeMatrix_0$, \cref{pdg1} simplifies to
\begin{equation}
\pdif{}{t}\vec{g}\rt=  - \rho\rt\Nabla\frac{\delta F}{\delta \rho\rt} - \DissipativeMatrix_0\vec{g}\rt + \vec{\mathfrak{f}}\rt .
\label{pdg2}%
\end{equation}
An adiabatic approximation, here corresponding to the assumption that $\rho$ is the only slow variable (such that $\vec{g}$ relaxes quickly and we can set $\partial_t \vec{g} = \vec{0}$), gives from \cref{pdg2}
\begin{equation}
\vec{g}\rt = \frac{1}{\DissipativeMatrix_0}\bigg(\! -\rho\rt\Nabla\frac{\delta F}{\delta \rho\rt} 
+\vec{\mathfrak{f}}\rt\bigg)\label{pdg3}.
\end{equation}
In this case, since we have started from the underdamped equations \eqref{pdg0} and \eqref{pdg1}, the adiabatic approximation is required to obtain a closed diffusive (first order in temporal derivatives) equation of motion for $\rho\rt$. Inserting \cref{pdg3} into \cref{pdg0} gives the DDFT equation
\begin{equation}
\pdif{}{t}\rho\rt = \frac{1}{m \DissipativeMatrix_0}\Nabla\cdot\bigg(\rho\rt \Nabla\frac{\delta F}{\delta \rho\rt} + \vec{\mathfrak{f}}\rt\bigg),
\end{equation}
which can also be written as a Fokker-Planck equation (see \cref{kawasaki}).

A different type of phenomenological derivation was presented by \citet{Evans1979} and later by \citet{DieterichFM1990} (similar arguments were employed by \citet{FraaijevVMPEHAGW1997}). Instead of the hydrodynamic equations \eqref{pdg0} and \eqref{pdg1}, the starting point is the generalized Fick's law
\begin{equation}
\vec{J}\rt=-\Gamma\rho\rt\Nabla\mu\rt
\label{generalizedfickslaw}
\end{equation}
with the current $\vec{J}$, where the gradient of the chemical potential $\mu$ is used as the thermodynamic driving force. As long as $\mu$ is not specified, the form \eqref{generalizedfickslaw} does not exclude, e.g., advection by a flow field, since $\mu$ may contain a velocity potential (this possibility is discussed in \cref{potentialflow}). To obtain DDFT, we need to assume that the function $\mu$ is a chemical potential that is constructed on the basis of equilibrium theory. In static DFT, $\mu$ is given by
\begin{equation}
\mu=\Fdif{F[\rho]}{\rho(\vec{r})}.
\end{equation}
Together with the continuity equation for conserved $\rho$
\begin{equation}
\pdif{}{t}\rho\rt + \Nabla\cdot\vec{J}\rt = 0,
\end{equation}
we obtain the DDFT equation 
\begin{equation}
\pdif{}{t}\rho\rt = \Gamma \Nabla\cdot \bigg(\rho\rt\Nabla\Fdif{F[\rho]}{\rho\rt}\bigg).
\end{equation}
This is a nonlinear diffusion equation that reduces to the ordinary diffusion equation \eqref{diffusionequation} in the ideal-gas limit.

\subsubsection{\label{langevin}Via Langevin formalism}
The original microscopic derivation of the deterministic DDFT for colloidal fluids by \citet{MarconiT1999,MarconiT2000} starts from Langevin equations. Without hydrodynamic interactions and inertial terms, the motion of the $i$-th particle is governed by
\begin{equation}
\tdif{\vec{r}_i(t)}{t}=-\Gamma\Nabla_{\vec{r}_i}\bigg(\underset{j\neq i}{\sum_{j=1}^{N}}U_2(\vec{r}_i - \vec{r}_j) + U_1(\vec{r}_i)\bigg) + \vec{\chi}_i(t),
\label{eq:langevinglgnherleitung}%
\end{equation}
where $\vec{r}_i(t)$ is the position of the $i$-th particle at time $t$, $U_2$ the pair-interaction potential, $U_1$ an external potential, and $\vec{\chi}_i(t)$ a noise term with the properties
\begin{align}
\braket{\vec{\chi}_i(t)}&= \vec{0}\label{noise1},\\
\braket{\vec{\chi}_i(t)\otimes\vec{\chi}_j(t')} &= 2D\Eins\delta_{ij}\delta(t-t').\label{noise2}
\end{align}
After some manipulation using stochastic calculus, one finds the dynamic equation 
\begin{equation}
\begin{split}
\pdif{}{t}\hat{\rho}\rt &= \vec{\nabla}\cdot(\sqrt{2\Gamma k_B T\hat{\rho}\rt}\vec{\eta}\rt) \\
&\quad\:\!+ \Gamma\vec{\nabla}\cdot\bigg(\hat{\rho}\rt\INT{}{}{^3r'}\hat{\rho}(\vec{r}',t)\vec{\nabla}U_2(\vec{r}-\vec{r}')\bigg) \\
&\quad\:\!+ D\vec{\nabla}^2\hat{\rho}\rt
\end{split}
\label{stochastic}%
\end{equation}
with the microscopic density $\hat{\rho}$ given by \cref{hatrho} and the vector noise $\vec{\eta}$ that has zero mean and is $\delta$-correlated in space and time. Equation \eqref{stochastic} is statistically equivalent to \cref{eq:langevinglgnherleitung} and identical to the result \eqref{dk1} derived by \citet{Dean1996}.

The next step is to introduce the \textit{ensemble average} as an average over all possible realizations of the noise, i.e., we consider an ensemble of microscopic states with the same initial conditions, but different realizations of the noise. For the averaged density $\rho\rt=\braket{\hat{\rho}\rt}$, we obtain from \cref{stochastic} the deterministic equation
\begin{equation}
\begin{split}
\pdif{}{t}\rho\rt&=\vec{\nabla}\cdot\bigg(D\vec{\nabla}\rho\rt + \Gamma\rho\rt\vec{\nabla}U_1(\vec{r}) \\
&\quad\:\!+ \Gamma\INT{}{}{^3r'}\braket{\hat{\rho}\rt\hat{\rho}(\vec{r}',t)}\Nabla U_2(\vec{r}-\vec{r}')\bigg). 
\label{deterministic}
\end{split}  
\end{equation}
In the case of interacting particles, this is not a closed equation due to the presence of the two-point correlation function $\braket{\hat{\rho}\rt\hat{\rho}(\vec{r}',t)}$. A dynamic equation for this quantity would depend on the three-point correlation function. An infinite hierarchy of correlation functions can be avoided by using approximations. 

It is at this point that we make the central adiabatic approximation, which is used here to close the governing equation \eqref{deterministic} for the ensemble-averaged one-body density $\rho\rt$. The central idea is to approximate the time evolution by a sequence of equilibrium states. In equilibrium, one can show that for each $\rho(\vec{r})$, there exists, given a fixed temperature, chemical potential, and interaction, a unique external potential such that $\rho(\vec{r})$ is the equilibrium distribution (see \cref{static}). Thermodynamic equilibrium implies that \cite{LovettMB1976}
\begin{equation}
\begin{split}
\frac{1}{\rho(\vec{r})}\INT{}{}{^3r'}\rho^{(2)}(\vec{r},\vec{r}')\Nabla U_2(\vec{r}-\vec{r}') = -k_B T\INT{}{}{^3r'}c^{(2)}(\vec{r},\vec{r}')\Nabla\rho(\vec{r}')=\Nabla\frac{\delta F_{\mathrm{exc}}[\rho]}{\delta \rho(\vec{r})}
\label{lovett}
\end{split}
\end{equation}
with the two-point density-density correlation (two-particle density) $\rho^{(2)}(\vec{r},\vec{r}')$, the direct pair-correlation function $c^{(2)}(\vec{r},\vec{r}')$, and the equilibrium excess free energy functional $F_{\mathrm{exc}}[\rho]$. We now make the \textit{approximation} of replacing the nonequilibrium correlation $\braket{\hat{\rho}\rt\hat{\rho}(\vec{r}',t)}$ in \cref{deterministic} by the equilibrium correlation $\rho^{(2)}(\vec{r},\vec{r}')$. By inserting \cref{lovett} into \cref{deterministic}, we then obtain the DDFT equation
\begin{equation}
\pdif{}{t}\rho\rt = \Gamma\vec{\nabla}\cdot\bigg(\rho\rt\vec{\nabla}\frac{\delta F[\rho]}{\delta\rho\rt}\bigg) 
\end{equation}
with the free energy \eqref{freeenergy}.

\subsubsection{\label{derivationsmolu}Via Smoluchowski formalism}
An alternative way for the derivation of DDFT is to start from a Smoluchowski equation, as done by \citet{ArcherE2004}. Their derivation also includes multibody interactions. Neglecting hydrodynamic interactions, the Smoluchowski equation governs the many-particle probability density $\SmoluchowskiDistribution$ as
\begin{equation}
\pdif{}{t}\SmoluchowskiDistribution(\{\vec{r}_k\},t)=\Gamma\sum_{i=1}^{N}\vec{\nabla}_{\vec{r}_i}\cdot(k_B T \vec{\nabla}_{\vec{r}_i} + \vec{\nabla}_{\vec{r}_i} U(\{\vec{r}_k\},t))\SmoluchowskiDistribution(\{\vec{r}_k\},t).
\label{smoluchowski}
\end{equation}
The one-particle number density $\rho\rt$ is introduced as
\begin{equation}
\rho(\vec{r}_1,t)=N\INT{}{}{^3 r_2}\dotsb \INT{}{}{^3 r_N}\SmoluchowskiDistribution(\{\vec{r}_k\},t).
\end{equation}
We integrate in \cref{smoluchowski} over the coordinates of all particles except for one and write $\vec{r}, \vec{r}', \vec{r}'',\dotsc$ instead of $\vec{r}_1, \vec{r}_2, \vec{r}_3,\dotsc$ for consistency with \cref{langevin}. Using $D=\Gamma k_B T$, this gives
\begin{equation}
\begin{split}
\pdif{}{t}\rho\rt &= D \Nabla^2 \rho\rt + \Gamma\Nabla\cdot(\rho\rt\Nabla U_1\rt)\\
&\quad\:\!+ \Gamma \vec{\nabla}\cdot\INT{}{}{^3r'}\braket{\hat{\rho}\rt\hat{\rho}(\vec{r}',t)}\vec{\nabla}U_2(\vec{r},\vec{r}')\\
&\quad\:\!+ \Gamma\vec{\nabla}\cdot\INT{}{}{^3r'}\INT{}{}{^3r''}\braket{\hat{\rho}\rt\hat{\rho}(\vec{r}',t)\hat{\rho}(\vec{r}'',t)}\vec{\nabla}U_3(\vec{r},\vec{r}',\vec{r}'')\\
&\quad\:\!+ \dotsb,  
\end{split}
\label{deterministicnbody}
\end{equation}
where, in contrast to \cref{deterministic}, $n$-body-interaction potentials $U_n$ are included.

Again, we face the problem that \cref{deterministicnbody} is not a closed equation for $\rho\rt$, since it involves the unknown $n$-body correlations. As in the derivation from the Langevin equations (see \cref{langevin}), a closure is possible using an adiabatic approximation: For an equilibrium fluid, we can relate the interparticle forces to the direct one-body correlation function $c^{(1)}(\vec{r})$ as 
\begin{equation}
\begin{split}
-k_B T \rho(\vec{r})\vec{\nabla}c^{(1)}(\vec{r})&=\INT{}{}{^3r'}\rho^{(2)}(\vec{r},\vec{r}')\vec{\nabla}U_2(\vec{r},\vec{r}') \\
&\quad\:\!+ \INT{}{}{^3r'}\INT{}{}{^3r''}\rho^{(3)}(\vec{r},\vec{r}',\vec{r}'')\vec{\nabla}U_3(\vec{r},\vec{r}',\vec{r}'') 
+ \dotsb,
\end{split}
\end{equation}
where $\rho^{(3)}$ is the three-particle density, and $c^{(1)}$ to the excess free energy $F_{\mathrm{exc}}$ as
\begin{equation}
k_B T c^{(1)}(\vec{r}) = -\frac{\delta F_{\mathrm{exc}}[\rho]}{\delta \rho(\vec{r})}.
\label{paircorfreeen}
\end{equation}
Making the approximation that these relations also hold out of equilibrium, which corresponds to replacing the nonequilibrium correlations in \cref{deterministicnbody} by their equilibrium counterparts, we again find
\begin{equation}
\pdif{}{t}\rho\rt = \Gamma\vec{\nabla}\cdot\bigg(\rho\rt\vec{\nabla}\frac{\delta F[\vec{\rho}]}{\delta \rho\rt}\bigg).
\label{ddfteq}
\end{equation}

\subsubsection{\label{derivationmz}Via projection operator formalism}
A third option for the derivation of DDFT is to use the Mori-Zwanzig projection operator formalism. This is a very general method for deriving theories for mean values and fluctuations of an arbitrary set of $\kappa$ observables $\{A_i(\vec{z})\}$ (\ZT{relevant variables}) that depend on the phase-space coordinates $\vec{z}$. (The formalism is also applicable to quantum systems, but we only require the classical case here.) The key idea is to project the complete microscopic dynamics of a system onto a subset of relevant variables $\{A_i(\vec{z})\}$. From the statistical point of view, this corresponds to replacing the microscopic probability density $\MicroscopicPhaseSpaceDistribution$ by a relevant probability density $\bar{\rho}$. The projection operator method exists in two forms. In the microcanonical framework or nonlinear Langevin theory \cite{MoriF1973}, one fixes the precise values of the relevant variables \cite{Kawasaki2000}. This gives a Fokker-Planck equation for the probability distribution of the relevant variables or, equivalently, a set of Langevin equations. In the canonical framework, on the other hand, one fixes the mean values and projects the full microscopic probability density onto the corresponding local equilibrium distribution. A DDFT derived using a microcanonical projection operator has a free energy that differs from that of DFT \cite{Kawasaki2006b}. For deriving a DDFT with the free energy functional of DFT, one requires a canonical method \cite{Yoshimori2005}. Additional difficulties can arise from the fact that the relevant variable of DDFT (the density operator \eqref{hatrho}) is a sum over Dirac delta distributions \cite{EspanolL2009,delTED2015}. A simple introduction to projection operators can be found in Ref.\ \cite{teVrugtW2019d}.

We start by explaining the canonical projection operator formalism introduced in Ref.\ \cite{teVrugtW2019}, which is a generalization of the method used by \citet{EspanolL2009} to derive DDFT. The projection operator $\ProjectionOperator(t)$ acting on a dynamical variable $Y$ is defined as \cite{Grabert1982}
\begin{equation}
\ProjectionOperator(t)Y = \Tr(\bar{\rho}(t)Y) + \sum_{i=1}^{\kappa}(A_i - a_i(t))\Tr\!\bigg(\pdif{\bar{\rho}(t)}{a_i(t)}Y\bigg),
\label{projectionoperator}
\end{equation}
where the relevant density has the grand-canonical form
\begin{equation}
\bar{\rho}(t)=\frac{1}{\Xi(t)}\exp\!\bigg(\! -\beta \bigg(H_N(t) -\mu N -\sum_{i=1}^{\kappa}a^\natural_i(t)A_i\bigg)\!\bigg)
\label{relevantdensity}
\end{equation}
with the normalization 
\begin{equation}
\Xi(t)=\Tr\!\bigg(\exp\!\bigg(\! -\beta \bigg(H_N(t) -\mu N -\sum_{i=1}^{\kappa}a^\natural_i(t)A_i\bigg)\!\bigg)\!\bigg),
\label{zvont}
\end{equation}
time-dependent $N$-particle Hamiltonian $H_N(t)$, and thermodynamic conjugates $\{a^\natural_i(t)\}$ (the superscript $\natural$ denotes a thermodynamic conjugation), which ensure that the macroequivalence condition $a_i(t)=\Tr(\bar{\rho}(t)A_i)$ is satisfied, where $a_i(t)$ is the mean value of the variable $A_i$. Defining the coarse-grained free energy as
\begin{equation}
F(t)=\Tr(\bar{\rho}(t)H_N(t)) + k_B T \Tr(\bar{\rho}(t)\ln(\bar{\rho}(t)))
\end{equation}
allows to express the thermodynamic conjugates as 
\begin{equation}
a^\natural_i(t) = \pdif{F(t)}{a_i(t)}.
\label{thermodynamicconjugate}
\end{equation}
If the relevant variable $A_i$ is a field (as in DDFT), the partial derivative in \cref{thermodynamicconjugate} becomes a functional derivative. Frequently, one also needs the complementary operator $\mathcal{Q}(t)=1-\ProjectionOperator(t)$. One can then show that the time evolution of the mean values is given by \cite{teVrugtW2019}
\begin{equation}
\dot{a}_i(t) = \organized_i(t) - \sum_{j=1}^{\kappa}\INT{0}{t}{t'}R_{ij}(t,t')\beta a^\natural_j(t') + \randomf_i(t,0)
\label{mzexact}
\end{equation}
with an overdot denoting a derivative with respect to time, the organized drift
\begin{equation}
\organized_i(t) = \Tr(\bar{\rho}(t)\ii\LS(t)A_i),
\end{equation}
the (classical) retardation matrix
\begin{equation}
R_{ij}(t,t')=\Tr(\bar{\rho}(t')(\mathcal{Q}(t')\mathcal{G}(t',t)\ii\LS(t)A_i)\ii\LS(t')A_j),
\end{equation}
the orthogonal dynamics propagator
\begin{equation}
\mathcal{G}(t',t)=\exp_R\!\bigg(\ii\TINT{t'}{t}{t''}\LS(t'')\mathcal{Q}(t'')\bigg),
\end{equation}
the mean random force
\begin{equation}
\randomf_i(t,0)=\braket{\mathcal{Q}(0)\mathcal{G}(0,t)\ii\LS(t)A_i},  
\label{eq:meanrandomforce}%
\end{equation}
the imaginary unit $\ii$, the Schr\"odinger picture Liouvillian $\LS(t)$, and the right-time-ordered exponential $\exp_R(\cdot)$. Although it is less well known, a difference between Heisenberg picture and Schr\"odinger picture also exists for classical systems. A corresponding discussion can be found in Refs.\ \cite{HolianE1985,teVrugtW2019,teVrugtW2019d}. If the Hamiltonian depends on time, the Liouvillian in the Schr\"odinger picture differs from its form in the Heisenberg picture, such that it is important to clarify which picture is used \cite{teVrugtW2019,teVrugtW2019d}. The mean random force \eqref{eq:meanrandomforce} vanishes if the initial distribution is given by $\bar{\rho}(0)$, which we assume in this review. To see this, note that \cref{eq:meanrandomforce} can be written as
\begin{equation}
\begin{split}
\randomf_i(t,0) =& \Tr(\MicroscopicPhaseSpaceDistribution(0)\mathcal{Q}(0)\mathcal{G}(0,t)\ii\LS(t)A_i)\\
=& \Tr(\delta \MicroscopicPhaseSpaceDistribution(0)\mathcal{G}(0,t)\ii\LS(t)A_i),
\label{randomfvanishes}
\end{split}    
\end{equation}
where $\delta \MicroscopicPhaseSpaceDistribution(t)=\MicroscopicPhaseSpaceDistribution(t)-\bar{\rho}(t)$ \cite{Grabert1978,Grabert1982}. To move from the first to the second line of \cref{randomfvanishes}, we have transposed $\mathcal{Q}$ in order to apply it to the phase-space distribution $\MicroscopicPhaseSpaceDistribution$. Although noise due to the irrelevant variables is, of course, present in general and causes dissipation, the \textit{mean} random force in \cref{mzexact} depends on whether the initial relevant density $\bar{\rho}(0)$ is a good approximation for the actual distribution. This can be thought of as a condition for the construction of $\bar{\rho}$ \cite{Grabert1982}, the form \eqref{relevantdensity} is then most appropriate if the system is initially in a state of constrained equilibrium \cite{Grabert1978}.

From now on, we also assume time-independent Hamiltonians (see Refs.\ \cite{teVrugtW2019,MeyerVS2019} for the more general case). 
In this case, we can simply write $L$ instead of $\LS$, since there is no difference between Schr\"odinger and Heisenberg picture Liouvillians. If the relevant variables relax slowly compared to microscopic degrees of freedom, one can approximate the exact evolution equation \eqref{mzexact} in the form (\ZT{Markovian approximation}) \cite{EspanolL2009}
\begin{equation}
\dot{a}_i(t) = \organized_i(t) - \sum_{j=1}^{\kappa}D_{ij}(t)\beta a^\natural_j(t)
\label{markov}
\end{equation}
with the diffusion tensor
\begin{equation}
D_{ij}(t) = \INT{0}{\infty}{t'}\Tr(\bar{\rho}(t)(\mathcal{Q}(t)\ii L A_j)e^{\ii L t'}(\mathcal{Q}(t)\ii L A_i)).  
\label{diffusiontensor}
\end{equation}

This general theory allows to derive DDFT if the relevant variable is the one-particle density \eqref{density}, such that \cref{markov} provides a dynamic equation for $\rho\rt = \braket{\hat{\rho}\rt}$. (The fact that $\hat{\rho}\rt$ is a field merely requires us to replace the sums in \cref{projectionoperator,relevantdensity,mzexact,markov} by integrals over space \cite{teVrugtW2019d}.) This was suggested by \citet{Kawasaki1998,Kawasaki2000}. \citet{Yoshimori1999} and \citet{Munakata2003} derived DDFT using the nonlinear Langevin equation method (see also Ref.\ \cite{Yoshimori2004}), which gives stochastic theories that are not based on the DFT free energy. Finally, \citet{Yoshimori2005} and \citet{EspanolL2009} derived DDFT using canonical methods given by the Kawasaki-Gunton \cite{KawasakiG1973} and Grabert \cite{Grabert1982} projection operators, respectively. Projection operators also allow to derive generalizations of standard DDFT (see \cref{po} for a discussion). For this derivation method, the crucial step is the Markovian approximation that allows us to move from \cref{mzexact} to \cref{markov}. In the derivation of DDFT, which starts from the underdamped Hamiltonian dynamics, it is this approximation that gives a memoryless overdamped dynamic equation. Physically, it is the assumption that $\rho\rt$ varies slowly on time scales on which the current correlation decays. This step leads to an overdamped dynamics, since the momentum is assumed to relax quickly.

We here present the derivation by \citet{EspanolL2009}, which is also discussed in Ref.\ \cite{teVrugtW2019d}. It is based on a grand-canonical framework, which is why it employs a density functional $\widehat{\Omega}$ rather than a free energy $F$. This is justified in the thermodynamic limit. (Note that this is a property of this particular derivation and not a general restriction of the projection operator method.) If the relevant variable $A$ is the density $\hat{\rho}$ given by \cref{hatrho}, the grand-canonical partition function \eqref{zvont} becomes 
\begin{equation}
\Xi(t) =\Tr\!\bigg(\exp\!\bigg(-\beta\bigg(H_N - \mu N -\beta \sum_{i=1}^{N}\rho^\natural(\vec{r}_i,t)\bigg)\!\bigg)\!\bigg)
\end{equation}
with the thermodynamic conjugate $\rho^\natural$. The partition function allows to introduce the grand-canonical potential
\begin{equation}
\Omega[\rho^\natural] = - k_B T\ln(\Xi(t)).
\end{equation}
Based on the density functional $\widehat{\Omega}$, introduced through the Legendre transformation 
\begin{equation}
\widehat{\Omega}[\rho] = \Omega[\rho^\natural] + \INT{}{}{^3r}\rho\rt \rho^\natural\rt,
\end{equation}
we can express the thermodynamic conjugate as
\begin{equation}
\rho^\natural\rt = \frac{\delta \widehat{\Omega}[\rho]}{\delta \rho\rt}.
\label{conjugate}
\end{equation}
The organized drift $\organized_i(t)$ in \cref{markov} vanishes and the diffusion tensor \eqref{diffusiontensor} can be expressed in terms of the currents
\begin{equation}
\vec{J}(\vec{r}) = \sum_{i=1}^{N}\vec{v}_i \delta(\vec{r}-\vec{r}_i)
\end{equation}
with the particle velocities $\{\vec{v}_i\}$ as
\begin{equation}
D_{ij} \to (\vec{\nabla}_{\vec{r}}\otimes\vec{\nabla}_{\vec{r}'}):\mathcal{D}(\vec{r},\vec{r}',t),
\end{equation}
where $:$ is the double tensor contraction and $\mathcal{D}(\vec{r},\vec{r}',t)$ is the tensor
\begin{equation}
\mathcal{D}(\vec{r},\vec{r}',t) = \INT{0}{\infty}{t'} \Tr\big(\bar{\rho}(t)\vec{J}(\vec{r}',0)\otimes\vec{J}(\vec{r}',t')\big).
\label{diffusiontensor2}
\end{equation}
Using \cref{conjugate,markov} and integration by parts, we obtain the dynamic equation \cite{EspanolL2009}
\begin{equation}
\pdif{}{t}\rho\rt = \frac{1}{k_B T}\vec{\nabla}_{\vec{r}}\cdot\INT{}{}{^3r'}\mathcal{D}(\vec{r},\vec{r}',t)\vec{\nabla}_{\vec{r}'}\frac{\delta \widehat{\Omega}[\rho]}{\delta \rho(\vec{r}',t)}.
\label{nonlocal}
\end{equation}
To obtain the standard local equation \eqref{trddft}, one assumes that positions and velocities are statistically independent and that positions evolve much slower than velocities. Both approximations are reasonable for a dilute colloidal suspension. In this case, the tensor \eqref{diffusiontensor2} can be approximated as
\begin{equation}
\mathcal{D}(\vec{r},\vec{r}',t) \approx D \Eins \rho\rt \delta(\vec{r}-\vec{r}'),
\label{localize}%
\end{equation}
where the diffusion constant $D$ can be expressed in terms of the velocity autocorrelation function by the usual Green-Kubo expression. Inserting \cref{localize} into the general form \eqref{nonlocal} gives the standard result
\begin{equation}
\pdif{}{t}\rho\rt = \frac{D}{k_B T}\Nabla \cdot \bigg(\rho\rt \Nabla \frac{\delta \widehat{\Omega}[\rho]}{\delta \rho\rt}\bigg). 
\label{localized}
\end{equation}
For higher colloidal densities, more sophisticated approximations for the diffusion tensor can be developed that include hydrodynamic interactions. This includes, as discussed by \citet{EspanolL2009}, the forms proposed in Refs.\  \cite{RoyallDSvB2007,RexL2008} (see \cref{hydro}).

The described derivation starts from the deterministic Hamiltonian dynamics rather than from the stochastic Langevin dynamics. Alternatively, a derivation of DDFT using projection operators is possible starting from the Smoluchowski equation, provided the projection operator method is adapted to stochastic microscopic dynamics \cite{EspanolV2002}. Since the procedure is also applicable to variables other than $\rho\rt$, it allows for the easy derivation of more general theories involving other variables. Note that, although \cref{mzexact} is formally exact, making a Markovian approximation here corresponds to the assumption that the particle density is the only relevant variable, i.e., that other variables such as the momentum relax quickly (overdamped limit). The projection operator formalism also allows to study the dynamics of fluctuations and correlation functions \cite{Grabert1978,teVrugtW2019}. For DDFT, fluctuations were considered by \citet{Yoshimori2005} and correlation functions by \citet{WittkowskiLB2012}.

\subsection{Accuracy}
\subsubsection{\label{limit}Limitations of DDFT}
Although DDFT is, in general, a very successful theory, it has certain limitations. These result from the approximations and assumptions involved in its derivation. Some of them can be avoided by using extensions of DDFT, which will be discussed in \cref{extensions}. We mostly confine ourselves to deterministic DDFT in this section, since stochastic DDFT is exact for Brownian particles if it is formulated as a theory for the density operator \eqref{hatrho}. An obvious drawback of a stochastic DDFT for the microscopic density is, of course, that it is impossible to solve in practice. 

In general, DDFT calculations involve three approximations \cite{SchindlerWB2019}: (i) using a grand-canonical rather than a canonical free energy, (ii) making an adiabatic approximation, and (iii) using approximate rather than exact grand-canonical functionals. A fourth problem is that it allows hard particles to pass through each other \cite{ReinhardtB2012,SchindlerWB2019}.

Approximation (i) arises from the problem that DDFT is a dynamical theory that has the form of a continuity equation which conserves the total number of particles. Hence, the canonical free energy functional $F$ should be used. However, this is not known, such that DDFT uses the grand-canonical functionals from DFT instead. This can lead to wrong predictions for systems that only have a few particles or large density gradients \cite{delasHerasBFS2016}. Consequences include unphysical self-interactions resulting from the coupling to a grand-canonical reservoir \cite{ReinhardtB2012}. Canonical approaches to DDFT have also been developed (see \cref{particleconserving}).

Approximation (ii) arises from the fact that DDFT is a closed equation of motion for the one-body density. In the microscopic derivation of DDFT (see \cref{langevin}), one arrives at \cref{deterministic}, which is \textit{not} closed as it also depends on the pair correlation. As a closure, one makes the adiabatic approximation. It states that the relaxation of the system is very slow such that it can be assumed that the pair correlation is always given by that of the corresponding equilibrium system (which then allows to relate the pair correlation to the functional derivative of the equilibrium free energy). This is, of course, not exactly true, such that predictions of DDFT for nonequilibrium pair correlations are not correct \cite{KohlIBML2012}. In other words, DDFT does not include \ZT{superadiabatic forces} \cite{FortinidlHBS2014} (see \cref{exact} for a discussion of these forces as well as of extensions of DDFT including them). The assumption that a system is always in a local equilibrium state often leads to relaxation times that are too short \cite{Kawasaki2006b}. Similar problems arise in the modeling of dense nonequilibrium steady states, such as in channel flow \cite{AlmenarR2011}. In particular, flow fields can induce distortions of the correlation function that DDFT does not capture \cite{ScacchiKB2016}. Moreover, memory effects are not included \cite{WittkowskiLB2012}.

The fact that in simple forms of DDFT the density is the only variable has a number of consequences. A rather obvious one is that properties of the system that are not captured by the density profile alone, such as temperature gradients, cannot be described. A less obvious consequence, discussed in \cref{extensions}, is that by only using the density as a relevant variable, one effectively assumes that the other variables characterizing the system relax very quickly compared to the density, which is a strong approximation \cite{EspanolL2009,HaatajaGL2010}. For simple DDFT to be accurate, it is therefore required that the actual currents are diffusive rather than convective \cite{TarazonaM2008}.

A related point is the fact that DDFT, being a dynamical theory for the one-body density only, cannot distinguish between two states that have the same one-body density but different pair-correlations. In equilibrium DFT, the one-body density gives, as discussed in \cref{static}, all relevant information about the system. For a system initially out of equilibrium, however, this is not the case due to an additional dependence on the initial condition \cite{ChanF2005,teVrugt2020}. Consequently, it is possible that a nonequilibrium system has an equilibrium density profile that minimizes the free energy functional, but a nonequilibrium correlation. Such problems can be relevant, e.g., for glassy systems \cite{HeinrichsDMF2004}.

Approximation (iii) is also related to the glass transition, since approximations for the free energy can lead to the existence of local minima \cite{MarconiT1999}. A poor approximation can lead to the observation of a dynamic arrest that disappears when using a more accurate free energy \cite{StopperRH2015,StopperMRH2015}. Finally, the inability to describe a strict particle order makes it problematic to apply DDFT to nonergodic systems where caging is important \cite{SchindlerWB2019}. Both aspects are discussed in \cref{glass}.

\subsubsection{\label{test}Tests of DDFT}
The success of DDFT results, of course, also from the fact that it has proven to be reliable. Usually, DDFT is tested by comparing its predictions to simulations of the underlying (Brownian) dynamics for which it is an approximation.

In Brownian dynamics (BD) simulations, the microscopic equations of motion for the individual particles are solved numerically. This allows to check whether the DDFT results for the same system are correct. The outcomes of BD simulations can also be used to analyze where shortcomings of DDFT, if existing, come from, e.g., whether they arise from a bad choice of the free energy functional or from a breakdown of the adiabatic approximation \cite{KruppaNML2012}. They also allow to test the validity of assumptions made in DDFT and related theories regarding fluid properties such as the viscosity \cite{MorcianoFNBYSGCK2017}.

Comparisons between DDFT and BD simulations are made frequently. Examples of studies combining DDFT and microscopic simulations consider active particles \cite{WensinkL2008,KrinningerS2019,SharmaB2017,PototskyS2012,PaliwalRvRD2018,HermannKdlHS2019}, advected DDFT \cite{RauscherDKP2007}, anisotropic pair correlations \cite{KohlIBML2012}, Brownian aggregation \cite{KelkarFC2014,KelkarFC2015}, canonical dynamics \cite{delasHerasBFS2016}, capillary collapse \cite{BleibelDO2011}, capillary interactions \cite{BleibelDOD2014}, cluster crystallization \cite{DelfauOLBH2016}, colloidal fluids in a cavity \cite{Archer2005,Archer2005b}, colloidal shaking \cite{KruppaNML2012}, colloids in a DNA solution \cite{GutscheKKRWH2008,GutscheEtAl2011}, demixing \cite{NunesGAdG2018}, diffusion-controlled reactions \cite{DzubiellaM2005}, drag forces \cite{LipsRM2019}, driven particles \cite{PennaDT2003,NunesAT2016,PototskyABMSM2010,PototskyASTM2011,RexLL2005}, dynamical correlations \cite{PennaT2006}, dynamic mode locking \cite{JuniperZSBALD2017}, flowing colloids \cite{ZimmermannSL2016}, hydrodynamic interactions \cite{RexL2008,RexL2009,BleibelDO2015,BleibelDO2016,GoddardNK2013,GoddardNK2016,GoddardNSPK2012,GoddardNSYK2013,Loewen2010b,PelaezUPDD2018,ZhaoWZDTM2014}, inhomogeneous polymer systems \cite{QiS2017}, lane formation \cite{WachtlerKK2016,ChakrabartiDL2003,ChakrabartiDL2004,Loewen2010c}, plasmas \cite{DiawM2015,DiawM2016}, power functional theory \cite{TreffenstadtS2019,delasHerasS2018,SchindlerS2016,delasHerasRS2019,KrinningerSB2016,StuhlmullerEdlHS2018,GeigenfeinddlHS2020}, protein-polyelectrolyte interaction \cite{XuALDB2018}, protein solvation \cite{MondalMB2017}, rheology \cite{ReinhardtWB2013}, sedimentation \cite{ArcherM2011,RoyallDSvB2007,SchmidtRvBD2008}, shear flow \cite{ScacchiKB2016}, solvation dynamics \cite{Yoshimori1DP998b}, stratification \cite{HowardNP2017,HowardNP2017b}, superadiabatic forces \cite{FortinidlHBS2014,BernreutherS2016}, thermostatted granular fluids \cite{MarconiTC2007}, the van Hove function \cite{HopkinsFAS2010,ArcherHS2007}, transport diffusivity \cite{Liu2016}, traveling band formation \cite{TaramaEL2019}, ultrasoft colloids \cite{RexLLD2006,DzubiellaL2003}, and vacancy diffusion \cite{vanTeeffelenAL2013}. In \vref{fig:trap}, a comparison between DDFT results and BD simulations for a system of driven colloids is shown. The agreement is very good, which confirms the accuracy of DDFT. 

Moreover, DDFT can be tested by comparing its predictions to experiments. This was done for, e.g., active particles \cite{EnculescuS2011}, Brownian hard disks \cite{StopperTDR2018}, charging processes \cite{QingLZTQMXZ2020}, colloids in a DNA solution \cite{GutscheKKRWH2008,GutscheEtAl2011,HartingZWH2009}, crystals \cite{YeNTZM2016,YeTZDM2016}, diffusion and hydrodynamic interactions \cite{BleibelDO2015}, dynamic mode locking \cite{JuniperZSBALD2017}, ion channels \cite{Gillespie2008,GillespieXWM2005}, nonequilibrium sedimentation of colloids \cite{RoyallDSvB2007,SchmidtRvBD2008}, particles in confinement \cite{LiuWDZM2020}, phase separation \cite{ZvyagolskayaAB2011}, Poisson-Nernst-Planck (PNP) theory \cite{EvertsSEvdHvdBvdR2017}, protein adsorption \cite{SatulovskyCS2000,AngiolettiBD2014}, protein-polyelectrolyte interaction \cite{XuALDB2018}, resistance nonadditivity  \cite{ZimmermannLKELS2017}, the van Hove function \cite{BiervRDvdS2008,GreletLBvRvdS2008}, and wetting \cite{YeTZDM2016}. Not discussed here due to lack of space is the large body of work that applies polymer DDFT (see \cref{pd}) to experimental results and employs it in technological applications. Among the many examples are Refs.\ \cite{XuWHH2008,XuLH2007b} considering microphase separation in polymers, Ref.\ \cite{KnollLHKSZM2004} discussing phase transitions in nanostructured fluids, and Ref.\ \cite{KnollHLKSZM2002} studying the phase behavior of thin films of block copolymers. Finally, an interesting opportunity for further tests of DDFT are experiments on diffusion in space \cite{BraibantiEtAl2019}.

A third option for testing DDFT is to compare it to exact results \cite{MarconiT1999,WittmannLB2020}, provided these are available.

\subsection{Further aspects}
\subsubsection{Terminology}
Terminological difficulties can arise since the name \ZT{DDFT} is not always used with the same meaning in the literature, and since DDFT also has other names. Very often, \ZT{DDFT} is used to refer to the theory derived by \citet{MarconiT1999}. However, stochastic theories \cite{Kawasaki1994} (see \cref{kawasaki}) and methods for polymers \cite{Fraaije1993} (see \cref{pd}) are also called \ZT{DDFT}. Some authors see phase field crystal models (see \cref{phasefield}) as a form of DDFT \cite{TothTPTG2010,TothTPTG2012,TegzeGTDP2011,KorbulyPHPAG2017,PodmaniczkyTPG2014,TothPG2015}, while others do not \cite{ArcherRRS2019}. Moreover, \ZT{dynamical density functional theory} is the name of a nucleation theory proposed by \citet{TalanquerO1994}. Finally, the abbreviation \ZT{DDFT} is also used in economics for \ZT{degree of dependence on foreign trade} \cite{TaoZCL2019} and in equine anatomy for \ZT{deep digital flexor tendon} \cite{BlundenDMS2006,MoraBGPRSM2014}.

On the other hand, various names are used for DDFT in the literature. The most common ones are \ZT{dynamical density functional theory}, \ZT{dynamic density functional theory}, and \ZT{time-dependent density functional theory}. Note that the name \ZT{time-dependent density functional theory} is not only used for the classical DDFT, but also for the quantum-mechanical method \cite{RungeG1984} and for lattice models \cite{ReinelD1996}. More exotic names for DDFT include \ZT{generalized diffusion approach} (GDA) \cite{FangSS2005} and \ZT{time-dependent classical density functional formalism} \cite{Tozzini2009}. For stochastic theories, \ZT{time-dependent density functional method} \cite{Yoshimori1999}, \ZT{stochastic density functional theory} \cite{DeanLMP2016}, and \ZT{Dean-Kawasaki equation} \cite{DasY2013} are alternatives to \ZT{DDFT} or \ZT{TDDFT}. The polymer DDFT by Fraaije and coworkers is referred to as \ZT{dynamic mean-field density functional method} \cite{FraaijevVMPEHAGW1997} or \ZT{dynamic density functional theory} \cite{Fraaije1993}. 

\begin{figure*}[!htb]
\centering
\TikzBilderNeuErzeugen{\newcommand{\ZU}{\\[-2pt]}
\newcommand{\XY}{\phantom{Xy}}
\tikzstyle{begriffsfeldI} = [rectangle, rounded corners, minimum size=5mm, align=center, draw=black, fill=yellow!30]
\tikzstyle{dummyfeld} = [rectangle, rounded corners, minimum size=5mm, align=center, draw=black, draw=none]
\tikzstyle{pfeil} = [thick,->,>=latex]
\tikzstyle{doppelpfeil} = [thick,implies-implies, double equal sign distance]
\tikzstyle{beschriftung} = []
\begin{tikzpicture}[node distance=18mm and 5mm]
\node(K0)[begriffsfeldI,minimum width=\linewidth]{Hamiltonian dynamics};
\node(K1)[begriffsfeldI, above=of K0]{Brownian dynamics};
\node(K2)[begriffsfeldI, above=of K1,fill=green!10]{Stochastic DDFT\\for microscopic density};
\node(K3)[dummyfeld, above=of K2]{\XY\\\XY\\\XY};
\node(K4)[begriffsfeldI, anchor=west,fill=green!10]at(K3.west-|K0.west){Deterministic DDFT\\for ensemble-averaged\\density};
\node(K5)[begriffsfeldI, anchor=east,fill=green!10]at(K3.east-|K0.east){Stochastic DDFT\\for coarse-grained\\density};
\node(K4b)[dummyfeld, above=of K5]{\XY};
\node(K6)[begriffsfeldI, above=of K4b,fill=red!10]{MCT for\\coarse-grained\\density};
\node(K7)[begriffsfeldI,fill=red!10]at(K4b.east-|K2.north){Microscopic MCT};
\draw[pfeil](K0.north) -- node[beschriftung,align=left,anchor=west]{coarse graining,\ZU{}overdamped\ZU{}dynamics}(K1.south);
\draw[doppelpfeil](K1.north) -- node[anchor=east, align=right]{statistically\ZU{}equivalent}(K2.south);
\draw[pfeil](K2.north) -- node[midway, sloped, below, align=center]{MCT approximation}(K7.south);
\draw[pfeil](K5.north) -- node[midway, sloped, below, align=center]{MCT approximation}(K6.south);
\draw[pfeil](K1.east) -- (K5.south|-K1.east) -- node[pos=1, sloped, above, anchor=south east]{coarse graining,} node[pos=1, sloped, below, anchor=north east]{approximation for interactions}(K5.south);
\draw[pfeil](K1.west) -- (K4.south|-K1.west) -- node[pos=1, sloped, above,anchor=south east]{reduction of degrees of freedom,} node[pos=1, sloped, below,anchor=north east]{adiabatic approximation} (K4.south);
\coordinate[](P1)at($(K4.south)!0.5!(K4.south west)$);
\coordinate[](P2)at($(K4.south)!0.5!(K4.south east)$);
\draw[pfeil](P1|-K0.north) -- node[midway, sloped, above, align=center]{slow density, ensemble average}(P1);
\draw[pfeil](K2.west) -- (P2|-K2.west) -- node[midway, anchor=west, align=left]{ensemble\ZU{}average,\ZU{}adiabatic\ZU{}approxi-\ZU{}mation}(P2);
\coordinate[](P3)at($(K5.south)!0.5!(K5.south west)$);
\coordinate[](P4)at($(K5.south)!0.5!(K5.south east)$);
\draw[pfeil](K2.east) -- (P3|-K2.east) -- node[anchor=east,align=right]{coarse graining,\ZU{}approximation\ZU{}for interaction\ZU{}term}(P3);
\draw[pfeil](P4|-K0.north) -- node[midway, sloped, below, align=center]{coarse graining, momentum relaxes fast}(P4);
\coordinate[](P5)at($(K7.south)!0.5!(K7.south west)$);
\coordinate[](P6)at($(K2.south)!0.5!(K2.south east)$);
\coordinate[](P7)at($(K1.east)!0.5!(K5.west)$);
\coordinate[](P8)at($(K2.north west)+(-0.25,0.25)$);
\coordinate[](P9)at($(K1.north)!0.25!(K2.south)$);
\coordinate[](P10)at(P8|-K0.north);
\coordinate[](P11a)at(P7|-K0.north);
\coordinate[](P11b)at($(K0.north)!0.5!(P4|-K0.north)$);
\coordinate[](P12)at(P11b|-P9);
\draw[pfeil](P10) -- (P8) -- (P5|-P8) -- node[midway, sloped, above, align=center]{projection operator\\formalism}(P5);
\draw[pfeil](P11b) -- (P12) -- (P6|-P12) -- node[midway, anchor=west, align=left]{momentum\ZU{}relaxes fast}(P6);
\end{tikzpicture}}%
\TikzBildEinfuegen{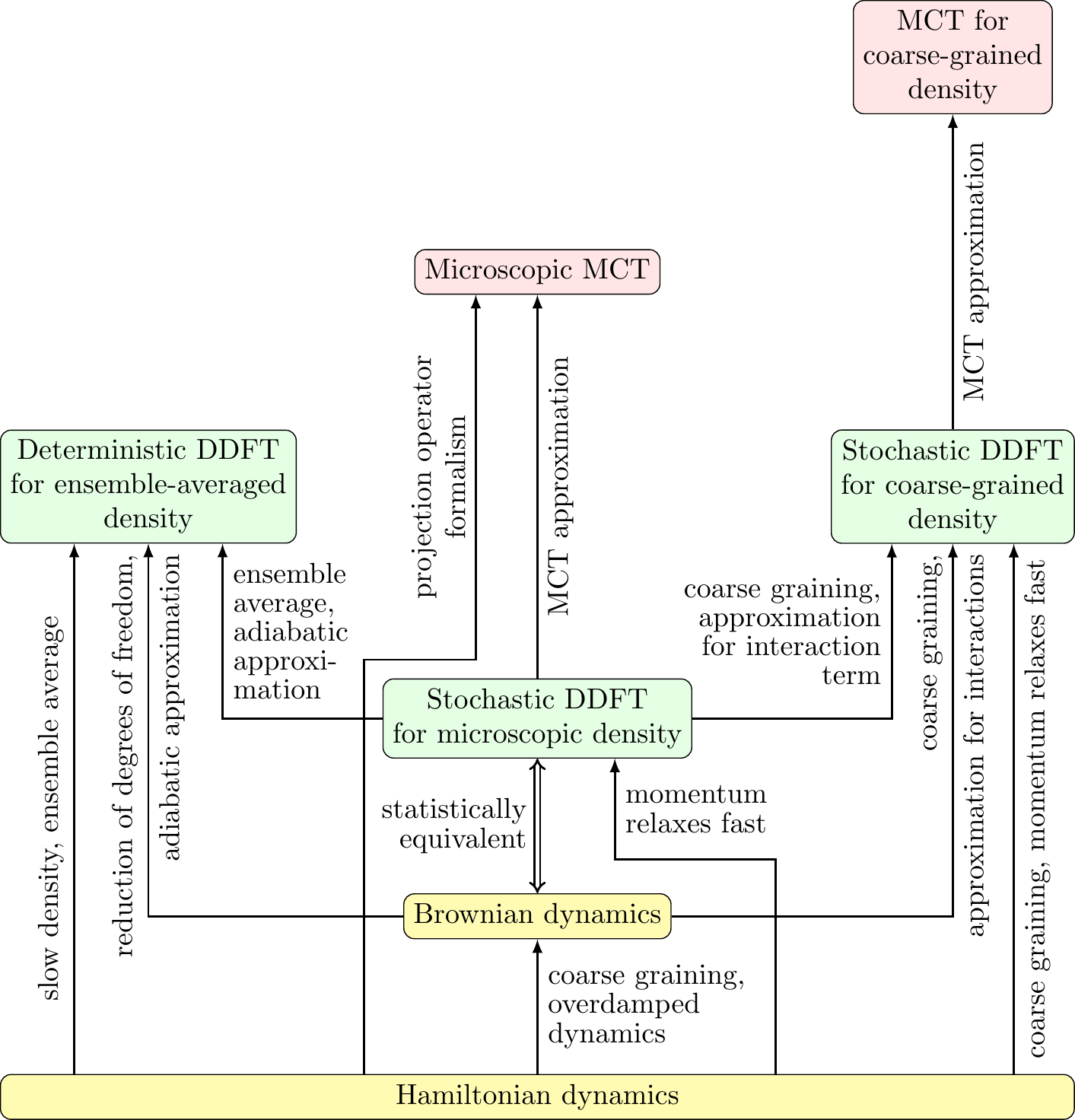}{}%
\caption{\label{fig:schematicglas}Relations of different forms of DDFT and MCT.}%
\end{figure*}
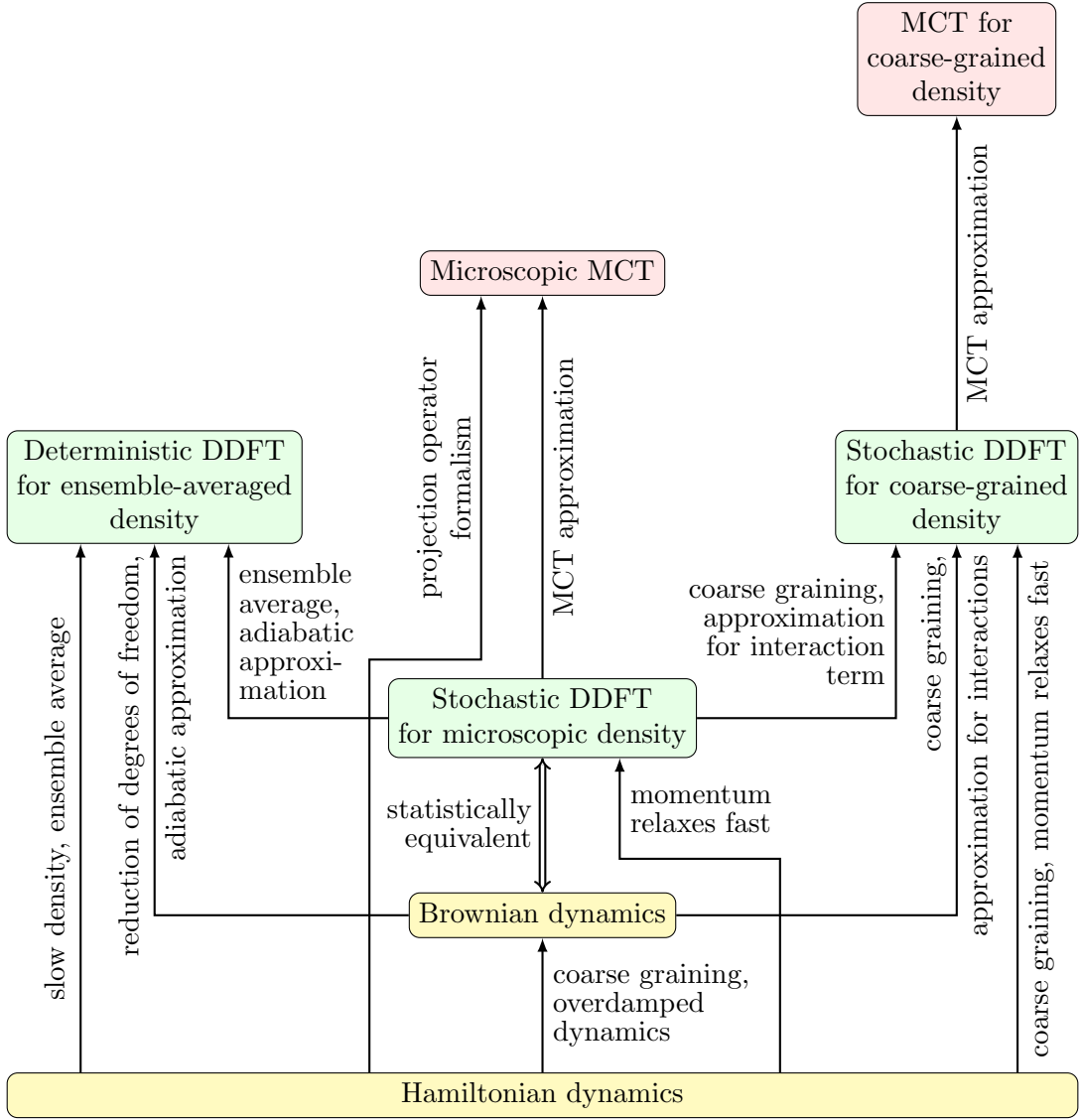
\FloatBarrier

\subsubsection{\label{noise}Deterministic or stochastic?}
An occasional source of confusion and controversy is the question whether DDFT should contain a noise term, i.e., whether \cref{trddft} or \cref{trddfts} is correct \cite{FrusawaH2000}. In fact, they are both correct, but have a different meaning \cite{ArcherR2004}. As discussed in \cref{traditional}, DDFT exists in two forms, as a deterministic and as a stochastic theory. Here, we will discuss how they are related. An illustration is given in \vref{fig:schematicglas}. The crucial point here is the interpretation of the density field $\rho$. While it always denotes an ensemble-averaged density in static DFT, its meaning in DDFT can vary.

We start with the deterministic DDFT derived by \citet{MarconiT1999} (on the left in \vref{fig:schematicglas}). Here, the density $\rho\rt$ is an \textit{ensemble average}, i.e., an average over all possible realizations of the noise. Physically, these correspond to fixed initial positions of the colloids, but different microscopic configurations of the bath (\ZT{Brownian ensemble}) \cite{MarconiT1999}. This average, given an initial state, is uniquely determined. Hence, there can be no noise terms \cite{ArcherR2004,ArcherE2004}. These would lead to the wrong equilibrium limit, since $F[\rho]$ already includes the fluctuations and adding them to the time evolution would correspond to overcounting them \cite{EmmerichEtAl2012}. In deterministic DDFT, the free energy functional $F$ has the same form as in DFT. This is the main advantage of this form, since, in constructing the free energy functional, one can make use of the well-developed theory of DFT \cite{MarconiT1999}. Some examples are discussed in \cref{fmt}.

There is, however, a second aspect that has to be taken into account: The exact description always gives a convex thermodynamic potential, such that through the dynamics the free energy approaches its unique (local and global) minimum. In practice, one will use approximate free energy functionals, which might have more than one local minimum. In this case, it is possible that through the time evolution the system gets trapped in a metastable state corresponding to a local minimum, such that phenomenological noise terms have to be added to push the system out of the metastable states towards its true equilibrium \cite{EmmerichEtAl2012,MarconiT1999}.

One can, however, also define $\rho\rt$ as a spatially or temporally coarse-grained density (on the right in \vref{fig:schematicglas}). In this case, the equations of motion for $\rho\rt$ are stochastic \cite{ArcherE2004,ArcherR2004}. The density $\rho\rt$ appearing in the deterministic equation can be interpreted as the probability density of finding a particle at time $t$ at position $\vec{r}$, whereas the density appearing in the stochastic equation is the number of particles in a small volume \cite{delaTorreE2011,delTED2015}. Coarse graining in the form of a spatial average, introduced by dividing space into small cells, is employed in the DDFT by \citet{Kawasaki1994} (see \cref{kawasaki}).

A further possibility for averaging in addition to noise, space, and time is an average over initial conditions. This possibility is discussed in Refs.\ \cite{BleibelDO2016,DonevVE2014}. Averaging over initial conditions leads to a smooth initial density profile $\rho(\vec{r},0)$ instead of a \ZT{spiky} distribution that has peaks at the positions of the particles. The time-dependent density profile $\rho\rt$ is a functional of the initial density profile and is thus different if such an average is taken \cite{BleibelDO2016}. This aspect can be relevant for phase separation, as discussed in \cref{ps}. One can combine this type of averaging with various types of DDFT. For example, \citet{BleibelDO2016} employ it in the context of deterministic DDFT to study capillary collapse, while \citet{DonevVE2014} discuss its possible application to a stochastic model.

Finally, a DDFT can be formulated as a theory for the density operator $\hat{\rho}\rt$ given by \cref{hatrho}, which is a sum of Dirac delta distributions (in the middle of \vref{fig:schematicglas}). This DDFT provides an exact description of a system of Brownian particles (a paradigmatic example is the result \eqref{dk} obtained by \citet{Dean1996}), or an approximate description of a system of Hamiltonian particles. A DDFT for the density operator is also stochastic. Note that the free energy entering Dean's DDFT is given by \cref{deanfreeenergy} and not by the DFT free energy \cite{ArcherR2004}. To emphasize this difference, the free energy \eqref{deanfreeenergy} is sometimes denoted as $H[\rho]$ rather than $F[\rho]$. Dean's stochastic DDFT can serve as a basis for the derivation of a deterministic DDFT based on ensemble averages and further approximations \cite{MarconiT1999} (see \cref{langevin}). Although it is exact, it might be insensitive to the conservation of particle order in 1D as it uses a symmetric pair-interaction potential (an effect described by \citet{SchindlerWB2019} for deterministic DDFT, see also \cref{glass}).

A detailed discussion of the relation between deterministic and stochastic DDFT was given by \citet{ArcherR2004} (see also Refs.\ \cite{Lutsko2010,Chavanis2019}).

Another way to understand the differences between deterministic and stochastic DDFT is to compare their microscopic foundations in the projection operator formalism (see \cref{po}). This method allows for a systematic derivation of field theories, such as DDFT, by projecting the microscopic dynamics onto the subdynamics of relevant variables. For the projection operator method, two forms can be distinguished, called \ZT{microcanonical} and \ZT{canonical} form \cite{Kawasaki2000,Kawasaki2006b}. In the microcanonical form, also known as \ZT{Zwanzig projection operator} or \ZT{nonlinear Langevin theory}, one projects onto states with fixed values of the relevant variables, i.e., the density $\rho\rt$ is specified \textit{exactly}. The free energy functional one gets is the \ZT{microcanonical free energy}. This name is chosen in analogy to the microcanonical ensemble, where, unlike in the canonical ensemble, one specifies the exact value of the energy rather than the mean value. Microcanonical methods lead to stochastic theories, where the equations derived in the projection operator framework contain a random noise term. This is not a problem in the equilibrium limit, since the free energy appearing in stochastic DDFT is different from the free energy of deterministic DDFT. In the microscopic derivation of stochastic DDFT, one gets a \textit{bare} free energy and \textit{bare} transport coefficients \cite{delTED2015}. These can be connected to the \textit{renormalized} free energy of DFT and the \textit{renormalized} transport coefficients of deterministic DDFT through a fluctuation renormalization \cite{Kawasaki2006b}. This procedure is discussed in Ref.\ \cite{Grabert1982}.

On the other hand, in a canonical theory, such as the \ZT{Robertson projection operator method} \cite{Robertson1966} or \ZT{Kawasaki-Gunton projection operator method} \cite{KawasakiG1973}, one fixes only the mean values of the relevant variables (such as the one-body density) rather than the exact values. This is analogous to the canonical ensemble in statistical mechanics, where fluctuations of the energy around its mean value are allowed. Here, the free energy is of the canonical form, which is the one appearing in DFT \cite{Yoshimori2005}. Hence, fluctuations are already included in the free energy functional. From this approach, one can derive the well-known form of deterministic DDFT \cite{EspanolL2009}. (Formally, neglecting the noise term in the averaged equations requires an additional assumption about initial conditions \cite{Yoshimori2005,teVrugtW2019d}.) It is also possible to interpolate between deterministic and stochastic theories using more general projection operators \cite{Kawasaki2006b}. A further discussion of the relation of deterministic and stochastic methods based on projection operators can be found in Ref.\ \cite{delTED2015}, where the problem is also related to numerical aspects.

Similar considerations as for DDFT apply to phase field crystal models \cite{Emmerich2011,HaatajaGL2010,TegzeGTPJANP2009,TahaDMEH2019} and Cahn-Hilliard models \cite{Duque2018}.

\subsubsection{\label{solvent}Role of the solvent}
In the context of versions and applications of DDFT, it is important to pay attention to the difference between \ZT{atomic/molecular} and \ZT{colloidal} fluids. A (one-component) atomic/molecular fluid (henceforth called a \ZT{simple} fluid) consists of a large number of identical particles (atoms or molecules) that interact through a certain two-body interaction potential. These interactions are governed by the deterministic laws of Newtonian mechanics. Coarse graining the microscopic description of fluids leads to the well-known equations of hydrodynamics. A colloidal fluid, on the other hand, consists of small \ZT{solvent particles} in which much larger \ZT{colloidal particles} are immersed. The exact microscopic description of colloidal fluids is challenging because, due to the huge difference in mass and size between the two particle types, very different time scales have to be considered. Typically, one models only the colloidal particles and describes their interaction using an effective theory. The solvent particles enter the dynamics through the noise term and the friction force. This gives a description in terms of Langevin equations or, equivalently, a Smoluchowski equation. Since not all microscopic degrees of freedom are considered on the colloid level of description, these equations are stochastic. The Smoluchowski equation can be derived by coarse graining from the exact microscopic Liouville equation that describes both solvent and colloidal particles \cite{Archer2005}. A useful method for connecting the mesoscopic dynamics of particles in a solvent to the microscopic level is the projection operator formalism (see \cref{po}), as shown by \citet{EspanolD2015}. \citet{DuranYGK2017} obtained the coarse-grained description of the $N$-body distribution for the colloids from the full microscopic dynamics using the projection operator method.

The original microscopic derivations of deterministic DDFT, based on Langevin equations \cite{MarconiT1999} and Smoluchowski equation \cite{ArcherE2004}, started from the overdamped stochastic dynamics and thus apply to colloidal fluids. In these theories, the density describes the number of colloidal particles per unit volume and not, as in hydrodynamics, the mass density. This is different in DDFT-based descriptions of atomic fluids. A DDFT for atomic fluids, which has as a microscopic starting point Newton's equation of motion rather than Langevin equations, was derived by \citet{Archer2006}. The resulting dynamic equation \eqref{atomic} includes inertial effects and is a differential equation of second order in time derivatives. In the limit of large collision frequencies, the usual first-order overdamped DDFT equation \eqref{trddft} is recovered. For the description in the projection operator formalism, the situation is more subtle. \citet{EspanolL2009} derived the more general nonlocal expression \eqref{nonlocal}, which only contains the number density as a relevant variable (and thus assumes all other variables to relax quickly). For the case of colloidal fluids, further approximations lead to the usual local form. A DDFT for simple fluids was derived with the projection operator formalism by \citet{CamargodlTDZEDBC2018}. Here, the momentum density is also used as a relevant variable. Moreover, standard DDFT for colloidal fluids can be recovered from a microscopic hydrodynamic theory for a binary mixture if one particle species (the colloids) is much heavier and more dilute than the other one (the solvent) \cite{Marconi2011}.

A further interesting topic is the dynamics of the solvent itself. Often, it is assumed that the solvent relaxes very rapidly compared to the colloids. On the other hand, nonequilibrium effects arise if the relaxation time scales are comparable, in particular if confinement prohibits easy relaxation of the solvent \cite{HaraY2015,Fuchizaki2015}. Solvation dynamics (rearrangement of solvent particles when a solute particle is present) can also be modeled in DDFT \cite{KarBC2008,HaraY2015}. 

\subsubsection{\label{thermo}Approach to equilibrium}
Among the central topics of nonequilibrium statistical mechanics is the question how and why systems approach a state of thermodynamic equilibrium \cite{Wallace2015,teVrugt2020}. The tendency to approach equilibrium is often associated with the second law of thermodynamics (see Ref.\ \cite{Uffink2001} for a detailed discussion) and has even been granted the status of a \ZT{minus first law of thermodynamics} by \citet{BrownU2001}. An early and important result in this area is the H-theorem derived by \citet{Boltzmann1872}. It states that a quantity known as the H-functional, which depends on the phase-space distribution and is identified with the negative entropy (\ZT{negentropy}), is monotonously decreasing \cite{BrownUM2009}. This has inspired a significant amount of work aiming at the derivation of other H-theorems. DDFT, which describes the way systems approach a state of thermodynamic equilibrium as described by DFT, is no exception.

In deterministic DDFT, one can prove that the free energy is monotonously decreasing. This was done for the standard DDFT \eqref{trddft} by \citet{Munakata1994} and \citet{MarconiT1999} and for the generalized form \eqref{nonlocal} by \citet{EspanolL2009}. From \cref{trddft}, one obtains
\begin{equation}
\tdif{F[\rho]}{t} = - \Gamma\INT{}{}{^3r}\rho\rt\bigg(\vec{\nabla}\frac{\delta F[\rho]}{\delta \rho\rt}\bigg)^2 \leq 0.
\label{htheorem}
\end{equation}
It can be shown from microscopic calculations that the mobility $\Gamma$ is positive, which completes the proof of this \ZT{H-theorem} of DDFT \cite{EspanolL2009}. The minimum of the free energy is reached for $\delta F / \delta \rho = \mu$ with a constant chemical potential $\mu$ \cite{Munakata1994}, which is in agreement with the variational principle \eqref{dft} of static DFT. Hence, \cref{trddft} indeed describes a monotonous approach towards the equilibrium state described by DFT. The precise speed of this relaxation is slightly overestimated due to the approximations made in the derivation of deterministic DDFT \cite{Kawasaki2006b}. An interesting consequence of the result \eqref{htheorem} is that the time evolution can be trapped in local minima of the free energy that arise from approximations made in its construction \cite{MarconiT1999}. This can be avoided by adding noise terms. A H-theorem for stochastic DDFT was derived by \citet{Munakata1994}. H-theorems can also be obtained for extended theories describing phase-space dynamics \cite{AneroE2007,Chavanis2011}. 

Having discussed \textit{how} systems approach equilibrium in DDFT, we can also address \textit{why} this is the case. A derivation of irreversible macrodynamics from reversible microdynamics requires a coarse-graining procedure \cite{DonevFVE2014}. The discussion of thermodynamic irreversibility has become a matter of intense discussion in philosophy of physics. This debate is concerned with (a) the location of irreversibility within thermodynamics, (b) the definition of \ZT{thermodynamic equilibrium}, (c) the justification of coarse graining, (d) the approach to equilibrium, and (e) the direction of the arrow of time \cite{teVrugt2020}. Being a paradigmatic example of a theory describing the irreversible approach to equilibrium based on a microscopic theory, DDFT can provide interesting insights into this debate. For example, the derivation by \citet{EspanolL2009} (see \cref{derivationmz}) starts from Hamilton's equations, which are reversible, and obtains from them the DDFT equation \eqref{localized}, which is irreversible. This is a consequence of coarse graining combined with a Markovian approximation \cite{teVrugtW2019d}. A discussion of philosophical aspects of DDFT can be found in Ref.\ \cite{teVrugt2020}.

\section{\label{extensions}Extensions of standard DDFT}
\begin{figure*}[htb]
\centering
\TikzBilderNeuErzeugen{\newcommand{\ZU}{\\[-2pt]}
\newcommand{\XY}{\phantom{Xy}}
\newcommand{\EA}{\phantom{$\bullet$ }}
\tikzstyle{begriffsfeld0} = [align=center,rectangle,fill=purple!20]
\tikzstyle{begriffsfeldI} = [rectangle, rounded corners, minimum size=5mm, align=center, draw=black, fill=yellow!30]
\tikzstyle{begriffsfeldII} = [rectangle, rounded corners, minimum size=5mm, align=left, draw=black, fill=gray!10, font=\scriptsize]
\tikzstyle{dummyfeld} = [rectangle, rounded corners, minimum size=5mm, align=center, draw=black, draw=none]
\tikzstyle{pfeil} = [thick,->,>=latex]
\tikzstyle{doppelpfeil} = [thick,implies-implies, double equal sign distance]
\tikzstyle{beschriftung} = []
\begin{tikzpicture}[node distance=25mm and 1.75mm]
\node(B0a)[align=center,rectangle,anchor=north,rotate=90,draw=none]at(0,0){\phantom{Steps in derivation of DDFT}};
\node(B1)[begriffsfeldI,anchor=west,right=of B0a.south]{Microscopic dynamics\\(Hamiltonian, Langevin)};
\node(K0)[begriffsfeldII,below=of B1]{Other or more general\\microscopic dynamics:\\[1mm]
$\bullet$ Active DDFT\\
$\bullet$ Quantum DDFT\\
$\bullet$ \dots};
\coordinate[](P0a)at(\textwidth,0);
\coordinate[](P0)at(P0a|-K0.north);
\node(K3)[begriffsfeldII, anchor=north east]at(P0){Larger configu-\\ration space:\\[1mm]
$\bullet$ Nonspherical\\\EA{}particles\\\EA{}($\varrho(\vec{r},\uu,t)$)\\
$\bullet$ Kinetic\\\EA{}theory\\\EA{}($\PSDistributionRVT(\vec{r},\vec{p},t)$)\\
$\bullet$ \dots};
\node(K2)[begriffsfeldII,left=of K3.north west, anchor=north east]{Additional variables:\\[1mm]
$\bullet$ Extended DDFT\\\EA{}(mixtures,\\\EA{} energy density)\\
$\bullet$ Functional\\\EA{}thermodynamics\\\EA{}(energy density)\\
$\bullet$ DDFT with\\\EA{}momentum\\\EA{}density\\
$\bullet$ \dots};
\coordinate[](P1)at($(K2.west)!0.5!(K3.east)$);
\node(B2)[begriffsfeldI,align=center,fill=green!10]at(P1|-B1){dynamic equation for $\rho(\vec{r},t)$};
\draw[pfeil](B1.south) -- node[beschriftung,align=left,anchor=west]{}(K0.north);
\draw[pfeil](B1.east) -- node[beschriftung,align=center,above]{approximations} node(N1)[beschriftung,align=center,below]{(adiabatic closure,\\approximate\\free energy, \dots)} (B2.west);
\node(K1)[begriffsfeldII,anchor=north]at(N1|-K0.north){Omission of certain\\approximations:\\[1mm]
$\bullet$ Particle-conser-\\\EA{}ving dynamics\\\EA{}(free energy)\\
$\bullet$ Power functional\\\EA{}theory (adiabatic\\\EA{}approximation)\\
$\bullet$ \dots};
\draw[pfeil](N1.south) -- node[beschriftung,align=left,anchor=west]{}(K1.north);
\coordinate[](P2)at($(B2.south west)!0.6666666666!(B2.south)$);
\coordinate[](P3)at($(B2.south)!0.3333333333!(B2.south east)$);
\coordinate[](P4)at($(P2)!0.5!(K2.north)$);
\coordinate[](P5)at($(P3)!0.5!(K3.north)$);
\draw[pfeil](P2) -- (P2|-P4) -| node[beschriftung,align=left,anchor=west]{}(K2.north);
\draw[pfeil](P3) -- (P3|-P5) -| node[beschriftung,align=left,anchor=west]{}(K3.north);
\node(B0b)[begriffsfeld0,anchor=north west,rotate=90,draw=black]at(B0a.north|-N1.south){Steps in derivation of DDFT};
\node(B0c)[begriffsfeld0,anchor=north east,rotate=90,draw=black]at(B0a.north|-K0.north){Corresponding extensions};
\end{tikzpicture}}%
\TikzBildEinfuegen{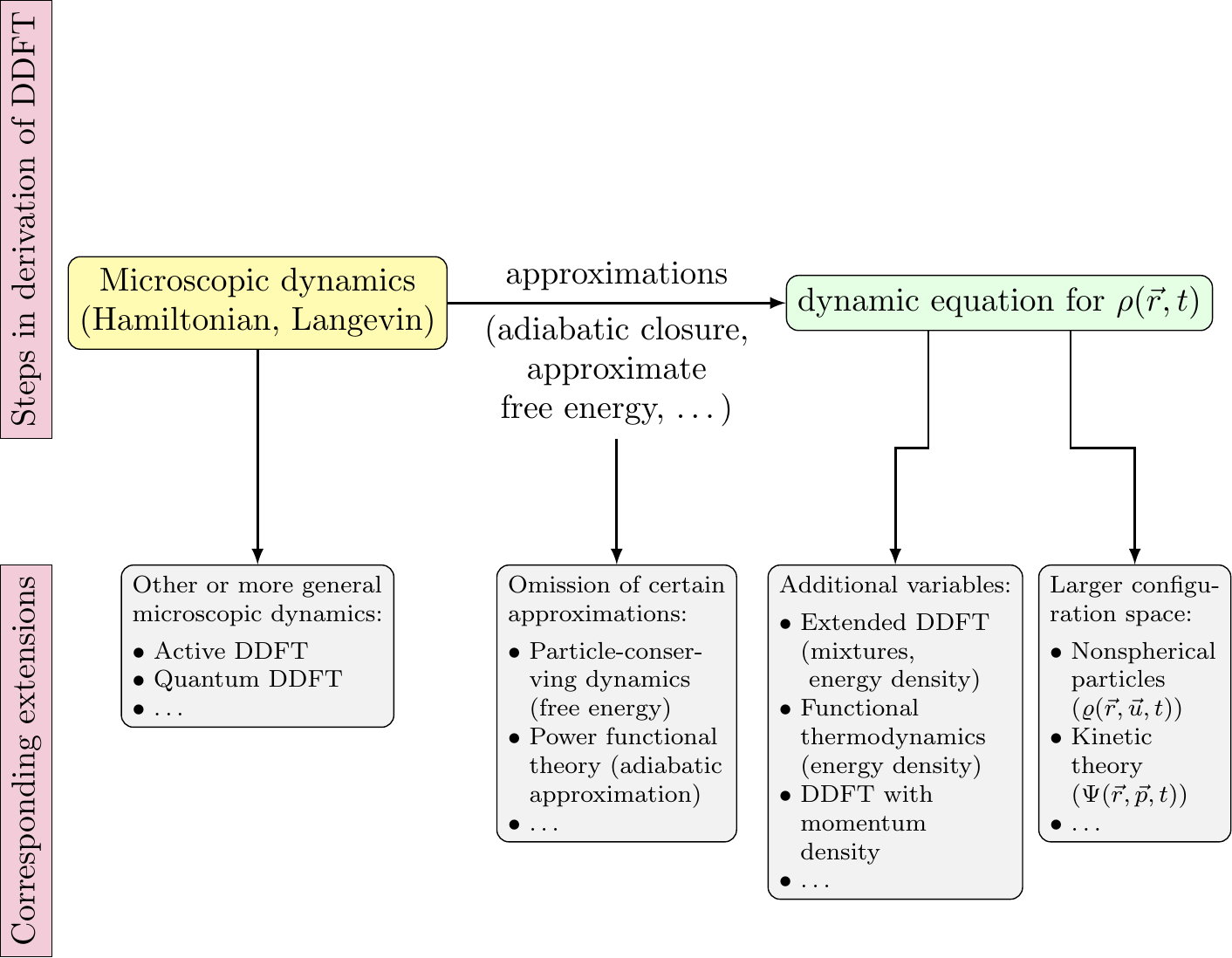}{}%
\caption{\label{fig:classificationextensions}Classification of extensions of standard DDFT.}%
\end{figure*}
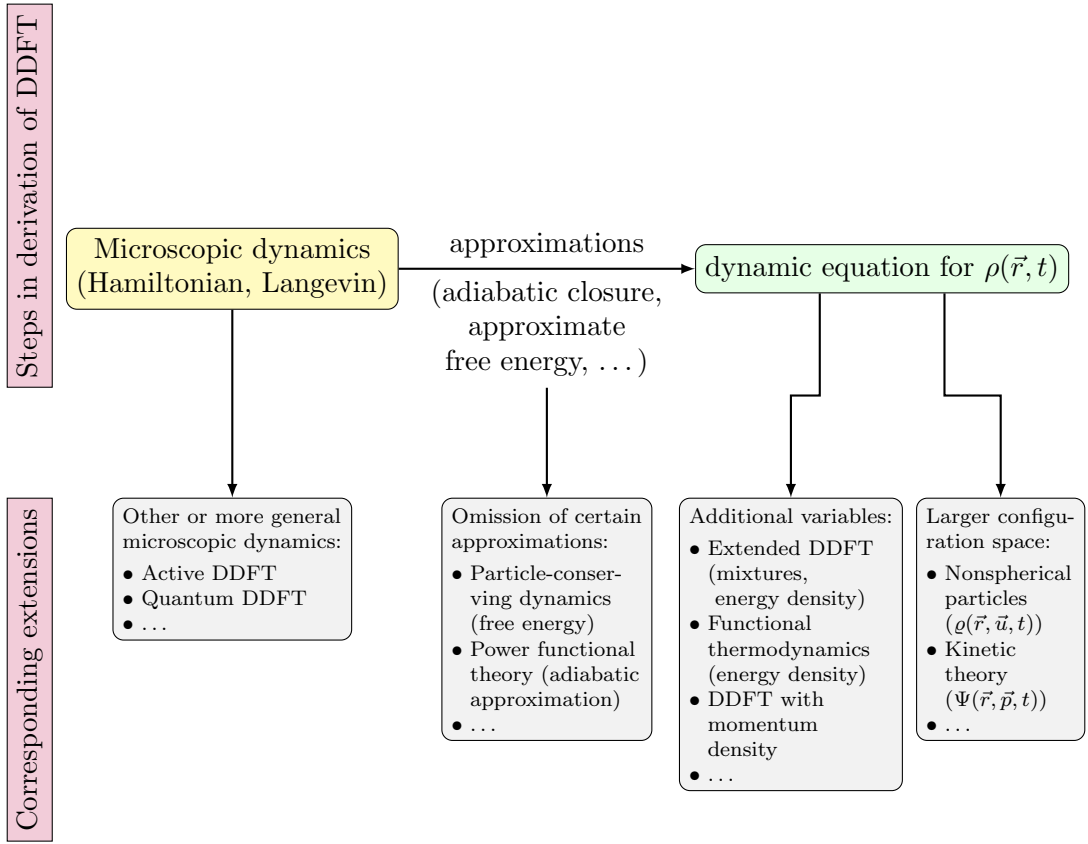

\begin{table}[htb]
\begin{tabularx}{\linewidth}{|p{0.38\linewidth}|X|}
\hline
\rowcolor{lightgray}\textbf{Extension} & \textbf{References}\\ \hline
Nonconstant diffusion coefficient & \cite{Liu2016,Lopez2006,LiuGHLH2017,RexL2008,RexL2009,Loewen2010b,Loewen2010W,Loewen2017,BleibelDO2015,Rauscher2010,YuJEAR2017,GoddardPK2012,GoddardMP2020,GoddardNSPK2012,GoddardNSPK2013,GoddardNSYK2013,GoddardNK2013,DuranGK2016,KikkinidesM2015,DonevVE2014,WittkowskiLB2012,YeDTZM2017,ZhaoWZDTM2014,JabeenYEAR2018,BleibelDO2016,PelaezUPDD2018,AerovK2015,GoddardNK2016,DufrecheJTB2008,HouYDZM2017,HouDZTM2017,YeNTZM2018,SunKK2019,MenzelSHL2016,HoellLM2017,HoellLM2018,HoellLM2019,Chavanis2019,Chavanis2015}\\ \hline 
Hydrodynamic interactions & \cite{RexL2008,RexL2009,Loewen2010b,Loewen2010W,Loewen2017,BleibelDO2015,Rauscher2010,YuJEAR2017,GoddardPK2012,GoddardMP2020,GoddardNSPK2012,GoddardNSPK2013,GoddardNSYK2013,GoddardNK2013,DuranGK2016,KikkinidesM2015,DonevVE2014,WittkowskiLB2012,YeDTZM2017,ZhaoWZDTM2014,JabeenYEAR2018,BleibelDO2016,PelaezUPDD2018,AerovK2015,GoddardNK2016,DufrecheJTB2008,HouYDZM2017,HouDZTM2017,YeNTZM2018,SunKK2019,MenzelSHL2016,HoellLM2017,HoellLM2018,HoellLM2019,RoyallDSvB2007,EspanolL2009}\\ \hline 
Source and sink terms & \cite{LoewenH2014,LiebchenL2019,WerkhovenSvR2019,WerkhovenESvR2018,ChauviereLC2012,AlSaediHAW2018,Lutsko2016,LutskoN2016,LiuL2020,teVrugtBW2020,MonchoD2020,GrawitterS2018,teVrugtBW2020b}\\ \hline
Potential flow & \cite{RauscherDKP2007,AlmenarR2011,Rauscher2010,PraetoriusV2011,GutscheKKRWH2008,ReinhardtSB2014,PuertasV2014,InoueY2014,InoueY2017,YuJEAR2017,FarokhiradRMMAER2019,MenzelSHL2016,HoellLM2017,HoellLM2018,HoellLM2019}\\ \hline 
Shear flow & \cite{ArakiM1995,YoshidaHM1996,RexLL2005,BraderK2011,ScacchiKB2016,AerovK2014,AerovK2015,MutchLAFE2013,SchmidtB2013,delasHerasS2018,StuhlmullerEdlHS2018,TreffenstadtS2019,KrugerSDRD2018,DuqueZumajoCdlTCE2019,KrugerB2011,ScacchiB2018,ScacchiAB2017,ScacchiB2018b,BraderS2013,ZvelindovskySvVMF1998,ZvelindovskySLA2004,ZvelindovskyS2003,NikoubashmanRP2013,LaicerCLT2005,JawalkarNR2008,JawalkarA2006,MuLZ2011,MuLW2011,RakkapaoV2014,MuHLS2008,XuLH2007,XuLH2007c,YeNTZM2019,DengPHL2016,HoellLM2019,ScacchiMA2020}\\ \hline
Mixtures & \cite{Munakata1989,Archer2005,Archer2005b,ZvyagolskayaAB2011,LichtnerAK2012,LichtnerK2014,LichtnerK2013,HoellLM2019,ChauviereLC2012,AlSaediHAW2018,SchindlerWB2019,ArcherHS2007,HopkinsFAS2010,StopperRH2015,StopperRH2016,BraderS2015,StopperMRH2015,StopperTDR2018,YeNTZM2019,BiervRDvdS2008,KohlIBML2012,OkamotoO2016,Lutsko2016,LutskoN2016,RoaSKD2018,LiuL2020,YuZWD2019,BottB2016,BobelBMBR2019,HuangEP2010,TahaDMEH2019,teVrugtBW2020,teVrugtBW2020b,RobbinsAT2011,ThieleVARFSPMBM2009,ArcherRT2010,ChalmersSA2017,NunesGAdG2018,ChakrabartiDL2003,ChakrabartiDL2004,ZakineFvW2018,ArcherWTK2014,ZhaoWZDTM2014,FangS2001,FangS2003,MalijevskyA2013,RothRA2009,ChandraB1991,YoshimoriDP1998,Yoshimori1991,MurataY2006,HowardNP2017,HowardNP2017b,GrawitterS2018,KruppaNML2012,GoddardNK2013,WittkowskiLB2012,MarconiM2011,MarconiM2011b,Marconi2011,ReinhardtB2012,GeigenfeinddlHS2020,KhelashviliWH2008,BlomP2004,DiawM2016,AngiolettiBD2014,AngiolettiBD2018,XuALDB2018,BabelEL2018,EvertsSEvdHvdBvdR2017,JiangCJW2014,GaoX2018,KondratK2013,JiangCJW2014b,LianZLW2016,MarconiM2014,RotenbergPF2010,WerkhovenESvR2018,WerkhovenSvR2019,ReindlB2013,LianL2018,MonteferranteMM2014,MarconiM2012,MarconiMP2013,JanssenvR2017,ZhanLZTXWKCJW2017,MonchoD2020,QingLZTQMXZ2020}\\ \hline 
Nonspherical particles & \cite{WensinkLMHWZKM2013,ChandraB1989,ChandraB1989b,VijayadamodarB1989,ChandraB1990,BagchiC1993,ChandraB1988,BagchiC1991,BurghardtB2006,MondalMB2017,WensinkL2008,WittkowskiL2011,DuranGK2016,BiervR2007,BiervR2008,GreletLBvRvdS2008,BiervRDvdS2008,HaertelL2010,HaertelBL2010,UematsuY2012,DuranYGK2017,KlopotekHGDSSO2017,GomesKPY2019,Loewen2010,MenzelOL2014,ChoudharyLEL2011,CugliandoloDLvW2015}\\ \hline 
Magnetic particles & \cite{LichtnerK2013,LichtnerAK2012,LichtnerK2014}\\ \hline
Active particles & \cite{WensinkL2008,WittkowskiL2011,MenzelSHL2016,HoellLM2019,WittmannB2016,FarageKB2015,WittmannMSSBM2017,WittmannMMB2017,PototskyS2012,MenzelOL2014,SharmaB2017,SharmaB2016,PaliwalRvRD2018,ZakineFvW2018,AroldS2019,KrinningerS2019,KrinningerSB2016,HermannKdlHS2019,MonchoD2020}\\ \hline 
Inertia & \cite{MarconiT2006,MarconiM2007,MarconiTCM2008,Archer2006,ChanF2005,Chavanis2011,GoddardNSPK2012,GoddardNSPK2013,GoddardNSYK2013,GoddardNK2013,DuranGK2016,BerryG2011,BaskaranBL2014,GaoCX2017,Schmidt2018}\\ \hline 
Momentum density & \cite{Archer2009,KikkinidesM2015,MarconiM2009,MarconiM2010,MarconiM2011b,Lutsko2010,Marconi2011,MarconiM2014,MickelJB2011,HughesB2012,DuqueZumajoCdlTCE2019,DuqueZumajodlTCE2019,ZhaoW2011,GoddardHO2020,MajaniemiG2007,GoddardNK2013,OkamotoO2016,BurghardtB2006,BousquetHMB2011,HughesBBRB2012,DiawM2017,RotenbergPF2010,WerkhovenESvR2018,WerkhovenSvR2019,BaskaranBL2014,DiawM2015,DiawM2016,PraetoriusV2015,AlandLV2011,AlandLV2012,AroldS2019,Lutsko2012,NakamuraY2009,PerezMadridRR2002,TothGT2013,ShangVC2011,BaskaranGL2016}\\ \hline 
Kinetic theory & \cite{AneroE2007,PerezMadridRR2002,MarconiM2014,Lutsko2010,MarconiM2007,MarconiTCM2008,MarconiTC2007,BaskaranBL2014,MarconiM2009,MarconiM2011,MarconiM2011b,Marconi2011,MonteferranteMM2014,MarconiM2012,GoddardHO2020}\\ \hline 
Fixed temperature gradients & \cite{LopezM2007,MarconiTCM2008}\\ \hline
Energy density & \cite{Schmidt2011,WittkowskiLB2012,WittkowskiLB2013,AneroET2013,ZhaoW2011,ZhaoW2011b,HuetterB2009,MarconiM2009,Marconi2011,GoddardHO2020}\\ \hline
Entropy density & \cite{Schmidt2011,WittkowskiLB2013,HuetterB2009}\\ \hline
Particle-conserving dynamics & \cite{delasHerasBFS2016,delasHerasS2014,SchindlerWB2019,WittmannLB2020}\\ \hline
Extensions of polymer DDFT & \cite{MauritsvVF1997,MauritsF1997,MauritsZSvVF1998,MauritsZF1998b}\\ \hline
Quantum mechanics & \cite{BurghardtB2006,BousquetHMB2011,HughesBBRB2012,Schmidt2015,BruttingTdlHS2019,DiawM2017}\\ \hline
\end{tabularx}
\vskip2mm
\begin{tabularx}{\linewidth}{|p{0.38\linewidth}|X|}
\hline
\rowcolor{lightgray} \textbf{Exact generalization} & \textbf{References}\\ \hline
Power functional theory & \cite{SchmidtB2013,BraderS2015,Schmidt2015,BruttingTdlHS2019,Schmidt2018,KrinningerS2019,BraderS2015b,delasHerasRS2019,BraderS2014,StuhlmullerEdlHS2018,TreffenstadtS2019,GeigenfeinddlHS2020,KrinningerSB2016,HermannKdlHS2019,JahreisS2020,delasHerasS2020} \\ \hline 
Projection operator formalism & \cite{KinjoH2007,Hyodo2012,Nakajima1958,Zwanzig1960,Mori1965,Kawasaki2000,teVrugtW2019,MeyerVS2019,Grabert1982,teVrugtW2019d,Yoshimori2005,Kawasaki2006b,DuranGK2016,WittkowskiLB2012,WittkowskiLB2013,Yoshimori1999,Munakata2003,EspanolL2009,MajaniemiPN2010,MurataY2006,Yoshimori2011,AneroET2013,CamargodlTDZEDBC2018,CamargodlTDBCE2019,Kawasaki2006,DuranYGK2017,DuqueZumajoCdlTCE2019,DuqueZumajodlTCE2019,MajaniemiG2007,MjaniemiNG2008,WalzF2010,RasSF2020,AneroE2007,EspanolD2015,delTED2015,Grabert1978,Robertson1966,EspanolV2002,Kawasaki1994}\\ \hline
Runge-Gross theorem & \cite{ChanF2005,RungeG1984}\\ \hline
\end{tabularx}
\vskip2mm
\caption{\label{tab:extensionoverview}Overview over extensions and exact generalizations of standard DDFT.}
\end{table}
In this section, we discuss extensions of standard DDFT. By an \ZT{extension of standard DDFT}, we denote any form of DDFT that allows to model aspects not covered by the theories presented in \cref{traditional} (deterministic and stochastic DDFT for simple and colloidal fluids and polymer DDFT). Such extensions exist in a large variety of forms. An overview is given in \cref{tab:extensionoverview}. As visualized in \vref{fig:classificationextensions}, these extensions can be distinguished by the way in which they modify the standard DDFT that is obtained by deriving an approximate transport equation for the one-body density $\rho\rt$ based on the microscopic dynamics of simple or colloidal fluids. Some extensions modify the microscopic dynamics, e.g., by replacing the Langevin equations for passive by those for active Brownian particles (see \cref{active}) or by considering quantum rather than classical systems (see \cref{qm}). Other extensions reduce the number of approximations in the derivation of the transport equation for $\rho\rt$, e.g., by not making an adiabatic approximation (see \cref{pft}) or by not using an approximate grand-canonical free energy functional (see \cref{particleconserving}). This type of extensions also involves formally exact approaches, such as power functional theory (PFT) \cite{SchmidtB2013}, which are discussed in \cref{exact}.

Finally, one can use an extended set of order parameters rather than just the density field $\rho\rt$. Some examples which we present in \cref{exact} are extended dynamical density functional theory (EDDFT) involving also the energy or entropy density, functional thermodynamics (FTD) involving also the energy density, and PFT involving also density currents or velocity gradients. Such extensions have two advantages. The first and obvious one is that by considering an additional field (e.g., the energy density) one has more detailed information about the system. A second and less obvious point is the range of applicability of the resulting theories. As an example, we take a derivation in the projection operator formalism (see \cref{derivationmz}): For a given set of relevant variables, one first rewrites the microscopic equations in an exact form, which is possible regardless of how they are chosen. In the second step, which is crucial for the dynamic equations to be useful, one makes a Markovian approximation corresponding to the assumption that the relevant variables are slow compared to the degrees of freedom that are projected out. If we now compare a DDFT with and without energy density as a relevant variable, the second one requires for a Markovian approximation that all irrelevant variables -- including the energy density -- relax quickly on relevant time scales, which for the energy means that the conjugate variable, which is the temperature, has equilibrated. In summary: A DDFT in which the number density is the only variable effectively needs to assume that the temperature is constant, while a DDFT with energy density does not. A slightly different form of this extension is to extend the configuration space on which the density is defined, such as by using a density $\varrho(\vec{r},\uu,t)$ rather than $\rho\rt$ to incorporate orientational degrees of freedom $\uu$ (see \cref{nonspherical}) or by including a momentum-dependence (see \cref{kinetic}). Many extensions combine a variety of these aspects.

In the following, we present a variety of extensions in (very roughly) increasing order of complexity. A simple modification of the standard DDFT equation \eqref{trddft} is to make the diffusion coefficient nonconstant (\cref{nonconstant}), which includes the case of hydrodynamic interactions (\cref{hydro}). More complex extensions involve adding additional terms to \cref{trddft}, such as source and sink terms (\cref{sourceandsink}) or advection terms (\cref{flow}). In addition, one can introduce additional order parameter fields, which in the simplest case are also density fields (\cref{mixtures}). Further possibilities for order parameter fields arise from additional degrees of freedom of the particles, such as orientation (\cref{odf}) -- giving rise to polarization and nematic order -- and momentum (\cref{moment}) -- giving rise to the momentum density. The latter extension naturally leads to the classical equations of hydrodynamics, which often also involve the energy density (\cref{noniso}). Finally, we turn to special types of extensions that involve an alternative approach to the construction of the free energy (\cref{particleconserving}) or are designed for specific applications such as polymer (\cref{expolymer}) and quantum (\cref{qm}) systems.
\FloatBarrier

\subsection{\label{nonconstant}Nonconstant diffusion coefficient}
While in standard DDFT, the diffusion coefficient $D$ is assumed to be a constant (and used as an input parameter), extended approaches allow it to vary. The most important case is the one of hydrodynamic interactions presented in \cref{hydro}. More generally, derivations in the projection operator formalism (see \cref{derivationmz,po}) give precise microscopic expressions for the diffusion coefficient that can, in certain cases, be approximated by a constant. \citet{Liu2016} calculated (nonconstant) diffusion coefficients based on an entropy scaling rule and applied this method to gas diffusion in porous media. A general discussion of density-dependent diffusion coefficients can be found in Ref.\ \cite{Lopez2006}. Moreover, a position-dependent diffusion coefficient can describe an inhomogeneous environment, e.g., in transport through membranes \cite{LiuGHLH2017}. Finally, the generalized Smoluchowski equation \eqref{generalizedsmoluchowski} discussed in \cref{kawasaki} has a modified mobility function $\Gamma g(\rho)$ instead of the standard form $\Gamma\rho$ \cite{Chavanis2019,Chavanis2015}.

\subsection{\label{hydro}Hydrodynamic interactions}
Hydrodynamic interactions are the most important reason for modifying the diffusion coefficient. A DDFT with hydrodynamics is a theory in which the mobility includes hydrodynamic interactions between the particles. In this case, solvent-mediated interactions lead to additional terms in the DDFT current. This results in a configuration-dependent diffusion tensor. \citet{RoyallDSvB2007} proposed a first simple treatment of hydrodynamic interactions in DDFT, using an equation by \citet{HayakawaI1995} for the dependence of the mobility $\Gamma$ on the colloid packing fraction. A microscopic theory including hydrodynamic interactions was derived by \citet{RexL2008} (see also Refs.\ \cite{RexL2009,Loewen2010b,Loewen2010W,Loewen2017}). In the Smoluchowski equation (in the simplest case given by \cref{smoluchowski}), they used a two-particle diffusion tensor
\begin{equation}
D_{ij}(\{\vec{r}_k\}) \approx D \Eins \delta_{ij}+D\bigg(\delta_{ij}\sum_{\begin{subarray}{c}l=1\\l \neq i\end{subarray}}^{N}\omega_{11}(\vec{r}_i - \vec{r}_l) + (1-\delta_{ij})\omega_{12}(\vec{r}_i - \vec{r}_j)\bigg) 
\end{equation}
with the single-particle diffusion constant $D$. The tensors $\omega_{11}$ (self term) and $\omega_{12}$ (distinct hydrodynamic tensor) are given by \cite{Dhont1996}
\begin{align}
\omega_{11}(\vec{r}_i - \vec{r}_j)&= A_\mathrm{s}(\norm{\vec{r}_i - \vec{r}_j})\frac{(\vec{r}_i - \vec{r}_j)\otimes(\vec{r}_i - \vec{r}_j)}{\norm{\vec{r}_i - \vec{r}_j}^2}\notag \\
&\quad\:\! + B_\mathrm{s}(\norm{\vec{r}_i - \vec{r}_j})\bigg(\Eins - \frac{(\vec{r}_i - \vec{r}_j)\otimes(\vec{r}_i - \vec{r}_j)}{\norm{\vec{r}_i - \vec{r}_j}^2} \bigg),\\
\omega_{12}(\vec{r}_i - \vec{r}_j)&=A_\mathrm{c}(\norm{\vec{r}_i - \vec{r}_j})\frac{(\vec{r}_i - \vec{r}_j)\otimes(\vec{r}_i - \vec{r}_j)}{\norm{\vec{r}_i - \vec{r}_j}^2}\notag \\
&\quad\:\! + B_\mathrm{c}(\norm{\vec{r}_i - \vec{r}_j})\bigg(\Eins - \frac{(\vec{r}_i - \vec{r}_j)\otimes(\vec{r}_i - \vec{r}_j)}{\norm{\vec{r}_i - \vec{r}_j}^2} \bigg) 
\end{align}
with the self-mobility functions $A_\mathrm{s}$ and $B_\mathrm{s}$ and the cross-mobility functions $A_\mathrm{c}$ and $B_\mathrm{c}$. On the Rotne-Prager level of approximation \cite{RotneP1969}, these are given by \cite{Dhont1996}
\begin{align}
A_\mathrm{s}(\norm{\vec{r}_i - \vec{r}_j})&= \mathcal{O}\bigg(\bigg(\frac{R}{\norm{\vec{r}_i - \vec{r}_j}}\bigg)^4\bigg),\\ 
B_\mathrm{s}(\norm{\vec{r}_i - \vec{r}_j})&= \mathcal{O}\bigg(\bigg(\frac{R}{\norm{\vec{r}_i - \vec{r}_j}}\bigg)^4\bigg),\\
A_\mathrm{c}(\norm{\vec{r}_i - \vec{r}_j})&=\frac{3}{2}\frac{R}{\norm{\vec{r}_i - \vec{r}_j}} - \bigg(\frac{R}{\norm{\vec{r}_i - \vec{r}_j}}\bigg)^3 + \mathcal{O}\bigg(\bigg(\frac{R}{\norm{\vec{r}_i - \vec{r}_j}}\bigg)^4\bigg),\\ 
B_\mathrm{c}(\norm{\vec{r}_i - \vec{r}_j})&= \frac{3}{4}\frac{R}{\norm{\vec{r}_i - \vec{r}_j}} +\frac{1}{2} \bigg(\frac{R}{\norm{\vec{r}_i - \vec{r}_j}}\bigg)^3 + \mathcal{O}\bigg(\bigg(\frac{R}{\norm{\vec{r}_i - \vec{r}_j}}\bigg)^4\bigg),
\end{align}
where $R$ is the radius of the particles (which are assumed to be spherical). If the hydrodynamic radius of the particles differs from the interaction radius, one has to use the hydrodynamic radius (see Ref.\ \cite{Dhont1996}). This was done in Refs.\ \cite{RexL2008,RexL2009}. The expression for $D_{ij}$ is valid for particles that are far away from each other. Using the method of reflections \cite{KimK1991,Dhont1996}, the mobility functions can be calculated up to arbitrary order.
By integrating over the Smoluchowski equation, similar to the procedure employed for standard DDFT (see \cref{derivationsmolu}), Rex and L\"owen obtained the result
\begin{equation}
\begin{split}
\frac{k_B T}{D}\pdif{}{t}\rho\rt &= \vec{\nabla}_{\vec{r}}\cdot\bigg(\rho\rt\vec{\nabla}_{\vec{r}}\frac{\delta F[\rho]}{\delta \rho\rt} \\
&\quad\:\!+ \INT{}{}{^3r'}\rho^{(2)}(\vec{r},\vec{r}',t)\omega_{11}(\vec{r}-\vec{r}')\cdot\vec{\nabla}_{\vec{r}}\frac{\delta F[\rho]}{\delta \rho\rt} \\
&\quad\:\!+ \INT{}{}{^3r'}\rho^{(2)}(\vec{r},\vec{r}',t)\omega_{12}(\vec{r}-\vec{r}')\cdot\vec{\nabla}_{\vec{r}'}\frac{\delta F[\rho]}{\delta \rho(\vec{r}',t)}\bigg),
\end{split}\label{eq:DDFTwithHI}%    
\end{equation}
which, in addition to the standard current of DDFT (first term on the right-hand side), also involves a current from the reflection of solvent flow by the thermodynamic force at position $\vec{r}$ (second term on the right-hand side) and from the solvent flow induced by the thermodynamic force at position $\vec{r}'$ (last term). The equation can be closed by using an Ornstein-Zernike relation for the two-particle density $\rho^{(2)}(\vec{r},\vec{r}',t)$ \cite{RexL2008}. A simplified form of \cref{eq:DDFTwithHI}, which uses the Oseen tensor for the hydrodynamic interactions and an ideal gas correlation, is presented in Ref.\ \cite{BleibelDO2015}. The mobility can get an explicit spatial dependence if the translational symmetry of the system is broken by interfaces. In this way, hydrodynamic interactions with walls can be described \cite{Rauscher2010,YuJEAR2017}.

Work on DDFTs including hydrodynamics has been performed in various directions. Derivations based on the projection operator formalism \cite{EspanolL2009,WittkowskiLB2012} naturally lead to dynamic equations with hydrodynamic interactions (as discussed in \cref{derivationmz}), such that they are included in EDDFT \cite{WittkowskiLB2012} (see \cref{eddft}). A rigorous derivation of a Smoluchowski equation for the one-particle distribution incorporating hydrodynamic interactions is given in Ref.\ \cite{GoddardPK2012}. \citet{GoddardMP2020} examined the well-posedness of hydrodynamic DDFT. The inertia of particles is included in hydrodynamic DDFTs in Refs.\ \cite{GoddardNSPK2012,GoddardNSPK2013,GoddardNSYK2013,GoddardNK2013,DuranGK2016,KikkinidesM2015}. \citet{DonevVE2014} obtained a hydrodynamic DDFT that also includes the fluctuations (averaging then recovers the result from \citet{RexL2008}). \citet{DuranGK2016} considered orientable particles with arbitrary shape. Another extension is the study of hydrodynamic interactions in multi-species systems \cite{WittkowskiLB2012,GoddardNK2013}. A DDFT for polymer chains with hydrodynamic interactions is presented in Refs.\ \cite{YeDTZM2017,ZhaoWZDTM2014}. In Ref.\ \cite{JabeenYEAR2018}, hydrodynamic interactions in a flowing hard-sphere suspension are studied using Monte Carlo simulations. Experimental results on the importance of hydrodynamic effects on short-time diffusion (which are compared to a DDFT without hydrodynamic interactions) can be found in Ref.\ \cite{StopperTDR2018}.

Regarding applications, the original hydrodynamic DDFTs were used to study sedimentation \cite{RoyallDSvB2007} and oscillations in optical traps \cite{RexL2008}. Effects of hydrodynamics in confined systems have also been studied in DDFT \cite{BleibelDO2015,BleibelDO2016,PelaezUPDD2018,AerovK2015,GoddardNK2016}. Moreover, hydrodynamic DDFT allows to study hydrodynamic interactions in ion diffusion \cite{DufrecheJTB2008}, wavenumber-dependent diffusion coefficients \cite{BleibelDO2016}, crystal growth \cite{HouYDZM2017,HouDZTM2017}, glassy water \cite{YeNTZM2018}, and velocities of particles subject to hydrodynamic stress \cite{SunKK2019}. Finally, hydrodynamic interactions are included in DDFT descriptions of microswimmers \cite{MenzelSHL2016,HoellLM2017,HoellLM2018,HoellLM2019}.

\subsection{\label{sourceandsink}Source and sink terms}
A further possibility for modifying the standard DDFT equation \eqref{trddft} is to add source and sink terms. Sources and sinks can be a result from influx or outflux, but in the case of mixtures (see \cref{mixtures}) also originate from a transformation of one particle type into another. If the system of particles is not closed, the number of particles is not necessarily conserved and particle inflow is possible. This was discussed by \citet{LoewenH2014} for the case of particles being injected into a confined system at position $\vec{r}$ with rate $\InjectionRate(t)$. In this case, a point source $\InjectionRate(t)\delta(\vec{r})$ has to be added to the DDFT equation \eqref{trddft}. If \cref{trddft} is linearized (assuming small density variations), an analytical solution of this problem is possible using Green's functions. In Ref.\ \cite{LiebchenL2019}, source terms are employed to describe diffusion of chemicals around a point source. Other cases in which source terms have to be added to the conserved DDFT equation include evaporating thin films \cite{RobbinsAT2011} (see \cref{thin}), surface charge densities \cite{WerkhovenSvR2019,WerkhovenESvR2018} (see \cref{electro}), biological dynamics \cite{ChauviereLC2012,AlSaediHAW2018} (see \cref{biology}), chemical reactions \cite{Lutsko2016,LutskoN2016,LiuL2020} (see \cref{chemical}), switching \cite{GrawitterS2018,MonchoD2020}, and epidemic spreading \cite{teVrugtBW2020,teVrugtBW2020b} (see \cref{disease}).

\subsection{\label{flow}Flow}
\subsubsection{\label{potentialflow}Potential flow}
In general, it cannot be assumed that the solvent in which the colloids are immersed is at rest. Thus, a natural generalization of DDFT is to allow for solvent flow with a velocity $\vec{v}\rt$. An additional modification of DDFT is the incorporation of \textit{shear} flow, which will be discussed separately (see \cref{shear}). 

Early models studied effects, such as colloids being dragged through a polymer solution, by shifts to a frame that is co-moving with a velocity $\vec{v}_{\mathrm{cm}}$ \cite{PennaDT2003} and did not consider perturbations of the flow \cite{KruegerR2007}. Essentially, this corresponds to assuming that the solvent simply flows through the colloid \cite{RauscherDKP2007}. However, as discussed in Refs.\ \cite{RauscherDKP2007,AlmenarR2011}, the colloids also influence the solvent. If a spherical obstacle of radius $\RObstacle$, such as a colloid, moves through a solution, the solvent particles, which are very small, can get very close. This leads to a modified flow field, which can for incompressible flow at low Reynolds number be calculated analytically as
\begin{equation}
\vec{v}(\vec{r})= \frac{3\RObstacle}{4r}\bigg(1+\frac{\RObstacle^2}{3r^2}\bigg)\vec{v}_{\mathrm{cm}} + \frac{3\RObstacle}{4r^3}(\vec{r}\cdot\vec{v}_{\mathrm{cm}})\vec{r}\bigg(1-\frac{\RObstacle^2}{r^2}\bigg)-\vec{v}_{\mathrm{cm}}
\label{eq:geschwindigkeitsfeld}%
\end{equation}
with $r=\norm{\vec{r}}$. Only for $r\to\infty$, \cref{eq:geschwindigkeitsfeld} can be approximated by $\vec{v} = -\vec{v}_{\mathrm{cm}}$. 

A DDFT for colloidal particles advected by a flow field was derived by \citet{RauscherDKP2007}, who did not take hydrodynamic interactions between solvent and colloidal particles into account. The starting point are the overdamped Langevin equations for the motion of the $i$-th particle in the presence of a flow field $\vec{v}\rt$, given by
\begin{equation}
\tdif{\vec{r}_i(t)}{t} = \vec{v}(\vec{r}_i,t) - \Gamma \vec{\nabla}_{\vec{r}_i}\bigg(U_1(\vec{r}_i)+\sum_{\begin{subarray}{c}j=1\\j \neq i\end{subarray}}^{N}U_2(\norm{\vec{r}_i - \vec{r}_j})\bigg) + \vec{\chi}_i(t).
\end{equation}
For these Langevin equations, one can obtain a corresponding Fokker-Planck equation from which the evolution of the averaged density $\rho$ is computed. If we have a potential flow, which is always the case when detailed balance holds, we can write
\begin{equation}
\vec{v}(\vec{r}) = - \Gamma \vec{\nabla}U_{\mathrm{vel}}(\vec{r})
\end{equation}
and define a modified external potential $U^\star (\vec{r})=U_1(\vec{r})+U_{\mathrm{vel}}(\vec{r})$ with the velocity potential $U_{\mathrm{vel}}$. This allows to approximate the interaction term in the usual way. The resulting \textit{advected DDFT} reads
\begin{equation}
\pdif{}{t}\rho\rt +\vec{\nabla}\cdot(\vec{v}\rt\rho\rt) = \Gamma \vec{\nabla}\cdot\bigg(\rho\rt\vec{\nabla}\frac{\delta F[\rho]}{\delta \rho\rt}\bigg).
\label{advected}
\end{equation}
As can be seen, this corresponds to a standard DDFT equation with $U_{1}$ replaced by $U^\star$.

In Ref.\ \cite{Rauscher2010}, this treatment is extended to hydrodynamic interactions with walls, incorporated by a position-dependent mobility, and hydrodynamic interactions among the particles (see \cref{hydro}).
\citet{PraetoriusV2011} used the advected DDFT to derive an advected PFC equation. A highly simplified model is applied to colloids in a DNA solution in Ref.\ \cite{GutscheKKRWH2008}. Moreover, advected DDFT is used in microrheology \cite{ReinhardtSB2014,PuertasV2014,InoueY2014,AlmenarR2011} (see also \cref{ps}). A possible setup is the flow of interacting particles around a fixed probe particle \cite{InoueY2014,InoueY2017}. Hydrodynamic lift forces are discussed in Refs.\ \cite{YuJEAR2017,FarokhiradRMMAER2019}. Microscopic considerations of the flow field are also important in the derivation of a hydrodynamic DDFT for microswimmers \cite{MenzelSHL2016,HoellLM2017,HoellLM2018,HoellLM2019} (see \cref{active}). Problems of the local equilibrium closure are considered in Ref.\ \cite{AlmenarR2011}. A review on nanoparticle flow can be found in Ref.\ \cite{RadhakrishnanFEA2019}. 

\subsubsection{\label{shear}Shear flow}
Investigating the properties of sheared simple and complex fluids is a central topic in soft matter physics. Therefore, effects of shear were considered already at early stages in the development of DDFT. The shear viscosity was calculated based on shear-induced deviations from the equilibrium profile \cite{ArakiM1995,YoshidaHM1996}. Soft colloids sheared by traveling waves were studied by \citet{RexLL2005}.

As discussed in Ref.\ \cite{BraderK2011}, there exist certain problems in describing shear flow using standard DDFT. To see this, we try to apply the governing equation of advected DDFT \eqref{advected} to a simple shear flow experiment with two parallel plane walls, where $y$ is the direction normal and $x$ the direction parallel to the walls. The flow velocity $\vec{v}$ has the form 
\begin{equation}
\vec{v}(\vec{r})=y\dot{\gamma}\vec{e}_{x}
\label{shearflow}
\end{equation}
with the shear rate $\dot{\gamma}$ and the unit vector in $x$ direction $\vec{e}_{x}$. The density will, because of translational symmetry, only depend on $y$. Unfortunately, the form \eqref{shearflow} implies
\begin{equation}
\vec{\nabla}\cdot(\vec{v}(\vec{r}) \rho\rt) = 0,
\end{equation}
such that \cref{advected} reduces to the standard DDFT equation without a flow field. However, it is certainly wrong that shear has no effects on the particle distribution. The source of the problem is the distortion of the correlation functions (which DDFT assumes to be as in equilibrium) due to the external flow \cite{ScacchiKB2016}. A DDFT-based description of distorted correlation functions for nonlinear rheology is proposed in Ref.\ \cite{ReinhardtWB2013}.

A solution to this problem has been suggested by \citet{BraderK2011}. The flow velocity is corrected by the mean-field term
\begin{equation}
\vec{v}_{\mathrm{fl}}\rt = \INT{}{}{^3r'}\rho(\vec{r}',t)\vec{K}^{\mathrm{s}}(\vec{r}-\vec{r}'),
\end{equation}
where $\vec{K}^{\mathrm{s}}(\vec{r}-\vec{r}')$ is the flow kernel accounting for the effects of shear flow. While this is a phenomenological correction, systematic derivations are possible in certain cases \cite{ScacchiKB2016}. A detailed discussion of shear DDFT can be found in Ref.\ \cite{AerovK2014} (see also Ref.\ \cite{AerovK2015} for the role of hydrodynamic interactions in shear flow). A review article on time-dependent shear flow can be found in Ref.\ \cite{MutchLAFE2013}.

Shear was studied in a variety of forms and extensions of DDFT. Since superadiabatic effects are important in rheology, it is reasonable to describe shear flow by PFT \cite{SchmidtB2013,delasHerasS2018}. Superadiabatic forces in sheared systems are discussed in the framework of PFT in Refs.\ \cite{StuhlmullerEdlHS2018,TreffenstadtS2019,JahreisS2020,delasHerasS2020}. \citet{KrugerSDRD2018} derived the stress tensor for stochastic DDFT, considering the shear viscosity and shear-stress fluctuations. Shear can also be described within DDFT extensions including the momentum density \cite{DuqueZumajoCdlTCE2019}. 

An application of DDFT with shear-flow corrections is the study of the influence of time-dependent shear on sedimentation \cite{KrugerB2011}. In Ref.\ \cite{ScacchiB2018}, shear DDFT is used to analyze the interplay between shear and crystallization. A linear stability analysis allows to determine the dispersion relation for the laning instability \cite{ScacchiAB2017}. Effects of shear flow on nonequilibrium phase transitions are considered in Ref.\ \cite{StopperR2018}. Shear flow can also have effects on the formation of low-density regions behind tracer particles \cite{ScacchiB2018b}. \citet{BraderS2013} discussed effects of shear flow on the time-dependent diffusion tensor. \citet{ScacchiMA2020} studied the influence of shear on length scales in soft matter. Polymer DDFT (see \cref{pd}) is also frequently applied to sheared systems \cite{ZvelindovskySvVMF1998,ZvelindovskySLA2004,ZvelindovskyS2003,NikoubashmanRP2013,LaicerCLT2005,JawalkarNR2008,JawalkarA2006,MuLZ2011,MuLW2011,RakkapaoV2014,MuHLS2008,XuLH2007,XuLH2007c,YeNTZM2019,DengPHL2016}. \citet{HoellLM2019} model a shear cell consisting of microswimmers trapped in passive colloidal particles. Note that not all applications of DDFT mentioned in this section employ the method by \citet{BraderK2011}.

\subsection{\label{mixtures}Mixtures}
DDFT has, very frequently, been applied to systems consisting of different particle types. The general idea, already employed by \citet{Munakata1989}, is straightforward: One uses a separate density field $\rho_i\rt$ for each particle species $i$, defines a free energy $F[\{\rho_i\}]$ that depends on all fields and includes the single-species dynamics as well as possible interactions, and then writes down the DDFT equation \eqref{trddft} for each particle type. The result is
\begin{equation}
\pdif{}{t}\rho_i\rt = \Gamma_i\Nabla\cdot\bigg(\rho_i\rt\Nabla\frac{\delta F[\{\rho_j\}]}{\delta \rho_i\rt}\bigg).
\label{ddftformixtures}
\end{equation}
Here, $\Gamma_i$ is the mobility of particle type $i$. Equation \eqref{ddftformixtures} is a special case of \cref{gradientdynamics}, which is the general form of a diffusive transport equation for a conserved order parameter known as \ZT{gradient dynamics} (this can be seen by writing \cref{ddftformixtures} in vector notation by introducing a vector $\vec{\rho}$ whose elements are the fields $\{\rho_i\}$). To obtain \cref{ddftformixtures} from \cref{gradientdynamics} for the case that the general conserved variable $a^\mathrm{c}$ from \cref{gradientdynamics} is given by $a^\mathrm{c} = \vec{\rho}$, we have to assume that the mobility matrix $\MobilityGradientDynamics$ appearing in \cref{gradientdynamics} has the elements ($\MobilityGradientDynamics)_{ij}=\Gamma_i\rho_i\delta_{ij}$ (no sum over $i$), i.e., that all off-diagonal coefficients are zero. This can change in two cases: First, hydrodynamic interactions (see \cref{hydro}) lead to a more complex diffusion tensor that contains off-diagonal terms. Second, incompressibility plays a role here. As discussed in \cref{pd}, polymer DDFT deals, in its simplest form, with incompressible mixtures. For a binary mixture, an incompressibility constraint leads from \cref{loco}, which is a standard DDFT equation for mixtures, to \cref{locoa,locob}, which, if written in the form \eqref{gradientdynamics}, would correspond to a mobility matrix with off-diagonal elements. This procedure was discussed by \citet{deGennes1980}. Mobility matrices with off-diagonal elements also arise in the thin-film hydrodynamics of incompressible binary mixtures, in a form that is similar to the one arising in incompressible polymer DDFT. \citet{XuTQ2015} compared this form to the standard form \eqref{ddftformixtures} of DDFT for mixtures without off-diagonal contributions and attributed this difference to the different choice of the frame of reference. While a detailed investigation of the relation between incompressible thin-film hydrodynamics and incompressible polymer theory is a task for future work, the similarity of the arguments employed in the derivations and of the form of the resulting theories makes it likely that they are connected. 

We now return to \cref{ddftformixtures}. The simplest case is that of a binary mixture. As demonstrated in Refs.\ \cite{Archer2005,Archer2005b,ZvyagolskayaAB2011}, this allows to model phase separation. A further extension is to give the particle species additional degrees of freedom, such as a magnetic moment \cite{LichtnerAK2012,LichtnerK2014,LichtnerK2013}. Mixtures of microswimmers are considered in Ref.\ \cite{HoellLM2019}, this allows to study the phase behavior of, e.g., pusher-puller mixtures or mixtures of active and passive particles. \citet{ChauviereLC2012} derived a DDFT for a system of multiple cell species and \citet{AlSaediHAW2018} modeled the competition of healthy and cancer cells using a two-species DDFT (see \cref{biology}). The particle-conserving dynamics (see \cref{particleconserving}) of binary mixtures is considered in Ref.\ \cite{SchindlerWB2019}. One can also apply DDFT for mixtures to one-component systems whose components can be in different states \cite{MonchoD2020}. A natural application of DDFT for mixtures is the calculation of the van Hove function \cite{ArcherHS2007,HopkinsFAS2010,StopperRH2015,StopperRH2016,BraderS2015,StopperMRH2015,StopperTDR2018,YeNTZM2019,BiervRDvdS2008} (see \cref{vh}).

Moreover, DDFT for binary mixtures has been applied to model anisotropic pair correlations \cite{KohlIBML2012}, bubble dynamics \cite{OkamotoO2016}, chemical reactions \cite{Lutsko2016,LutskoN2016,RoaSKD2018,LiuL2020}, CO$_2$ dissolving in poly (lactic acid) melt \cite{YuZWD2019}, colloid-polymer mixtures \cite{StopperRH2016}, demixing on a sphere \cite{BottB2016,BobelBMBR2019}, derivation of a PFC model for binary mixtures \cite{HuangEP2010,TahaDMEH2019}, disease spreading \cite{teVrugtBW2020,teVrugtBW2020b}, evaporating nanoparticle suspensions \cite{RobbinsAT2011,ThieleVARFSPMBM2009,ArcherRT2010,ChalmersSA2017}, field-driven demixing \cite{NunesGAdG2018}, instabilities in driven systems \cite{ChakrabartiDL2003,ChakrabartiDL2004}, particles switching between two states \cite{ZakineFvW2018}, pattern formation \cite{ArcherWTK2014}, polymer-particle suspensions \cite{ZhaoWZDTM2014}, protein adsorption \cite{FangS2001,FangS2003}, sheared polymer nanocomposites \cite{YeNTZM2019}, sedimentation \cite{MalijevskyA2013}, selectivity effects \cite{RothRA2009}, solvation dynamics \cite{ChandraB1991,YoshimoriDP1998,Yoshimori1991,MurataY2006}, stratification \cite{HowardNP2017,HowardNP2017b}, surfactant mixtures \cite{GrawitterS2018}, and the Brazil nut effect \cite{KruppaNML2012},

A more general case is a multicomponent suspension, as discussed by \citet{WittkowskiLB2012} and \citet{GoddardNK2013}. Both articles include a treatment of hydrodynamic interactions. As an additional order parameter field, they use the energy density \cite{WittkowskiLB2012} and the velocity field for each particle species \cite{GoddardNK2013}, respectively. The momentum density of multicomponent mixtures is also used as a dynamical variable in Refs.\ \cite{MarconiM2011,MarconiM2011b,Marconi2011}. Grand-canonical artefacts and canonical corrections in multicomponent DDFT are discussed in Ref.\ \cite{ReinhardtB2012}. A PFT for multicomponent mixtures was derived by \citet{BraderS2015} and applied in Ref.\ \cite{GeigenfeinddlHS2020}. Polymer DDFT (see \cref{pd}) is typically set up as a multicomponent theory. 

An application of DDFT for mixtures is the description of cell membranes \cite{KhelashviliWH2008,BlomP2004}. Moreover, \citet{DiawM2016} modeled an $N$-component astrophysical plasma using DDFT. Mixtures of various protein species are considered in Refs.\ \cite{AngiolettiBD2014,AngiolettiBD2018,XuALDB2018}. Finally, electrochemical systems containing ions of different charge \cite{BabelEL2018,EvertsSEvdHvdBvdR2017,JiangCJW2014,GaoX2018,KondratK2013,JiangCJW2014b,LianZLW2016,MarconiM2014,RotenbergPF2010,WerkhovenESvR2018,WerkhovenSvR2019,ReindlB2013,LianL2018,MonteferranteMM2014,MarconiM2012,MarconiMP2013,JanssenvR2017,ZhanLZTXWKCJW2017,QingLZTQMXZ2020} are an important application of DDFT for mixtures (see \cref{electro}). An option for future work is the investigation of systems with nonreciprocal interactions \cite{BartnickHIL2015}.

\subsection{\label{odf}Orientational degrees of freedom}
Up to now, we have focused on passive spherical particles that (in the overdamped case) are fully described by their center-of-mass positions. However, real particles often have additional (e.g., orientational) degrees of freedom, which can be incorporated in extensions of DDFT. Orientation can be taken into account by using an extended configuration space, i.e., using a density $\varrho(\vec{r},\uu,t)$ that also depends on the orientation $\uu$, or by additional order parameter fields such as a polarization field. Sources of orientational degrees of freedom are geometrical anisotropy (\cref{nonspherical}), magnetic moments (\cref{magnetic}), and self-propulsion (\cref{active}).

\subsubsection{\label{nonspherical}Nonspherical particles}
The standard DDFT formalism describes the dynamics of particles with (effective) spherical symmetry, since it is assumed in the microscopic derivations that the state of a Brownian particle is (in the overdamped limit) characterized by its position. However, standard DDFT can also be extended to nonspherical particles. For this purpose, we additionally need to specify the orientation of a particle. If the particle has an axis of continuous rotational symmetry (\ZT{uniaxial particle}), this can be done by a vector $\uu$ which points in the direction of this axis, or, for a spherical active particle, in the direction of self-propulsion. The orientation vector $\uu$ is parametrized by one angle $\varphi$ in 2D and by two angles $\varphi$ and $\vartheta$ in 3D \cite{teVrugtW2020b}, which corresponds to the use of polar coordinates and spherical coordinates, respectively. If the particle has a more general shape, its orientation is fixed by the rotation that transforms from the laboratory-fixed to the body-fixed frame of reference. This rotation can be parametrized in 2D by one angle $\varphi$ \cite{teVrugtW2020b} and in 3D by the three Euler angles $\varphi$, $\vartheta$, and $\winkel$ \cite{teVrugtW2020b}. In the convention employed in Refs.\ \cite{GrayG1984,teVrugtW2020b,WittkowskiL2012,WittkowskiL2011}, which is equivalent to the second convention of Ref.\ \cite{Schutte1976}, the first two Euler angles coincide with the angles $\varphi$ and $\vartheta$ of the spherical coordinates. Theoretical descriptions of anisotropic particles are reviewed in Ref.\ \cite{WensinkLMHWZKM2013}.

Orientational degrees of freedom were considered already at early stages of the development of DDFT. A Smoluchowski equation for orientational dynamics was obtained by \citet{CalefW1983}, whose work was extended by \citet{ChandraB1989}. In the study of solvation \cite{ChandraB1989b}, diffusion \cite{VijayadamodarB1989}, and orientational relaxation \cite{ChandraB1990,BagchiC1993,ChandraB1988}, the DDFT equation 
\begin{equation}
\pdif{}{t}\varrho(\vec{r},\uu,t)= D \Nabla_{\vec{r}}\cdot\bigg(\varrho(\vec{r},\uu,t)\Nabla_{\vec{r}}\frac{\delta F}{\delta \varrho(\vec{r},\uu,t)}\bigg) + D_R\Nabla_{\uu} \cdot\bigg(\varrho(\vec{r},\uu,t)\Nabla_{\uu} \frac{\delta F}{\delta \varrho(\vec{r},\uu,t)}\bigg)
\label{eq:rhorut}%
\end{equation}
for the one-body density $\varrho$ that measures the probability of finding a particle with position $\vec{r}$ and orientation $\uu$ at time $t$ was used. Here, $D$ and $D_R$ are the translational and rotational diffusion coefficients of a free particle, respectively, and $\Nabla_{\uu}$ denotes the rotation operator. We here adapt the notation of Refs.\ \cite{BagchiC1991,WittkowskiL2012} and use $\Nabla_{\uu}$ for the rotation operator. An alternative, employed, e.g., in Ref.\ \cite{RexWL2007}, is to use $\Nabla_{\uu}$ for a partial derivative with respect to the elements of the orientation vector $\uu$, in analogy to $\Nabla_{\vec{r}}$. In this case, the rotation operator is given by $\uu \times \Nabla_{\uu}$. A review on orientational relaxation dynamics is given in Ref.\ \cite{BagchiC1991} and a review on early forms of DDFT (including the method from Chandra and Bagchi) can be found in Ref.\ \cite{Yoshimori2004}. The DDFT \eqref{eq:rhorut} can also be extended to mixtures \cite{ChandraB1991}. Applications of this form of DDFT include the calculation of orientational relaxation times \cite{Bagchi2001,JanaB2012}, mixed quantum-classical dynamics \cite{BurghardtB2006} (see \cref{qm}), and protein solvation \cite{MondalMB2017}.

From a microscopic point of view, uniaxial particles were considered by \citet{RexWL2007} (see also Refs.\ \cite{Loewen2010b,Loewen2010W,Loewen2017}). In a DFT (or DDFT) for uniaxial particles, the orientation is incorporated by using a density field $\varrho(\vec{r},\uu)$ rather than simply $\rho(\vec{r})$, such that $\varrho(\vec{r},\uu)$ measures the probability of finding a particle at position $\vec{r}$ with orientation $\uu$ \cite{SittaSWL2016,SittaSWL2018}. The derivation of a dynamic equation for $\varrho(\vec{r},\uu)$ then proceeds along the same lines as the derivation from the Smoluchowski equation outlined in \cref{derivationsmolu}. For nonspherical particles, the Smoluchowski equation reads 
\begin{equation}
\begin{split}
\pdif{}{t}\SmoluchowskiDistribution(\{\vec{r}_k\},\{\uu_k\},t)
&=\sum_{i=1}^{N}\bigg(\vec{\nabla}_{\vec{r}_i}\cdot \bigg( D(\uu_i) \bigg(\vec{\nabla}_{\vec{r}_i} + \frac{1}{k_B T}\vec{\nabla}_{\vec{r}_i} U(\{\vec{r}_k\},\{\uu_k\},t) \bigg)\!\bigg)\\
&\quad\:\! + D_R \Nabla_{\uu_i}\cdot\bigg(\Nabla_{\uu_i} + \frac{1}{k_B T}\Nabla_{\uu_i} U(\{\vec{r}_k\},\{\uu_k\},t)\bigg)\!\bigg)\SmoluchowskiDistribution(\{\vec{r}_k\},\{\uu_k\},t)\big).
\label{smoluchowskiorientation}    
\end{split}\raisetag{3.5em}%
\end{equation}
Compared to \cref{smoluchowski}, the microscopic density $\SmoluchowskiDistribution$ now also depends on the orientations $\{\uu_k\}$ of the individual particles, the translational diffusion constant is now a tensor
\begin{equation}
D(\uu) = D_{\parallel}\uu\otimes\uu + D_{\perp}(\Eins-\uu\otimes\uu)
\end{equation}
with parallel and perpendicular translational diffusion coefficients $D_{\parallel}$ and $D_{\perp}$, respectively, and identity matrix $\Eins$, and we have added a term for the rotation with the coefficient of rotational diffusion $D_R$. The generalized force balance 
\begin{equation}
\begin{split}
\frac{k_B T}{\varrho(\vec{r},\uu)}\vec{\nabla}_{\vec{r}}\varrho(\vec{r},\uu) +\vec{\nabla}_{\vec{r}}U_1(\vec{r},\uu) = -\vec{\nabla}_{\vec{r}}\frac{\delta F_{\mathrm{exc}}[\varrho]}{\delta\varrho(\vec{r},\uu)}
\end{split}
\end{equation}
of thermodynamic equilibrium used analogously in the standard derivations is now supplemented by a generalized torque balance 
\begin{equation}
\frac{k_B T}{\varrho(\vec{r},\uu)}\Nabla_{\uu}\varrho(\vec{r},\uu) + \Nabla_{\uu}U_1(\vec{r},\uu)=-\Nabla_{\uu}\frac{\delta F_{\mathrm{exc}}[\varrho]}{\delta\varrho(\vec{r},\uu)}.
\end{equation}
Making an adiabatic approximation in the usual way, this gives the DDFT for uniaxial particles 
\begin{equation}
\begin{split}
\pdif{}{t}\varrho(\vec{r},\uu,t)
&=\vec{\nabla}_{\vec{r}}\cdot \bigg(D(\uu) \varrho(\vec{r},\uu,t)\vec{\nabla}_{\vec{r}}\frac{\delta F[\varrho]}{\delta \varrho(\vec{r},\uu,t)}\bigg)\\
&\quad\:\! + D_R \Nabla_{\uu}\cdot\bigg(\varrho(\vec{r},\uu,t)\Nabla_{\uu}\frac{\delta F[\varrho]}{\delta \varrho(\vec{r},\uu,t)}\bigg),
\label{uniaxialddft}%
\end{split}
\end{equation}
where the free energy $F$ still has the form \eqref{freeenergy} (the ideal contribution is now that of an ideal rotor gas). 

For a further analysis of the field $\varrho(\vec{r},\uu,t)$, it is typically helpful to perform an expansion of the orientational dependence in symmetric traceless tensors. The general expansion for a function $f(\uu)$ in 3D reads \cite{teVrugtW2020b}
\begin{equation}
f(\uu)=\SUM{l=0}{\infty} \SUM{i_{1},\dotsc,i_{l}=1}{3} \ffX{3}{l}{i_{1}\dotsb i_{l}} u_{i_{1}}\!\dotsb u_{i_{l}}
\label{fu}%
\end{equation}
with the coefficients
\begin{equation}
\ffX{3}{l}{i_{1}\dotsb i_{l}} = A^{(3\mathrm{D})}_{l} \INT{S_{2}}{}{\Omega} f(\uu) \TT{3}{l}{i_{1}\dotsb i_{l}},
\label{eq:fdD}%
\end{equation}
the normalization $A^{(3\mathrm{D})}_{l} = (2l+1)/(4\pi)$, the integration over the unit sphere $\TINT{S_{2}}{}{\Omega}\!$, and
the tensor polynomials
\begin{equation}
\TT{3}{l}{i_{1}\dotsb i_{l}} = \frac{(-1)^{l}}{l!} \partial_{i_{1}}\!\dotsb\partial_{i_{l}}\frac{1}{r}\bigg\rvert_{\vec{r}=\uu},
\end{equation}
where $\partial_i$ denotes the spatial derivative with respect to the $i$-th element of $\vec{r}$.
Note that different expansions have to be used in 2D or for asymmetric particles whose one-body distribution depends on three angles. (A detailed overview over various types of orientational expansions is given in Ref.\ \cite{teVrugtW2020b}.) If we perform the expansion \eqref{fu} for the function $\varrho(\vec{r},\uu,t)$ up to second order, we find
\begin{equation}
\varrho(\vec{r},\uu,t) = \frac{1}{4\pi}\rho\rt + \sum_{i=1}^{3}P_i\rt u_i + \sum_{i,j=1}^{3}Q_{ij}\rt u_i u_j 
\end{equation}
with
\begin{align}
\rho(\vec{r},t) &= \INT{S_2}{}{\Omega}\varrho(\vec{r},\uu,t),\\
P_i(\vec{r},t) &= \frac{3}{4\pi}\INT{S_2}{}{\Omega}\varrho(\vec{r},\uu,t) u_i,\\
Q_{ij}(\vec{r},t) &= \frac{15}{8\pi}\INT{S_2}{}{\Omega}\varrho(\vec{r},\uu,t) \Big(u_i u_j - \frac{1}{3}\delta_{ij}\Big),
\end{align}
where $\vec{P}(\vec{r},t)$ is the local polarization and $Q(\vec{r},t)$ the local nematic tensor. By inserting this expansion into \cref{uniaxialddft} and converting the functional derivative with respect to $\varrho(\vec{r},\uu,t)$ into derivatives with respect to the fields $\rho\rt$, $\vec{P}\rt$, and $Q\rt$, one obtains a set of coupled dynamic equations for these fields. A detailed calculation following this procedure can be found in Ref.\ \cite{WittkowskiLB2011b}. (The orientational expansion method is not only used in DDFT, but also in other soft and active matter theories \cite{BickmannW2020,AchimWL2011,WittkowskiSC2017}.) Alternatively, one could directly derive the equations of motion for the order parameter fields from the projection operator formalism.

The DDFT for nonspherical particles can be generalized further in two ways. First, we have assumed that the particles have an axis of continuous rotational symmetry. This includes apolar and polar particles, which both can be described using an orientation vector $\uu$ (apolarity is reflected by a symmetry $\varrho(\uu) = \varrho(-\uu)$ \cite{teVrugtW2020b}). The more general case would be a particle with arbitrary shape, whose orientation, as mentioned above, is described by a rotation parametrized by the three Euler angles $\varphi$, $\vartheta$, and $\winkel$. For convenience, we summarize the Euler angles in a vector $\vec{\varpi}=(\varphi,\vartheta,\winkel)^{\mathrm{T}}$. Second, we can consider active particles, as done, e.g., by \citet{WensinkL2008} for active rods. These have, due to their self-propulsion, always an intrinsic orientation. Self-propulsion is an additional extension, since it leads to a change of the dynamical structure of DDFT (see \cref{active}). Both cases were considered by \citet{WittkowskiL2011}, whose result is discussed in \cref{active}. A theory involving hydrodynamic interactions, inertia, and particles of general shape was derived using projection operator methods by \citet{DuranGK2016}.  

\citet{BiervR2007,BiervR2008} studied platelike colloids in a phenomenological model. Self-diffusion in smectic and nematic liquid crystals is studied in Refs.\ \cite{BiervRDvdS2008,GreletLBvRvdS2008}. Non-Gaussian diffusion in such systems is observed in Ref.\ \cite{BiervRDvdS2008} as a result of temporary cages. Similar behavior is also found in simulations of equilibrium smectic phases of hard rods \cite{PattiEvRD2009,PattiEvRD2010} and relaxation of spherocylinders \cite{PattiBvRD2011,BelliPvRD2010}. 

The application of FMT (see \cref{fmt}) to hard spherocylinders in DDFT is discussed in Refs.\ \cite{HaertelL2010,HaertelBL2010}. \citet{UematsuY2012} studied polarization relaxation using a DDFT with orientational dynamics. In Ref.\ \cite{CugliandoloDLvW2015}, Dean's stochastic DDFT equation \eqref{dk} is extended to dipolar particles. \citet{DuranYGK2017} derived a general fluctuating DDFT for particles with arbitrary shape that also includes momentum and angular momentum density as relevant variables. The growth of hard-rod monolayers is analyzed in Ref.\ \cite{KlopotekHGDSSO2017}, including a consideration of nematic order. In Ref.\ \cite{GomesKPY2019}, the nucleation of hard squares is studied.

DDFTs for nonspherical particles can also be used to obtain PFC equations that include orientational dynamics \cite{Loewen2010,MenzelOL2014} or anisotropic particle interaction potentials \cite{ChoudharyLEL2011} (see \cref{phasefield}). Finally, the methods presented in this section are also relevant for magnetic particles \cite{LichtnerK2013,LichtnerAK2012} (see \cref{magnetic}).

\subsubsection{\label{magnetic}Magnetic particles}
In order to describe continuum systems with magnetic order, DDFT has also been extended to systems of particles with a (classical) spin $\vec{\sigma}_{\mathrm{s}}$. These theories have been used to study phase separation in fluids containing magnetic particles \cite{LichtnerK2013,LichtnerAK2012,LichtnerK2014}. The spin, which is another degree of freedom of the particles in addition to the position, is normalized and described by an orientation specified by the spherical coordinate angles $\vartheta$ and $\varphi$ or an orientation vector $\uu$. Thus, the description of magnetic particles in DDFT is completely analogous to the description of nonspherical particles discussed in \cref{nonspherical}. For the spin contribution to the interaction potential, one frequently uses the Heisenberg model 
\begin{equation}
U_2(\norm{\vec{r}-\vec{r}'},\uu,\uu') = \coupling(\norm{\vec{r}-\vec{r}'})\vec{\sigma}_{\mathrm{s}}\cdot\vec{\sigma}_{\mathrm{s}}',  
\end{equation}
where the sign of the coupling $\coupling$ determines whether ferromagnetic or antiferromagnetic ordering is preferred.

The stochastic dynamics of magnetic dipoles is considered in Ref.\ \cite{CugliandoloDLvW2015}. A DFT for ferrogels (magnetic colloids in a polymer matrix) was derived in Refs.\ \cite{CremerHML2017,GohWML2019}. Moreover, the DFT of ferrofluids is discussed in Refs.\ \cite{CerdaKH2008,KantorovichCH2008,IvanovK2004,PyanzinaKCIH2009,KantorovichIRTS2015,KantorovichIRTS2013}. Methods of this type provide a possible starting point for future work on applications of DDFT to magnetism.

\subsubsection{\label{active}Active particles}
DDFT can also be applied to study systems of active particles. Field theories for active matter can, in general, be classified into \textit{dry models}, which model active particles without explicitly considering the solvent, and \textit{wet models}, where hydrodynamics is also considered. General reviews on active matter can be found in Refs.\ \cite{GompperEtAl2020,WensinkLMHWZKM2013,Menzel2015,MarchettiJRLPRS2013,Klapp2016,Ramaswamy2010,Ramaswamy2017,BechingerdLLRVV2016,NeedlemanD2017,CavagnaG2014,CichosGMV2020,ShaebaniWWGR2020,CatesT2015}. A particular challenge in DDFT is that active particles are far-from-equilibrium systems whose dynamics cannot be assumed to be driven by the gradient of the equilibrium free energy. Possible solutions are to add to the standard DDFT for passive systems (with passive free energy) an active term or to derive an \ZT{effective} free energy that describes (a steady state of) the active system. A DDFT model for dry active particles of the former type was obtained for uniaxial particles by \citet{WensinkL2008} and for general shapes by \citet{WittkowskiL2011}. These models only hold for the case of weak activity. 

We start by discussing the DDFT for active particles with general shape by \citet{WittkowskiL2011}. The density is now a function $\varrho(\vec{r},\vec{\varpi},t)$ and the Smoluchowski equation reads
\begin{equation}
\begin{split}
\pdif{}{t}\SmoluchowskiDistribution(\{\vec{r}_k\},\{\vec{\varpi}_k\},t)
&=\sum_{i=1}^{N}
\begin{pmatrix}
\vec{\nabla}_{\vec{r}_i}\\
\Nabla_{\vec{\varpi}_i}
\end{pmatrix}
\cdot\Bigg(\!
\begin{pmatrix}
D_i^{\mathrm{TT}}(\vec{\varpi}_i) &  D_i^{\mathrm{TR}}(\vec{\varpi}_i)\\
D_i^{\mathrm{RT}}(\vec{\varpi}_i) &  D_i^{\mathrm{RR}}(\vec{\varpi}_i)
\end{pmatrix}
\\
&\quad\:\!\Bigg(\frac{1}{k_B T}
\begin{pmatrix}
\vec{\nabla}_{\vec{r}_i}\\
\Nabla_{\vec{\varpi}_i}
\end{pmatrix}
U(\{\vec{r}_k\},\{\vec{\varpi}_k\},t) - \frac{1}{k_B T}
\begin{pmatrix}
\vec{F}_{\mathrm{A},i}(\vec{r}_i,\vec{\varpi}_i,t)\\
\vec{T}_{\mathrm{A},i}(\vec{r}_i,\vec{\varpi}_i,t)
\end{pmatrix}
\\
&\quad\:\!+
\begin{pmatrix}
\vec{\nabla}_{\vec{r}_i}\\
\Nabla_{\vec{\varpi}_i}
\end{pmatrix}
\!\Bigg)\!\Bigg)\SmoluchowskiDistribution(\{\vec{r}_k\},\{\vec{\varpi}_k\},t)
\end{split}    
\end{equation}
with the angular gradient operator for particles with arbitrary shape 
\begin{equation}
\Nabla_{\vec{\varpi}}=
\begin{pmatrix}
- \cos(\varphi)\cot(\vartheta)\partial_{\varphi} - \sin(\varphi)\partial_{\vartheta} + \cos(\varphi)\csc(\vartheta)\partial_{\winkel}\\
-\sin(\varphi)\cot(\vartheta)\partial_{\varphi} + \cos(\varphi)\partial_{\vartheta} + \sin(\varphi)\csc(\vartheta)\partial_{\winkel}\\
\partial_{\varphi}
\end{pmatrix}
\end{equation}
and the derivative operators $\partial_\varphi=\partial/\partial\varphi$, $\partial_\vartheta=\partial/\partial\vartheta$, and $\partial_{\winkel}=\partial/\partial\winkel$. In the Smoluchowski equation,
we have used a very general diffusion matrix with elements $D_i^{\mathrm{TT}}(\vec{\varpi}_i)$, $D_i^{\mathrm{TR}}(\vec{\varpi}_i)$, $D_i^{\mathrm{RT}}(\vec{\varpi}_i)$, and $D_i^{\mathrm{RR}}(\vec{\varpi}_i)$ that involves couplings between translation and rotation, in addition to the purely translational and rotational diffusion, but no hydrodynamic interactions. The precise form of the elements depends on the particle shape. (Various shapes are discussed in Ref.\ \cite{VossW2018}, see Ref.\ \cite{VossJW2019} for the accompanying software.)
Furthermore, the Smoluchowski equation contains the active force $\vec{F}_{\mathrm{A},i}(\vec{r}_i,\vec{\varpi}_i,t) = R^{-1}(\vec{\varpi}_i)\vec{F}_{\mathrm{A},0,i}(\vec{r}_i,t)$ and active torque $\vec{T}_{\mathrm{A},i}(\vec{r}_i,\vec{\varpi}_i,t) = R^{-1}(\vec{\varpi}_i)\vec{T}_{\mathrm{A},0,i}(\vec{r}_i,t)$ with the rotation matrix $R(\vec{\varpi})$, the propulsion force $\vec{F}_{\mathrm{A},0,i}(\vec{r}_i,t)$, and the propulsion torque $\vec{T}_{\mathrm{A},0,i}(\vec{r}_i,t)$, where the latter two can be space- and time-dependent.
After integrating over positional and orientational degrees of freedom of all particles except for the first one and performing an adiabatic approximation, one arrives at the generalized DDFT equation 
\begin{equation}
\begin{split}
\pdif{}{t}\varrho(\vec{r},\vec{\varpi},t) &= \frac{1}{k_B T}
\begin{pmatrix}
\vec{\nabla}_{\vec{r}}\\
\Nabla_{\vec{\varpi}}
\end{pmatrix}
\cdot\Bigg(\!
\begin{pmatrix}
D^{\mathrm{TT}}(\vec{\varpi}) &  D^{\mathrm{TR}}(\vec{\varpi})\\
D^{\mathrm{RT}}(\vec{\varpi}) &  D^{\mathrm{RR}}(\vec{\varpi})
\end{pmatrix}\\
&\quad\,
\Bigg(\varrho(\vec{r},\vec{\varpi},t)\Bigg(\!
\begin{pmatrix}
\vec{\nabla}_{\vec{r}}\\
\Nabla_{\vec{\varpi}}
\end{pmatrix}
\frac{\delta F[\rho]}{\delta \varrho(\vec{r},\vec{\varpi},t)} - 
\begin{pmatrix}
\vec{F}_{\mathrm{A}}(\vec{r},\vec{\varpi},t)\\
\vec{T}_{\mathrm{A}}(\vec{r},\vec{\varpi},t)
\end{pmatrix}
\!\Bigg)\!\Bigg)\!\Bigg),
\end{split}
\label{arbitraryshapeddft}%
\end{equation}
where we have omitted the index 1 for the quantities $\vec{r}$, $\vec{\varpi}$, $D^{\mathrm{TT}}(\vec{\varpi})$, $D^{\mathrm{TR}}(\vec{\varpi})$, $D^{\mathrm{RT}}(\vec{\varpi})$, $D^{\mathrm{RR}}(\vec{\varpi})$, $\vec{F}_{\mathrm{A}}$, and $\vec{T}_{\mathrm{A}}$.
(In principle, the elements $D^{\mathrm{TT}}(\vec{\varpi})$, $D^{\mathrm{TR}}(\vec{\varpi})$, $D^{\mathrm{RT}}(\vec{\varpi})$, and $D^{\mathrm{RR}}(\vec{\varpi})$ of the short-time diffusion tensor could also be space-dependent \cite{WittkowskiLB2011}.)

\citet{MenzelSHL2016} derived a DDFT for hydrodynamically interacting microswimmers. Here, the individual swimmers are modeled as spherical particles that exert a force on the surrounding fluid at two force centers located along an orientation vector $\uu$. This leads to self-propulsion. Assuming low Reynolds number, the fluid flow velocity $\vec{v}\rt$ is governed by Stokes' equation
\begin{equation}
-\ShearViscosity\vec{\nabla}^2\vec{v}\rt + \vec{\nabla}p\rt = \sum_{i=1}^{N}\vec{f}_i(\vec{r}_i,\uu_i,t)
\end{equation}
with the shear viscosity $\ShearViscosity$, pressure $p\rt$, and force density $\vec{f}_i(\vec{r}_i,\uu_i,t)$ exerted by the $i$-th swimmer. Since the swimmers influence the fluid flow by the force they apply, but at the same time are dragged along by the fluid, they self-propel and interact hydrodynamically.

The derivation of the DDFT then proceeds in a similar way as in the treatment of nonspherical particles discussed in \cref{nonspherical}, since the one-body density for active particles also needs to take into account the orientation. With the translational velocity $\vec{v}_i$ and angular velocity $\vec{\omega}_i$ of the $i$-th particle resulting from the flow field, the continuity equation for the $N$-body density $\SmoluchowskiDistribution(\{\vec{r}_k\},\{\uu_k\},t)$ is given by
\begin{equation}
\pdif{}{t}\SmoluchowskiDistribution(\{\vec{r}_k\},\{\uu_k\},t) = - \SUM{i=1}{N}\vec{\nabla}_{\vec{r}_i}\cdot(\vec{v}_i\SmoluchowskiDistribution(\{\vec{r}_k\},\{\uu_k\},t)) 
+ \Nabla_{\uu_{i}}\cdot(\vec{\omega}_i\SmoluchowskiDistribution(\{\vec{r}_k\},\{\uu_k\},t)).
\end{equation}
The hydrodynamics is contained in the expressions for the velocities $\vec{v}_i$ and $\vec{\omega}_i$, which can be expressed in terms of the hydrodynamic force and torque, respectively. By integrating over the coordinates of $N-1$ particles and applying the usual adiabatic approximation, one can then derive a DDFT for the microswimmer system. (The exact equations are rather lengthy and can be found in Ref.\ \cite{MenzelSHL2016}.) This approach has also been extended towards circle swimmers \cite{HoellLM2017}, polar ordering \cite{HoellLM2018}, and multi-species microswimmer suspensions \cite{HoellLM2019}. It also holds, being based on an equilibrium approximation (as standard DDFT), only for weak activity. 

Another DDFT for active particles was obtained, based on effective free energies for active steady states rather than passive free energies, by \citet{EnculescuS2011}. \citet{WittmannB2016} constructed a DDFT-based effective equilibrium description that predicts certain phase transitions of active Brownian particles (ABPs). It is based on the work in Refs.\ \cite{PototskyS2012,FarageKB2015} and reads
\begin{equation}
\pdif{}{t}\rho\rt=\frac{D}{k_B T}\Nabla\cdot\bigg(\rho\rt\Nabla\frac{\delta F_{\mathrm{eff}}[\rho]}{\delta\rho\rt}\bigg)
\label{effective}
\end{equation}
with the translational diffusion coefficient $D$. Note that this type of active DDFT differs in an interesting way from the type for which \cref{arbitraryshapeddft} is a paradigmatic example. Equation \eqref{arbitraryshapeddft} provides a dynamic equation for $\varrho(\vec{r},\vec{\varpi},t)$ rather than $\rho\rt$, i.e., the orientational degrees of freedom are considered explicitly, and it incorporates the activity using an additional term in the final DDFT equation. In contrast, \cref{effective} looks exactly like the standard DDFT equation \eqref{trddft}. The activity is incorporated in \cref{effective} by the fact that an \textit{effective} free energy $F_{\mathrm{eff}}$ is used. It is obtained by replacing the actual external potential by an effective external potential that consists of the actual external potential $U_1$ and a contribution that results from integrating an approximate expression for the steady-state active force. Effective equilibrium states are also considered in Refs.\ \cite{WittmannMSSBM2017,WittmannMMB2017}. A DDFT for active particles driven by colored noise was derived by \citet{WittmannMMB2017}. 

\citet{PototskyS2012} derived a DDFT for active particles by mapping them onto a passive system with a modified potential. \citet{MenzelOL2014} obtained an active PFC model from an active DDFT (see \cref{phasefield}). \citet{SharmaB2017} have developed a DDFT for ABPs with spatially inhomogeneous activity (building up on work in Ref.\ \cite{SharmaB2016}). \citet{PaliwalRvRD2018} constructed a chemical-potential-like function for ABPs in a steady state to obtain a DDFT-like model. Activity-driven spins were modeled by \citet{ZakineFvW2018}. \citet{KrinningerS2019} derived a PFT for ABPs (see \cref{nonspft}). \citet{AroldS2019} proposed a DDFT for active particles with inertia that includes the polarization as a dynamical variable. A reaction DDFT (see \cref{chemical}) was used by \citet{MonchoD2020} to model active switching.

\subsection{\label{moment}Momentum}
A further microscopic degree of freedom is the momentum, which is of importance if the particles are not overdamped. Just like the description of particles with orientational degrees of freedom requires a density $\varrho(\vec{r},\uu,t)$ rather than $\rho\rt$ in the microscopic description, momentum degrees of freedom require an extension to a phase-space distribution function $\PSDistributionRVT(\vec{r},\vec{v},t)$ depending on the velocity $\vec{v}$ of the particles. Moreover, as in the case of orientation, an expansion of this extended density is possible that leads to an infinite hierarchy of order parameter fields. At lowest orders, this gives rise to the one-body density $\rho\rt$ and the momentum density $\vec{g}\rt$.

\subsubsection{\label{inertia}Inertia}
The effects of inertia in the DDFT of (one-dimensional) colloidal fluids were studied by \citet{MarconiT2006}. The analysis is based on the phase-space distribution function $\PSDistributionRVT(\vec{r},\vec{v},t)$, whose dynamics follows the (1D) Fokker-Planck equation
\begin{equation}
\begin{split}
&\pdif{}{t}\PSDistributionRVT(x,v,t) + \bigg(v\pdif{}{x} + \frac{F}{m}\pdif{}{v}\bigg)\PSDistributionRVT(x,v,t) =  \gamma\bigg(\pdif{}{v}v+\frac{T}{m}\pdif{^2}{v^2}\bigg)\PSDistributionRVT(x,v,t)+C(x,v,t,\PSDistributionRVT_2),
\end{split}
\end{equation}
where $C$ is a collision operator depending on the two-body distribution $\PSDistributionRVT_2$ that can be approximated by the assumption of molecular chaos. The dynamic equation is then rewritten in terms of nondimensionalized parameters. This allows for a time-scale separation in terms of a dimensionless parameter $\gamma_{\mathrm{nd}}$ (proportional to $\gamma$) measuring the relaxation (see also \cref{multi}). At lowest nontrivial order, one recovers the usual overdamped DDFT. These results were generalized to arbitrarily many dimensions in Ref.\ \cite{MarconiM2007}. The described method for incorporating inertia into DDFT can be used to systematically obtain corrections to the overdamped case \cite{MarconiTCM2008}.

A DDFT for atomic fluids, where, unlike in colloidal fluids, inertia always plays a role, has been derived by \citet{Archer2006} (general fluids were also treated by \citet{ChanF2005}, see \cref{existence}). Starting from Newton's equation of motion and performing an ensemble average, one finds for the one-body density $\rho\rt$ the exact equation of motion
\begin{equation}
\begin{split}
\pdif{^2}{t^2}\rho\rt + B\rt &= \frac{k_B T}{m}\vec{\nabla}^2\rho\rt
+\frac{1}{m}\vec{\nabla}\cdot\INT{}{}{^3r'} \braket{\hat{\rho}\rt\hat{\rho}(\vec{r}',t)} \vec{\nabla}U_2(\vec{r}-\vec{r}') \\
&\quad\:\! +\frac{1}{m}\vec{\nabla}\cdot(\rho\rt\vec{\nabla}U_1\rt)
\label{inertiaexact}
\end{split}\raisetag{2em}%    
\end{equation}
with the term 
\begin{equation}
B\rt=\frac{1}{(2\pi)^3}\INT{}{}{^3k}k^2\bigg\langle\sum_{i=1}^{N}\alpha_i(t)e^{\ii \vec{k}\cdot \vec{r}_i}\bigg\rangle e^{-\ii\vec{k}\cdot\vec{r}},   
\end{equation}
where $m$ is the mass of the particles, $\vec{r}_i$ is the position of the $i$-th particle, and $\alpha_i(t)$ is the deviation of $(\dot{\vec{r}}_i(t)\cdot\vec{k})^2/k^2$ with the wavevector $\vec{k}$ from its equilibrium mean value. Equation \eqref{inertiaexact} is closed with two approximations: First, the two-particle correlations are, as in standard DDFT, replaced by their equilibrium values, which are related to the functional derivative of the equilibrium free energy. Second, the term $B\rt$, which vanishes in equilibrium, is assumed to have the form $\nu \dot{\rho}\rt$ with an undetermined frequency $\nu$. This gives the result
\begin{equation}
\pdif{^2}{t^2}\rho\rt + \nu \pdif{}{t}\rho\rt = \frac{1}{m}\vec{\nabla}\cdot\bigg(\rho\rt\vec{\nabla}\frac{\delta F[\rho]}{\delta \rho\rt}\bigg). 
\label{atomic}%
\end{equation}

DDFTs with inertia can also be shown to possess a H-theorem \cite{Chavanis2011}. They can be extended towards hydrodynamic interactions \cite{GoddardNSPK2012,GoddardNSPK2013,GoddardNSYK2013,GoddardNK2013,DuranGK2016} and orientational degrees of freedom \cite{DuranGK2016}. Moreover, a DDFT with inertia has been developed for active particles \cite{AroldS2019}. One can also incorporate inertia in PFT \cite{Schmidt2018} (see \cref{newtonianpft}). Applications of inertial DDFT include the study of sound waves \cite{Archer2006} (see \cref{sound}), mode coupling theory \cite{Archer2006} (see \cref{mct}), glassy behavior \cite{BerryG2011}, granular systems \cite{MarconiTCM2008,MarconiTC2007}, freezing \cite{BaskaranBL2014}, and dielectricity \cite{GaoCX2017}.

\subsubsection{\label{momentum}Momentum density}
An important extension of DDFT are more general theories that describe not only the density $\rho\rt$, but also the momentum density $\vec{g}\rt$. These theories typically relate $\rho\rt$ to the flux of $\vec{g}\rt$ by the continuity equation
\begin{equation}
\pdif{}{t}\rho\rt + \frac{1}{m}\vec{\nabla}\cdot \vec{g}\rt = 0.
\end{equation}
Furthermore, they give a separate equation governing the momentum density, which has the form of a (generalization of the) Navier-Stokes equation \cite{BurghardtB2006}
\begin{equation}
\pdif{}{t}\vec{g}\rt= -\rho\rt\vec{\nabla}\frac{\delta F[\rho]}{\delta \rho\rt} + \vec{\DissipativeContribution}[\vec{g}]
\end{equation}
with a dissipative contribution $\vec{\DissipativeContribution}$. These theories then contain standard DDFT as a limiting case and can thus be seen as generalizations of DDFT towards hydrodynamics. (Historically, however, they also formed a basis for the derivation of overdamped DDFT \cite{Munakata1989}, see \cref{phenomenological}.) Eliminating the velocity or momentum density field from the coupled equations gives a second-order equation of motion for the one-body density \cite{Archer2009,KikkinidesM2015}.

Frequently, such theories are derived from kinetic theories that model the phase-space distribution $\PSDistributionRVT(\vec{r},\vec{v},t)$ rather than just the one-particle density $\rho\rt$ \cite{MarconiM2009,MarconiM2010,MarconiM2011b,BurghardtB2006,HughesB2012,Lutsko2010,Marconi2011,MarconiM2014}. These results contain DDFT as the high-friction limit. Such methods allow to relate the well-known continuum theories of fluid mechanics to the microscopic particle dynamics, which is important, e.g., for modeling flow in microchannels \cite{MickelJB2011}. Moreover, they can be seen as arising from a truncated expansion of the full phase-space dynamics in moments of the momentum (that can also be stopped at higher orders, in this case the kinetic energy density tensor, which contains the kinetic pressure tensor, is also among the dynamical variables) \cite{HughesB2012}. Theories including the momentum density can also be derived using the projection operator formalism by projecting the microscopic dynamics onto mass and momentum densities as relevant variables. This is done for simple fluids by \citet{CamargodlTDZEDBC2018}. A further exploration of this approach can be found in Refs.\ \cite{CamargodlTDBCE2019,DuqueZumajoCdlTCE2019,DuqueZumajodlTCE2019}. 

One can further generalize DDFT with momentum density, e.g., by using additional order parameter fields. Dynamic equations that include, in addition to mass and momentum density, also the energy density, are derived in Refs.\ \cite{HuetterB2009,MarconiM2009,Marconi2011,ZhaoW2011,GoddardHO2020}. \citet{MajaniemiG2007} coupled mass and momentum density to a dynamic equation for a displacement field. Mixtures are considered in Refs.\ \cite{Marconi2011,GoddardNK2013,OkamotoO2016}. One can also derive the coupled equations for mass and momentum density from quantum mechanics \cite{BurghardtB2006,BousquetHMB2011,HughesBBRB2012}. \citet{DiawM2017} applied DDFT to quantum hydrodynamics (see \cref{qm}).

\citet{Archer2009} derived a theory for mass density and flow field that contains as limiting cases both overdamped DDFT and the Euler equation. \citet{RotenbergPF2010} coupled a DDFT-like equation for charged systems to the Navier-Stokes equation. A coupling to the Navier-Stokes equation is also possible for the PNP equations \cite{WerkhovenESvR2018,WerkhovenSvR2019} (see \cref{electro}). Another theory combining mass and momentum density is the \ZT{kinetic density functional theory} derived in Ref.\ \cite{BaskaranBL2014}, which is used to model the liquid-solid transition. The momentum density is also included in DDFT-based theories of plasma dynamics \cite{DiawM2015,DiawM2016} (see \cref{plasma}). \citet{PraetoriusV2015} derived, starting from DDFT, a PFC model coupled to a dynamic equation for the flow field. A PFC equation coupled with a Navier-Stokes-Cahn-Hilliard equation is derived in Ref.\ \cite{AlandLV2011} (see also Ref.\ \cite{AlandLV2012}). \citet{AroldS2019} included the momentum density in an active matter model. The momentum density is considered in the context of stochastic theories in Refs.\ \cite{Lutsko2012,NakamuraY2009,PerezMadridRR2002,TothGT2013,ShangVC2011}. Numerical methods for equations of this type can be found in Ref.\ \cite{BaskaranGL2016}. 

\subsubsection{\label{kinetic}Kinetic theory}
For the description of inertia and momentum transport, it is important to include not only the positions, but also the momenta or velocities of the particles in a dynamical theory. Hence, a natural extension of DDFT is to consider the phase-space distribution $\PSDistributionRVT(\vec{r},\vec{v},t)$ rather than the spatial distribution $\rho\rt$. A first theory of this type, which is only applicable to dilute gases, has already been developed by \citet{Boltzmann1872}. More recently and more generally, a dynamic theory for $\PSDistributionRVT(\vec{r},\vec{v},t)$ has been derived using the projection operator formalism by \citet{AneroE2007}. \citet{PerezMadridRR2002} have obtained a Fokker-Planck equation for the phase-space distribution functional. A review of kinetic theory can be found in Ref.\ \cite{MarconiM2014}.

Starting from a phase-space theory, it is possible to derive DDFT as an overdamped limit, based on the assumption that the velocity very quickly reaches its steady state \cite{Lutsko2010}. Phase-space methods also give inertial corrections to DDFT \cite{MarconiM2007,TarazonaM2008}. The relation to overdamped DDFT can be obtained using a multiple-time-scale analysis \cite{MarconiTCM2008,MarconiTC2007} (see \cref{multi}). A particularly important application of kinetic theory is the microscopic description of fluid dynamics \cite{MarconiM2014}. Kinetic theory also allows to obtain a \ZT{kinetic density functional theory} for the description of the solid-liquid phase transition \cite{BaskaranBL2014}. Moreover, a phase-space description connects DDFT to the Boltzmann theory \cite{MarconiM2009}. From the phase-space description, one can derive the equations of hydrodynamics \cite{MarconiM2010}. This is also possible for multicomponent systems \cite{MarconiM2011,MarconiM2011b,Marconi2011,MarconiM2014,MonteferranteMM2014,MarconiM2012} and granular media \cite{MarconiTC2007,GoddardHO2020}.

\subsection{\label{noniso}Nonisothermal systems}
While standard DDFT assumes the system to have a fixed temperature, extensions to nonisothermal systems have also been derived. This topic has a clear connection to the case of momentum discussed in \cref{moment}, since a natural dynamical variable for nonisothermal systems is the energy density. In many hydrodynamic theories, it is used as an order parameter field in addition to mass and momentum density, which allows to describe nonisothermal systems. The reason why the energy density is important is that energy -- like mass and momentum -- is a conserved quantity, such that the energy density can often be assumed to be a slow variable. As can be shown using projection operator methods \cite{Grabert1982} (see \cref{po}), this implies that the energy density is important for the dynamics of the system.

\subsubsection{Fixed temperature gradients}
Treatments of nonisothermal systems can either consider externally fixed temperature gradients (this is the simpler case) or couple DDFT to an additional dynamic equation for the local temperature. \citet{LopezM2007} derived a DDFT-like transport equation for particles in a nonuniform heat bath with temperature distribution $T(x)$ (in 1D), which reads
\begin{equation}
\pdif{}{t}\rho(x,t)=\vec{\nabla}\cdot\bigg(\vec{\nabla}(D(x)\rho(x,t)) + \frac{1}{\gamma}\rho(x,t)\vec{\nabla}\bigg(\frac{\delta F_{\mathrm{exc}}}{\delta \rho(x,t)} + U_1(x)\bigg)\!\bigg)
\end{equation}
with $D(x)=k_B T(x)/\gamma$. The derivation is sketched in \cref{multi}. This result is also given by \citet{MarconiTCM2008}.

\subsubsection{\label{nonisothermal}Energy density}
A description in terms of a single order parameter field (such as the one-body density in DDFT) implies that all other degrees of freedom (such as the local temperature) relax very quickly. Hence, in order to extend DDFT to nonisothermal systems, one requires an additional order parameter field. Natural candidates are the local energy density \cite{WittkowskiLB2012,AneroET2013} and the local entropy density \cite{Schmidt2011}.  

Microscopic derivations of DDFTs for nonisothermal systems can be performed in the projection operator formalism (see \cref{po}) by projecting onto the energy density as a second relevant variable in addition to the number density. This is the idea behind EDDFT \cite{WittkowskiLB2012,WittkowskiLB2013} and FTD \cite{AneroET2013}, both of which are extensions of DDFT towards nonisothermal systems. They are discussed in \cref{eddft,ftd}, respectively. Moreover, an extended theory involving mass, momentum, and energy density as well as a dynamic equation for the two-body correlation was derived by \citet{ZhaoW2011} (see also Ref.\ \cite{ZhaoW2011b}). The dynamics of the energy density is also considered in Refs.\ \cite{HuetterB2009,MarconiM2009,Marconi2011}. Coupled equations for heat and mass transport (the latter formally identical to DDFT) were obtained by \citet{KocherP2019} from nonequilibrium thermodynamics. The temperature of granular fluids was studied by \citet{GoddardHO2020}.

\subsubsection{\label{entropydensity}Entropy density}
The entropy density was considered by \citet{Schmidt2011} based on a phenomenological generalization of an equilibrium formalism: An energy functional $E[\rho,\mathfrak{s}]$ is defined that depends not only on the number density $\rho(\vec{r})$ but also on the entropy density $\mathfrak{s}(\vec{r})$. It satisfies the equilibrium Euler-Lagrange equations
\begin{align}
\frac{\delta E[\rho,\mathfrak{s}]}{\delta \rho(\vec{r})}\bigg|_{\rho_{\mathrm{eq}},\mathfrak{s}_{\mathrm{eq}}} 
&= \mu - U_1(\vec{r}),\\
\frac{\delta E[\rho,\mathfrak{s}]}{\delta \mathfrak{s}(\vec{r})}\bigg|_{\rho_{\mathrm{eq}},\mathfrak{s}_{\mathrm{eq}}} 
&= T
\end{align}
with the equilibrium number density $\rho_{\mathrm{eq}}(\vec{r})$, equilibrium entropy density $\mathfrak{s}_{\mathrm{eq}}(\vec{r})$, chemical potential $\mu$, and temperature $T$. Assuming the Gibbs-Duhem relation (solved for $\dif \mathfrak{s}$)
\begin{equation}
\dif \mathfrak{s} = \frac{1}{T}\dif e - \frac{\mu}{T}\dif\rho
\label{gdh}%
\end{equation}
to be valid, one can calculate the rate of change of the entropy as
\begin{equation}
\pdif{}{t}\mathfrak{s}\rt=\frac{1}{T\rt}\pdif{}{t}e\rt- \frac{\mu\rt}{T\rt}\pdif{}{t}\rho\rt,
\label{dotentropy}
\end{equation}
where $e$ is the energy density. One then identifies temperature $T$ and chemical potential $\mu$ as the driving forces for changes of energy density $e$ and number density $\rho$, respectively. Inserting these driving forces into a general continuity equation and expressing temperature and chemical potential as functional derivatives of $E$ with respect to entropy and number density, respectively, then gives a dynamic equation for the entropy density. 

A similar approach is used by \citet{WittkowskiLB2013} based on a microscopic derivation in EDDFT. It is based on the framework of linear irreversible thermodynamics discussed in \cref{net}. Here, \cref{gdh} is written for a general set of conserved and nonconserved variables $\{a_i\}$ rather than just for $\rho\rt$. This leads to a generalization of \cref{dotentropy} given by
\begin{equation}
\pdif{}{t}\mathfrak{s}\rt=\frac{1}{T\rt}\pdif{}{t}e\rt-\sum_{i=1}^{\kappa}\frac{1}{T\rt}\frac{\delta F[\{a_i\}]}{\delta a_i\rt}\pdif{}{t} a_i\rt.  
\end{equation}
The projection operator formalism allows to obtain the transport equations for these variables in terms of their thermodynamic conjugates. Since the entropy is a nonconserved variable, its dynamics can be expressed in the general form
\begin{equation}
\pdif{}{t}\mathfrak{s}\rt=\Nabla\cdot\vec{J}_{\mathfrak{s}}\rt - \EntropyProduction\rt
\end{equation}
with the entropy current $\vec{J}_{\mathfrak{s}}\rt$ and the entropy production
\begin{equation}
\EntropyProduction\rt=\frac{2\mathfrak{r}\rt}{T\rt},
\end{equation}
where $\mathfrak{r}\rt$ is the dissipation function (see \cref{net}). Note that all these approaches are based on the local validity of the laws of thermodynamics, which do not hold for systems far from equilibrium.

Finally, \citet{HuetterB2009} employed (in nonequilibrium thermodynamics) an entropy density $\mathfrak{s}\rt$ that is a functional of mass and energy density and obtained a dynamic equation for $\mathfrak{s}$. Entropy production has also been discussed in the context of PFC models \cite{ChengKLE2012}.

\subsection{\label{particleconserving}Particle-conserving dynamics}
We now turn to a different type of extension, which does not aim to incorporate additional order parameter fields. Instead, it corrects an inaccuracy arising from one of the approximations involved in DDFT (see \cref{limit}): Although the canonical and the grand-canonical ensemble are equivalent in the thermodynamic limit, inaccuracies can arise when treating systems with a finite -- and, in particular, small -- fixed number of particles in the grand-canonical ensemble \cite{GonzalezWRVE1997,WhiteGRV2000}. While DFT is a grand-canonical theory, DDFT is not completely consistent in this regard. In principle, it is a canonical theory that is based on the free energy $F$ and conserves the particle number as it has the form of a continuity equation. However, in practice it uses thermodynamic forces based on a grand-canonical free energy. This can lead to differences between DFT and DDFT, e.g., in the prediction of nucleation pathways \cite{LichtnerAK2012}. Agreement between DFT and DDFT in the equilibrium case is only guaranteed in the thermodynamic limit \cite{MalijevskyA2013}. A comparison of canonical and grand-canonical results is made in Ref.\ \cite{YuJEAR2017} (see also Ref.\ \cite{FarokhiradRMMAER2019}) for particles confined to a cylinder, where in the canonical case a Lagrange multiplier is used. 

To overcome problems of the grand-canonical ensemble, the method of particle-conserving dynamics (PCD) \cite{delasHerasBFS2016} uses a canonical framework for DDFT. We start by discussing the equilibrium case. While the canonical free energy functional is not explicitly available, one can convert between the grand-canonical partition function $\Xi$ and the canonical partition function $\KanonischeZustandssumme_N$ for a system with $N$ particles using \cite{delasHerasS2014}
\begin{equation}
\Xi = \sum_{N=0}^{\infty}e^{\beta\mu N}\KanonischeZustandssumme_N.
\end{equation}
Moreover, the grand-canonical density $\rho_\mu(\vec{r})$ corresponding to the chemical potential $\mu$ is related to the canonical density $\rho_N(\vec{r})$ for a system with $N$ particles by
\begin{equation}
\rho_\mu(\vec{r})= \sum_{N=0}^{\infty}p_N(\mu)\rho_N(\vec{r}),  
\end{equation}
where 
\begin{equation}
p_N(\mu) = \frac{1}{\Xi}e^{\beta\mu N}\KanonischeZustandssumme_{N} 
\end{equation}
is the probability of finding $N$ particles for a given chemical potential $\mu$. Starting from these relations and the grand-canonical Euler-Lagrange equation, an iterative scheme can be developed that gives the external potential $U_{\mathrm{pcd}}(\vec{r})$ creating a density $\rho(\vec{r})$ in \textit{canonical} equilibrium. This leads to the canonical intrinsic free energy 
\begin{equation}
F_N[\rho]= - k_B T \ln(\KanonischeZustandssumme_N) - \INT{}{}{^3r}\rho(\vec{r}) U_{\mathrm{pcd}}(\vec{r}).
\end{equation}
The dynamic equation of the canonical density $\rho_N$ is given by
\begin{equation}
\pdif{}{t}\rho_N\rt = D\Nabla\cdot\bigg(\Nabla\rho_N\rt - \frac{1}{k_B T}\vec{f}_N\rt - \frac{1}{k_B T}\rho_N\rt(\vec{X}\rt - \Nabla U_1\rt)\bigg),
\end{equation}
where $\vec{f}_N$ is the internal force density arising from interactions, which in DDFT is approximated by the adiabatic force $\vec{f}^{\mathrm{ad}}_N$, and $\vec{X}$ is a nonconservative force acting on the particles. Obtaining the dynamics within the DDFT approximation then requires to find an expression for $\vec{f}^{\mathrm{ad}}_N$. Various options are discussed in Ref.\ \cite{delasHerasBFS2016}. One of them is 
\begin{equation}
\vec{f}^{\mathrm{ad}}_N\rt =- \rho_N^{\mathrm{ad}}\rt\vec{\nabla}\frac{\delta F_N^{\mathrm{exc}}[\rho]}{\delta \rho\rt}\bigg|_{\rho\rt = \rho_N^{\mathrm{ad}}\rt}
\end{equation}
with the one-body density $\rho_N^{\mathrm{ad}}\rt=\rho_N\rt$ of the adiabatic reference system and the canonical excess free energy $F_N^{\mathrm{exc}}$. 

Extensions of PCD have also been derived. In Ref.\ \cite{SchindlerWB2019}, PCD was generalized to binary mixtures. Here, it was shown that PCD does not conserve particle order for hard rods in 1D. To solve this problem, the theory of order-preserving dynamics (OPD) was developed based on PCD in Ref.\ \cite{WittmannLB2020} (see \cref{glass}).

\subsection{\label{expolymer}Extensions of DDFT for polymer melts}
Extensions have also been developed for the polymer DDFT discussed in \cref{pd}. We present these extensions here in a separate section, since, historically, this form of DDFT has had a separate development.

A useful approximation is provided by the method of \ZT{external potential dynamics} (EPD), which was introduced by \citet{MauritsF1997}. As discussed in \cref{pd}, the dynamic equations governing polymer dynamics are typically nonlocal. An example is the Rouse dynamics, which is given by \cref{rouse}. In EPD, it is possible to rewrite this nonlocal equation in a local form while keeping the nonlocal coupling, provided that a certain approximation can be made. What is exploited here is the existence of a bijective relation between density fields $\rho_i$ and external potentials $U_i$. The approximation is that, for the two-body correlator $P_{ij}$ in \cref{rouse}, we can write 
\begin{equation}
\Nabla_{\vec{r}} P_{ij}(\vec{r},\vec{r}',t)= - \Nabla_{\vec{r}'}P_{ij}(\vec{r},\vec{r}',t).
\end{equation}
This allows to write 
\begin{equation}
\Nabla_{\vec{r}}\cdot\INT{}{}{^3r'}P_{ij}(\vec{r},\vec{r}',t)\Nabla_{\vec{r}'}\mu_j(\vec{r}',t) = \INT{}{}{^3r'}P_{ij}(\vec{r},\vec{r}',t)\Nabla_{\vec{r}'}^2\mu_j(\vec{r}',t).
\label{gauss}
\end{equation}
The relation 
\begin{equation}
k_B T\frac{\delta \rho_i(\vec{r},t)}{\delta U_j(\vec{r}',t)} = - P_{ij}(\vec{r},\vec{r}',t)    
\end{equation}
gives by applying the chain rule
\begin{equation}
\begin{split}
\pdif{}{t}\rho_i\rt &=\sum_{j=1}^{\NumberOfParticleTypes}\INT{}{}{r'}\frac{\delta \rho_i(\vec{r},t)}{\delta U_j(\vec{r}',t)} \pdif{}{t}U_j(\vec{r}',t)\\ 
&=-\frac{1}{k_B T}\sum_{j=1}^{\NumberOfParticleTypes}\INT{}{}{r'}P_{ij}(\vec{r},\vec{r}',t) \pdif{}{t}U_j(\vec{r}',t).
\label{chain}
\end{split}
\end{equation}
Combining \cref{gauss,chain,rouse} leads to 
\begin{equation}
\pdif{}{t}U_i\rt = -\frac{D_{\mathrm{ro}}}{k_B T}\Nabla^2\mu_i\rt,
\end{equation}
such that the nonlocal equation \eqref{rouse} has been replaced by a local equation. This allows for an efficient numerical solution \cite{MauritsF1997,HeS2006,HeS2008,HeuserSS2017,QiZY2010}. A more exact scheme, along with a comparison to BD simulations, can be found in Ref.\ \cite{QiS2017}.

An extension towards compressible copolymer melts was developed in Ref.\ \cite{MauritsvVF1997}. Hydrodynamic effects were considered by \citet{MauritsZSvVF1998}, who added a streaming term to \cref{loco}. This gives
\begin{equation}
\pdif{}{t}\rho_i\rt = \frac{D_{\mathrm{lo}}}{k_B T}\Nabla\cdot(\rho_i\rt\Nabla\mu_i\rt) - \Nabla\cdot(\rho_i\rt\vec{v}\rt),
\end{equation}
where the velocity field $\vec{v}$ can be approximated using Darcy's law. Moreover, polymer DDFT was extended to include viscoelastic effects in Ref.\ \cite{MauritsZF1998b}. 

\subsection{\label{qm}Quantum mechanics}
As discussed in \cref{qtddft}, density functional methods can also be applied to quantum systems in the form of TDDFT. This is an area on its own, which is beyond the scope of this review. Nevertheless, there are certain theories that relate quantum and classical DDFT methods. Examples of such methods are presented here. A further method of this type is quantum PFT, which is presented in \cref{quantumpft}.

One way to obtain a relation between classical and quantum many-particle systems is the use of Wigner functions \cite{Wigner1932}. These allow for an exact reformulation of quantum mechanics as a dynamical theory on classical phase space, where the Wigner function $\mathcal{W}(x,p)$, which is related to the statistical operator $\hat{\rho}_{\mathrm{s}}$ by the transformation (in 1D)
\begin{equation}
\mathcal{W}(x,p) = \frac{1}{2\pi\hbar}\INT{}{}{x'}\bigg\langle{x-\frac{1}{2}x'|\hat{\rho}_{\mathrm{s}}|x+\frac{1}{2}x'\bigg\rangle e^{\ii\frac{x' p}{\hbar}}}
\label{Wigner}%
\end{equation}
with the reduced Planck constant $\hbar$, replaces the classical distribution function. Similar transformations exist for orientational degrees of freedom \cite{teVrugtW2019c}. From the quantum Liouville equation, an equation of motion for the Wigner function can be derived that is governed by a generalized Poisson bracket. 

For the description of mixed quantum-classical dynamics, one performs the transformation \eqref{Wigner} only for a subset of the degrees of freedom of the complete system. The result is a partially Wigner transformed equation of motion. Introducing a classical limit by linearizing the Poisson bracket operator in $\hbar$ (formally justified by the small mass ratio between quantum and classical parts), one obtains a hybrid dynamic equation describing classical systems coupled to quantum systems that contains the classical Liouville equation as a limiting case \cite{KapralC1999,NielsenKC2001,BurghardtB2006}. Further background information can be found in Refs.\ \cite{BurghardtP2004,BurghardtMPCB2004,HughesPPB2007}.

This method can be used to obtain a mixed quantum-classical (T)DDFT that includes the momentum density (and the polarization field, if orientational relaxation is of interest) as dynamical variables \cite{BurghardtB2006}. The kinetic pressure can, in certain cases, be written as the functional derivative of an entropy functional. On this basis, a free energy functional can be defined that then governs the time evolution of density and current. This relates these theories to a DDFT with an equation for the current \cite{BousquetHMB2011}. They can be applied, e.g., to nonpolar solvation dynamics \cite{HughesBBRB2012}.

Moreover, classical DDFT has been applied to quantum hydrodynamics \cite{DiawM2017}. The general transport equations for a dense electron plasma are closed adiabatically by assuming for the free energy the equilibrium form. All thermodynamic information is encoded in the free energy functional. The resulting theory has a close connection to Bohmian quantum hydrodynamics.

\section{\label{exact}Exact approaches generalizing DDFT}
\subsection{General aspects}
Although DDFT is a highly successful theory, it is nevertheless only approximately valid. Various limitations are presented in \cref{limit}. Often, these arise from the \textit{adiabatic approximation} (see \cref{derivation}), which is a central step in the derivation of DDFT. As discussed \cref{extensions}, one can derive extended forms of DDFT to obtain a larger domain of applicability.

For going beyond standard DDFT, it is often helpful to use formalisms that provide an \textit{exact} description of the underlying many-body dynamics. While the resulting equations of motion can typically not be solved exactly, they are important for at least two reasons: First, studying how DDFT can be obtained from them as a limiting case gives conceptual insights into the approximations involved in the derivation of DDFT and the corresponding limitations. Second, the exact equations of motion can serve as a starting point for the development of new approximate transport equations that improve on the limitations of DDFT. An overview over the literature on this topic is given in \cref{tab:extensionoverview}.

Forces that are not incorporated in deterministic DDFT are known as \textit{superadiabatic forces}. (Stochastic DDFT is not necessarily based on the same type of approximation as deterministic DDFT and can therefore include, e.g., memory effects not present in the deterministic formalism. An obvious example is Dean's DDFT, which is formally exact and therefore contains all superadiabatic forces.) The importance of superadiabatic forces was discussed by \citet{FortinidlHBS2014}, who developed a numerical scheme for their determination. It is based on calculating the exact force integral as well as the fictitious \ZT{adiabatic} potential that generates the instantaneous nonequilibrium density. The difference between the exact and the adiabatic force integral is then the superadiabatic contribution. BD simulations show differences to DDFT results, resulting from neglecting the superadiabatic forces in the latter theory. The numerical scheme is applied to states with a variety of initial conditions in Ref.\ \cite{BernreutherS2016}. Problems arising from the adiabatic approximation involve wrong estimates of relaxation times and difficulties in the description of glass transition and shear flow \cite{SchmidtB2013}. The case of shear was considered in Refs.\ \cite{JahreisS2020,delasHerasS2020}, where superadiabatic forces were further classified into viscous and structural forces. Superadiabatic forces also arise in quantum mechanics \cite{BruttingTdlHS2019}.

A particularly important superadiabatic effect is \textit{memory}. While it is typically not present on the level of single-particle dynamics (Langevin or Hamiltonian equations), reduced descriptions generally lead to a history dependence. This can be shown using the projection operator formalism  \cite{teVrugtW2019d,TreffenstadtS2019}. A procedure for incorporating memory into gradient dynamics (in addition to the ones discussed in this section) was suggested by \citet{GalenkoJ2005} and \citet{KoideKR2006}. It is based on replacing the instantaneous current of the system by a time convolution to take the finite propagation speed of signals into account. This allows to study the effects of memory and local nonequilibrium on spinodal decomposition \cite{KoideKR2007,GalenkoV2007,GalenkoL2008,GalenkoL2008b}. As discussed by \citet{Archer2006,Archer2009}, applying this procedure to DDFT gives a telegrapher's equation that is identical in form to the DDFT for atomic fluids with inertia presented in \cref{inertia}. The same result is achieved when starting from a damped Euler equation \cite{Chavanis2008}. A further discussion of memory effects can be found in Ref.\ \cite{RussoDYK2020}.

Superadiabatic effects are also relevant for the calculation of correlation functions, as discussed in Ref.\ \cite{BraderS2014}. A nonequilibrium Ornstein-Zernike relation relates the superadiabatic current to the direct time-correlation function. Hence, superadiabatic forces take memory effects into account \cite{BraderS2013,SchindlerS2016}. In Refs.\ \cite{BraderS2015,SchindlerS2016}, the superadiabatic/non-Markovian dynamics of the van Hove function is discussed. \citet{TreffenstadtS2019} describe superadiabatic forces in a Brownian liquid, where a resulting memory-induced motion reversal is observed.

Superadiabatic forces can be studied using exact theories. A theory of this type is the stochastic DDFT by \citet{Dean1996}, which has already been discussed in \cref{traditional} and which forms the starting point of the derivation of deterministic DDFT by \citet{MarconiT1999}. In this section, we present three additional exact approaches. The first one is \textit{power functional theory} (\cref{pft}), which was derived by \citet{SchmidtB2013} as an exact description of Brownian dynamics and which has subsequently been extended to Newtonian mechanics \cite{Schmidt2018} (see \cref{newtonianpft}) and quantum mechanics \cite{Schmidt2015,BruttingTdlHS2019} (see \cref{quantumpft}). It is based on a variational principle that is formulated for the current rather than for the density. DDFT can be obtained by setting the \textit{excess power dissipation} (see \cref{standardpft}) to zero. Consequently, studying the excess power dissipation allows to gain insights into superadiabatic forces.

The second one is the \textit{projection operator formalism} (\cref{projectionoperatorformalism}), developed by \citet{Nakajima1958}, \citet{Zwanzig1960}, and \citet{Mori1965}. It allows to derive exact equations of motion for the density -- and for any other dynamical variable or set of variables -- by projecting the full microscopic dynamics onto the subset of variables that is of interest. These equations do, in general, contain a memory term. In the derivation of DDFT using this method (see \cref{derivationmz}), memory effects are dropped (\textit{Markovian approximation}). The projection operator method therefore allows to extend DDFT by enlarging the set of relevant variables or by keeping memory effects. Moreover, it is very important for understanding the relation of stochastic and deterministic forms of DDFT. Applications of the projection operator formalism range far beyond soft matter physics and include fields such as solid state theory \cite{KakehashiF2004}, spin relaxation theory \cite{KivelsonO1974, Bouchard2007,teVrugtW2019}, and particle physics \cite{Koide2002,HuangKKR2011}.

A third approach starts from the \textit{Runge-Gross theorem} (\cref{rungegross}). This result, derived by \citet{RungeG1984}, is the basis of quantum time-dependent density functional theory (TDDFT), a formally exact extension of quantum DFT towards time-dependent systems. \citet{ChanF2005} derived a similar result for classical systems. Since it is nonconstructive, this result has not led to significant further developments in DDFT. Nevertheless, it is often considered an important result as an existence proof \cite{GoddardNK2013,AlmenarR2011,RexLL2005,RothRA2009}.

\subsection{\label{pft}Power functional theory}
\subsubsection{\label{standardpft}Standard power functional theory}
Power functional theory (PFT) is an important exact generalization of DDFT. Unlike standard DDFT, it is based on the one-body current $\vec{J}\rt$. A \textit{power functional} that depends on both $\rho\rt$ and $\vec{J}\rt$ is introduced. The equations of motion are then obtained by a variation of the power functional. 

We start by outlining the derivation of the standard form of PFT following Ref.\ \cite{SchmidtB2013}: One considers a system of $N$ overdamped Brownian particles, where the $i$-th particle obeys the Langevin equation
\begin{equation}
\gamma \vec{v}^L_i(\{\vec{r}_k\},t) = \vec{F}_i(\{\vec{r}_k\},t) + \vec{\chi}_i(t)
\label{langevinpft}
\end{equation}
with the friction constant $\gamma$, the single-particle velocities $\{\vec{v}^L_i\}$ that depend on the particle positions $\{\vec{r}_i\}$ and time $t$, the forces $\{\vec{F}_i\}$, and the noises $\{\vec{\chi}_i\}$ described by \cref{noise1,noise2}. Introducing the $N$-body density $\SmoluchowskiDistribution(\{\vec{r}_k\},t)$, the dynamics can equivalently be rewritten in the form of the Smoluchowski equation
\begin{equation}
\pdif{}{t}\SmoluchowskiDistribution(\{\vec{r}_k\},t) = - \sum_{i=1}^{N}\vec{\nabla}_{\vec{r}_i}\cdot\vec{J}_i(\{\vec{r}_k\},t)
\label{smoluchowskipft}
\end{equation}
with the current of particle $i$
\begin{equation}
\begin{split}
\vec{J}_i(\{\vec{r}_k\},t) &= \SmoluchowskiDistribution(\{\vec{r}_k\},t)\gamma^{-1}(\vec{F}_i(\{\vec{r}_k\},t)-k_B T \Nabla_{\vec{r}_i} \ln(\SmoluchowskiDistribution(\{\vec{r}_k\},t)))\\
&=\SmoluchowskiDistribution(\{\vec{r}_k\},t)\gamma^{-1}\vec{F}^{\mathrm{tot}}_i(\{\vec{r}_k\},t) \\
&=\SmoluchowskiDistribution(\{\vec{r}_k\},t)\vec{v}_i(\{\vec{r}_k\},t),
\end{split}
\end{equation}
where we have identified the total force $\vec{F}^{\mathrm{tot}}_i=\vec{F}_i - k_B T\Nabla_{\vec{r}_i} \ln(\SmoluchowskiDistribution)$ on and the deterministic velocity $\vec{v}_i = \gamma^{-1}\vec{F}^{\mathrm{tot}}_i$ of the $i$-th particle. The deterministic velocity $\vec{v}_i$ is equivalent to the stochastic velocity $\vec{v}^L_i$ appearing in the Langevin equation \eqref{langevinpft} as far as average values are concerned. This allows to introduce the one-body current (which can also be calculated using the stochastic velocity $\vec{v}^L_i$) as
\begin{equation}
\vec{J}\rt=\bigg\langle\sum_{i=1}^{N}\vec{v}_i(\{\vec{r}_k\},t)\delta(\vec{r}-\vec{r}_i)\bigg\rangle .
\end{equation}

We now have the necessary ingredients to set up PFT. The starting point is the generating functional 
\begin{equation}
\begin{split}
\mathcal{R}_t[\SmoluchowskiDistribution,\{\tilde{\vec{v}}_i\}] &=\INT{}{}{^3r_1}\dotsb\INT{}{}{^3r_N}\SmoluchowskiDistribution(\{\vec{r}_k\},t)
\bigg(\sum_{i=1}^{N} \gamma\Big(\frac{1}{2}\tilde{\vec{v}}_i(\{\vec{r}_k\},t) -\vec{v}_i(\{\vec{r}_k\},t)\Big) \\ 
&\quad\:\! \cdot\tilde{\vec{v}}_i(\{\vec{r}_k\},t)+ \dot{U}_1(\vec{r}_i,t)\bigg)
\label{generatingfunctional}%
\end{split}\raisetag{3em}%
\end{equation}
with the time derivative of the external potential $\dot{U}_1= \partial U_1 / \partial t$ and the trial velocity functions $\{\tilde{\vec{v}}_i\}$. (We follow the convention of Ref.\ \cite{SchmidtB2013} and use a subscript $t$ to denote time-dependence.) The actual velocity $\vec{v}_i$ can then be determined from the variation
\begin{equation}
\frac{\delta\mathcal{R}_t[\SmoluchowskiDistribution,\{\tilde{\vec{v}}_i\}]}{\delta \tilde{\vec{v}}_i(\{\vec{r}_k\},t)}=\vec{0}
\end{equation}
performed at fixed $\SmoluchowskiDistribution$ and $t$. Based on the generating functional \eqref{generatingfunctional}, the \textit{free power functional} is defined as
\begin{equation}
R_t[\rho,\vec{J}]= \min_{\{\tilde{\vec{v}}_i\}\to\rho,\vec{J}}\mathcal{R}_t[\SmoluchowskiDistribution,\{\tilde{\vec{v}}_i\}],
\end{equation}
which denotes a minimization with respect to the trial velocities $\{\tilde{\vec{v}}_i\}$ with the constraint that a certain one-body density and current are realized (Levy method \cite{Levy1979}). After removing the dependence on the distribution function $\SmoluchowskiDistribution$ by using the many-body continuity equation \eqref{smoluchowskipft}, we arrive at the simple variational principle 
\begin{equation}
\frac{\delta R_t[\rho,\vec{J}]}{\delta \vec{J}\rt}=\vec{0}.
\label{variationalprinciple}
\end{equation}
The power functional depends on density and current and also on the system's history, i.e., it contains memory effects. In the variation, $\rho(\vec{r},t')$ for $t' \leq t$ and $\vec{J}(\vec{r},t')$ for $t<t'$ are held constant.

By separating external contributions from the complete functional, one can obtain the intrinsic free power
\begin{equation}
W_t[\rho,\vec{J}] = R_t[\rho,\vec{J}] + \INT{}{}{^3r}(\vec{X}\rt - \Nabla U_1\rt)\cdot\vec{J}\rt - \dot{U}_1 \rt\rho\rt,
\end{equation}
where $\vec{X}\rt$ contains all nonconservative forces. The intrinsic free power is further split up as
\begin{equation}
W_t[\rho,\vec{J}] = P_t[\rho,\vec{J}] + \INT{}{}{^3r}\vec{J}\rt\cdot\Nabla\bigg(\frac{\delta F}{\delta\rho\rt }- U_1\rt\bigg)
\end{equation}
with the power dissipation functional $P_t$ describing irreversible loss of energy and a term for reversible contributions. The power dissipation can, moreover, be split into two contributions as
\begin{equation}
P_t[\rho,\vec{J}] = P_t^{\mathrm{id}}[\rho,\vec{J}] +  P_t^{\mathrm{exc}}[\rho,\vec{J}]
\end{equation}
with the ideal gas contribution
\begin{equation}
P_t^{\mathrm{id}}[\rho,\vec{J}] = \INT{}{}{^3r}\frac{\gamma\vec{J}\rt^2}{2\rho\rt} 
\label{idealgascontribution}
\end{equation}
and the excess dissipation $P_t^{\mathrm{exc}}[\rho,\vec{J}]$ resulting from particle interactions. If evaluated at the DDFT current $\vec{J} = - \Gamma \rho \Nabla \delta F / \delta \rho$, \cref{idealgascontribution} leads to the DDFT dissipation functional \eqref{ddftdissipation} (see \cref{net}). The variational principle \eqref{variationalprinciple} gives the fundamental equation of motion
\begin{equation}
\frac{\gamma\vec{J}\rt}{\rho\rt} + \frac{\delta P_t^{\mathrm{exc}}[\rho,\vec{J}]}{\delta \vec{J}\rt} = - \vec{\nabla}\frac{\delta F[\rho]}{\delta\rho\rt} + \vec{X}\rt.
\label{fundamentaleqpft}
\end{equation}
If one sets the excess dissipation to zero, one recovers DDFT as a special case. Theories that include nonadiabatic contributions can be obtained by including effects of particle interactions on the power dissipation. In practice, the application of PFT requires an approximation for $P_t^{\mathrm{exc}}$ \cite{SchmidtB2013}. Moreover, an accurate DDFT for the considered system is needed \cite{WittmannLB2020}, since the adiabatic DDFT term $\Nabla \delta F / \delta \rho$ in \cref{fundamentaleqpft} also has to be approximated.

In Ref.\ \cite{delasHerasS2018}, a reformulation of PFT based on velocity gradients $\vec{\nabla}\otimes\vec{v}$ is developed. A change of variables allows to express the power functional in terms of $\vec{\nabla}\otimes\vec{v}$. This is motivated by the need to construct approximations for the excess power functional $P_t^{\mathrm{exc}}$. For example, the assumption that it can, to lowest order, be written as a nonlocal bilinear in velocity gradients gives the form
\begin{equation}
\begin{split}
P_t^{\mathrm{exc}}[\rho,\vec{J}]&=k_B T\INT{}{}{^3r}\INT{}{}{^3r'}\INT{0}{t}{t'}\rho\rt(\vec{\nabla}\otimes\vec{v}\rt) \\
&\quad\:\! :\pftkernel(\vec{r}-\vec{r}',t-t'):(\vec{\nabla}\otimes\vec{v}(\vec{r}',t'))\rho(\vec{r}',t')
\end{split}
\end{equation}
with the fourth-rank tensor $\pftkernel$ 
the velocity gradients are contracted with. With an additional Markovian approximation and the assumption of spatial locality, this allows to find, for the superadiabatic forces, the form familiar from the Stokes equation in hydrodynamics.

PFT has been generalized towards mixtures \cite{BraderS2015}, Newtonian mechanics \cite{Schmidt2018} (see \cref{newtonianpft}), quantum mechanics \cite{Schmidt2015,BruttingTdlHS2019} (see \cref{quantumpft}), and active particles \cite{KrinningerS2019} (see \cref{nonspft}). Methods for calculating the power dissipation are presented in Ref.\ \cite{BraderS2015b}. An iterative method for relating the one-body density and current to the external force generating them is described in Ref.\ \cite{delasHerasRS2019}. Applications of PFT include the calculation of correlation functions \cite{BraderS2014}, the van Hove function \cite{BraderS2015}, shear \cite{StuhlmullerEdlHS2018,TreffenstadtS2019}, demixing \cite{GeigenfeinddlHS2020}, and phase diagrams \cite{KrinningerSB2016,HermannKdlHS2019}. A further discussion of superadiabatic forces can be found in Refs.\ \cite{FortinidlHBS2014,BernreutherS2016,JahreisS2020,delasHerasS2020}. 

\subsubsection{\label{nonspft}Nonspherical and active particles}
A PFT for ABPs, which also takes orientational degrees of freedom into account, was derived by \citet{KrinningerS2019}. The generating functional, which for translational motion is given by \cref{generatingfunctional}, has the form 
\begin{equation}
\begin{split}
\mathcal{R}_t[\SmoluchowskiDistribution,\{\tilde{\vec{v}}_i ,\tilde{\vec{v}}^\omega_i \}] &= \INT{}{}{^3r_1}\INT{S_2}{}{\Omega_1}\dotsb\INT{}{}{^3r_N}\INT{S_2}{}{\Omega_N}\SmoluchowskiDistribution(\{\vec{r}_k\},\{\uu_k\},t)\\
&\quad\!\:\Bigg(\sum_{i=1}^{N}  \gamma\Big(\frac{1}{2}\tilde{\vec{v}}_i(\{\vec{r}_k\},\{\uu_k\},t) -\vec{v}_i(\{\vec{r}_k\},\{\uu_k\},t)\Big)\cdot\tilde{\vec{v}}_i(\{\vec{r}_k\},\{\uu_k\},t)
\\
&\quad\!\:+\sum_{i=1}^{N} \gamma^\omega\Big(\frac{1}{2}\tilde{\vec{v}}^\omega_i(\{\vec{r}_k\},\{\uu_k\},t) -\vec{v}^\omega_i(\{\vec{r}_k\},\{\uu_k\},t)\Big)\cdot\tilde{\vec{v}}^\omega_i(\{\vec{r}_k\},\{\uu_k\},t) \Bigg).
\end{split}\raisetag{5.9em}%
\end{equation}
Compared to \cref{generatingfunctional}, this generating functional includes the rotational velocities $\{\vec{v}^\omega_i\}$, related to the total torque, and the corresponding trial fields $\{\tilde{\vec{v}}^\omega_i\}$ and friction $\gamma^\omega$. The continuity equation for the one-body density
\begin{equation}
\pdif{}{t}\varrho(\vec{r},\uu,t) = -\vec{\nabla}_{\vec{r}}\cdot\vec{J}(\vec{r},\uu,t) - \Nabla_{\uu}\cdot\vec{J}^\omega(\vec{r},\uu,t)
\end{equation}
contains two currents $\vec{J}$ and $\vec{J}^\omega$ for translational and rotational motion, respectively. Then, the variational equation \eqref{variationalprinciple}, which is a generalized force balance, is supplemented by the generalized torque balance 
\begin{equation}
\frac{\delta R_t[\rho,\vec{J},\vec{J}^\omega]}{\delta \vec{J}^\omega(\vec{r},\uu,t)}=\vec{0}.
\end{equation}
The phase diagram of active particles is moreover calculated using PFT in Refs.\ \cite{KrinningerSB2016,HermannKdlHS2019}.

\subsubsection{\label{newtonianpft}Newtonian mechanics}
Inertia can also be incorporated in PFT \cite{Schmidt2018}. Here, the variational principle of PFT (see \cref{standardpft}) is formulated for $\dot{\vec{J}}$ rather than for $\vec{J}$. Denoting positions and momenta of the individual particles by $\{\vec{r}_i\}$ and $\{\vec{p}_i\}$, respectively, and introducing the trial acceleration fields $\{\vec{a}_i\}$ defined on phase space, one defines the functional 
\begin{equation}
\mathcal{G}_t = \INT{}{}{^{3}r_1}\INT{}{}{^{3}p_1}\dotsi\INT{}{}{^{3}r_N}\INT{}{}{^{3}p_N} \sum_{i=1}^{N}\frac{(\vec{F}_i - m\vec{a}_i)^2}{2m}\MicroscopicPhaseSpaceDistribution(\{\vec{r}_i\},\{\vec{p}_i\},t) 
- \INT{}{}{^3r}\frac{m}{2\braket{\hat{\rho}}}\braket{\dot{\hat{\vec{J}}}}
\end{equation}
with the force $\vec{F}_i$ exerted on particle $i$, the phase-space distribution function $\MicroscopicPhaseSpaceDistribution$, the many-body density $\hat{\rho}$ given by \cref{hatrho}, and the total time derivative of the microscopic current $\dot{\hat{\vec{J}}}$. A functional of the three fields $\rho\rt$, $\vec{J}\rt$, and $\dot{\vec{J}}\rt$ can be obtained by the constrained search
\begin{equation}
G_t[\rho,\vec{J},\dot{\vec{J}}] = \min_{\{\vec{a}_i\}\to\rho,\vec{J},\dot{\vec{J}}}\mathcal{G}_t.
\end{equation}
Assuming physical values of $\rho$ and $\vec{J}$, the functional is minimized by the true value of $\dot{\vec{J}}$, i.e.,
\begin{equation}
\frac{\delta G_t[\rho,\vec{J},\dot{\vec{J}}]}{\delta \dot{\vec{J}}\rt}=\vec{0}.    
\end{equation}
As in usual PFT, the functional $G_t$ can be split into an ideal, external, and excess part. Contributions that go beyond the Brownian case are contained in a \ZT{superpower functional}.

\subsubsection{\label{quantumpft}Quantum mechanics}
PFT has also been extended to quantum systems \cite{Schmidt2015}. Since the Schr\"odinger equation is not overdamped, one needs to take inertia into account. Thus, the formalism is analogous to the Newtonian PFT with inertia presented in \cref{newtonianpft} (which, interestingly, has been developed after quantum PFT). The variational principle is then based on a functional $G_t$ that also depends on the time derivative $\dot{\vec{J}}$ of the current. In the quantum case, the force $\vec{F}$ and the current $\vec{J}$ are replaced by the corresponding operators. Moreover, quantum systems are described using the many-particle wave function $\varPsi$ rather than the phase-space distribution $\SmoluchowskiDistribution$. Hooke's helium model atom was explored using this method in Ref.\ \cite{BruttingTdlHS2019}.

\subsection{\label{projectionoperatorformalism}Projection operator formalism}
\subsubsection{\label{po}Overview}
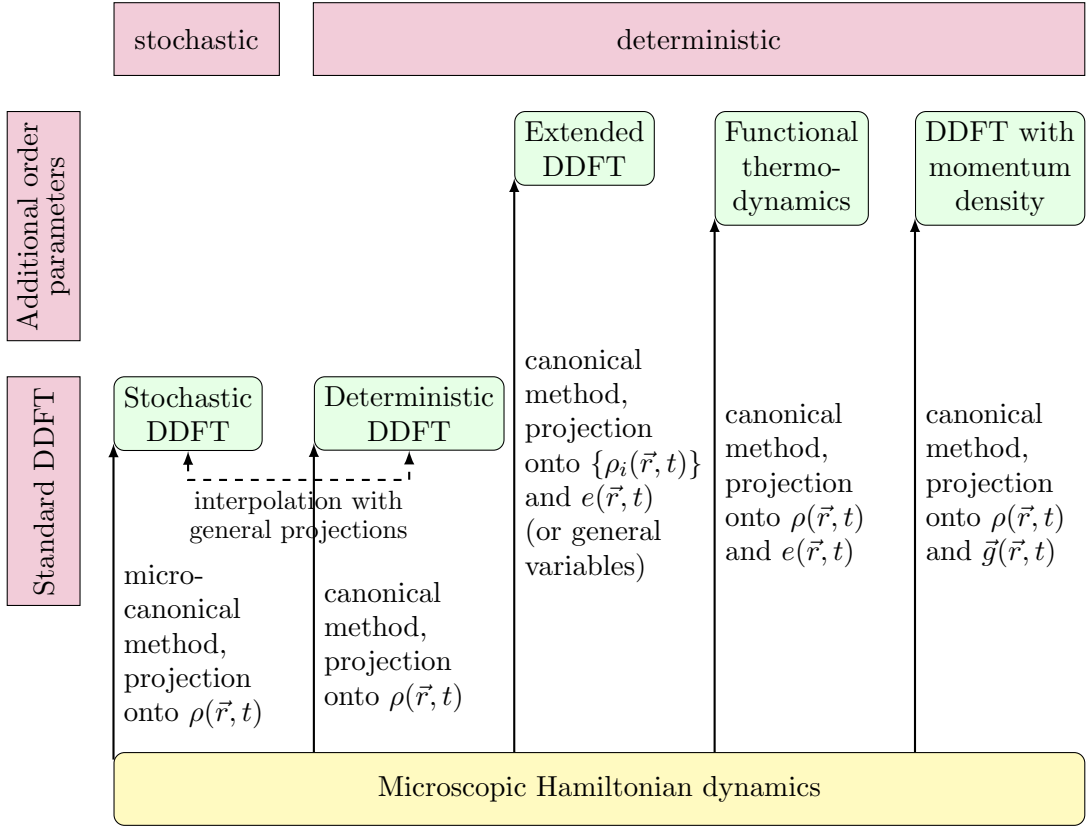
\begin{figure*}[htb]
\centering
\TikzBilderNeuErzeugen{\newcommand{\ZU}{\\[-2pt]}
\newcommand{\XY}{\phantom{Xy}}
\newcommand{\EA}{\phantom{$\bullet$ }}
\newcommand{\Abstand}{0.0325\textwidth}
\newcommand{\Mindesthoehe}{0.0675\textwidth}
\tikzstyle{begriffsfeld0} = [align=center,rectangle,fill=purple!20]
\tikzstyle{begriffsfeldI} = [rectangle, rounded corners, minimum size=\Mindesthoehe, align=center, draw=black, fill=yellow!30]
\tikzstyle{begriffsfeldII} = [rectangle, rounded corners, minimum size=\Mindesthoehe, align=center, draw=black, fill=green!10]
\tikzstyle{begriffsfeldIII} = [rectangle, rounded corners, minimum size=5mm, align=left, draw=black, fill=gray!10, font=\scriptsize]
\tikzstyle{dummyfeld} = [rectangle, rounded corners, minimum size=5mm, align=center, draw=black, draw=none]
\tikzstyle{pfeil} = [thick,->,>=latex]
\tikzstyle{pfeilII} = [thick,<->,>=latex]
\tikzstyle{doppelpfeil} = [thick,implies-implies, double equal sign distance]
\tikzstyle{beschriftung} = []
\begin{tikzpicture}
\coordinate[](P1)at(\textwidth,0);
\node(B1)[begriffsfeldI,minimum width=0.9\textwidth,anchor=north east]at(P1){Microscopic Hamiltonian dynamics};
\node(K1)[begriffsfeldII,anchor=north west,yshift=50mm]at(B1.north west){Stochastic\\DDFT};
\newdimen\yI
\pgfextracty{\yI}{\pgfpointanchor{K1}{north}}
\coordinate[](PBL1)at(0,\yI);
\node(BL1)[begriffsfeld0,rotate=90,anchor=north east,draw=black,minimum height=\Mindesthoehe]at(PBL1){Standard DDFT};
\node(BL2)[begriffsfeld0,rotate=90,anchor=north west,draw=black,align=center,minimum height=\Mindesthoehe,xshift=\Abstand]at(PBL1.north east){Additional order\\[-3pt]parameters};
\node(K5)[begriffsfeldII,anchor=north east]at(B1.east|-BL2.east){DDFT with\\momentum\\density};
\newdimen\xI
\newdimen\xII
\newdimen\xneu
\newdimen\yII
\newdimen\differenz
\pgfextractx{\xI}{\pgfpointanchor{B1}{west}}
\pgfextractx{\xII}{\pgfpointanchor{K5}{west}}
\differenz=\xII
\advance\differenz by -\xI
\divide\differenz by 4
\xneu=\differenz
\advance\xneu by \xI
\coordinate[](K2X)at(\xneu,\yI);
\advance\xneu by \differenz
\pgfextracty{\yII}{\pgfpointanchor{BL2}{east}}
\coordinate[](K3X)at(\xneu,\yII);
\advance\xneu by \differenz
\coordinate[](K4X)at(\xneu,\yII);
\node(K2)[begriffsfeldII,anchor=north west]at(K2X){Deterministic\\DDFT};
\newdimen\breiteIII
\pgfextractx{\breiteIII}{\pgfpointdiff{\pgfpointanchor{K1}{west}}{\pgfpointanchor{K2}{west}}}
\advance\breiteIII by -\Abstand
\node(BL3)[begriffsfeld0,anchor=south west,draw=black,align=center,minimum height=\Mindesthoehe,yshift=\Abstand,minimum width=\breiteIII]at(K1.west|-BL2.north east){stochastic};
\newdimen\breiteIVa
\newdimen\breiteIVb
\pgfextractx{\breiteIVb}{\pgfpointanchor{B1}{east}}
\pgfextractx{\breiteIVa}{\pgfpointanchor{K2}{west}}
\advance\breiteIVb by -\breiteIVa
\node(BL4)[begriffsfeld0,anchor=north east,draw=black,align=center,minimum height=\Mindesthoehe,minimum width=\breiteIVb]at(B1.east|-BL3.north){deterministic};
\node(K3)[begriffsfeldII,anchor=north west]at(K3X){Extended\\DDFT};
\node(K4)[begriffsfeldII,anchor=north west]at(K4X){Functional\\thermo-\\dynamics};
\coordinate[yshift=1mm](PK1)at(K1.south west);
\coordinate[yshift=1mm](PK2)at(K2.south west);
\coordinate[yshift=1mm](PK3)at(K3.south west);
\coordinate[yshift=1mm](PK4)at(K4.south west);
\coordinate[yshift=1mm](PK5)at(K5.south west);
\coordinate[yshift=-1em](NH)at(K1.south);
\draw[pfeilII,dashed](K1.south) -- (NH) -- node[beschriftung,align=center,anchor=north,font=\small](NHII){interpolation with\\general projections}(K2.south|-NH) -- (K2.south);
\coordinate[yshift=0.5ex](NHIII)at($(NHII.south)!0.5!(B1.north)$);
\draw[pfeil]([yshift=-1mm]K1.west|-B1.north) -- (K1.west|-PK1);
\draw[] node[beschriftung,align=left,anchor=west]at(K1.west|-NHIII){micro-\\canonical\\method,\\projection\\onto $\rho(\vec{r},t)$};
\draw[pfeil](K2.west|-B1.north) -- (K2.west|-PK2);
\draw[] node[beschriftung,align=left,anchor=west]at(K2.west|-NHIII){canonical\\method,\\projection\\onto $\rho(\vec{r},t)$}(K2.west|-PK2);
\draw[pfeil](K3.west|-B1.north) -- node[beschriftung,align=left,anchor=west]{canonical\\method,\\projection\\onto $\{\rho_i(\vec{r},t)\}$\\and $e(\vec{r},t)$\\(or general\\variables)}(K3.west|-PK3);
\draw[pfeil](K4.west|-B1.north) -- node[beschriftung,align=left,anchor=west]{canonical\\method,\\projection\\onto $\rho(\vec{r},t)$\\and $e(\vec{r},t)$}(K4.west|-PK4);
\draw[pfeil](K5.west|-B1.north) -- node[beschriftung,align=left,anchor=west]{canonical\\method,\\projection\\onto $\rho(\vec{r},t)$\\and $\vec{g}(\vec{r},t)$}(K5.west|-PK5);
\end{tikzpicture}}%
\TikzBildEinfuegen{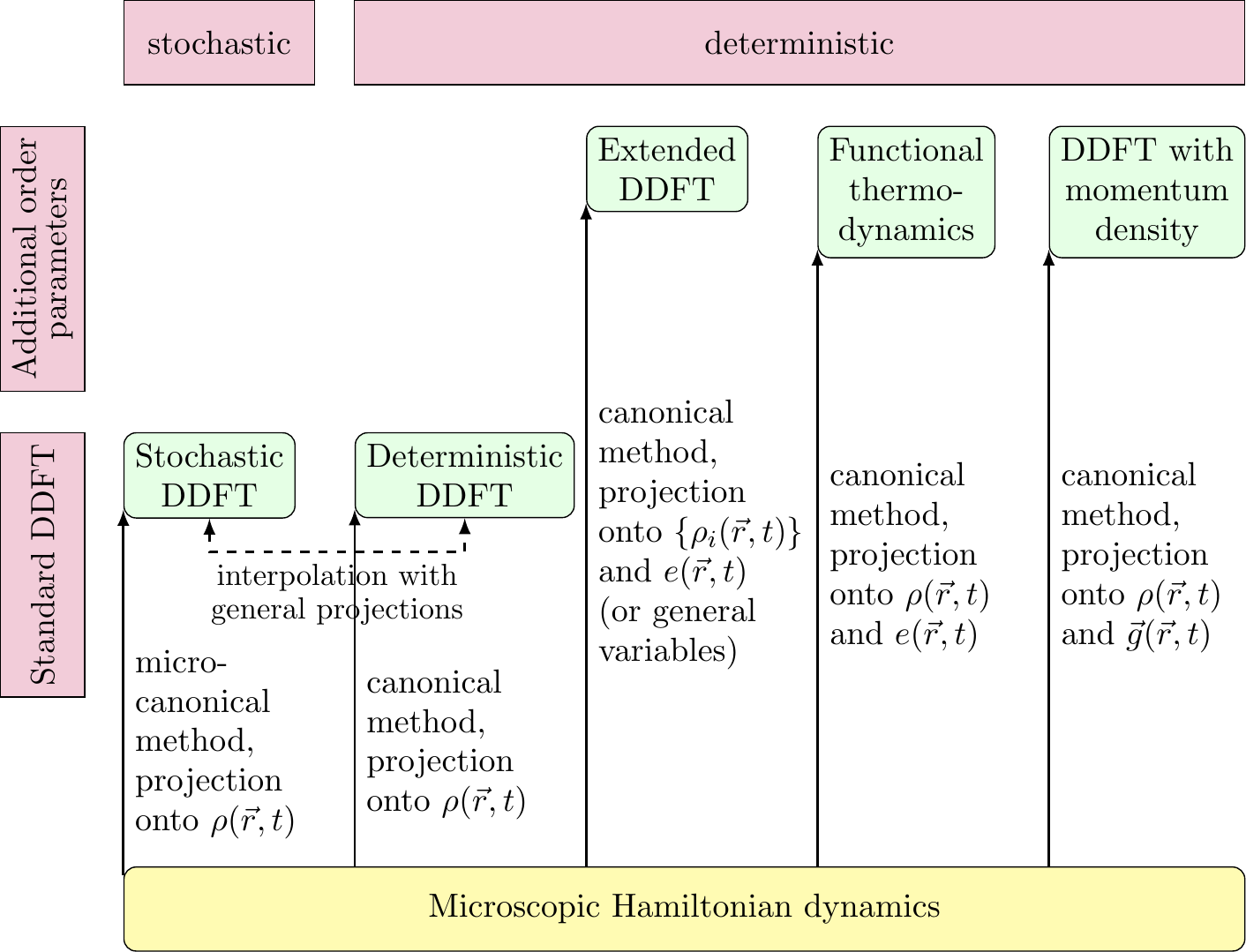}{}%
\caption{\label{fig:derivationviaMZF}Derivation of DDFT using the projection operator formalism.}%
\end{figure*}
Projection operator methods are another central tool in the context of DDFT. These allow, in general, to derive coarse-grained equations of motion \cite{KinjoH2007,Hyodo2012} for a set of relevant variables by projecting the complete microscopic dynamics onto the subdynamics of the relevant variables. Based on the original work by \citet{Nakajima1958}, \citet{Zwanzig1960}, and \citet{Mori1965}, a large class of such methods has been developed. These methods all share the same general structure, but differ in certain details, such as the definition of the projection operator or the domain of applicability. 
 
The role of projection operators in the derivation of DDFT is visualized in \vref{fig:derivationviaMZF}. They can be classified based on whether they fix the value (\ZT{microcanonical} case) or the average (\ZT{canonical} case) of the relevant variables \cite{Kawasaki2000} (left of \vref{fig:derivationviaMZF}, see \cref{noise} for a discussion of this point). Microcanonical methods allow to derive stochastic DDFT. The derivation of a deterministic DDFT with a free energy from DFT requires a canonical method \cite{Yoshimori2005}. In these theories, the projection operator is generally time-dependent. They were developed by \citet{Robertson1966} and \citet{KawasakiG1973}, and subsequently generalized towards the dynamics of fluctuations \cite{Grabert1978} and time-dependent Hamiltonians \cite{teVrugtW2019,MeyerVS2019}. A general overview is given in the textbook by \citet{Grabert1982} and a simple introduction in Ref.\ \cite{teVrugtW2019d}. Various names are used in the literature on DDFT, such as \ZT{Kawasaki-Gunton operator method} \cite{Yoshimori2005,Kawasaki2006b}, \ZT{Zwanzig projection operator} \cite{DuranGK2016}, \ZT{Mori-Zwanzig-Forster technique} \cite{WittkowskiLB2012,WittkowskiLB2013}, and \ZT{Mori-Zwanzig formalism} \cite{teVrugtW2019}. (Some of these names, such as \ZT{Kawasaki-Gunton method}, refer to specific forms.)

As discussed in \cref{derivationmz}, projection operators can be used to derive DDFT by projecting onto the one-body density as a relevant variable \cite{Yoshimori1999,Munakata2003,Yoshimori2005,EspanolL2009,EspanolV2002,Kawasaki1994}. Additional projections allow to go from DDFT to phase field crystal models \cite{MajaniemiPN2010}. When using the Mori theory, which is a simple form of the projection operator formalism, one obtains a linear model that has different properties than full DDFT \cite{MurataY2006}. \citet{Yoshimori2011} used projection operators to extend DDFT to a rigid interaction site model.  

By choosing additional relevant variables, extensions of DDFT to further order parameter fields can be (and have been) derived (right of \vref{fig:derivationviaMZF}). This is done in extended dynamical density functional theory (EDDFT) for several concentration fields for the different species of colloidal particles in mixtures, the energy density \cite{WittkowskiLB2012}, and the entropy density \cite{WittkowskiLB2013} (see \cref{eddft}) and in functional thermodynamics (FTD) for the energy density \cite{AneroET2013} (see \cref{ftd}). For fluids, the momentum density can also be included \cite{CamargodlTDZEDBC2018,CamargodlTDBCE2019,Kawasaki2006,DuranYGK2017,DuqueZumajoCdlTCE2019,DuqueZumajodlTCE2019,MajaniemiPN2010}. Additional order parameters, such as an elastic strain field, can account for translational symmetry breaking in solids \cite{MajaniemiPN2010,MajaniemiG2007,MjaniemiNG2008}. In Refs.\ \cite{WalzF2010,RasSF2020}, projection operators were used to calculate elastic properties of solids (see \cref{sound}). In principle, arbitrary variables can be incorporated in DDFT, although for obtaining a Markovian equation it is required that these variables are slow compared to the microscopic dynamics. Moreover, additional degrees of freedom can be incorporated by using a density $\PSDistributionRVT(\vec{r},\vec{p},t)$ that also depends on the momentum $\vec{p}$ \cite{AneroE2007} or a density $\varrho(\vec{r},\vec{\varpi},t)$ with the orientation $\vec{\varpi}$ \cite{DuranGK2016}.

In addition, projection operator methods allow for important conceptual insights into DDFT. For example, they allow to connect deterministic and stochastic forms of DDFT, thereby clarifying the origin and necessity of noise terms \cite{Kawasaki2006,Kawasaki2006b} (see \cref{noise}). This is also useful for fluctuating hydrodynamics \cite{DuranYGK2017,EspanolD2015}. Furthermore, projection operators allow for an exact definition of the free energy functional \cite{Yoshimori2005}. Finally, an interesting connection between projection-operator-based coarse graining and numerical discretization has been found by \citet{delTED2015}.

\subsubsection{Projection-operator-based extensions of standard DDFT}
\subsubsubsection{\label{eddft}Extended dynamical density functional theory}
As discussed in \cref{po}, the Mori-Zwanzig projection operator formalism allows to derive equations of motion for arbitrary dynamical variables, which for slow variables can, in most cases, be approximated by memoryless dissipative transport equations. Since standard DDFT can be derived by projecting the microscopic dynamics onto the number density as a relevant variable (see \cref{derivationmz}), an extended dynamical density functional theory (EDDFT) can be derived by projecting onto a larger number of variables. 

The projection operator method allows, for a set of relevant variables with mean values $\{a_i\rt\}$, to introduce a free energy $F[\{a_i\}]$ such that the dynamics is governed by the thermodynamic conjugates $\delta F / \delta a_i\rt$. This allows to derive the general transport equations \eqref{dotac} for conserved and \eqref{dotanc} for nonconserved variables (see below), along with microscopic expressions for the current $\vec{J}_i$ and the quasi-current $\Phi_i$, respectively, that depend on the thermodynamic conjugates \cite{WittkowskiLB2012}. For close-to-equilibrium systems, the current and quasi-current can be obtained from a dissipation functional \cite{WittkowskiLB2013} (see \cref{net}). Moreover, since the projection operator formalism is also a standard method for the derivation of MCT, this approach allows for insights into the relation of EDDFT and MCT \cite{WittkowskiLB2012}.

\citet{WittkowskiLB2012} derived an EDDFT for particle mixtures by choosing the concentration of each particle species $\rho_i\rt$ and the energy density $e\rt$ as relevant variables. The resulting theory is an extension of DDFT to colloidal mixtures and temperature gradients. In general, the expressions for the currents also contain hydrodynamic interactions. Moreover, the entropy density (see \cref{entropydensity}) is incorporated as a relevant variable into EDDFT in Ref.\ \cite{WittkowskiLB2013} within the framework of linear irreversible thermodynamics, assuming that a local formulation of the first law of thermodynamics holds. 

For a general set of ${\kappa}$ conserved relevant variables $\{A_i\rt\}$ that are fields on spacetime, the extended DDFT equations for the corresponding mean values $\{a_i\rt\}$ read \cite{WittkowskiLB2012}
\begin{equation}
\pdif{}{t}a_i\rt = - \Nabla_{\vec{r}}\cdot\Tr(\bar{\rho}(t)\Stromoperator{\vec{J}}_i(\vec{r},0)) + \sum_{j=1}^{\kappa}\frac{1}{k_B T}\Nabla_{\vec{r}}\cdot\INT{}{}{^3r'} D^{(ij)}(\vec{r},\vec{r}',t)\Nabla_{\vec{r}'}\frac{\delta F}{\delta a_j(\vec{r}',t)}
\label{eddftequation}
\end{equation}
with the microscopic current $\Stromoperator{\vec{J}}_i$ for the relevant variable $A_i$ defined by the relation
\begin{equation}
\ii L A_i\rt = - \Nabla\cdot\Stromoperator{\vec{J}}_i\rt, 
\end{equation}
where in classical mechanics the Liouvillian $L$ is defined with the Poisson bracket $\{\cdot,\cdot\}$, 
and the diffusion tensor
\begin{equation}
D^{(ij)}_{ml}(\vec{r},\vec{r}',t) = \INT{0}{\infty}{t'}\Tr\big(\bar{\rho}(t)(\mathcal{Q}(t)\Stromoperator{J}_{j,l}(\vec{r}',0))
e^{\ii L t'}(\mathcal{Q}(t)\Stromoperator{J}_{i,m}(\vec{r},0))\big),
\end{equation}
where $\bar{\rho}(t)$ is the relevant probability density, $\mathcal{Q}=1-\ProjectionOperator$ is the complementary projection operator (see \cref{derivationmz}), and $\Stromoperator{J}_{i,j}=(\Stromoperator{\vec{J}}_i)_j$. For real-valued relevant variables with definite time-reversal signature, Onsager's principle demands that $D^{(ij)}_{kl} = D^{(ji)}_{lk}$. A notable aspect of the general EDDFT equation \eqref{eddftequation} is the presence of the organized drift term $\Nabla_{\vec{r}}\cdot\Tr(\bar{\rho}(t)\Stromoperator{\vec{J}}_i(\vec{r},0))$. In usual DDFT, this drift vanishes by symmetry, which is a manifestation of the fact that DDFT describes overdamped dynamics. For general dynamical variables, this is, however, not guaranteed. In this case, the drift term corresponds to the nondissipative part of the dynamics. 

A notable feature of the EDDFT approach is that it, being based on the projection operator formalism, also provides dynamic equations for the correlation functions. For this reason, it is a natural framework for analyzing the relation of DDFT to MCT (see Ref.\ \cite{WittkowskiLB2012} and \cref{mct}).

\subsubsubsection{\label{ftd}Functional thermodynamics}
Functional thermodynamics (FTD), derived by \citet{AneroET2013}, is another extension of DDFT towards nonisothermal systems. As in EDDFT, the complete microscopic dynamics of a system is, using the projection operator formalism, projected onto the number density and the energy density as relevant variables. A conceptual aim of this framework is to study the relation of different \ZT{levels of description} of a certain system, which are characterized by the choice of the relevant variables. Each level possesses its own entropy functional, which can be connected to that of other levels using a bridge theorem. This generalizes the theorem by \citet{Mermin1965} on the relation between density, external potentials, and free energy functional exploited in DFT (see \cref{static}).

\subsection{\label{rungegross}Runge-Gross theorem}
\subsubsection{\label{qtddft}Quantum time-dependent density functional theory}
Besides its importance in classical statistical physics, DFT is also an important tool of many-body quantum mechanics. Quantum DFT also has a dynamical extension, which in contrast to its classical counterpart is usually referred to as \ZT{time-dependent density functional theory} (TDDFT). (In this review, we use \ZT{DDFT} to refer to the classical and \ZT{TDDFT} to refer to the quantum-mechanical method. Note that \ZT{TDDFT} is sometimes also used in the literature to refer to classical DDFT.) Reviews of TDDFT can be found in Refs.\ \cite{MarquesG2004,BurkeWG2005,CasidaH2012,NakatsukasaMMY2016,Ullrich2011,vanLeeuwen2001,SantoroJ2016,NiehausM2010,BottiSDSR2007}. Quantum DFT is reviewed in Refs.\ \cite{Jones2015,KuemmelK2008,JonesG1989,JainSP2016,MardirossianH2017}. A discussion of the relation of classical and quantum DFT can be found in Ref.\ \cite{ZhaoLCLLH2015}. TDDFT is based on the \textit{Runge-Gross theorem}, which is a formally exact result that, in a very similar form, can also be derived for classical systems \cite{ChanF2005}. From the latter result, one can obtain classical DDFT as a limiting case. Here, we will first give a brief introduction to quantum DFT and TDDFT. The relation to classical DDFT is discussed in \cref{existence}.

We follow Refs.\ \cite{UllrichY2014,Jones2015}. The state of a system of $N$ electrons can be described using the $N$-particle wave function $\varPsi(\vec{r}_1,\dotsc,\vec{r}_N)$, which satisfies the time-independent Schr\"odinger equation
\begin{equation}
\hat{H}\varPsi = E\varPsi
\label{schrodinger}%
\end{equation}
with the Hamiltonian $\hat{H}$ and energy eigenvalue $E$. The Hamiltonian can be written as
\begin{equation}
\hat{H} = \hat{T} + \hat{V}_{\mathrm{ee}} + \hat{V}_{\mathrm{ext}}
\end{equation}
with the kinetic energy $\hat{T}$, the electron-electron interaction potential $\hat{V}_{\mathrm{ee}}$, and an external potential $\hat{V}_{\mathrm{ext}}$. In position space, these are given by
\begin{align}
\hat{T}&=\sum_{i=1}^{N}-\frac{\hbar^2}{2m_e}\Nabla_{\vec{r}_i}^2,\\
\hat{V}_{\mathrm{ee}} &= \sum_{\begin{subarray}{c}i,j=1\\i\neq j\end{subarray}}^{N} \frac{Q_e^2}{8\pi\epsilon_0}\frac{1}{\norm{\vec{r}_i - \vec{r}_j}},\\
\hat{V}_{\mathrm{ext}} &= \sum_{i=1}^{N}U_1(\vec{r}_i)
\end{align}
for a system of $N$ electrons with mass $m_e$ and charge $Q_e$. Here, $\hbar=h/(2\pi)$ is the reduced Planck constant and $\epsilon_0$ is the vacuum permittivity. The ground state of the system is the state $\varPsi$ with the lowest energy $E$. In practice, \cref{schrodinger} is impossible to solve for all but the simplest systems. A more efficient description of many-particle systems is possible in DFT: Just as classical DFT has allowed us to use the density $\rho$ instead of the full phase-space distribution $\MicroscopicPhaseSpaceDistribution$, quantum DFT allows to work with the electron density \cite{Jones2015}
\begin{equation}
n(\vec{r}) = N\INT{}{}{^3r_2}\dotsb\INT{}{}{^3r_N}\abs{\varPsi(\vec{r}_1,\dotsc,\vec{r}_N)}^2
\end{equation}
instead of the full $N$-body wavefunction. Since $n(\vec{r})$ is a function of 3 rather than $3N$ variables, this leads to a significant computational simplification. The reason why this is possible is that, as shown by \citet{HohenbergK1964}, the external potential generating a certain density $n(\vec{r})$ is unique up to a constant. Moreover, since the external potential determines the Hamiltonian which then fixes the ground-state wavefunction via the Schr\"odinger equation \eqref{schrodinger}, the wavefunction is a functional of $n(\vec{r})$. The same then holds for the internal energy of the system. One can therefore introduce the universal functional \cite{HohenbergK1964}
\begin{equation}
F_{\mathrm{univ}}[n] = \braket{\varPsi|\hat{T} + \hat{V}_{\mathrm{ee}}|\varPsi},
\end{equation}
which is valid for any external potential. Here, $\braket{\cdot}$ denotes a quantum-mechanical expectation value. \citet{HohenbergK1964} showed that the ground-state energy can then be found by minimizing the functional
\begin{equation}
E = F_{\mathrm{univ}}[n] + \INT{}{}{^3r} n(\vec{r}) U_1(\vec{r}).  
\end{equation}
The minimization is over all nondegenerate densities that are $v$-representable, i.e., that are ground-state densities of an external potential \cite{Jones2015}. Moreover, the original derivation by \citet{HohenbergK1964} is restricted to nondegenerate ground states. A generalization was derived by \citet{Levy1979}. For the practical application of DFT, approximation methods were developed by \citet{KohnS1965}: We consider a noninteracting reference system that has the same density as the interacting system. For the noninteracting system, the density can be written in terms of single-particle orbitals $\{\zeta_i(\vec{r})\}$ as
\begin{equation}
n_{\mathrm{gs}}(\vec{r}) = \sum_{i=1}^{N}\abs{\zeta_i(\vec{r})}^2,
\label{groundstate}
\end{equation}
where the single-particle orbitals satisfy the Kohn-Sham equation
\begin{equation}
\bigg(\! -\frac{\hbar^2}{2m_e}\Nabla^2 + u(\vec{r})\bigg)\zeta_i(\vec{r}) = E_i \zeta_i(\vec{r})
\end{equation}
with the single-particle energies $\{E_i\}$. The potential $u(\vec{r})$ is the potential which produces the ground-state density $n_{\mathrm{gs}}(\vec{r})$ in the noninteracting system and thus has a functional dependence on $n$. Once the potential $u(\vec{r})$ is known, the ground-state energy $E_{\mathrm{gs}}$ can be calculated immediately. To determine this potential, we write it as
\begin{equation}
u(\vec{r}) = U_1(\vec{r}) + u_{\mathrm{H}}(\vec{r}) + u_{\mathrm{xc}}(\vec{r})
\label{quantumpotential}
\end{equation}
with the Hartree potential
\begin{equation}
u_{\mathrm{H}}(\vec{r}) = \INT{}{}{^3r'}\frac{Q_e^2}{4\pi\epsilon_0} \frac{n(\vec{r}')}{\norm{\vec{r}-\vec{r}'}}
\end{equation}
and the exchange correlation potential $u_{\mathrm{xc}}(\vec{r})$, which is not known exactly and has to be approximated. 

TDDFT extends quantum DFT by using a generalization of the Hohenberg-Kohn theorem which is known as the Runge-Gross theorem \cite{RungeG1984}. It states that for two $N$-electron systems that start from the same initial state $\varPsi_0$ and are subject to two different time-dependent potentials, the time-dependent densities will be different (unless the potentials only differ by a time-dependent function). The proof requires that the potential can be Taylor-expanded in time. As in DFT, one therefore has a one-to-one correspondence between the density and the external potential. The main difference to the time-independent case is that this correspondence holds for a fixed initial state $\varPsi_0$. Consequently, the external potential has functional dependence on both $n$ and $\varPsi_0$, where the latter dependence vanishes if the system starts in the ground state. As in time-independent DFT, one can introduce a noninteracting reference system in an external potential $u\rt$ that reproduces the density $n\rt$ of the interacting system. The electrons in the noninteracting system satisfy the time-dependent Kohn-Sham equations
\begin{equation}
\ii\hbar\pdif{}{t}\zeta_i\rt = \bigg(\! -\frac{\hbar^2}{2m_e}\Nabla^2 + u\rt\bigg)\zeta_i\rt.
\end{equation}
Again, the potential $u\rt$ is decomposed in the form \eqref{quantumpotential}, this time with an exchange correlation potential that depends on the density and on the initial state of the interacting system and the Kohn-Sham system.

\subsubsection{\label{existence}Existence proof for DDFT}
A TDDFT has been derived for classical systems by \citet{ChanF2005}, who used an analogue of the Runge-Gross theorem to establish an invertible mapping between the density $\rho\rt$ and the phase-space distribution function. For Hamiltonian systems, this allows to obtain a stationary action principle. The hydrodynamic behavior can, in analogy to quantum mechanics, be obtained from a Kohn-Sham-like noninteracting reference system. This framework allows to relate classical DDFT to quantum TDDFT, since it is derived along the same lines as the quantum mechanical formalism, but at the same time contains classical DDFT as a limiting case. In a sense, one has thereby derived classical DDFT from the corresponding quantum-mechanical formalism (by formal analogy).

The proof by \citet{ChanF2005} is often thought of as an existence proof for a classical DDFT, but, unfortunately, it is nonconstructive \cite{GoddardNK2013,AlmenarR2011,RexLL2005,RothRA2009}. It concerns a system whose microscopic dynamics is described by Hamilton's equations. For a fixed initial phase-space distribution, solving the deterministic Liouville equation gives a unique mapping between the phase-space distribution and the external potential. It is then shown that different potentials (that do not only differ by a purely time-dependent function) lead to different one-body currents, which allows, through the continuity equation, to obtain a unique mapping between the time-dependent one-body density and the time-dependent external potential for a given initial distribution. This mapping also exists if the microscopic dynamics is Brownian rather than Hamiltonian. Most of the derivation in Ref.\ \cite{ChanF2005} is, however, restricted to Hamiltonian systems. One can introduce a noninteracting reference system that at each time reproduces the exact one-body density. The excess potential can be obtained as the functional derivative of an \ZT{excess action functional} that in the simplest (adiabatic) case can be approximated using the equilibrium excess free energy. This allows to obtain the standard DDFT equation \eqref{trddft} \cite{ChanF2005}. In this approximation, memory effects are dropped.

\section{\label{theories}Theories related to DDFT}
In this section, we discuss the relation of DDFT to various other theories used in the theoretical description of soft matter systems. This is important for understanding the precise location of DDFT in the landscape of soft matter theory. On the one hand, the development of DDFT has historically been influenced by work on related methods, such as mode coupling theory (MCT) \cite{Kawasaki1994} or the Cahn-Hilliard equation \cite{Cahn1965,CahnH1958,Evans1979}. On the other hand, DDFT has influenced the development of other theories such as phase field crystal (PFC) models \cite{ElderKHG2002,ElderG2004,BerryGE2006,ElderPBSG2007,vanTeeffelenBVL2009}. In addition to the historical importance, such a discussion also provides conceptual insights: Studying the relation of DDFT to MCT allows to understand to which extent DDFT can and cannot describe glassy systems, and comparing DDFT to general models of the gradient dynamics type is helpful for work on spinodal decomposition and relaxation dynamics.

We start with a discussion of MCT (\cref{mct}), which has been important at early stages of the historical development of DDFT and continues to be important for the study of the glass transition. Then, we continue with theories of the gradient dynamics type, beginning with a general discussion of nonequilibrium thermodynamics (\cref{net}), which has also (in particular in the form of the Cahn-Hilliard equation) influenced DDFT. We subsequently turn to two younger theories (\cref{phasefield,nescgle}) that have been influenced by DDFT. Here, PFC models (\cref{phasefield}), which also have a gradient dynamics form, are of particular interest. Afterwards, we discuss nonequilibrium self-consistent generalized Langevin equations (\cref{nescgle}), which are based on nonequilibrium thermodynamics. Finally, we present dynamic mean-field theory (\cref{dmft}), which is a discrete model. The latter two approaches are not frequently discussed in the literature on DDFT, but are nevertheless of conceptual interest and therefore presented here for completeness.

\subsection{\label{mct}Mode coupling theory}
Mode coupling theory (MCT), reviewed in Refs.\ \cite{Das2004,Janssen2018}, is the most widely used theory for the description of the glass transition \cite{BengtzeliusGS1984,Leutheusser1984,Goetze2009} (although it has its origin in the theory of critical dynamics \cite{Fixman1962,Kawasaki1966,KadanoffS1968}, see Ref.\ \cite{Kawasaki1998} for a discussion). Historically, MCT was very important for the development of DDFT, in particular in the forms proposed by \citet{KirkpatrickW1987} and \citet{Kawasaki1994}. There is a significant amount of literature discussing the relation of DDFT and MCT, in which, remarkably, two different views are expressed: In treatments of stochastic DDFT, MCT is typically presented as a limiting case of DDFT. DDFT is considered a more general theory that models aspects of the dynamics of supercooled liquids and glasses not captured by MCT \cite{Kawasaki1994,Kawasaki2009,FuchizakiK2002}. The literature on deterministic DDFT, on the other hand, considers DDFT and MCT to be complementary theories that arise as different limiting cases of a general exact treatment \cite{BraderS2013,WittkowskiLB2012}. After giving a brief introduction to MCT, we here discuss how this apparent contradiction can be resolved. Finally, we explain how MCT can be derived from stochastic DDFT.

The object of interest in MCT is the normalized density correlator $\phi(\vec{k},t)$, which is defined as \cite{Kawasaki2009} 
\begin{equation}
\phi(\vec{k},t) = \frac{\braket{\rho_{\vec{k}}(t)\rho_{-\vec{k}}(0)}}{\braket{\rho_{\vec{k}}(0)\rho_{-\vec{k}}(0)}}
\label{correlator}%
\end{equation}
with the ensemble average $\langle\cdot\rangle$ and the Fourier-transformed reduced density
\begin{equation}
\rho_{\vec{k}}(t)= \INT{}{}{^3r}(\rho\rt-\rho_0)e^{-\ii\vec{k}\cdot\vec{r}}.
\end{equation}
The correlation approximately follows the closed equation of motion \cite{Kawasaki2009,Archer2006} 
\begin{equation}
\ddot{\phi}(\vec{k},t)+ \nu\dot{\phi}(\vec{k},t)+ \frequenz^2(\vec{k})\phi(\vec{k},t) =-\INT{0}{t}{t'}\memory(\vec{k},t')\dot{\phi}(\vec{k},t-t')
\label{mcteqofmotion}
\end{equation}
with the friction coefficient $\nu$, squared frequency $\frequenz^2(\vec{k})=k_BT k^2/(mS(\vec{k}))$, and memory kernel
\begin{equation}
\memory(\vec{k},t)=\frac{k_B T \rho_0 }{2(2\pi)^3mk^2}\INT{}{}{^3k'}\big(\vec{k}\cdot\vec{k}'c(\vec{k}') + \vec{k}\cdot(\vec{k}-\vec{k}') c(\vec{k}-\vec{k}')\big)^2S(\vec{k}')S(\vec{k}-\vec{k}')\phi(\vec{k}',t)\phi(\vec{k}-\vec{k}',t),
\label{mctexpression}%
\end{equation}
where $S(\vec{k})$ is the static structure factor and $c(\vec{k})$ is the Fourier-transformed direct pair-correlation function. Note that the explicit form of the MCT equation can be different and depends on the system (e.g., on whether we consider simple or colloidal fluids) \cite{WittkowskiLB2012}.

Using certain approximations, this result can be derived using the projection operator formalism (see \cref{po}). The derivation can be found in Ref.\ \cite{Janssen2018}. Starting from the microscopic dynamics, a formally exact equation of motion for the correlation function is obtained. This equation contains a memory term, in which the memory kernel is the autocorrelation of the random force. The random force corresponds to the part of the complete dynamics that is orthogonal to the relevant slow variables given by the collective density modes. Next, one makes the approximation that the random force is dominated by products of two density modes, which corresponds to the simplest nontrivial form. (Note that, due to the nature of the linear projection operator, \textit{products} of the density modes are contained in the random force if we project onto the density modes.) Finally, it is assumed that the four-point density correlation function in the memory kernel, which one gets as a result, can be approximated by a product of two-point correlation functions. This gives the \ZT{MCT approximation} for the memory kernel \cite{RadonsJH2005}. (Note that the name \ZT{MCT approximation} is used with different meanings in the literature. Here, we use it to denote the approximations required for obtaining the MCT expression \eqref{mctexpression} from the full density dynamics described by stochastic DDFT, as discussed below and visualized in \vref{fig:schematicglas}.) 

The success of MCT thus shows that, if the density is a slow variable, the same can be assumed about its products (which appear in the random force and therefore in the memory kernel). This suggests that, in a system with the density as a slow variable, a more general theory can be derived by projecting the microscopic dynamics of the system onto the set of all nonlinear functions of the density \cite{Kawasaki2009}, which is precisely what is done in the derivation of (stochastic) DDFT in the (microcanonical) projection operator formalism \cite{Munakata2003}. The assumption of a slow density is motivated by the fact that, in dense liquids, particle motion is restricted, whereas momentum can be transferred very quickly \cite{FuchizakiK2002}.
 
Physically, MCT has the interesting feature that it predicts a transition from an ergodic to a nonergodic state. This is related to its role as a theory of the glass transition: Glasses form when a liquid is cooled below its freezing temperature, but does not freeze into a crystal, which would be the thermodynamic equilibrium state (\ZT{equilibration}), but into some other state \cite{Kawasaki2009}. Systems in glassy states have a local structure that looks like that of a fluid. However, the viscosity is much larger. The formation of glasses has a kinetic rather than a thermodynamic origin \cite{Loewen1994a}, which is reflected by the fact that their dynamical behavior depends strongly on their history and the way they were prepared (\ZT{aging}). A simple picture for the origin of glass formation is that, in dense supercooled liquids, particles get trapped in cages and thus cannot move to their equilibrium positions. These geometrical constraints, which prevent the system from accessing the complete phase space, make the system nonergodic.

However, the actual physics of the glass transition is more complicated. Real materials are always subject to thermal fluctuations and these fluctuations have the consequence that there is no ideal ergodic-to-nonergodic transition. To overcome this problem, extensions of MCT have been developed (see Ref.\ \cite{Das2004} for an overview). The shortcomings of the simple MCT motivated the development of the stochastic DDFT by \citet{Kawasaki1994} (see \cref{kawasaki}), in which thermal fluctuations are explicitly included. Stochastic DDFT allows for the derivation of MCT as a limiting case. Details are discussed below.

A somewhat different view on the relation of DDFT and MCT can be found in the literature on deterministic DDFT, where DDFT and MCT are viewed as different limiting cases of a general exact description. For example, \citet{WittkowskiLB2012} compared, within the framework of EDDFT (see \cref{eddft}), the derivation of DDFT and MCT via the projection operator formalism (see \cref{derivationmz}) and noted that the derivations involve different approximations. In (E)DDFT, it is assumed that the ensemble-averaged one-body density (and possible additional variables) follow closed memoryless transport equations (Markovian approximation). In MCT, on the other hand, one makes a linearization, but does not assume the dynamics to be Markovian (since MCT contains a memory term). A similar conclusion was reached by \citet{BraderS2013} within the framework of PFT (see \cref{pft}), who argued that DDFT is obtained by neglecting superadiabatic contributions to the power dissipation, whereas MCT corresponds to the assumption that these contributions have a certain form.

While this is certainly a topic that deserves further investigation, it should be noted here that the views \ZT{stochastic DDFT allows to derive MCT} and \ZT{deterministic DDFT and MCT are complementary} do not necessarily have to be in conflict. To see this, consider the DDFT equation \eqref{dk} obtained by \citet{Dean1996}. This is an \textit{exact} description of a system of Brownian particles, from which the deterministic DDFT can, as shown by \citet{MarconiT1999}, be derived with additional approximations. On the other hand, since MCT for colloidal fluids can also be derived from the system of Langevin equations that is equivalent to \cref{dk}, it follows that MCT and deterministic DDFT are complementary approaches that both arise as limiting cases of the (exact) stochastic DDFT by Dean. 

An additional point that needs to be taken into account is that, as discussed in \cref{kawasaki,noise}, stochastic DDFT can be formulated for the exact microscopic density operator \eqref{hatrho} (defined as a sum over Dirac delta distributions) or for a coarse-grained density. This also affects an MCT derived from these theories. If the stochastic DDFT describes a coarse-grained density, this will also be the density that enters the correlator \eqref{correlator} (an example is the derivation by \citet{Archer2006} presented below). \citet{Kawasaki1994} explicitly compares his derivation of MCT from stochastic DDFT to the derivation of MCT from the Smoluchowski equation by \citet{SzamelL1991} and notes as a crucial difference that the former employs a continuum description. 
A derivation of an MCT for the correlator \eqref{correlator} with $\rho$ given by the microscopic density operator $\hat{\rho}$ is possible by starting from a DDFT for the microscopic density \cite{KimKJvW2014}. The relations between the different forms of DDFT and MCT are visualized in \vref{fig:schematicglas}.

The problems regarding the relation to MCT discussed in the literature on deterministic DDFT do not necessarily apply to derivations from stochastic DDFT. As discussed by \citet{WittkowskiLB2012}, Archer's result concerns the temporally coarse-grained density, while the derivation of MCT from the projection operator formalism starts from a microscopic theory. In the literature on deterministic DDFT, \ZT{MCT} typically denotes a dynamic equation for the correlator of the microscopic density while \ZT{DDFT} denotes an adiabatic transport equation for the ensemble-averaged density. This then leads to an incompatibility that is not necessarily present if \ZT{MCT} and \ZT{DDFT} are understood in a different way.

One of the problems in combining DDFT and MCT, arising in the microscopic derivation via the projection operator formalism, is that DDFT neglects memory effects while MCT does not \cite{WittkowskiLB2012}. As discussed in detail in the textbook on projection operators by \citet{Grabert1982}, the question whether or not memory effects in the transport equations for the macroscopic variables can be neglected depends on how large the set of macroscopic variables is. In particular, a Fokker-Planck equation (as arising in stochastic DDFT) for a dynamical variable can be Markovian even if the equation for its mean value (as given by deterministic DDFT) contains memory effects. In the derivation of such a Fokker-Planck equation, one projects onto all nonlinear functions of the relevant variables and assumes these to be slow \cite{Grabert1982}. In contrast, the derivation of MCT uses only the density itself as a relevant variable, such that quadratic functions of the density contribute to the memory kernel. If these are also slow, it is, of course, inappropriate to ignore the memory effects, while it still can be appropriate for a Fokker-Planck equation that includes the quadratic functions.

Regarding the PFT-based argument, it is important to note that not all forms of stochastic DDFT employ the adiabatic approximation of deterministic DDFT. Since, by the terminology of PFT, everything that is not included in the adiabatic dynamics of deterministic DDFT is a \ZT{superadiabatic force}, it is thus possible that stochastic DDFT allows to model phenomena that would be classified as \ZT{superadiabatic} in PFT. Consequently, the positions of \citet{Kawasaki1994}, \citet{Archer2006}, \citet{WittkowskiLB2012}, and \citet{BraderS2013} are compatible, even though they superficially appear to be incompatible.

In addition, DDFT and MCT can be combined in the microscopic calculation of certain properties of complex fluids. This has been done, e.g., for ionic systems \cite{ChandraB1999,ChandraB2000,ChandraB2000b,ChandraB2000c,DufrecheJTB2008,DufrecheB2002,Bagchi2014,RoyYB2015}, friction in supercooled liquids \cite{BhattacharyyaB1997,Bagchi1994}, and orientational relaxation \cite{Bagchi2001,JanaB2012}. We discuss the case of ions in \cref{electro}.

The aforementioned derivation of MCT from stochastic DDFT can be done in at least three ways. First, it was, with certain approximations, derived by \citet{Kawasaki1994} using projection operators (see also Ref.\ \cite{Kawasaki1995b}) via a similar procedure as in the derivation by \citet{SzamelL1991} from the Smoluchowski equation. Second, it can, with certain approximations, be obtained by renormalized perturbation theory at one-loop order \cite{KimK2008}. Mode coupling effects here take the form of a lifetime renormalization \cite{Das2004,Das2011}. The full one-loop-order result, however, goes beyond MCT, and additional approximations are required to recover it \cite{KimKJvW2014} (see \cref{renormalization}). A third derivation, performed by \citet{Archer2006} using ideas by \citet{Kawasaki1995} and \citet{ZaccarelliFDGSTD2002}, employs DDFT for atomic fluids (see \cref{inertia}) and is presented here. It is less rigorous than the other types of derivation, but also much simpler. Starting from a DDFT for the coarse-grained density, it gives an MCT equation for the correlator of the coarse-grained density.

One starts from the exact microscopic dynamics and then performs a temporal rather than a spatial coarse graining. The temporally coarse-grained density is
\begin{equation}
\rho\rt=\INT{-\infty}{\infty}{t'}K^{\mathrm{t}}(t-t')\hat{\rho}(\vec{r},t'),
\end{equation}
where $K^{\mathrm{t}}(t)$ is a normalized function that has finite width. After temporal coarse graining, one obtains an equation of the form \eqref{inertiaexact}. Making the approximation $B\approx\nu\dot{\rho}$ leads to an equation that looks like \cref{atomic}. (Strictly speaking, a dynamic equation for the coarse-grained density should contain thermal noise \cite{ArcherR2004}. However, thermal noise is not relevant for MCT \cite{Kawasaki1995}.) Performing a Taylor expansion of the free energy in $\Delta\rho\rt=\rho\rt-\rho_0$ and then a spatial Fourier transformation gives
\begin{equation}
\ddot{\rho}_{\vec{k}}(t) + \nu\dot{\rho}_{\vec{k}}(t) + \frequenz^2(\vec{k})\rho_{\vec{k}}(t)=\QuelltermMCT(\vec{k},t),
\label{rhoq}%
\end{equation}
where $\rho_{\vec{k}}(t)$ is the Fourier transformation of $\Delta\rho\rt$, $\frequenz^2(\vec{k}) = k_B T k^2/(m S(\vec{k}))$ with the static structure factor $S(\vec{k})=1/(1-\rho_0 c(\vec{k}))$, $c(\vec{k})$ is the Fourier transformation of the direct correlation function, and 
\begin{equation}
\QuelltermMCT(\vec{k},t) = \frac{k_B T}{(2\pi)^3m}\INT{}{}{^3k'}\vec{k}\cdot\vec{k}'\rho_{\vec{k}'}(t)c(\vec{k}')\rho_{\vec{k}-\vec{k}'}(t).
\label{hatr}%
\end{equation}
This is compared to the general form \eqref{mcteqofmotion} that can be obtained from the projection operator formalism. The memory kernel $\memory(\vec{k},t)$ depends on the time correlation of the random force. Assuming that one can use $\QuelltermMCT$, given by \cref{hatr}, (rather than the exact random force obtained from the projection operator formalism) for calculating the memory kernel $\memory(\vec{k},t)$, and making the usual factorization approximation (see \cref{mct}), gives
\begin{equation}
\memory(\vec{k},t)=\frac{k_B T \rho_0 }{2(2\pi)^3mk^2}\INT{}{}{^3k'}(\vec{k}\cdot\vec{k}'c(\vec{k}') + \vec{k}\cdot(\vec{k}-\vec{k}') c(\vec{k}-\vec{k}'))^2S(\vec{k}')S(\vec{k}-\vec{k}')\phi(\vec{k}',t)\phi(\vec{k}-\vec{k}',t),
\end{equation}
which is the MCT expression \eqref{mctexpression} (although derived here for the correlator of the coarse-grained density). Another discussion of the relation between DDFT and MCT can be found in Ref.\ \cite{Archer2009}.

\subsection{\label{net}Nonequilibrium thermodynamics}
The general structure of DDFT is shared by a larger class of close-to-equilibrium dynamical theories known as \ZT{gradient dynamics}. Here, the time evolution of (a set of) conserved or nonconserved variables is driven by a thermodynamic potential. A general form of gradient dynamics for a conserved variable $a^\mathrm{c}$ is
\begin{equation}
\pdif{}{t}a^\mathrm{c}\rt=\Nabla\cdot\bigg(\MobilityGradientDynamics(a^\mathrm{c}\rt)\Nabla\frac{\delta F[a^\mathrm{c}]}{\delta a^\mathrm{c}\rt}\bigg)
\label{gradientdynamics}
\end{equation}
with the positive mobility function $\MobilityGradientDynamics$ \cite{Thiele2018,ThieleFHEKA2019,LyMTG2020,LyTCG2019,ThieleH2019}. If $a^\mathrm{c}$ is a vector containing multiple variables, $\MobilityGradientDynamics$ is a matrix that is positive definite \cite{XuTQ2015,TrinschekJT2018,Thiele2018}. The DDFT equation \eqref{trddft} is a special case of \cref{gradientdynamics} corresponding to $a^\mathrm{c} = \rho$ and $\MobilityGradientDynamics=\Gamma\rho$. Theories of this form can, in nonequilibrium thermodynamics, be derived from Onsager's principle \cite{Onsager1931,Onsager1931b}. Onsager's linear thermodynamics can also be used to obtain the DDFT equation \cite{ChavanisS2011b}. A further discussion of Onsager's variational principle can be found in Refs.\ \cite{Doi2011,Chavanis2008b,Uneyama2020}. The structure of gradient dynamics thus relates DDFT to other nonequilibrium theories \cite{XuTQ2015,KraaijLMP2018,HonischLHTG2015,Thiele2011,ThieleAP2016}. 

A very simple example is the Cahn-Hilliard equation describing the dynamics of a conserved order parameter field $\FieldCahnHilliard\rt$, which is obtained from \cref{gradientdynamics} by using a constant mobility function $\MobilityCahnHilliard$ and a free energy \cite{CahnH1958}
\begin{equation}
F = \INT{}{}{^3r}\bigg(\frac{\CPCahnHilliard}{2}(\Nabla \FieldCahnHilliard)^2 + \FEDCahnHilliard(\FieldCahnHilliard)\bigg),
\label{cahnhilliardfreeenergy}
\end{equation}
where $\CPCahnHilliard > 0$ is a constant. The free energy \eqref{cahnhilliardfreeenergy} consists of a term penalizing gradients and the free energy density $\FEDCahnHilliard$ of a homogeneous solution, which is often approximated as a fourth-order polynomial \cite{WilczekG2014,LyTCG2019}. Combining \cref{gradientdynamics,cahnhilliardfreeenergy} gives the dynamic equation \cite{Cahn1965}
\begin{equation}
\pdif{}{t}\FieldCahnHilliard = \MobilityCahnHilliard \Nabla^2\bigg(\frac{\partial \FEDCahnHilliard(\FieldCahnHilliard)}{\partial \FieldCahnHilliard}-\CPCahnHilliard\Nabla^2\FieldCahnHilliard  \bigg). 
\label{cahnhilliardequation}
\end{equation}
Equation \eqref{cahnhilliardequation} is a simple model equation for spinodal decomposition \cite{Cahn1965}. As is easily confirmed by a linear stability analysis (see \cref{linear}), a homogeneous state $\FieldCahnHilliard\rt = \ConstantCahnHilliard$, where $\ConstantCahnHilliard$ is a constant, becomes unstable for $(\partial^2 \FEDCahnHilliard / \partial \FieldCahnHilliard^2)|_{\FieldCahnHilliard = \ConstantCahnHilliard} < 0$. This result has inspired later work on spinodal decomposition in DDFT \cite{Evans1979,ArcherE2004}, which is discussed in \cref{linear,ps}. Similar methods have also gained importance for other types of soft and active matter models \cite{BickmannW2019b,BickmannW2020,WittkowskiTSAMC2014,TiribocchiWMC2015}.

Within linear irreversible thermodynamics, equations of motion can be derived from a dissipation functional. This was discussed by \citet{WittkowskiLB2013} for the case of (extended) DDFT. For general conserved variables $a^{\mathrm{c}}_i\rt$ and nonconserved variables $a^{\mathrm{n}}_i\rt$, the \textit{thermodynamic forces} are given by
\begin{align}
\vec{a}_i^{\mathrm{c}\sharp}\rt&=-\Nabla\frac{\delta F[\{a_k\}]}{\delta a^{\mathrm{c}}_i\rt},\label{acsharp}\\
a_i^{\mathrm{n}\sharp}\rt&=\frac{\delta F[\{a_k\}]}{\delta a^{\mathrm{n}}_i\rt}.
\end{align}
Here, the superscript $\sharp$, not to be confused with the related symbol $\natural$ introduced in \cref{derivationmz}, denotes a thermodynamic force, which has a different form for conserved and nonconserved variables. The equations of motion for the relevant variables are given by
\begin{align}
\pdif{}{t}a_i^{\mathrm{c}}\rt + \Nabla\cdot(\vec{J}_{i,R}\rt + \vec{J}_{i,D}\rt) &= 0,\label{dotac}\\
\pdif{}{t}a_i^{\mathrm{n}}\rt + \Phi_{i,R}\rt + \Phi_{i,D}\rt &= 0\label{dotanc}
\end{align}
with the currents $\vec{J}_i$ and quasi-currents $\Phi_i$ that have a reversible and a dissipative contribution, denoted by subscripts $R$ and $D$, respectively. From the dissipation functional
\begin{equation}
\mathfrak{R}[\{a_i, a_i^{\sharp}\}]=\INT{}{}{^3r}\mathfrak{r}\rt,
\end{equation}
giving the energy dissipated per unit time and defined by the integral over the dissipation function $\mathfrak{r}\rt$, one can obtain the dissipative currents and quasi-currents as
\begin{align}
\vec{J}_{i,D}\rt &= \frac{\delta \mathfrak{R}[\{a_i, a_i^{\sharp}\}]}{\delta \vec{a}_i^{\mathrm{c}\sharp}\rt},\label{jid}\\
\Phi_{i,D}\rt &= \frac{\delta \mathfrak{R}[\{a_i, a_i^{\sharp}\}]}{\delta a_i^{\mathrm{n}\sharp}\rt}.
\end{align}
General expressions for the dissipation function are derived in Ref.\ \cite{WittkowskiLB2013}. In particular, the standard DDFT equation \eqref{trddft} can be obtained from the dissipation functional
\begin{equation}
\mathfrak{R}[\rho,\vec{\rho}^\sharp]=\INT{}{}{^3r}\frac{1}{2}\Gamma\rho\rt(\vec{\rho}^\sharp\rt)^2,
\label{ddftdissipation}
\end{equation}
which, together with \cref{acsharp,dotac,jid}, constitutes a reformulation of DDFT. The thermodynamic force, defined by \cref{acsharp}, is given by $\vec{\rho}^\sharp = - \Nabla \delta F /\delta\rho$. A further discussion of the relation of DDFT-based models to linear irreversible thermodynamics can be found in Ref.\ \cite{WittkowskiLB2011b}.

\subsection{\label{phasefield}Phase field crystal models}
Phase field crystal (PFC) models are another example of theories with a gradient dynamics structure. They arise as a limiting case of DDFT upon assuming a constant mobility and making certain approximations for the free energy \cite{ArcherRRS2019} (see below for a detailed discussion). Note that the line of demarcation between DDFT and PFC is drawn in different ways by different authors, sometimes models with a nonconstant mobility are also called \ZT{PFC models} \cite{Loewen2010}. In standard PFC models, a system is described in terms of a dimensionless conserved order parameter field $\psi\rt$ governed by the dynamic equation
\begin{equation}
\pdif{}{t}\psi = D \Nabla^2\frac{\delta F}{\delta\psi}.
\end{equation}
PFC models were first proposed phenomenologically \cite{ElderKHG2002,ElderG2004,BerryGE2006} and then connected to DFT \cite{ElderPBSG2007} and DDFT \cite{vanTeeffelenBVL2009}. One can give PFC models a microscopic foundation by deriving them as an approximation to DDFT. This allows to connect the parameters of PFC models to the parameters of the microscopic dynamics \cite{BackofenEV2014,MkhontaEH2016} and to give a physical interpretation to the PFC order parameter field $\psi$ \cite{RobbinsATK2012} by identifying it as a dimensionless form of the deviation of the density $\rho$ from a reference value. In this sense, PFC models are an intermediate theory between atomistic models (such as DDFT) and macroscopic continuum theories \cite{SalvalaglioAHVEV2020,SalvalaglioBVE2017}. PFC models are less accurate, but also simpler. (Some authors classify PFC models as a form of DDFT \cite{TothTPTG2010}. We do not adapt this terminology here.)

Applications of PFC models, such as pattern formation or solidification, frequently overlap with those of DDFT. This allows for a comparison of both methods. An example is the study of vacancy diffusion in colloidal crystals by \citet{vanTeeffelenAL2013}, who compared results of both DDFT and PFC models to BD simulations. It was found that DDFT correctly predicts the temperature dependence of the diffusion constant, whereas modifications are required in PFC models.

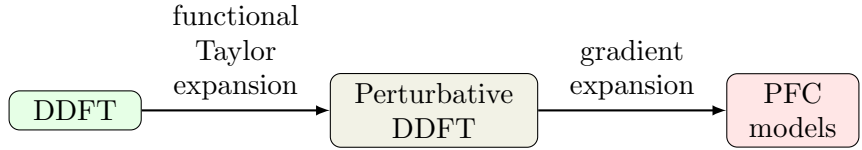
\begin{figure}[htb]
\centering
\TikzBilderNeuErzeugen{\tikzstyle{begriffsfeldI} = [rectangle, rounded corners, minimum width=5mm, minimum height=5mm, text width=15mm, text centered, draw=black, fill=green!10]
\tikzstyle{begriffsfeldII} = [rectangle, rounded corners, minimum width=5mm, minimum height=5mm, text width=25mm, text centered, draw=black, fill=green!50!red!10]
\tikzstyle{begriffsfeldIII} = [rectangle, rounded corners, minimum width=5mm, minimum height=5mm, text width=15mm, text centered, draw=black, fill=red!10]
\tikzstyle{arrow} = [thick,->,>=latex]
\begin{tikzpicture}[node distance=25mm]
\node(K0)[begriffsfeldI]{DDFT};
\node(K1)[begriffsfeldII, right=of K0]{Perturbative DDFT};
\node(K2)[begriffsfeldIII, right=of K1]{PFC models};
\draw[arrow](K0) -- node[anchor=south, text width=20mm, text centered]{functional Taylor expansion} (K1);
\draw[arrow](K1) -- node[anchor=south, text width=20mm, text centered]{gradient expansion} (K2);
\end{tikzpicture}}%
\TikzBildEinfuegen{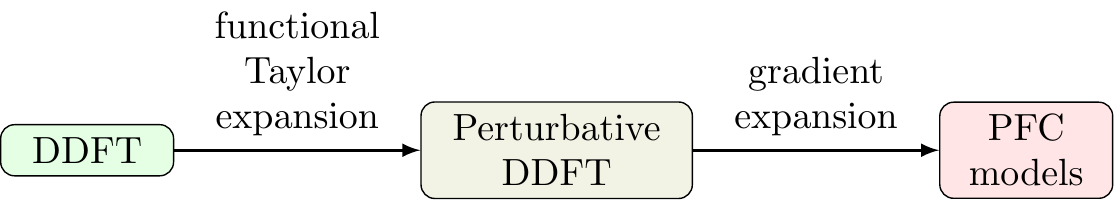}{}%
\caption{\label{fig:PFC}Derivation of PFC models from DDFT.}%
\end{figure}
The derivation of PFC models is visualized in \vref{fig:PFC}. Following Ref.\ \cite{EmmerichEtAl2012}, we introduce the order parameter field of PFC models $\psi$ as a dimensionless parametrization of deviations from a constant reference density $\rho_0$ in the form
\begin{equation}
\rho(\vec{r},t) = \rho_0(1+\psi(\vec{r},t)).
\end{equation}
This parametrization is inserted into the free energy \eqref{freeenergy}, which allows for three main approximations \cite{ArcherRRS2019}:
\begin{enumerate}
\item Assuming small deviations from the reference density, a functional Taylor expansion of $F$ in $\psi$ is performed.
\item A Ramakrishnan-Yussouff or random phase approximation is made for the free energy functional.
\item The functional is made local using a gradient expansion.
\end{enumerate}
\FloatBarrier
The Taylor expansion of the ideal gas free energy gives (dropping irrelevant constants)
\begin{equation}
F_{\mathrm{id}}[\psi] = \rho_0k_B T \INT{}{}{^3r}\bigg(\psi + \frac{\psi^2}{2}-\frac{\psi^3}{6}+\frac{\psi^4}{12}\bigg),
\label{fidpfc}%
\end{equation}
where the expansion is performed up to fourth order to allow for the formation of stable crystals. The excess free energy has, after performing the Ramakrishnan-Yussouff approximation, the form of an integral over a convolution. To make this functional local, we assume translational and rotational invariance to write the direct pair-correlation function as
\begin{equation}
c^{(2)}(\vec{r}_1,\vec{r}_2) = c^{(2)}(\norm{\vec{r}_1-\vec{r}_2}) = c^{(2)}(r)
\end{equation}
and then, exploiting the symmetries, perform a Taylor expansion of the Fourier transformed direct pair-correlation function $c(\vec{k})$ around the wavevector $\vec{k}=\vec{0}$ in the form
\begin{equation}
c(\vec{k}) = c_0 + c_2k^2 + c_4k^4
\end{equation}
with expansion coefficients $c_0$, $c_2$, and $c_4$.
Again, we have stopped the expansion at fourth order. After an inverse Fourier transformation, the expansion in wavevectors $\vec{k}$ becomes an expansion in gradients. Omitting irrelevant constants, the excess free energy becomes
\begin{equation}
F_{\mathrm{exc}}[\psi] = - \frac{\rho_0}{2}k_B T\INT{}{}{^3r}(\mathcal{A}_1 \psi^2 + \mathcal{A}_2 \psi\vec{\nabla}^2 \psi + \mathcal{A}_3\psi\vec{\nabla}^4\psi),
\label{fexcpfc}%
\end{equation}
where the expansion coefficients $\{\mathcal{A}_i\}$ are moments of the direct pair-correlation function. Combining \cref{fidpfc,fexcpfc} and rescaling gives the complete PFC free energy
\begin{equation}
F_{\mathrm{pfc}}[\psi]=\rho_0k_B T\INT{}{}{^3r}\bigg(\mathcal{A}_1'\psi^2 + \mathcal{A}_2'\psi\Nabla^2\psi+ \mathcal{A}_3'\psi\Nabla^4\psi - \frac{\psi^3}{6} + \frac{\psi^4}{12}\bigg)
\end{equation}
with the rescaled expansion coefficients $\{\mathcal{A}_i'\}$. Making the further approximation that the mobility is constant and equal to $D \rho_0$ gives the equation of motion
\begin{equation}
\pdif{}{t}\psi= D\Nabla^2\frac{\delta F_{\mathrm{pfc}}[\psi]}{\delta\psi}
\end{equation}
for the PFC model.

The approximations made in the derivation of PFC models from DDFT have some consequences for the structure and predictions of the models. These were discussed in detail by \citet{ArcherRRS2019}. They identify dropping a term of the form 
\begin{equation}
\vec{\nabla}\cdot(\psi\vec{\nabla}(\mathcal{L}\psi))
\label{essentialterm}%
\end{equation}
with the nonlocal operator $\mathcal{L}$ defined by $\mathcal{L}\psi\rt = - \psi\rt+\rho_0\TINT{}{}{^3r'}c^{(2)}(\vec{r},\vec{r}')\psi(\vec{r}',t)$, along with the assumption of constant mobility and an expansion of the logarithm in the ideal gas free energy, as the central approximation that sets the border between DDFT and PFC models. The reason is that, after making the substitution $\rho=\rho_0(1+\psi)$, performing a functional Taylor expansion of the free energy, truncating at fourth order, and taking zero-wavelength components of third- and fourth-order correlation functions (following the derivation of PFC models by \citet{HuangEP2010}), one obtains 
\begin{equation}
\pdif{}{t}\psi = -\Nabla^2(\mathcal{L}\psi + C_a\psi^2 + C_b\psi^3 + C_c\psi^4) - \vec{\nabla}\cdot(\psi\vec{\nabla}(\mathcal{L}\psi)),
\label{ddft2}%
\end{equation}
where $C_a$, $C_b$, and $C_c$ are constants. Dropping the term \eqref{essentialterm} in \cref{ddft2} gives an equation of the form
\begin{equation}
\pdif{}{t}\psi = \Nabla^2\frac{\delta \FpfcII[\psi]}{\delta \psi}
\label{ddft2mod}%
\end{equation}
with some functional $\FpfcII[\psi]$. Deriving an equation of the form \eqref{ddft2mod} from DDFT requires the assumption of a constant mobility. However, this assumption is incompatible with an ideal gas free energy of the form \eqref{idealfreeenergy}, since it would lead to
\begin{equation}
\pdif{}{t}\rho\rt = D\Nabla\cdot\bigg(\frac{1}{\rho\rt}\Nabla\rho\rt\bigg)
\end{equation}
instead of the usual diffusion equation \eqref{diffusionequation} in the noninteracting case (the factor $\rho$ from the mobility would usually cancel the $1/\rho$ resulting from the logarithm). Hence, dropping \eqref{essentialterm} is necessarily connected to the constant mobility assumption and the expansion of the logarithm. This expansion does not come without a price: It leads to a second spinodal point in the phase diagram, such that PFC models have an unphysical phase behavior. It predicts that structures formed at high densities melt again to a uniform state at even higher densities.

In Ref.\ \cite{OettelDBNS2012}, DFT and PFC approaches are compared at various stages of the approximation using hard-sphere simulations. \citet{TupperG2008} interpreted PFC models as the result of coarse graining molecular dynamics in time. A further discussion of the coarse-graining chain leading from microscopic dynamics to DDFT and then to PFC models using the projection operator method (see \cref{po}) can also be found in Ref.\ \cite{MajaniemiPN2010}. The derivation of PFC models is moreover presented in Refs.\ \cite{BackofenRV2007,EmmerichGL2011,ArcherRTK2012}. 

In Refs.\ \cite{HuangEP2010,TahaDMEH2019}, a binary PFC model is derived from DDFT. \citet{Loewen2010} has derived a PFC model for liquid crystals (consisting of particles with orientational degrees of freedom) from the corresponding DDFT. The derivation of an eighth-order PFC model can be found in Ref.\ \cite{JaatinenA2010}. \citet{PraetoriusV2011} derived an advected PFC model from an advected DDFT. The dissipative dynamics of orientable particles was moreover derived by \citet{WittkowskiLB2011b} based on DDFT and compared to macroscopic Ginzburg-Landau theories, building upon the DFT-based discussions in Refs.\ \cite{WittkowskiLB2010,WittkowskiLB2011}. The nematic tensor $Q_{ij}$ can also be included as an order parameter in PFC models \cite{PraetoriusVWL2013,TangPBVYW2015}. An \ZT{anisotropic phase field crystal model} for condensed matter systems built from oriented nonspherical particles, proposed by \citet{PrielerHLVHE2009} (see also Ref.\ \cite{PrielerLE2010}), was derived from DDFT by \citet{ChoudharyLEL2011}. 

Finally, PFC models can also be developed for active particles. The first theory of this type, proposed by \citet{MenzelL2013}, was derived from active DDFT (see \cref{active}) by \citet{MenzelOL2014}. It reads
\begin{align}
\pdif{}{t}\psi &= \vec{\nabla}^2\frac{\delta F}{\delta \psi} - v_0 \vec{\nabla}\cdot\vec{P},\label{activepfc1}\\
\pdif{}{t}\vec{P} &= \vec{\nabla}^2\frac{\delta F}{\delta \vec{P}} - D_R \frac{\delta F}{\delta \vec{P}}- v_0 \vec{\nabla}\psi\label{activepfc2}
\end{align}
with the self-propulsion velocity $v_0$, polarization $\vec{P}\rt$ arising from an orientational expansion (see \cref{nonspherical}), and rotational diffusion constant $D_R$. Microscopically, the system of active particles is described by a DDFT for the orientation-dependent density $\varrho(\vec{r},\uu,t)$ (see \cref{nonspherical,active}). The derivation of \cref{activepfc1,activepfc2} from the microscopic DDFT for active particles employs a Cartesian expansion of the density $\varrho$ in the form \eqref{fu} (see \cref{nonspherical}), which gives $\psi$ and $\vec{P}$ as zeroth and first orders, respectively. In future work, this derivation could be extended to active PFC models on a sphere \cite{PraetoriusVWL2018} starting from a DDFT for particles on a spherical surface \cite{BottB2016}. A further discussion of active PFC models can be found in Refs.\ \cite{AlaimoPV2016,AlaimoV2018,OphausGT2018,OphausKGT2020}.

\subsection{\label{nescgle}Nonequilibrium self-consistent generalized Langevin equations}
The theory of nonequilibrium self-consistent generalized Langevin equations (NE-SCGLE), derived by \citet{RamirezGonzalezM2010}, is a nonequilibrium generalization of the self-consistent generalized Langevin equations (SCGLE) \cite{YeomansM2001} for colloids. It is derived from a nonequilibrium extension of the Onsager theory \cite{Medina2009} and describes the relaxation dynamics of colloidal systems based on the diffusion equation
\begin{equation}
\pdif{}{t}\rho\rt = \frac{D}{k_B T}\Nabla\cdot\big( b\rt\rho\rt\Nabla\mu(\vec{r},[\rho])\big)
\label{nescglee}
\end{equation}
for the local concentration $\rho$ with the local reduced mobility $b\rt$ describing local frictional effects. The standard DDFT equation \eqref{trddft} is recovered when setting $b\rt = 1$. Equation \eqref{nescglee} is coupled to the equation 
\begin{equation}
\begin{split}
\pdif{}{t}\covariance(\vec{r},\vec{r}',t) &=D\Nabla_{\vec{r}} \cdot\bigg(\rho\rt b\rt\Nabla_{\vec{r}}
\INT{}{}{^3r''}\varepsilon(\vec{r},\vec{r}'',[\rho])\covariance(\vec{r}'',\vec{r}',t)\bigg)\\
&\quad\:\!+D\Nabla_{\vec{r}'}\cdot\bigg(\rho(\vec{r}',t) b(\vec{r}',t)\Nabla_{\vec{r}'}
\INT{}{}{^3r''}\varepsilon(\vec{r}',\vec{r}'',[\rho])\covariance(\vec{r}'',\vec{r},t)\bigg)\\
&\quad\:\!-2D\Nabla_{\vec{r}}\cdot\big(\rho\rt b\rt\Nabla_{\vec{r}}\delta(\vec{r}-\vec{r}')\big)
\end{split}    
\end{equation}
for the covariance $\covariance(\vec{r},\vec{r}',t)$ with the thermodynamic matrix 
\begin{equation}
\varepsilon(\vec{r},\vec{r}',[\rho]) = \frac{1}{k_B T}\frac{\delta\mu(\vec{r},[\rho])}{\delta \rho(\vec{r}',t)}.
\end{equation}
NE-SCGLE theory has also been extended to multicomponent systems \cite{SanchezLOGMN2014}. It is mainly used for glass-forming liquids, allowing to describe both equilibration \cite{SanchezRM2013b} and aging \cite{RamirezGonzalezM2010b} and in particular the crossover from the equilibration to the aging regime \cite{SanchezRM2013,MendozaLSRPM2017}. Moreover, it has been applied to arrested spinodal decomposition \cite{OlaisLM2015}.

\subsection{\label{dmft}Dynamic mean field theory}
A discrete theory that has a close relationship to DDFT is \textit{dynamic mean field theory} (DMFT) or \textit{mean-field kinetic theory} (not to be confused with the \textit{mean-field dynamical density functional theory} \cite{DzubiellaL2003} or \textit{dynamic mean-field density functional theory} \cite{FraaijevVMPEHAGW1997}). This method, based on work by \citet{Martin1990} and \citet{Penrose1991}, describes the dynamics of particles on a lattice. It has been interpreted as a form of DDFT \cite{Monson2008,EdisonM2010}, which is motivated by formal analogies discussed below. The \ZT{time-dependent density functional theory} for lattice gases \cite{ReinelD1996,FischerRDGM1998,KesslerDFGM2002,GouyetPDM2003,DierlME2011,DierlME2012,HeinrichsDMF2004,DierlEM2013}, which has been developed as a time-dependent extension of classical DFT, can also be seen as a form of DMFT \cite{EinaxSDN2010}. Applications of DMFT include fluids in porous materials \cite{Monson2008,Monson2008b,Monson2009,Monson2011,EdisonM2013,KierlikLRT2011,LeoniKRT2011,RathiKFM2019}, wetting \cite{EdisonM2009,EdisonM2010,EdisonM2014}, evaporation \cite{CasselmanDM2015}, diffusion \cite{MatuszakAD2004,MatuszakAD2006}, and gelation \cite{WitmanW2006}. Moreover, DMFT has been extended to hydrodynamic interactions \cite{KikkinidesM2015}. A review of lattice methods of this type can be found in Ref.\ \cite{GouyetPDM2003}. Here, we focus on the relation to DDFT, which is discussed for DMFT in Ref.\ \cite{Monson2008}:

Lattice DFT can be used to describe equilibrium states of fluids based on a free energy $F(\{\rho_k\})$ that depends on the mean density at the individual lattice sites in the mean-field approximation. We denote by $\rho_i$ the mean density at lattice site $i$ \cite{NieswandMD1993} and assume a 1D lattice for simplicity. The exchange dynamics conserves the total particle number (Kawasaki dynamics \cite{Kawasaki1966b}), such that the dynamics is given by
\begin{equation}
\pdif{}{t}\rho_i(t) = - \sum_{j=i\pm 1}J_{ij}(t)
\label{kawasakidynamics}
\end{equation}
with $J_{ij}$ being the net flux from site $i$ to site $j$. In the mean-field approximation, the flux is given by
\begin{equation}
J_{ij}=w_{ij}\rho_i(1-\rho_j) - w_{ji}\rho_j(1-\rho_i)
\end{equation}
with the transition probability $w_{ij}$. Defining the (symmetric) mobility as
\begin{equation}
M_{ij} = - \frac{J_{ij}}{\mu_j - \mu_i}
\end{equation}
with the chemical potential $\mu_i$ corresponding to $\rho_i$, the flux can be written as
\begin{equation}
J_{ij}= - M_{ij}(\mu_j - \mu_i).
\label{latticeflux}
\end{equation}
Inserting \cref{latticeflux} into \cref{kawasakidynamics} gives
\begin{equation}
\begin{split}
\pdif{}{t}\rho_i &= \sum_{a=\pm 1}M_{i,i+a}(\mu_{i+a}-\mu_{i})\\& =\frac{1}{2}\sum_{a=\pm 1}\big(M_{i,i+a}(\mu_{i+a}-\mu_{i}) + M_{i,i-a}(\mu_{i-a}-\mu_{i})\big)
\label{symmetric},
\end{split}
\end{equation}
where we have assumed the lattice to have a center of symmetry. With the forwards/backwards-difference operators $\DifferenceOperator_{\pm a}$, defined as
\begin{equation}
\DifferenceOperator_{\pm a}f_i =\pm f_{i \pm a} \mp f_i,
\end{equation}
\cref{symmetric} can be written as 
\begin{equation}
\pdif{}{t}\rho_i = \frac{1}{2}\sum_{a=\pm 1}\DifferenceOperator_{-a}M_{i,i+a}\DifferenceOperator_{a}\mu_i
= \frac{1}{2}\sum_{a=\pm 1}\DifferenceOperator_{-a}M_{i,i+a}\DifferenceOperator_{a}\pdif{F}{\rho_i},
\end{equation}
which has the form of a DDFT equation (the free energy $F$ is defined in discretized form).

\section{\label{methods}Analyzing a DDFT}
After having presented the construction of DDFT (\cref{traditional}), its extensions (\cref{extensions,exact}), and its relation to other theories (\cref{theories}), we now proceed to the more practical problems of solving the dynamic equation(s) of DDFT. This will be (roughly) done in order of increasing generality: First, steady solutions are discussed in \cref{steadycurrents}. Second, we introduce dynamics in \cref{perturbation} by explaining systematic perturbative expansions. Third, numerical methods are presented in \cref{numeric}, which allow to study the full dynamics of DDFT.

\subsection{\label{steadycurrents}Steady solutions}
In many cases, the simplest way to study a dynamic equation is to consider its steady (in the simplest case stationary) solutions. In the case of the deterministic DDFT equation \eqref{trddft}, which describes the ensemble-averaged one-body density (see \cref{noise}), the stationary solution -- at least for an isolated system -- is given by the solutions of the DFT equation \eqref{dft}. This reflects the fact that, as discussed in \cref{thermo}, DDFT describes the approach to thermodynamic equilibrium. DFT, which is discussed in detail in \cref{static}, describes the equilibrium state of the ensemble-averaged density and is therefore the equilibrium solution of deterministic DDFT. For a stochastic DDFT such as \cref{eq:kawasaki}, which does not describe the ensemble-averaged density, the stationary solution is given by an equilibrium distribution of the form $P[\rho] \propto \exp(-\beta F[\rho])$ \cite{Dean1996}.

However, this is not the most general case, since driven systems do not generally approach equilibrium. In driven systems, steady states can therefore arise that do not correspond to equilibrium states. A simple and illustrative example was discussed by \citet{PennaT2003}: Consider (in 1D) a potential of the form $V(x,t) = V(x-vt)=V(\bar{x})$ with a speed $v$ and the position in a co-moving frame $\bar{x}=x-vt$. We write the density $\rho$ as $\rho(x-vt)$ and look for a stationary solution in the co-moving frame. In this case, using $\partial_t\rho = - v\partial_x \rho$, \cref{trddft} can be integrated twice to give 
\begin{equation}
\frac{\delta F[\rho]}{\delta \rho(\bar{x})} +\frac{v}{\Gamma}(\bar{x}-\bar{x}_0) 
- \frac{\IntegrationConstantJ}{\Gamma}\INT{\bar{x}_0}{\bar{x}}{\bar{x}'}\frac{1}{\rho(\bar{x}')} = \mu,
\label{steadystate}
\end{equation}
where $\bar{x}_0$ is the (arbitrary) integration boundary, $\mu$ an integration constant related to this boundary, and $\IntegrationConstantJ$ an integration constant with the dimension of a current. For $v=0$, $\IntegrationConstantJ$ has to vanish and \cref{steadystate} reduces to the DFT equation \eqref{dft}, such that the only solution for the stationary density is the equilibrium density. In this case, $\mu$ is the chemical potential of DFT. This, however, is not the case for $v\neq 0$. In Ref.\ \cite{TarazonaM2008}, a similar method is applied to a phase-space description, which shows that, although there is a general agreement with the DDFT results, inertia can lead to additional effects such as the formation of a wake. 

More generally, a steady state of a DDFT with a time-dependent external force does not have to be a stationary state, i.e., the system might relax not to a time-independent state described by static DFT but to, e.g., an oscillatory state \cite{RexL2008,RexLL2005}.

\subsection{\label{perturbation}Perturbative approaches}
\subsubsection{Linearization}
\subsubsubsection{\label{linear}Linear stability analysis}
A useful analytical method for the analysis of phase separation in DDFT is the \textit{linear stability analysis}. Its application to spinodal decomposition in DDFT was first suggested by \citet{Evans1979}. Here, we present it following Ref.\ \cite{ArcherE2004}: One writes the density as
\begin{equation}
\rho\rt = \rho_0 + \Delta\rho\rt,
\label{deviation}%
\end{equation}
where $\rho_0$ is the density of the homogeneous reference state and $\Delta\rho\rt$ is the deviation. If the homogeneous state is unstable, then a deviation will grow in time. This can be determined by inserting the parametrization \eqref{deviation} into \cref{trddft}. If we assume that the deviation is small, i.e., if we consider the initial stage of decomposition, then we can linearize around the homogeneous state to find
\begin{equation}
\frac{1}{\Gamma k_B T}\pdif{}{t}\Delta\rho\rt = \vec{\nabla}^2\Delta\rho\rt - \rho_0\vec{\nabla}^2\INT{}{}{^3r'}c^{(2)}(\norm{\vec{r}-\vec{r}'};\rho_0)\Delta\rho(\vec{r}',t),
\label{linearizedequation}
\end{equation}
where the direct pair-correlation function $c^{(2)}(\vec{r},\vec{r}')$ takes the form $c^{(2)}(\norm{\vec{r}-\vec{r}'};\rho_0)=c^{(2)}(r,\rho_0)$ for a homogeneous fluid of spherical particles. We introduce the Fourier transformed density deviation as
\begin{equation}
\rho_{\vec{k}}(t) = \INT{}{}{^3r}e^{\ii\vec{k}\cdot\vec{r}}\Delta\rho\rt
\end{equation}
with the wavevector $\vec{k}$, where the notation $\rho_{\vec{k}}(t)$ with $\vec{k}$ as a subscript (which we use only for the density) is adapted from the MCT literature. 
A Fourier transformation of \cref{linearizedequation} gives
\begin{equation}
\frac{1}{\Gamma k_B T}\pdif{}{t}\rho_{\vec{k}}(t)=(-k^2+\rho_0k^2 c(k))\rho_{\vec{k}}(t)
\end{equation}
with the wavenumber $k=\norm{\vec{k}}$ and Fourier-transformed direct pair-correlation function $c(k)=\INT{}{}{^3r}\exp(\ii\vec{k}\cdot\vec{r})c^{(2)}(r,\rho_0)$. This is a linear differential equation for $\rho_{\vec{k}}(t)$ that has the solution
\begin{equation}
\rho_{\vec{k}}(t) = \rho_{\vec{k}}(0)\exp(\lambda(k)t)
\end{equation}
with the initial deviation $\rho_{\vec{k}}(0)$ and the dispersion relation $\lambda(k)=-\Gamma k_B Tk^2(1-\rho_0 c(k))$. If $\lambda(k)>0$ for some value of $k$, the corresponding Fourier component will grow exponentially in time, such that the homogeneous state is unstable. For $\lambda(k) <0$, small perturbations decay exponentially and the homogeneous state is linearly stable. (The applicability of deterministic DDFT, which assumes local stability, to spinodal decomposition, which involves local instability, has been doubted by \citet{Kawasaki2006b,Kawasaki2000}. \citet{ArcherE2004} avoid this problem with the following argument: $\lambda(k)$ is always negative for an equilibrium fluid outside the spinodal. A careful treatment is required inside the spinodal. In Ref.\ \cite{ArcherE2004}, $c(k)$ is defined in terms of the excess free energy, constructed by a mean-field treatment of attractive interactions. This permits positive values of $\lambda(k)$ inside the spinodal.)

This method can be extended to more complicated situations, such as mixtures of two particle species \cite{ArcherWTK2014}. In this case, one needs to linearize two coupled differential equations, such that finding the dispersion relation becomes an eigenvalue problem. Moreover, in DDFTs for active particles (see \cref{active}) that include an orientational dependence, a linear stability analysis can be used to test the stability of an isotropic state against orientational order \cite{HoellLM2018,HoellLM2019}. Mathematical details on linear stability analysis can be found in Ref.\ \cite{GoddardMP2020}. Note that linear stability analysis does not allow to predict the resulting final state. This requires a solution of the full nonlinear problem, which in most cases is only possible numerically.

Linear stability analysis has found a large number of applications in DDFT. Examples include the determination of the dispersion relation for front-speed calculation \cite{ArcherRTK2012,ArcherWTK2014,ArcherWTK2016,teVrugtBW2020}, traveling waves \cite{TaramaEL2019}, spinodal decomposition of magnetic fluids \cite{LichtnerK2013}, spinodal decomposition in a fluid with anisotropic diffusion \cite{VuijkBS2019}, spinodal and freezing modes \cite{PototskyASTM2011,PototskyTA2014}, the McKean-Vlasov equation \cite{CarrilloGPS2019}, the Dean-Kawasaki equation \cite{DelfauOLBH2016}, solvent-density modes \cite{BiswasC2007,KarBC2008}, capillary interactions \cite{BleibelDOD2014}, lane formation \cite{WachtlerKK2016,ScacchiAB2017}, quasicrystal formation \cite{ArcherRK2013,ArcherRK2015,RatliffASR2019,WaltersSAE2018}, phase behavior of thin films \cite{RobbinsAT2011}, orientational order of microswimmers \cite{HoellLM2018,HoellLM2019}, actively switching particles \cite{ZakineFvW2018}, dynamics of cancer cells \cite{AlSaediHAW2018}, and epidemic outbreaks \cite{teVrugtBW2020}.

\subsubsubsection{\label{front}Front propagation}
If a fluid reference state has been found to be unstable against perturbations, it remains to be discussed how the occurring solidification front propagates. This can be calculated based on the \ZT{marginal stability hypothesis} \cite{DeeL1983,BenJacobDK1985}, which is applied to DDFT in Refs.\ \cite{ArcherRTK2012,ArcherWTK2014,ArcherWTK2016,teVrugtBW2020}. We here present the method in 1D (see Ref.\ \cite{ArcherWTK2014} for a discussion of the 2D case). Suppose we have, by means of a linear stability analysis, obtained a dispersion $\lambda(k)$. If a solidification front propagates with velocity $v$, we can transform the linearized equation of motion for the perturbations to the co-moving frame and obtain from that the dispersion relation $\lambda_v(k) = \ii vk + \lambda(k)$. In the co-moving frame, the growth rate at the leading edge of the solidification front should be zero (since it would leave the frame otherwise). Hence, we get the equations
\begin{align}
\ii v + \tdif{\lambda(k)}{k} &= 0,\\
\Rea(\ii vk + \lambda(k)) &= 0.
\end{align}
They can be solved for the unknown complex wavenumber $k$ and the unknown velocity $v$. The selected wavelength is not necessarily that of the resulting equilibrium lattice, which can lead to disorder \cite{ArcherRTK2012,ArcherWTK2014}.

Front propagation is also studied in PFC models (see \cref{phasefield}) \cite{GalenkoSE2015,vanTeeffelenBVL2009,JaatinenA2010} and Allen-Cahn equations \cite{StegemertenGT2020}. In Refs.\ \cite{vanTeeffelenBVL2009,JaatinenA2010}, front propagation in DDFT is discussed with an emphasis on how the results compare to PFC calculations.

\subsubsection{\label{multi}Separation of time scales}
Since standard DDFT is the overdamped limit of more general kinetic theories involving inertia, systematic expansions of such theories in the friction coefficient can be used to obtain DDFT as a limiting case, but also to find inertial corrections. This is possible using the \textit{multiple-time-scale method}. The general procedure of deriving overdamped dynamics and corrections from phase-space theories was pioneered by \citet{Titulaer1980,Titulaer1978} and \citet{Wilemski1976}. It was applied to DDFT in Refs.\ \cite{LopezM2007,MarconiTCM2008,MarconiM2007,MarconiT2006,MarconiTC2007}. Moreover, \citet{GoddardPK2012} employed methods of this form for a mathematically rigorous derivation of DDFT. Time-scale separation also plays a role in the study of pattern formation in weakly nonlinear theory \cite{ArcherRTK2012} and of the violation of the fluctuation-response relation in nonequilibrium steady states \cite{YamadaY2015}. Amplitude equations, the derivation of which is also based on a multiple-scale expansion, are obtained for a binary phase field crystal (PFC) model derived from DDFT in Refs.\ \cite{HuangEP2010,ElderHP2010}. 

Following Ref.\ \cite{LopezM2007}, we consider systems with a position-dependent temperature $T(\vec{r})$. One starts from the nondimensionalized Kramers equation for the phase-space distribution $\PSDistributionRVT$
\begin{equation}
\gamma_{\mathrm{nd}} \KramersOperator \PSDistributionRVT(\vec{r},\vec{v},t)=\bigg(\pdif{}{t} + \vec{v}\cdot\vec{\nabla}_{\vec{r}} 
+ \vec{F}\cdot\vec{\nabla}_{\vec{v}}\bigg)\PSDistributionRVT(\vec{r},\vec{v},t)
\label{kramers}
\end{equation}
with the friction parameter $\gamma_{\mathrm{nd}}$, velocity $\vec{v}$, force $\vec{F}$, and Fokker-Planck operator 
\begin{equation}
\KramersOperator = \vec{\nabla}_{\vec{v}}\cdot(T(\vec{r})\vec{\nabla}_{\vec{v}}+\vec{v}).
\end{equation}
Equation \eqref{kramers} can be expanded in the eigenfunctions of $\KramersOperator$, which are related to the Hermite polynomials. We replace the physical time scale $t$ by an infinite series of auxiliary time scales $(\tau_0,\tau_1,\dotsc)$ with $\tau_n = \gamma_{\mathrm{nd}}^{-n}t$ and the function $\PSDistributionRVT(\vec{r},\vec{v},t)$ by an auxiliary function $\PSDistributionRVT_a(\vec{r},\vec{v},\{\tau_n\})$, which depends on the auxiliary time scales that are now treated as independent. The time derivative becomes
\begin{equation}
\pdif{}{t} = \pdif{}{\tau_0} + \frac{1}{\gamma_{\mathrm{nd}}}\pdif{}{\tau_1} + \frac{1}{\gamma_{\mathrm{nd}}^2}\pdif{}{\tau_2} + \dotsb.
\end{equation}
We now expand the auxiliary function in powers of $\gamma_{\mathrm{nd}}^{-1}$ and then each term in eigenfunctions of $\KramersOperator$. By inserting all results into \cref{kramers} and identifying orders of $\gamma_{\mathrm{nd}}^{-1}$, one obtains a hierarchy of equations for the amplitudes (prefactors arising in the expansion of the auxiliary function in $\gamma_{\mathrm{nd}}^{-1}$ and eigenfunctions). Compared to simple perturbation theory, this method avoids secular growth. After some calculation, the lowest nontrivial order gives the usual overdamped DDFT (here with temperature gradient), while higher orders lead to inertial corrections. Physically, the multiple-time-scale analysis separates the fast time scale on which the momenta relax from the longer time scales on which the positions relax. The density $\rho\rt$ then \ZT{enslaves} all other dynamical variables and follows a closed equation of motion \cite{MarconiM2007,MarconiTC2007}.

\subsubsection{\label{renormalization}Renormalized perturbation theory}
Renormalized perturbation theory has also been applied to DDFT. We discuss it here following Ref.\ \cite{KimKJvW2014}. The Martin-Siggia-Rose (MSR) procedure \cite{MartinSR1973,Janssen1976} allows to write a stochastic differential equation in the form of a field theory based on an action functional \cite{VelenichCCK2008}, in close analogy to methods used in quantum field theory. This method became a popular tool for the study of the glass transition \cite{DasM1986}. A matter of debate is whether an ergodic-to-nonergodic (ENE) transition can be found. An ENE transition, which corresponds to the long-time limit of the density correlation function being nonzero, is predicted by simple MCT \cite{BidhoodiD2015}.

In particular, a path-integral formalism can be developed for stochastic DDFT. This was done by \citet{KawasakiM1997}. We present the method following Ref.\ \cite{KimKJvW2014}. The action $\mathcal{S}$ reads
\begin{equation}
\mathcal{S}[\rho,\widehat{\rho}] = \INT{}{}{^3r}\INT{}{}{t}\bigg(\ii \widehat{\rho}\rt\bigg(\pdif{}{t}\rho\rt - \Gamma\Nabla\cdot\bigg(\rho\Nabla\frac{\delta F}{\delta \rho\rt}\bigg)\!\bigg) -D\rho(\Nabla\widehat{\rho}\rt)^2\bigg).
\label{action}
\end{equation}
Here, $\widehat{\rho}$ denotes a real auxiliary field and the last term arises from the noise. The action \eqref{action} is invariant under two types of time-reversal (TR) transformations that, however, are nonlinear. An example is the U transformation
\begin{align}
\rho(\vec{r},-t)&\to\rho\rt,\\
\widehat{\rho}(\vec{r},-t)&\to-\widehat{\rho}\rt + \frac{\ii}{k_BT}\frac{\delta F}{\delta \rho\rt}.
\end{align}
A loop-expansion of the action \eqref{action} leads to problems with the fluctuation-dissipation relation \cite{MiyazakiR2005}. The reason for this is that the action \eqref{action} can be decomposed into a Gaussian and a non-Gaussian part, which are not separately invariant under the TR transformation. \citet{AndreanovBL2006} solved this problem by introducing two auxiliary fields $\theta$ and $\widehat{\theta}$. The action then becomes a functional $\mathcal{S}[\rho,\widehat{\rho},\theta,\widehat{\theta}]$. Using the new fields, the U transformation can be linearized. Under the linear transformation, Gaussian and non-Gaussian parts of the new action are separately invariant. Renormalized perturbation theory can then preserve the fluctuation-dissipation relation at each order.

Using this idea, \citet{KimK2007,KimK2008,KimK2009} developed a renormalized perturbation theory from stochastic DDFT. A derivation of MCT at one-loop order by \citet{KimK2008} turned out to lack a term that was identified by \citet{KimKJvW2014}. While MCT can be recovered by an additional approximation, the full result of Ref.\ \cite{KimKJvW2014} contains no ENE transition. This conclusion is also supported by the nonperturbative analysis of \citet{BidhoodiD2015}. \citet{JacquinKKvW2015} studied a modified stochastic model B equation, which, despite having the same equilibrium state as \cref{dk}, showed (even in the mode coupling approximation) no sign of glassy behavior. The case of Brownian particles with momentum was considered by \citet{Das2020}, who argued that an ENE transition is not supported by the fluctuation-dissipation relation also in this case. In Ref.\ \cite{KimFK2020}, a further TR-symmetry-preserving perturbative scheme was developed.

An introduction to this topic is given in Ref.\ \cite{Das2011} (here, the results from Ref.\ \cite{KimKJvW2014} are not yet taken into account). Using a Cole-Hopf transformation, a reformulation of the action is possible \cite{AndreanovBBL2006}. A diagrammatic derivation of mode coupling results can also be found in Ref.\ \cite{Szamel2007}. Feynman rules and exact results for the case of noninteracting particles can be found in Ref.\ \cite{VelenichCCK2008}. One can also include the momentum density \cite{NishinoH2008,Yeo2009,BidhoodiD2016} (in Ref.\ \cite{KawasakiM1997}, the DDFT action was obtained by eliminating the momentum density). Smoluchowski dynamics is considered in Refs.\ \cite{Mazenko2010,Mazenko2011,McCowan2015} and Newtonian dynamics in Refs.\ \cite{DasM2012,DasM2013}. In Ref.\ \cite{GrzeticWS2014}, an MSR integral for polymer dynamics is derived. Orientational degrees of freedom were included in Ref.\ \cite{CugliandoloDLvW2015}. Moreover, renormalization group theory in combination with (hydrodynamic) DDFT is applied to the glass transition of water in Ref.\ \cite{YeNTZM2018}. 

\subsection{\label{numeric}Numerical solutions}
Since the DDFT equation \eqref{trddft} is a nonlinear partial differential equation, it is, in general, impossible to solve it analytically. Hence, it is important to have numerical methods available that can be used to solve it. In general, DDFT has computational advantages compared to Brownian dynamics (BD) simulations, since the computational cost does not depend on the density \cite{MalijevskyA2013}. The computational cost of a BD simulation increases with the number of particles $N$, whereas the computational cost of a DDFT calculation increases with the desired resolution. As a consequence, for systems with a large number of particles and no small-scale pattern formation, continuum simulations based on numerically solving a DDFT equation are typically much more efficient than BD simulations. Compared to dissipative particle dynamics \cite{HoogerbruggeK1992,KoelmanH1993,MoeendarbaryNZ2009,EspanolW2017,ElleroE2018}, which is another method for simulating fluids on the mesoscale, DDFT is applicable down to microscopic scales \cite{PagonabarragaF2001}. In Ref.\ \cite{QiS2017b} (see also Refs.\ \cite{QiBS2013,QiBRS2016} for earlier work), a hybrid particle-continuum simulation method is presented, where the level of resolution switches adaptively between BD and DDFT field propagation. 

A significant amount of work has been done in numerical mathematics on the solution of partial differential equations of the gradient-dynamics type (see \cref{net}), which DDFT belongs to. However, only a few of these methods are specifically designed for DDFT. Many of them address PFC models \cite{BaskaranHLWWZ2013,DongFWWZ2018,DehghanM2016,HuWWL2009,LiK2017,LiLXK2019,WangW2011,WiseWL2009,LiNZG2018,PraetoriusV2015b}, the Allen-Cahn and Cahn-Hilliard equations \cite{GuanLW2017,GuanLWW2014,GuanWW2014,LiKW2017,LiQK2018,YangZ2019,DuJLQ2018}, the Swift-Hohenberg equation \cite{SuFYL2019}, or the Poisson-Nernst-Planck (PNP) equations \cite{XuCMYL2014}. For PFC models, \citet{BackofenRV2007} presented a semi-implicit integration scheme in order to avoid the problems of explicit (time-step restrictions) and implicit (nonlinear equations) methods. Due to the close mathematical relation of these models to DDFT, it is likely that similar approaches can also be useful here. 

Moreover, DDFT can benefit from methods developed for static DFT. Some of these methods have already been applied to DDFT (such as numerical continuation), whereas others have potential for future applications (such as machine learning). A variety of numerical strategies for DFT are discussed in Ref.\ \cite{FrinkSSWF2002}, including real-space numerics, fast Fourier transformations, parallelization, preconditioning, and pseudo-arc-length continuation. Numerical methods for polymer DFT can be found in Ref.\ \cite{FrischknechtWSCDM2002}. More recently, machine-learning methods have been used to approximate classical density functionals in DFT \cite{LinO2019}. Machine learning is already successfully applied in quantum DFT (see Refs.\ \cite{SchlederPACF2019,CarleoCCDSTVZ2019} for reviews). 

A variety of authors have discussed the discretization and integration of DDFT, focusing on a variety of specific problems. For stochastic DDFT, a finite-volume method was developed by \citet{RussoPDYCK2019}. A simple finite-difference integration scheme for deterministic DDFT can be found in Ref.\ \cite{ChalmersSA2017}. Discretization of deterministic DDFT is discussed in Ref.\ \cite{StopperRH2016}. Differential equations arising in DDFT can be stiff for fine grids. Since implicit Runge-Kutta algorithms, which are more efficient than simple time-stepping methods in such situations, cause difficulties due to the large dimensionality and nonlinearity of the problem, specialized explicit Runge-Kutta methods are employed \cite{ReinhardtWB2013}. A discussion of the finite-element discretization of DDFT, with an emphasis on the relation of stochastic and deterministic theories, the introduction of thermal fluctuations, and the relation to the physics of coarse graining, can be found in Ref.\ \cite{delTED2015}. In Ref.\ \cite{CarrilloKPS2018}, finite-volume schemes for general transport equations that include DDFT as a limiting case are presented.

Nonlocal terms resulting from the excess free energy are particularly difficult to treat numerically. Convolution terms on Cartesian grids can be evaluated efficiently using fast-Fourier-transformation methods. However, these are difficult to apply on general meshes \cite{ReinhardtSB2014}. Since hard spherical disks cause geometrical difficulties when using the standard method of uniform Cartesian grids, partially refined grids are useful \cite{ReinhardtWB2013}. A spectral method for both DFT and DDFT is presented in Ref.\ \cite{YatsyshinSK2012}. It involves choosing a discretization scheme where the mesh is dense close to walls, where larger density variations are expected. A Clenshaw-Curtis quadrature is used to evaluate the DFT convolution integrals. \citet{NoldGYSK2017} developed a pseudo-spectral method for the evaluation of the nonlocal terms. This scheme is discussed for a variety of contexts in DFT and FMT and also applied to DDFT where it is checked for mass conservation. On the surface of a sphere, the convolutions can be efficiently computed using expansions in spherical harmonics \cite{BottB2016}.

For DDFT calculations in higher dimensions, which can become rather slow due to the large number of lattice points required, parallelization is important. The parallelization of DFT calculations is discussed in Refs.\ \cite{StopperR2017,FrinkS2000}. \citet{FraaijeE1995} presented parallelized DDFT algorithms in FORTRAN.

Another useful method in DDFT is numerical continuation \cite{EngelnkemperGUWT2019,DoedelFSCKW2007,DijkstraEtAl2014,DoedelKK1991,DoedelKK1991b,ThieleKG2014,WillersTALK2020}, which allows to track solutions of an equation as a function of a control parameter. Numerical continuation is discussed in the context of DDFT in Refs.\ \cite{KruegerR2007,ArcherRRS2019,LichtnerPK2012}. It was used by \citet{GomesKPY2019} to determine bifurcation diagrams for the McKean-Vlasov equation (a DDFT-type equation). \citet{PototskyTA2014} employed continuation methods to calculate bifurcations of clusters. In Ref.\ \cite{YatsyshinSK2012}, an algorithm for the numerical continuation of DFT is presented.

DDFT (in the form discussed in \cref{pd}) also forms the basis of the MesoDyn software \cite{AltevogtEFMvV1999,FraaijeZS2004}, which can be used to simulate polymer materials. The time scales involved in phase separation and pattern formation of block copolymers are typically too large to be accessible by molecular dynamics simulations \cite{FraaijeZS2004}, in particular for industrial applications in soft nanoscience and material design. On the other hand, the time scales in industrial processing are much shorter than those required for full relaxation to thermodynamic equilibrium, such that the nonequilibrium behavior is important \cite{LiXWS2007}. DDFT-based methods allow for a simulation of the time-dependent thermodynamic behavior on the mesoscale \cite{FraaijeZS2004}. A discussion of various multiscale simulation methods for materials can be found in Refs.\ \cite{Hyodo2002,PosoccoPF2014}.

For kinetic extensions of DDFT (see \cref{kinetic}), the use of lattice Boltzmann methods \cite{ChenD1998,ShanC1993,ShanC1994,Succi2001,MelchionnaM2007,MarconiM2013,BernaschiMS2019} has been suggested by \citet{MarconiM2009,MarconiM2010,MarconiM2014}. First, the phase-space distribution function and the collision operator are expanded in Hermite polynomials. Second, the moments are evaluated using Gauss-Hermite quadratures. Third, the distribution function is propagated using a forward Euler procedure. The lattice Boltzmann method has certain advantages over the solution of the coupled hydrodynamic equations, such as an easier handling of boundary conditions \cite{MarconiM2010}. \citet{BaskaranGL2016} developed implicit finite-difference methods for a DDFT with momentum density that takes the form of a compressible Navier-Stokes equation.

A numerical iterative scheme allowing to determine the external force field that generates a certain density and velocity profile can be found in Ref.\ \cite{delasHerasRS2019}. The existence of a mapping between motion and force field is implied by power functional theory (PFT) (see \cref{pft}). In the context of PFT, numerical techniques have moreover been developed to sample the one-body current from BD simulations, where one has to take the stochastic nature of the underlying dynamics into account \cite{delasHerasRS2019,KrinningerSB2016}.

Finally, Monte Carlo (MC) simulations are often combined with DDFT. We only discuss a few examples here. The stochastic DDFT \eqref{eq:kawasaki} by Kawasaki (see \cref{kawasaki}) can be solved by mapping it onto a kinetic lattice gas model. This model can then be solved using ordinary MC techniques \cite{FuchizakiK1999,KawasakiFM1997,FuchizakiK1998b}. Dynamic MC simulations can be employed to test approximations made in DDFT \cite{PiazzaDDDF2013}. The scheme for the determination of superadiabatic forces presented in Ref.\ \cite{FortinidlHBS2014} requires an adiabatic reference potential, which can be obtained from MC simulations \cite{FortinidlHBS2014,BernreutherS2016}. In Ref.\ \cite{YuJEAR2017}, stationary density profiles obtained from a DDFT for flow-driven hard spheres are compared to MC results. Moreover, DDFT can help to incorporate hydrodynamic effects into MC simulations \cite{JabeenYEAR2018}.

\section{\label{applications}Applications}
Having discussed how to analyze a DDFT equation, we finally turn to applications of DDFT. This is a very broad and diverse field, which we present here ordered by systems DDFT is applied to (\cref{systems}) and by phenomena that are described by DDFT (\cref{phenomena}). In \cref{systems}, we start with the systems introduced in \cref{traditional}, namely colloidal fluids (\cref{colloid}), atomic and molecular fluids (\cref{amf}), and polymers (\cref{polymer}), for which standard DDFT has been developed. We then address thin films (\cref{thin}) and glassy systems (\cref{glass}), which are composed of the same materials, but have special properties that affect their description in DDFT. Different materials come into play when considering granular media (\cref{granular}) and plasmas (\cref{plasma}). Driven (\cref{driven}) and active (\cref{activesm}) soft matter are also nonstandard cases, since in these systems typical close-to-equilibrium approximations of DDFT tend to break down. Finally, we present electrochemical (\cref{electro}) and biological (\cref{biology}) systems, which are particularly complex. For phenomena (\cref{phenomena}), we again start with the most basic aspects, which are relaxation (\cref{relax}) and phase separation (\cref{ps}). This naturally leads to pattern formation (\cref{pattern}), which can be a consequence of phase separation, and nucleation and solidification (\cref{nucandsol}), which is a form of relaxation (to a particular thermodynamic equilibrium state) and can also lead to pattern formation. We then turn to specific applications where pattern formation is one (though not the only one) relevant effect, namely chemical reactions (\cref{chemical}), disease spreading (\cref{disease}), and feedback control (\cref{fc}). Finally, we discuss sound waves (\cref{sound}). A further important aspect are \ZT{theoretical applications}, which involve the derivation of new dynamic equations from DDFT. Important examples include MCT and PFC models, which have both been explained in previous sections (\cref{mct,phasefield}, respectively), and the dynamics of the van Hove function, which is discussed in \cref{vh}.

\subsection{\label{systems}Systems}
\subsubsection{\label{colloid}Colloidal fluids}
Colloidal fluids are a standard system of application for DDFT. This case is discussed extensively in \cref{traditional}, in particular in \cref{solvent}.
\begin{figure}
\includegraphics[width=\linewidth]{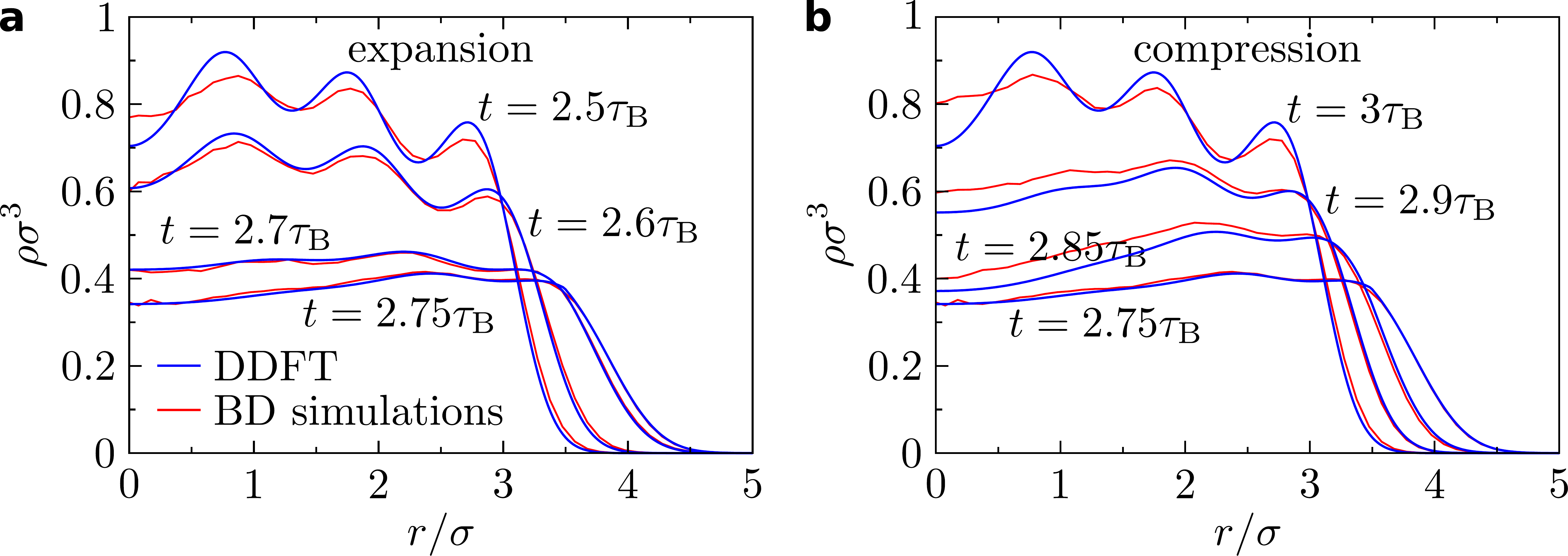}
\caption{\label{fig:trap}Colloidal spheres with diameter $\sigma$ in a periodically switching optical trap. Density profiles $\rho(r,t)$ obtained by DDFT and BD simulations are shown at different times $t$, where $\tau_{\mathrm{B}}=\sigma^2/D$ denotes the Brownian time. Adapted from Fig.\ 2 from Ref.\ \protect\cite{RexL2008}.}
\end{figure}

An example is presented in \vref{fig:trap}, which is adapted from Ref.\ \cite{RexL2008} and shows the time evolution of the density profile of colloidal spheres in a periodically switching optical trap. The steady state is a driven breathing mode. Results from BD simulations are also shown. While there are small deviations resulting from superadiabatic contributions, the general agreement between theory and simulation is very good.

\subsubsection{\label{amf}Atomic and molecular fluids}
Atomic and molecular fluids are another standard system of application for DDFT and discussed in \cref{traditional,inertia}.

\subsubsection{\label{polymer}Polymers}
\begin{table}[htb]
\begin{tabularx}{\linewidth}{|l|X|}
\hline
\rowcolor{lightgray}\textbf{Topic} & \textbf{References}\\ \hline
General theory & \cite{KawasakiS1987,KawasakiS1988,KawasakiS1989,MauritsF1997b,Kawakatsu1997,Binder1983,ManthaQS2020} \\ \hline
Original derivations & \cite{Fraaije1993,FraaijevVMPEHAGW1997}\\ \hline
Extensions:&\\ 
$\bullet$ External potential dynamics & \cite{MauritsF1997}\\ 
$\bullet$ Compressibility & \cite{MauritsvVF1997}\\ 
$\bullet$ Viscoelasticity & \cite{MauritsZF1998b}\\ \hline
Software:&\\ 
$\bullet$ MesoDyn &\cite{AltevogtEFMvV1999,FraaijeZS2004,Nicolaides2001}\\ 
$\bullet$ Hybrid BD-DDFT simulations & \cite{QiS2017b,SevinkCF2013,QiBS2013,QiBRS2016,SevinkF2014,SevinkF2008}\\ 
$\bullet$ Other aspects & \cite{vanVlimmerenF1996,FraaijeE1995}\\ \hline
Applications:&\\ 
$\bullet$ Phase behavior & \cite{FraaijeNRG2011,KyrylyukF2004,ZvelindovskySvVMF1998,MauritsZSvVF1998,PloshnikLHBMFAS2013,SevinkFH2002,MauritsZF1998,HondaK2007,HondaK2007b,TanSNWGY2009,HaoMSNL2014,KnollHLKSZM2002,XiaSQZY2005,NeratovaKK2010,SevinkZvVMF1999,SunHL2005,HorvatLSZM2004,KomarovVCK2010,KyrylyukLS2005,ZhangYW2006,KnollLHKSZM2004,SpyriouniV2001,YangYZY2008,MuLW2013,LiuLKPR2012,JawalkarRHSA2007,LiuHPGR2012,ChiangraengLN2019,MuLW2012,JawalkarASNA2005,MuLW2015,LamG2003,MaitiWG2005,TsarkovaHKZSM2006,Kawakatsu2004,MoritaKD2001,CoslanichFCMS2004,CoslanichFFS2004,XuLH2007b,XuWHH2008,DongZCSXW2010,ZhangYXL2007}\\ 
$\bullet$ Sheared systems & \cite{ZvelindovskySvVMF1998,ZvelindovskySLA2004,ZvelindovskyS2003,NikoubashmanRP2013,LaicerCLT2005,JawalkarNR2008,JawalkarA2006,MuLZ2011,MuLW2011,RakkapaoV2014,MuHLS2008,XuLH2007,XuLH2007c,YeNTZM2019}\\ 
$\bullet$ Pattern formation & \cite{KyrylyukCF2005,SevinkZFH2001,KyrylyukSZF2003,FraaijeZSM2000,vanVlimmerenMZSF1999,HeHLP2002,KomarovVCKK2010,TanZDSY2013,HuangHHL2003,LudwigsBVRMK2003,LuHL2004,ScherdelSM2007,Berendsen1999,NeratovaPK2010,MichielsenDR2001,KyrylyukF2004b,BootsdRNBPWDWF2014,SevinkF2008b,SevinkZF2003}\\ 
$\bullet$ Electric fields & \cite{KyrylyukF2005,KyrylyukF2006,KyrylyukZSF2002,ZhangLWX2017}\\ 
$\bullet$ Adsorption/desorption & \cite{MilchevEB2010}\\ 
$\bullet$ Thin films &  \cite{HorvatLSZM2004,TsarkovaHKZSM2006,MoritaKD2001,ParadisoFFF2012,SevinkF2008b,KnollHLKSZM2002,KnollLHKSZM2004,LudwigsBVRMK2003}\\ 
$\bullet$ Mechanical properties & \cite{DengZHLH2011,DengHXLLH2014,DengHLXLL2014,DengPHL2016} \\ 
$\bullet$ Response to nanoparticles & \cite{LaicerMT2007,LaicerCLT2005} \\ \hline
Reviews:&\\
$\bullet$ Polymer DDFT & \cite{SevinkF2008b,SevinkZF2003}\\
$\bullet$ Polymer simulation & \cite{LangnerS2012,NikoubashmanRP2013,XuLZWT2016,ZhanweiXJZZ2011,YuliangFPH2006,Schmid2010,FredricksonGD2002,GlotzerP2002,GartnerJ2019,McGrotherGL2004,Nicolaides2001,LiuHCXH2011,Theodorou2003,ZhangLWX2017}\\
\hline
\end{tabularx}
\vskip2mm
\caption{\label{tab:polymer}Overview over the literature on polymer DDFT.}
\end{table}
Polymers are the third standard system of application introduced in \cref{traditional}. They were among the first systems studied using DDFT and the theory presented in \cref{pd} is now one of the most widely used methods in the simulation of polymer systems. Applications include phase behavior \cite{FraaijeNRG2011,KyrylyukF2004,ZvelindovskySvVMF1998,MauritsZSvVF1998,PloshnikLHBMFAS2013,SevinkFH2002,MauritsZF1998,HondaK2007,HondaK2007b,TanSNWGY2009,HaoMSNL2014,KnollHLKSZM2002,XiaSQZY2005,NeratovaKK2010,SevinkZvVMF1999,SunHL2005,HorvatLSZM2004,KomarovVCK2010,KyrylyukLS2005,ZhangYW2006,KnollLHKSZM2004,SpyriouniV2001,YangYZY2008,MuLW2013,LiuLKPR2012,JawalkarRHSA2007,LiuHPGR2012,ChiangraengLN2019,MuLW2012,JawalkarASNA2005,MuLW2015,LamG2003,MaitiWG2005,TsarkovaHKZSM2006,Kawakatsu2004,MoritaKD2001,CoslanichFCMS2004,CoslanichFFS2004,XuLH2007b,XuWHH2008,DongZCSXW2010,ZhangYXL2007}, systems under shear \cite{ZvelindovskySvVMF1998,ZvelindovskySLA2004,ZvelindovskyS2003,NikoubashmanRP2013,LaicerCLT2005,JawalkarNR2008,JawalkarA2006,MuLZ2011,MuLW2011,RakkapaoV2014,MuHLS2008,XuLH2007,XuLH2007c,YeNTZM2019}, pattern formation and self-organization \cite{KyrylyukCF2005,SevinkZFH2001,KyrylyukSZF2003,FraaijeZSM2000,vanVlimmerenMZSF1999,HeHLP2002,KomarovVCKK2010,TanZDSY2013,HuangHHL2003,LudwigsBVRMK2003,LuHL2004,ScherdelSM2007,Berendsen1999,NeratovaPK2010,MichielsenDR2001,KyrylyukF2004b,BootsdRNBPWDWF2014,SevinkF2008b,SevinkZF2003}, effects of electric fields \cite{KyrylyukF2005,KyrylyukF2006,KyrylyukZSF2002,ZhangLWX2017}, adsorption/desorption kinetics \cite{MilchevEB2010}, thin films \cite{HorvatLSZM2004,TsarkovaHKZSM2006,MoritaKD2001,ParadisoFFF2012,SevinkF2008b,KnollHLKSZM2002,KnollLHKSZM2004,LudwigsBVRMK2003}, mechanical properties \cite{DengZHLH2011,DengHXLLH2014,DengHLXLL2014,DengPHL2016}, and the response of materials to embedded nanoparticles \cite{LaicerMT2007,LaicerCLT2005}. 

DDFT has gained importance in industrial applications due to its applicability as a simulation technique. In particular, the MesoDyn method \cite{AltevogtEFMvV1999,FraaijeZS2004,Nicolaides2001} is widely used for polymer simulation. Moreover, DDFT can be combined with BD in hybrid particle-field simulations \cite{QiS2017b,SevinkCF2013} (see also Refs.\ \cite{QiBS2013,QiBRS2016,SevinkF2014,SevinkF2008}). Some comments on parallelization can be found in Ref.\ \cite{FraaijeE1995}. Various alternative approaches are discussed in Refs.\ \cite{GrootM1998,FengLH2002,FrischknechtWSCDM2002,ChoW2006,FredricksonO2014,GrzeticWS2014}. For general reviews of (co)polymer modeling and simulation, see Refs.\ \cite{LangnerS2012,NikoubashmanRP2013,XuLZWT2016,ZhanweiXJZZ2011,YuliangFPH2006,Schmid2010,FredricksonGD2002,GlotzerP2002,GartnerJ2019,McGrotherGL2004,Nicolaides2001,LiuHCXH2011,Theodorou2003,ZhangLWX2017}. A fairly complete overview over the literature on polymer DDFT is given in \cref{tab:polymer}.

Since polymer DDFT is a large topic on its own that has been reviewed elsewhere \cite{SevinkF2008b,SevinkZF2003}, we only discuss a few studies here as examples. All of them consider polymer thin films, which have increasing importance for nanotechnology. \citet{KnollHLKSZM2002} used DDFT for computer simulations of copolymer thin films. The phase diagram was obtained both theoretically and experimentally. Wetting layers and lamella formation were observed. \citet{KnollLHKSZM2004} compared simulation data from DDFT to experiments on nanostructured fluids. These also show a lamella phase, as do the thin films of ABC triblock copolymer, whose self-assembly was studied theoretically and experimentally by \citet{LudwigsBVRMK2003}. All simulations reported here employ the MesoDyn software.

Other applications of DDFT to polymers \cite{RauscherDKP2007,KruegerR2009} are based on the methods presented in \cref{marconi,kawasaki}. Depending on the effects one is interested in, the polymers can be assumed to be interacting or noninteracting \cite{PennaDT2003}. Applications include diffusion in colloid-polymer mixtures \cite{StopperRH2016}, encapsulation kinetics of hollow hydrogels consisting of polymeric shells \cite{MonchoGAAD2019}, and polymer conformation \cite{TakahashiM1997,Munakata2001}. Polymer chains are described using a DDFT with hydrodynamics in Refs.\ \cite{ZhaoWZDTM2014,YeDTZM2017}. Cahn-Hilliard-type models can also be applied to polymer systems \cite{DiazPZPS2020}. Further applications of DDFT include thin films (see \cref{thin}) and biology (see \cref{biology}). 

\subsubsection{\label{thin}Thin films}
While they also fall into the category \ZT{fluids and polymers} discussed in \cref{colloid,amf,polymer}, thin films deserve a separate discussion due to their theoretical and practical importance. Polymer thin films have already been addressed in \cref{polymer}. Here, we give a broader overview. The wetting behavior of droplets and thin films is often modeled through phenomenological equations based on long-wave hydrodynamics or gradient dynamics (see Refs.\ \cite{Thiele2018,WilczekTGKCT2015,Thiele2014,Thiele2012} for reviews), which exist in deterministic and stochastic forms \cite{DuranGKP2019}. DDFT provides an alternative based on microscopic particle dynamics (see Ref.\ \cite{ThieleVARFSPMBM2009} for a review). It allows for a description of various important effects, such as contact angle hysteresis, capillary-driven motion \cite{MickelJB2011}, wetting of confined colloidal fluids \cite{Archer2005}, and nanodroplet coalescence \cite{NiuLLH2019}. Interesting insights into wetting behavior can already be gained using static DFT \cite{YinSA2019,YatsyshinDK2018,HughesTA2017,YatsyshinK2016,YatsyshinK2016b,YatsyshinSK2015,HughesTA2014,HughesTA2015,LiW2008,YatsyshinSK2015b}. Numerical aspects of wetting in (D)DFT are discussed in Ref.\ \cite{YatsyshinSK2012}. 

Frequently, one studies solvents in which nanoparticles are immersed. Dewetting processes are more complex in such systems, since various transport and evaporation processes are involved \cite{ArcherRT2010}. As in the stationary case \cite{HughesTA2015}, the system is often modeled as a lattice. Consequently, DMFT (see \cref{dmft}) is a method used in this field \cite{EdisonM2009,EdisonM2010,EdisonM2013,CasselmanDM2015}. The sites of the lattice can be unoccupied, occupied by liquid, or occupied by a nanoparticle \cite{RobbinsAT2011}. The free energy governing the DDFT can be obtained by a mapping from the lattice Hamiltonian or through gradient expansions of the free energy functionals for continuous systems \cite{ThieleVARFSPMBM2009}. The mobility of the nanoparticles depends on the density of the solvent, since they are transported through the film and cannot move on the dry substrate. Moreover, due to evaporation processes governed by a difference between the chemical potential of the liquid on the surface and that in a reservoir, the DDFT equation gets an additional nonconserved term that has an Allen-Cahn-type form \cite{RobbinsAT2011}. Denoting the densities of liquid and nanoparticles by $\rho_{\mathrm{l}}$ and $\rho_{\mathrm{n}}$, respectively, this gives
\begin{align}
\pdif{}{t}\rho_{\mathrm{n}}\rt&=\Nabla\cdot\bigg(M_{\mathrm{n}}(\rho_{\mathrm{n}},\rho_{\mathrm{l}})\Nabla\frac{\delta F[\rho_{\mathrm{n}},\rho_{\mathrm{l}}]}{\delta\rho_{\mathrm{n}}\rt}\bigg),\\
\pdif{}{t}\rho_{\mathrm{l}}\rt &= M_{\mathrm{l}}^{\mathrm{c}}\Nabla\cdot\bigg(\rho_{\mathrm{l}}\rt\frac{\delta F[\rho_{\mathrm{n}},\rho_{\mathrm{l}}]}{\delta\rho_{\mathrm{l}}\rt}\bigg) -M_{\mathrm{l}}^{\mathrm{n}}\frac{\delta F[\rho_{\mathrm{n}},\rho_{\mathrm{l}}]}{\delta\rho_{\mathrm{l}}\rt},
\end{align}
where the mobility $M_{\mathrm{n}}$ of the nanoparticles depends on the local densities, whereas the mobilities $M_{\mathrm{l}}^{\mathrm{c}}$ and $M_{\mathrm{l}}^{\mathrm{n}}$ for the conserved and nonconserved part of the liquid motion, respectively, are constant. DDFT models allow to analyze a variety of effects in the context of dewetting and evaporation, such as spinodal decomposition, nucleation, and fingering \cite{RobbinsAT2011,ThieleVARFSPMBM2009,ArcherRT2010}. Moreover, a thermodynamic equivalent of the coffee-ring stain effect can be found \cite{ChalmersSA2017}. A review is given by Ref.\ \cite{ZhongCD2015}. Simulations and experiments are described in Ref.\ \cite{CrivoiD2013}.

Standard DDFT is also applied to study drying mixtures of colloids and polymers. In particular, one is here interested in stratification. It is found, in agreement with molecular dynamics simulations, that initially mixed fluids stratify with smaller colloids on top \cite{HowardNP2017}, while in colloid-polymer mixtures the result depends on the polymer length and radius of gyration \cite{HowardNP2017b}. As a mechanism for stratification in drying films, diffusiophoresis has been proposed \cite{SearW2017}. Another application of DDFT in the context of thin films is the dynamics of surfactants \cite{ThieleAP2012,AlandV2012}. Their density can be described by a diffusion-advection-reaction equation, where the diffusive current of DDFT is combined with an advection term from the Marangoni flux and reaction terms for ad-/desorption and photoisomerization \cite{GrawitterS2018}. \citet{YeTZDM2016} studied the oleophobicity of polymer films. Thin films have also been considered in the context of the polymer DDFT presented in \cref{pd} \cite{HuininkBvDS2000,HorvatLSZM2004,TsarkovaHKZSM2006,MoritaKD2001,ParadisoFFF2012}, some examples are discussed in \cref{polymer}. Finally, DDFT can also be applied to wetting and drying in active systems \cite{WittmannB2016}.

\subsubsection{\label{glass}Glassy systems}
When entering a glassy state, soft matter systems exhibit a variety of interesting nonstandard properties. Like thin films (see \cref{thin}), glasses are therefore discussed separately. However, the application of DDFT to glasses is more controversial and therefore requires a particularly careful treatment.

The dynamics of supercooled liquids and glasses is often modeled using MCT, whose formalism and relation to DDFT are discussed in detail in \cref{mct}. It models glassy behavior as an ergodic-to-nonergodic (ENE) transition. The applicability of DDFT in this field is a question that is discussed very controversially. On the one hand, stochastic DDFT was specifically designed as a generalization of MCT for the treatment of supercooled liquids \cite{Kawasaki1994,KawasakiF1998,KawasakiK1996,Kawasaki1998b}, DDFT was used to investigate the origin of nonergodicity in MCT \cite{Kawasaki2003}, MCT was derived from DDFT \cite{Kawasaki1994,Archer2006}, and DDFT reproduces glass-like relaxation behavior \cite{HopkinsFAS2010}. On the other hand, it was argued that DDFT cannot describe caging and nonergodicity effects relevant for glasses \cite{ReinhardtB2012,SchindlerWB2019}, renormalized perturbation theory showed no ENE transition \cite{KimKJvW2014}, and the application to glasses seems incompatible with the fact that the derivation of DDFT involves neglecting memory effects \cite{WittkowskiLB2012}.

To resolve these issues, it is important to distinguish between three different questions:
\begin{enumerate}
\item Can DDFT reproduce MCT?
\item Can DDFT describe the physics of (real) glasses?
\item Can DDFT describe nonergodic systems?
\end{enumerate}

The subtle relation of DDFT and MCT is discussed in \cref{mct}. Stochastic DDFT was developed in the context of MCT by \citet{KirkpatrickW1987} and \citet{Kawasaki1994}. Based on stochastic DDFT, one can derive an equation of motion for the correlator of the density fluctuations that has the form of an MCT equation. This corresponds to the microscopic MCT if the DDFT describes the microscopic density, whereas a coarse-grained stochastic DDFT leads to an MCT-like equation for the fluctuations of the coarse-grained density. A derivation of the latter type \cite{Archer2006} is presented in \cref{mct}. Other ways of deriving MCT include projection operators \cite{Kawasaki1994} and renormalized perturbation theory \cite{KimKJvW2014} (see \cref{renormalization}). Deterministic DDFT and MCT are complementary approaches that correspond to making different approximations to the full microscopic dynamics \cite{WittkowskiLB2012,BraderS2013}.

The second issue was the original motivation for stochastic DDFT: The standard description of cage effects in simple MCT as an ENE transition does not apply to real materials due to hopping effects resulting from thermal fluctuations (although the deviations from MCT due to thermal fluctuations are comparatively small in dense colloidal liquids, since colloids are very large). These fluctuations are included in stochastic DDFT \cite{FuchizakiK2002}. In dense liquids, momentum and energy can be transferred on time scales much shorter than those on which the particle configuration can change, such that it is natural to use the density as the only slow variable in such a theory. \citet{FuchizakiK1998} showed that stochastic DDFT is capable of describing mode coupling mechanisms. Stochastic DDFT gives access to the later stages of freezing in a supercooled liquid, where the system relaxes from the glassy minima by thermally activated hopping \cite{FuchizakiK1999}. For practical purposes, the description can be simplified by using a lattice model \cite{FuchizakiK1998,FuchizakiK1999}. Another important method in this context is renormalized perturbation theory (see \cref{renormalization}). A brief overview over both MCT and stochastic DDFT is given by Ref.\ \cite{Kawasaki2009}. Further discussions can be found in Refs.\ \cite{Das2011,BhattacharyyaBW2005}.

With deterministic DDFT, a description of dense atomic liquids is also possible \cite{Archer2006}. A specific problem is, however, that if a system gets trapped in a local minimum, deterministic DDFT will predict that it is trapped forever \cite{Yoshimori2005}. A further problem regarding dense \textit{colloidal} suspensions is that hydrodynamic interactions are important here, which affects both deterministic and stochastic methods \cite{FuchizakiK2002}. Within the derivation of DDFT from the projection operator method by \citet{EspanolL2009} (see \cref{derivationmz}), this is reflected by the fact that one obtains the general nonlocal equation \eqref{nonlocal}, which can be approximated by the local form \eqref{trddft} if one assumes a dilute colloidal suspension. At finite colloidal densities, hydrodynamic interactions become important and they are naturally included in extensions of standard DDFT that are based on the projection operator formalism, such as EDDFT \cite{WittkowskiLB2012} (see \cref{eddft}).

DDFT has been applied to glassy behavior in a variety of contexts: DDFT can be used to calculate friction forces in supercooled liquids \cite{BhattacharyyaB1997,Bagchi1994}. The density of the glass transition was estimated by a calculation of the time required for crossing an energy barrier by \citet{ArcherHS2007}. \citet{BerryG2011} employed a stochastic DDFT with inertia to model the glass transition, showing agreement with MCT and indications of fragile behavior. Disorder can be found to arise from the wavelengths selected by front propagation \cite{ArcherRTK2012} (see \cref{front}).

The main problem, however, arises from the third question: As shown by \citet{ReinhardtB2012}, DDFT allows hard particles to pass through each other, which makes its application to nonergodic states problematic. This was studied in detail for a 1D system by \citet{SchindlerWB2019}. A system of hard rods in 1D is nonergodic because it has a strict particle order, i.e., if particle 1 is initially on the left of particle 2, it always will be there. However, when labeling the particles and describing their dynamics using an adiabatically exact canonical DDFT for mixtures, one finds that, in contrast to the BD description, the particle order is not conserved. \citet{SchindlerWB2019} inferred that caging effects as arising in glasses are not incorporated in standard DDFT. As a solution, they suggested the use of an asymmetric interaction potential. Based on this idea, \citet{WittmannLB2020} derived a statistical-mechanical theory for ordered ensembles. This forms the basis of order-preserving dynamics (OPD), a modified DDFT which takes the conservation of particle order into account.

For stochastic DDFT, in particular in the formalism by \citet{Dean1996}, the existence of an ENE transition is a widely debated and investigated topic, in particular in the context of the path integral formalism (see \cref{renormalization}). It was discussed both whether such a transition exists at one-loop order \cite{Mazenko2011} and whether higher-order loops can affect the result \cite{KimK2008}. Results by \citet{KimKJvW2014} indicate that such a transition does not exist, even though MCT can be recovered by further approximations. 

What is in need of explanation is then the fact that DDFT nevertheless sometimes gives useful results when applied to glasses and MCT. Two aspects play a role here. First, approximations can make an important difference. As already pointed out by \citet{MarconiT1999}, the exact free energy of deterministic DDFT has a global minimum at equilibrium (corresponding to DFT), but approximate free energy functionals can have local minima. These local minima, in which the deterministic dynamics can get trapped forever if an approximate free energy functional is used, correspond to very long but finite relaxation times in the exact description. Such minima have been found by \citet{ArcherHS2007} and were used to estimate transitions to nonergodic regimes resulting from the system being unable to escape from these minima. In treatments with more accurate free energy functionals, these minima do no longer arise \cite{StopperRH2015,StopperMRH2015}. The dynamic arrest observed in Refs.\ \cite{HopkinsFAS2010,StopperMRH2015} is no longer present if more accurate FMT-type free energies are used, such that \citet{StopperRH2015,StopperMRH2015} attributed it to a too strongly simplified free energy functional. In stochastic DDFT, the treatment in renormalized perturbation theory by \citet{KimKJvW2014} (see \cref{renormalization}) showed no ENE transition. Nevertheless, making additional approximations allows to recover MCT. A second point is that a lack of nonergodicity in DDFT only makes it inapplicable to glasses insofar they are actually nonergodic. As indicated above, the argument behind the application of stochastic DDFT to glasses is precisely the fact that, in the long run, thermal fluctuations will lead to a restoration of ergodicity and to the approach to an equilibrium state. From this perspective, the absence of a nonergodic transition is an advantage rather than a problem.

Ideas from DDFT have also been employed in the study of activated hopping in glassy liquids. This is done in the theory by \citet{Schweizer2005}, which he derived building up on earlier work in Refs.\ \cite{SchweizerS2003,SaltzmanS2003}. The idea is to use the single-particle radial displacement $r(t)$ as a slow dynamical variable. At short times, the particle moves diffusively. A caging force $-\partial F_{\mathrm{eff}}/\partial r$, derived from an effective free energy $F_{\mathrm{eff}}$ that is based on ideas from DFT, but has no rigorous equilibrium meaning, captures the tendency to localize at higher volume fractions which is described by MCT \cite{Schweizer2005}. Finally, thermal noise leads to ergodicity-restoring activated hopping. A more systematic derivation of the resulting dynamic equation \eqref{schweizerhopping} proceeds as follows: Starting from the single-particle Langevin equations, a formally exact stochastic theory for the single-particle density operator is obtained. Multiplying by $\vec{r}^2$ (motivated by using the local displacement as an order parameter), integrating, and averaging gives a dynamic equation for the squared displacement of a tagged particle $r^2(t)$. This equation involves the two-body correlation, which is then related to the functional derivative of a free energy as in the standard DDFT approximation. After using the chain rule, one finds 
\begin{equation}
\xi_s \frac{\dif r(t)}{\dif t} = - \pdif{F_{\mathrm{eff}}(r(t))}{r}+\chi(t)
\label{schweizerhopping}
\end{equation}
with the short-time friction constant $\xi_s$, the effective free energy
\begin{equation}
F_{\mathrm{eff}}(r(t)) = -3k_B T\ln(r(t))-\frac{k_B T}{2}\INT{}{}{^3k}\rho\frac{c(k)^2}{(2\pi)^3}S(k)e^{-\frac{1}{3}k^2r(t)^2},
\end{equation}
where $\rho$ is a constant number density, $c(k)$ is the Fourier-transformed direct pair-correlation function, and $S(k)$ is the dimensionless collective structure factor, and the zero-mean delta-correlated noise $\chi$. This method can be used to describe hopping, diffusion, and relaxation in glassy fluids \cite{SaltzmanS2006,SaltzmanS2008,SaltzmanYS2008,YangS2011,SchweizerY2007,ZhangS2010,MirigianS2014,TripathyS2009,SaltzmanS2006b,ViehmanS2008,PhanS2018,SussmanS2011,YangS2010,KobelevS2005,GhoshS2019,SchweizerS2016,ZhangS2018}. Reviews are given by Refs.\ \cite{ChenSS2009,ChenSS2010}.

Aging effects relevant for the glass transition can also be described by nonequilibrium generalized Langevin equations (see \cref{nescgle}) that include DDFT as a special case \cite{MendozaLSRPM2017,SanchezLOGMN2014,SanchezRM2013}. \citet{BraderS2013} derived from an excess dissipation functional a generalized theory including both deterministic DDFT and MCT as limiting cases. The relation of memory functions to power functionals is further explored in Ref.\ \cite{BraderS2014}. Glass-like relaxation in gels can be observed in DMFT \cite{WitmanW2006}.

\subsubsection{\label{granular}Granular media}
We now turn to systems composed of different materials, starting with granular media (such as sand or powder). Granular gases consist of particles exhibiting inelastic collisions. A dynamical theory for a thermostatted one-dimensional granular fluid is derived in Ref.\ \cite{MarconiTC2007} using multiple-time-scale methods (see \cref{multi}). It has the form of a DDFT equation with additional terms from inelastic collisions. The discussion is extended in Ref.\ \cite{MarconiTCM2008}. More recently, a DDFT for granular fluids was derived by \citet{GoddardHO2020}, which couples dynamic equations for mass, momentum, and energy density. The dissipative effects of inelastic collisions are accounted for by moments of a collision operator.

\subsubsection{\label{plasma}Plasmas}
Plasmas are, despite having physical properties very different from those of colloidal fluids, also studied using methods from the field of DDFT. The hydrodynamics of plasmas can be described using the one-particle density defined on phase space, which is governed by a dynamic equation that depends on the two-particle density (which, in turn, depends on the three-particle density and so on). A partial closure of this Bogoliubov-Born-Green-Kirkwood-Yvon (BBGKY) hierarchy can be obtained from DDFT, which relates the two-particle density to the free energy functional. Combined with relaxation closures for stress tensor and heat flow, this allows to derive closed hydrodynamic equations for plasmas \cite{DiawM2015}. When extended to mixtures, such approaches allow to model astrophysical plasmas in white dwarfs and neutron stars \cite{DiawM2016}. Moreover, these approaches can also be extended to quantum mechanics \cite{DiawM2017} (see \cref{qm}).

\subsubsection{\label{driven}Driven soft matter}
Systems can deviate from those considered in \ZT{standard} applications of DDFT not only because they are composed of different materials, but also because they are driven away from equilibrium. This can be a consequence of external drives or of activity (the latter case is discussed in \cref{activesm}). Here, we consider driven soft matter, which is a field of application for both DDFT and extensions such as PFT \cite{StuhlmullerEdlHS2018}. This includes, in particular, systems that are subject to a time-dependent external potential \cite{DzubiellaL2003}. A review of instabilities in driven soft matter (including DDFT applications) can be found in Ref.\ \cite{Loewen2010c}. More general aspects of colloids in external fields are reviewed in Ref.\ \cite{Loewen2001}. Examples of DDFT studies of driven soft matter include colloids that are dragged through a polymer solution at a constant rate \cite{PennaDT2003}, colloids in time-dependent optical traps exhibiting breathing modes \cite{RexL2008}, attractive particles that are subject to an ac or dc drive \cite{PototskyABMSM2010,PototskyASTM2011}, flow through constrictions \cite{ZimmermannSL2016}, ions in narrow confinement \cite{BabelEL2018}, and flow-driven particles \cite{ScacchiB2018,YuJEAR2017}. Moreover, driven nonequilibrium systems have been mapped to equilibrium systems that can then be studied using equilibrium DFT \cite{DwandaruS2007}. One can also study stochastic driving, such as by a thermostat \cite{MarconiTC2007}.

Early studies were interested in potentials of the form $U_1(x-vt)$ that move at a constant rate, a scenario that has been studied for flow in narrow channels \cite{PennaT2003}, dragged particles \cite{PennaDT2003}, and sheared soft colloids \cite{RexLL2005}. As discussed in \cref{steadycurrents}, steady currents can arise as solutions of the DDFT equation in this scenario \cite{PennaT2003}. Additional effects can be found if the inertia of the particles is taken into account \cite{TarazonaM2008}. Comparisons with BD simulations show that DDFT can also be applied to more complex time-dependent potentials such as oscillating cavities \cite{RexLLD2006}. A further extension are driven systems with orientational degrees of freedom \cite{HaertelBL2010}. Synchronized driven dipoles have also been described in an equilibrium DFT \cite{JaegerK2011}.

DDFT has also been used to model pattern formation in driven soft matter. An example is nonequilibrium lane formation \cite{WachtlerKK2016,ScacchiAB2017}. Laning has been incorporated in DDFT by adding currents corresponding to the external drive \cite{ChakrabartiDL2003,ChakrabartiDL2004}. It can also arise as a consequence of shear \cite{KrugerB2011}. In Ref.\ \cite{GeigenfeinddlHS2020}, laning is studied using PFT. Periodic external fields can lead to band formation \cite{NunesAT2016}. Moreover, time-dependent potentials are a way to design spatiotemporal patterns in driven colloidal systems \cite{AraujoZKT2017}. Instabilities resulting in time-dependent density profiles are a possible consequence of external drives \cite{PototskyASTM2011}. Driven tracer particles in colloidal systems have been observed to induce local phases such as colloid-poor bridges and cavitation bubbles \cite{ScacchiB2018b}. New types of dynamical instabilities have been found to arise as a consequence of time-dependent external fields combined with intrinsic phase separation \cite{LichtnerK2014}. Finally, DDFT has been applied to synchronization and dynamic mode-locking in colloidal particles that are subject to a modulated driving force \cite{JuniperZSBALD2017}.

A further application of DDFT is the study of depletion forces in driven systems \cite{KruegerR2007}. Depletion forces in nonequilibrium systems can be shown to have very different properties than their equilibrium counterparts \cite{DzubiellaLL2003}. An interesting consequence of depletion forces is the \ZT{Brazil nut effect}, where heavy particles float on top of lighter particles \cite{KruppaNML2012}.

There are, however, certain difficulties in applying DDFT to driven systems, since the presence of external drives can distort the correlation functions from their equilibrium form \cite{ScacchiKB2016}. Improved theories have also been derived \cite{KohlIBML2012}. \citet{LipsRM2019} found results of DDFT to agree with those small-driving approximations, which (like DDFT) underestimate mean interaction forces in systems driven by drag forces. Moreover, simple advected DDFT equations cannot capture effects of shear flow, since for symmetry reasons they are solved by the unsheared density field. Modifications are required to capture effects of shear flow \cite{AerovK2014,BraderK2011,KrugerB2011} (see \cref{shear} for a discussion).

\subsubsection{\label{activesm}Active soft matter}
DDFT for active particles (see \cref{active}) is used to describe a variety of effects in active soft matter. Many studies are interested in confined systems, such as self-propelled rods in a confining channel \cite{WensinkL2008}, microswimmers confined to a plane \cite{MenzelSHL2016}, circle swimmers in a circular trap \cite{HoellLM2017}, and ABPs in a 2D trap \cite{PototskyS2012}. A typical effect is the accumulation at boundaries and surfaces \cite{WittmannB2016}, e.g., in the form of transient clusters \cite{WensinkL2008}. Active DDFTs based on an \ZT{effective} description resembling a passive system allow to investigate steady states \cite{WittmannB2016,WittmannMMB2017} and in particular to apply concepts of pressure \cite{WittmannMMB2017} and chemical potential \cite{PaliwalRvRD2018} from equilibrium thermodynamics in the active case. Moreover, DDFT allows to describe the spatial dependence of the density \cite{SharmaB2017} and is thus suitable for describing pattern formation in active matter \cite{ZakineFvW2018} (see also \cref{pattern}). For example, \citet{AroldS2019} considered laning in underdamped active systems. The DDFT for microswimmers has been applied to study stability against orientational order and collective motion as well as behavior of pushers and pullers \cite{HoellLM2018,HoellLM2019}, localizing effects in circle swimmers \cite{HoellLM2017}, and the formation of \ZT{fluid pump states} \cite{MenzelSHL2016}. Reaction DDFT can model the phase behavior of actively switching soft colloids \cite{MonchoD2020}. \citet{MenzelOL2014} used active DDFT to derive an active PFC model. Finally, active PFT is used to find steady-state sum rules \cite{KrinningerS2019} and nonequilibrium phase behavior \cite{HermannKdlHS2019,KrinningerSB2016}.

\subsubsection{\label{electro}Electrochemical systems}
DDFT has also become an important method in electrochemistry, e.g., in the description of electrolyte conductivity \cite{DemeryD2016} or charging processes in capacitors \cite{KondratK2013,KongLW2015}, as it allows for a microscopic description of ion interactions \cite{PeraudNCBDG2016}. In this field, two lines of work can be distinguished. The first one is by Chandra, Bagchi, and coworkers, who applied DDFT to obtain dynamic equations for ions \cite{ChandraJB1996,ChandraB1999} and to study viscosity, friction, and conductivity in electrolyte solutions \cite{ChandraB2000,ChandraB2000b,ChandraB2000c,DufrecheJTB2008,DufrecheB2002,ChandraBB1999,RoyYB2015,Bagchi1991,BiswasB1997,BanerjeeB2019} (see Ref.\ \cite{NaegeleHBCA2013} for a review). Here, a combination of DDFT and MCT is used. Ion conductance is driven by ion diffusion, which is related to ion friction. MCT can be used to calculate ion friction and self-diffusion coefficients. DDFT provides the force \cite{ChandraB1999} 
\begin{equation}
\vec{F}_{\mathrm{s}}\rt = k_B T \rho_{\mathrm{s}}\rt \Nabla \sum_{i=1}^{\NumberOfSpecies}\INT{}{}{^3r'}c^{(2)}_{\mathrm{s}i}(\vec{r},\vec{r}')\Delta\rho_{i}(\vec{r}',t)
\end{equation}
with the density of the tagged ion $\rho_{\mathrm{s}}$, number of ion species $\NumberOfSpecies$, direct pair-correlation function $c^{(2)}_{\mathrm{s}i}$ corresponding to the correlation between species $i$ and the tagged ion, and density fluctuations $\Delta\rho_{i}$ of ion species $i$. The friction $\delta\xi_{\mathrm{s}}$ is then obtained from the Kirkwood formula \cite{ChandraB2000b,ChandraB1999} (see also Refs.\ \cite{Kirkwood1946,Kirkwood1946b})
\begin{equation}
\delta \xi_{\mathrm{s}} = \frac{1}{3k_B T}\INT{}{}{^3r}\langle\vec{F}_{\mathrm{s}}\rt\cdot\vec{F}_{\mathrm{s}}(\vec{r},0)\rangle.
\label{kirkwoodformula}%
\end{equation}
DDFT is also used here to calculate van Hove functions \cite{ChandraB1999} (see \cref{vh}).

The second line of work concerns the Poisson-Nernst-Planck (PNP) equations, which are a very popular model for ion transport. They are given by \cite{GaoX2018} 
\begin{align}
\pdif{}{t}\rho_i\rt &= \Gamma\vec{\nabla}\cdot\big(k_B T\vec{\nabla}\rho_i\rt + \rho_i\rt(Q_i \vec{\nabla}\psi\rt 
+ \vec{\nabla}U_1(\vec{r}))\big),\label{eq:PNPeqI}\\
\vec{\nabla}^2 \psi\rt &= - \frac{1}{\epsilon_0\epsilon_{\mathrm{r}}}\sum_{i=1}^{\NumberOfSpecies}Q_i\rho_i\rt \label{eq:PNPeqII}
\end{align}
and describe the density $\rho_i$ of ion species $i$ as well as the electrostatic potential $\psi$, where $Q_i$ is the charge of an ion of species $i$ and $\epsilon_{\mathrm{r}}$ is the dielectric constant of the solvent. 
These equations describe diffusion of ions in the presence of an electrostatic potential $\psi$, which is determined by the configuration of the ions via the Poisson equation. A microscopic derivation can be found in Ref.\ \cite{SchussNE2001}. The dynamic equation for the ions can be obtained from DDFT by including an electrostatic contribution in the free energy \cite{EvertsSEvdHvdBvdR2017,WerkhovenSvR2019,WerkhovenESvR2018,ReindlB2013,MarconiMP2013}. In Refs.\ \cite{MarconiM2014,MonteferranteMM2014,MarconiM2012}, the PNP equations are obtained from multicomponent kinetic theory. The PNP equations can also be coupled to a flow field $\vec{v}\rt$, which gives the Poisson-Nernst-Planck-Navier-Stokes equations \cite{PagonabarragaRF2010,WerkhovenESvR2018,WerkhovenSvR2019}. A further hydrodynamic theory for complex charged fluids can be found in Ref.\ \cite{RotenbergPF2010}. However, the standard PNP equations do not include excluded-volume effects and are therefore valid only in the dilute limit \cite{KilicBA2007,KilicBA2007b,XuCMYL2014,BertiFGBESF2014,JanssenB2018}. A phenomenological solution to this problem was proposed by \citet{BazantSK2011}. 

To give a more accurate description, DDFT is used to model the ions as charged hard spheres \cite{JiangCJW2014b}. The resulting theory is an extension of the PNP equations that includes a hard-sphere contribution in the excess free energy, which is typically obtained from FMT \cite{JiangCJW2014,GaoX2018}. Dispersion forces can also be incorporated, leading to a nonmonotonic time evolution of the surface charge of electric double layers \cite{LianZLW2016}. In addition, excluded-volume effects can be shown to affect the dielectric constant \cite{GaoCX2017}. Impedance resonance can be studied by analyzing the effects of a time-dependent external electric field on the ions in DDFT \cite{BabelEL2018}. Further applications include reversible heating \cite{JanssenvR2017}, viscosity effects \cite{GaoX2018}, ion dynamics in nanopores \cite{LianL2018,LianSLLW2019,ZhanLZTXWKCJW2017}, and colloid-interface interactions \cite{EvertsSEvdHvdBvdR2017}. \citet{LeeKVG2015} mentioned the assumption that the ion density is low compared to the solvent density as a limitation of DDFT. Finally, the dynamics of the surface charge density field in an ion system can, by introducing an effective potential, be written as a (stochastic) DDFT \cite{MamasakhlisovNP2008}. Stochastic DDFT has also been applied to fluctuations of counterions near a charged plate \cite{Frusawa2020}. In Refs.\ \cite{WerkhovenSvR2019,WerkhovenESvR2018}, a PNP model for the surface charge density involving source terms is derived using DDFT. Effects of kinetic dielectric decrement were studied in Ref.\ \cite{QingLZTQMXZ2020}. A remaining challenge is the accurate prediction of relaxation times \cite{LianJLvR2019}. Applications of DFT to capacitive systems are reviewed in Ref.\ \cite{Haertel2017} and the theory of solid electrolytes is reviewed in Ref.\ \cite{Dieterich1990} (both reviews include some remarks on DDFT).

The PNP equations are also used in biology to model ion channels (see Refs.\ \cite{Eisenberg2002,RamirezAACA2006,Boda2014} for an overview). To account for interactions of the ions, the PNP equations can be supplemented by a chemical potential that is derived from a DFT free energy \cite{GillespieNE2002} (see also Ref.\ \cite{GillespieNE2003} for the DFT of charged hard spheres), giving a DDFT-like equation. The PNP equations extended in this way then provide a model of calcium channels, which are relevant for muscle contraction \cite{Gillespie2008,GillespieXWM2005}. Similar methods are applied in studying transport through membranes \cite{Magnico2012} and nanopores \cite{TagliazucchiRS2011}. 

\subsubsection{\label{biology}Biological systems}
Biological systems, which are particularly complex, combine many of the topics presented so far: Biological microswimmers are a prototypical example of active matter \cite{MenzelSHL2016}, transport through ion channels is a case of electrochemical transport \cite{Gillespie2008}, polymers are important in biology \cite{MilchevEB2010}, and biofilms can be described using thin-film models \cite{TrinschekJLT2017}. A good example for a biological application of DDFT is the study of proteins \cite{Tozzini2009}, in particular protein adsorption. Here, one analyzes how proteins adsorb on a surface where polymers are chemically attached. This is important, e.g., in the development of biomimetic materials, where one wishes to avoid adsorption on artificial organs \cite{CarignanoS2000}, or for drug delivery, where adsorption of the drug carriers in the blood stream needs to be delayed so that they can reach their target \cite{FangSS2005}. Moreover, protein adsorption is central in various fields of biotechnology \cite{AngiolettiBD2014}. Experimental results on protein adsorption (with comparison to theory) can be found in Ref.\ \cite{OberleCADB2015}. One is interested in both minimizing the equilibrium adsorption and in the control of the adsorption kinetics. 

With DDFT, it is possible to obtain construction guidelines for the prevention of adsorption \cite{FangSS2005}. In the simplest case, one assumes the solvent and the polymers to relax very quickly and uses a DDFT equation to model the dynamics of the protein density. The free energy, whose functional derivative gives the chemical potential determining the protein adsorption, depends on the configuration of polymers and solvent \cite{SatulovskyCS2000}. More complex approaches include various types of proteins and allow the proteins to be in various configurations. Intermolecular repulsion can be taken into account through an additional volume constraint \cite{FangS2001}. Extended models allow to incorporate effects of electrostatic interactions and salt concentrations on the adsorption behavior \cite{FangS2003}. The free energy of protein adsorption models includes, in addition to the ideal gas term and the external potential, the adsorption free energy for protein-specific protein-surface interactions, a term for electrostatic interactions, and the usual excess free energy for protein-protein interactions \cite{AngiolettiBD2014}. Mixtures of various protein types can also be studied \cite{FangS2003,FangS2001}. An interesting application of DDFT is the Vroman effect \cite{VromanA1969}, where proteins show a nonmonotonic adsorption behavior (typically, smaller proteins adsorb first and are then replaced by larger ones) \cite{FangS2001,AngiolettiBD2018}. DDFT-based predictions agree with experiments much better than predictions from simple diffusion equations \cite{AngiolettiBD2014}. Reviews are given by Refs.\ \cite{WeiBADWNLBH2014,WeiBADWNLBH2014b,Angioletti2017,XuALDB2018,DograBCCGBCW2019,LongoPS2019,ChakrabortyEM2020}.

A further biological application is the study of cancer growth. A DDFT model for cancer has been developed by \citet{ChauviereLC2012} and further explored by \citet{AlSaediHAW2018}. Here, cells are modeled as interacting Brownian particles. The derivation incorporates the possibility of different cell species and phenotypes. A central property of cells that is absent for other Brownian particles is that they can reproduce and die. Hence, an additional term for this nonconserved part of the dynamics has to be added. For describing tumor growth, the DDFT is coupled to a reaction-diffusion equation for the density of a nutrient, whose availability determines the birth and death rates \cite{ChauviereLC2012}. By considering a system of two cell species, the competition of healthy and cancer cells can also be studied \cite{AlSaediHAW2018}. A general overview over models for cancer growth is given in Ref.\ \cite{WangBKCD2015}. Similarly, the dynamics of cell colonies can also be studied via PFC models that can be derived from DDFT \cite{AlandHLV2015}.

Another biophysical application of DDFT is protein diffusion in cell membranes. The membrane is a mixture of various types of lipids, which rearrange if a charged macromolecule adsorbs. This dynamics can be modeled in DDFT \cite{KhelashviliWH2008}. DDFT also serves as a basis for lipid bilayer models \cite{BlomP2004} and plays a role in biological solvation dynamics \cite{SenPBB2006}. Moreover DDFT-like methods have been applied in the derivation of a dynamical model for Mongolian gazelles \cite{MartinezCMOL2013}. A DDFT model for disease spreading \cite{teVrugtBW2020} is discussed in \cref{disease}. Finally, DDFT is used to study ion channels \cite{Gillespie2008,GillespieXWM2005} (see \cref{electro} for details).

\subsection{\label{phenomena}Phenomena}
\subsubsection{\label{relax}Relaxation}
As discussed in \cref{thermo}, DDFT is a theory that describes the diffusive relaxation of colloidal systems towards an equilibrium state corresponding to the minimum of the free energy. Hence, a natural application of DDFT is to model relaxation dynamics, i.e., the way in which a system moves to an equilibrium state after initially being out of equilibrium. For nonspherical particles, DDFT can be used to describe orientational relaxation, which was one of the main applications of early forms of DDFT \cite{ChandraB1990,BagchiC1993,BagchiC1991} (see \cref{nonspherical}). 

Relaxation dynamics was studied by \citet{BiervR2007} for platelike colloids. Here, a fluid is perturbed by a laser beam from an initial isotropic state. Depending on laser power and initial chemical potential, different relaxation paths are observed. The same model is applied in Ref.\ \cite{BiervR2008} to a system connecting two reservoirs. Relaxing binary fluids in a cavity were studied by \citet{Archer2005}. For certain systems, relaxation can also have the form of a gravitational collapse \cite{BleibelDOD2014,BleibelDO2016,ChavanisS2011b,BleibelDO2011}. \citet{UematsuY2012} analyzed the polarization relaxation of molecular liquids in external fields. \citet{BierA2013} compared relaxation processes of interfaces and bulk phases. In Ref.\ \cite{Bier2015}, nonequilibrium surface tension is considered. If the equilibrium state is uniform, interfaces and thus surface tension are nonequilibrium phenomena. Their time evolution can be calculated in DDFT (see Ref.\ \cite{LamorgeseM2016} for another discussion of nonequilibrium surface tension). Relaxation of colloid-polymer mixtures was studied by \citet{StopperRH2016}. \citet{KrugerD2017} used DDFT as a comparison to test the relaxation dynamics of a theory of density fluctuations. \citet{NunesGAdG2018} analyzed the relaxation of transient demixed states in a binary mixture. As a direction for future research, the study of nonequilibrium fluctuations in DDFT was suggested by \citet{GiavazziSVC2016}.

Studying relaxation dynamics is also interesting for relating DDFT to glassy dynamics as described by MCT. In general, the van Hove function provides a good measure for relaxation to equilibrium \cite{BraderS2014}. Its calculation is a typical application of DDFT (see \cref{vh}). For glasses, a two-stage relaxation of the self-part of the van Hove function with a diverging relaxation time is characteristic. As shown in Ref.\ \cite{HopkinsFAS2010}, this relaxation behavior can be recovered in DDFT. \citet{BerryG2011} found agreement between stochastic DDFT and MCT. \citet{ReinhardtB2012} have, however, argued that DDFT is not capable of describing the nonergodic transition of simple MCT. (This point is discussed in \cref{glass}.) Moreover, deviations from the true relaxation behavior can result from neglecting memory effects and superadiabatic forces \cite{YoshimoriDP1998,FortinidlHBS2014,BraderS2014}. The validity of DDFT in the description of relaxational dynamics was explored by \citet{PennaT2006} by comparisons to BD simulations.

\subsubsection{\label{ps}Phase separation}
Phase separation in colloidal fluids can occur through nucleation (if they are quenched near the binodal) or spinodal decomposition (if they are quenched inside the binodal). The investigation of spinodal decomposition was important for the early stages in the development of DDFT and its 
predecessors \cite{Evans1979,EvansTdG1979,Cahn1965,Abraham1976}. In Ref.\ \cite{ArcherE2004}, spinodal decomposition is studied based on DDFT. A linear stability analysis (see \cref{linear}) can be used to model the early stages of phase separation. Linear stability analysis can also reveal the existence of different ways for the homogeneous state to become unstable (spinodal and freezing modes) \cite{PototskyASTM2011,PototskyTA2014}. For the intermediate stages, a nonlinear theory is required. A comparison between DDFT and experiments was made by \citet{ZvyagolskayaAB2011}.

In Ref.\ \cite{Archer2005}, DDFT is applied to Gaussian core model (GCM) binary fluids exhibiting liquid-liquid phase separation. Both bulk phase separation and microphase separation occur. A symmetric GCM fluid is considered in Ref.\ \cite{Archer2005b}. If the different particle types have symmetric properties, the ensemble-averaged one-body profile is the same for both of them if the confining potentials are identical. DDFT is able to predict the phase separation if the symmetry of the confining potentials is broken. In addition, DDFT can be applied to gas-liquid phase separation. This gives insights into the influence of phase separation on sedimentation \cite{ArcherM2011}. Colloids with capillary interactions can be found to have a phase diagram that involves not only spinodal decomposition, but also a gravitation-like collapse \cite{BleibelDOD2014,BleibelDO2011}. Such a collapse was studied for a uniform disklike initial distribution by \citet{BleibelDO2016}. A detailed discussion  of the gravitational collapse of Brownian particles can be found in Refs.\ \cite{ChavanisS2011b,Chavanis2011b}. For binary fluids in a gravitational field, the interplay between thermodynamics and sedimentation can lead to a rich phase diagram \cite{MalijevskyA2013}. In studies of 1D hard-sphere fluids, conditions for the coexistence of 1D and quasi-0D states can be found to be analogous to conditions for phase separation \cite{Frusawa2014}. Coarsening modes (Ostwald ripening and translational modes) are studied in Ref.\ \cite{PototskyTA2014}. DDFT models including hydrodynamics can also exhibit phase separation, as shown for a van der Waals fluid in Ref.\ \cite{KikkinidesM2015}. In dilute binary mixtures, phase separation can lead to the emergence of bubbles that show damped oscillations \cite{OkamotoO2016}. Nonequilibrium phase separation under shear is discussed in Ref.\ \cite{StopperR2018}. For otherwise identical particles with different Stokes coefficients, external fields can lead to demixing \cite{NunesGAdG2018}. DDFT is also applied to phase separation of particles on a sphere \cite{BottB2016}. Finally, fluid demixing on spherical surfaces can be analyzed using Minkowski functionals \cite{BobelBMBR2019}. 

Local phase transitions are considered in Refs.\ \cite{ReinhardtSB2014,ScacchiB2018b}. Behind a driven tracer particle, a colloid-poor phase emerges that eventually forms a bubble \cite{ScacchiB2018b}. Cavitation, which is a local phase separation, results from the pressure being higher in front of a solid particle being driven through a fluid than behind it \cite{ReinhardtSB2014}. Microrheology, including DDFT applications, was reviewed in Ref.\ \cite{PuertasV2014}.

Phase separation is also modeled using DDFT in more complex fluids. A mixture of particles with opposing adsorption preferences can show a colloidal-liquid-colloidal-liquid demixing transition driven by critical fluctuations \cite{ZvyagolskayaAB2011}. Magnetic particles, with have the magnetic moment as an additional degree of freedom, can also show phase separation \cite{LichtnerAK2012}. Spinodal decomposition in a magnetic fluid is considered in Ref.\ \cite{LichtnerK2013}. In Ref.\ \cite{VuijkBS2019}, spinodal decomposition is studied in a system with an anisotropic diffusion tensor. Microphase separation in polymer systems is discussed in Refs.\ \cite{XuLH2007,XuWHH2008,DengZHLH2011,DengPHL2016}. Regarding nonequilibrium phase separation, DDFT is used in the context of lane formation \cite{WachtlerKK2016,ChakrabartiDL2004,ChakrabartiDL2003,AroldS2019,ScacchiAB2017,GeigenfeinddlHS2020} (see \cref{pattern}). A study of nonequilibrium phase separation in PFT can be found in Ref.\ \cite{KrinningerSB2016} (see also Ref.\ \cite{KrinningerSB2017}). In Refs.\ \cite{WittmannB2016,WittmannMMB2017}, DDFT was applied to phase separation and phase equilibria in active systems. Phase coexistence of ABPs was moreover studied by \citet{PaliwalRvRD2018} using a chemical-potential-like function. Actively switching colloids were studied in Ref.\ \cite{MonchoD2020}.

Polymer DDFT (see \cref{pd}) is a very frequently used method in the description of phase separation in polymer systems, in particular copolymer melts \cite{FraaijeNRG2011,KyrylyukF2004,ZvelindovskySvVMF1998,MauritsZSvVF1998,PloshnikLHBMFAS2013,SevinkFH2002,MauritsZF1998,HondaK2007,HondaK2007b,TanSNWGY2009,HaoMSNL2014,KnollHLKSZM2002,XiaSQZY2005,NeratovaKK2010,SevinkZvVMF1999,SunHL2005,HorvatLSZM2004,KomarovVCK2010,KyrylyukLS2005,ZhangYW2006,KnollLHKSZM2004,SpyriouniV2001,YangYZY2008,MuLW2013,LiuLKPR2012,JawalkarRHSA2007,LiuHPGR2012,ChiangraengLN2019,MuLW2012,JawalkarASNA2005,MuLW2015,LamG2003,MaitiWG2005,TsarkovaHKZSM2006,Kawakatsu2004,MoritaKD2001,CoslanichFCMS2004,CoslanichFFS2004,XuLH2007b,XuWHH2008,DongZCSXW2010,ZhangYXL2007}.

When studying phase separation in DDFT, the problem of averaging explained in \cref{noise} is of potential importance. This was discussed in detail by \citet{BleibelDO2016} who studied the capillary collapse of a uniform disk of particles and compared the time evolutions predicted by DDFT with and without an average over the initial distribution of the particles. The results do not agree, since the random initial distribution of particles that is captured only by a density profile \textit{not} averaged over initial conditions can lead to the transient formation of dense regions. This aspect can be relevant for spinodal decomposition, since even for an unstable bulk state an average over initial conditions may lead to a constant density profile and then to $\partial_t\rho = 0$. The typical solution for this problem, which is to add a small perturbation to the constant density profile (see \cref{linear}), does not always give accurate results \cite{BleibelDO2016}.

\subsubsection{\label{pattern}Pattern formation}
Pattern formation, which can be a consequence of phase separation, is one of the central effects studied in soft matter theory. As a dynamical theory for the spatial distribution of the particles, DDFT is also well suited to describe pattern formation. A good example is lane formation, which is a nonequilibrium phase transition \cite{LoewenD2003}. Laning can arise, e.g., in colloidal mixtures where different particle types are driven in different ways by an external force \cite{ChakrabartiDL2003}. It can be analyzed using linear stability analysis \cite{ChakrabartiDL2003,ChakrabartiDL2004,ScacchiAB2017}. Moreover, lane formation can occur on short time scales in systems that are subject to time-dependent shear \cite{KrugerB2011}. Laning can also be observed in active matter \cite{AroldS2019}. Superadiabatic effects in lane formation are considered in Ref.\ \cite{GeigenfeinddlHS2020}. DDFT has also been used to describe reentrance effects \cite{ChakrabartiDL2004} and fluids with attractive interactions \cite{WachtlerKK2016}. A particularly frequently used method for studying pattern formation is polymer DDFT \cite{KyrylyukCF2005,SevinkZFH2001,KyrylyukSZF2003,FraaijeZSM2000,vanVlimmerenMZSF1999,HeHLP2002,KomarovVCKK2010,TanZDSY2013,HuangHHL2003,LudwigsBVRMK2003,LuHL2004,ScherdelSM2007,Berendsen1999,NeratovaPK2010,MichielsenDR2001,KyrylyukF2004b,BootsdRNBPWDWF2014,SevinkF2008b,SevinkZF2003}.

Pattern formation can also occur in evaporation/dewetting processes in thin films (see also \cref{thin}). Here, effects such as fingering can occur as a consequence of the interplay between transport and evaporation processes. Suspensions of nanoparticles in fluids are modeled by a DDFT for mixtures that also contains nonconserved terms for evaporation. Tools employed here are both linear stability analysis and the numerical solution of the full nonlinear DDFT \cite{RobbinsAT2011,ThieleVARFSPMBM2009,ArcherRT2010}.

Recently, DDFT has been applied to quasicrystal formation in soft matter \cite{RatliffASR2019,WaltersSAE2018}. Self-organization into quasicrystals requires the presence of two different dominant wavenumbers (although other wavenumbers can be important too \cite{RatliffASR2019}). DDFT allows, by means of a linear stability analysis, to identify important wavenumbers, giving insights into which length scales are important. \citet{ArcherRK2013,ArcherRK2015} found quasicrystal formation in DDFT to be the result of linear growth of one wavenumber followed by the nonlinear selection of another one. A strategy for finding quasicrystals was proposed in Ref.\ \cite{ScacchiSBA2020}. Quasicrystals have also been studied in PFC models \cite{AchimSL2014,SubramanianAKR2016,SubramanianAKR2018}.

In colloidal systems, DDFT is used to study, e.g., the design of spatiotemporal patterns by means of time-dependent external potentials \cite{AraujoZKT2017} or traveling band formation \cite{TaramaEL2019}. Stripe formation in magnetic fluids can occur as a consequence of intrinsic phase separation combined with external drives \cite{LichtnerK2014}. In a DDFT model for epidemic spreading (see \cref{disease}), the separation of healthy and infected persons can be observed \cite{teVrugtBW2020}. Further applications of DDFT to pattern formation include patterns behind fronts \cite{ArcherRTK2012}, crystallization on structured surfaces \cite{NeuhausSL2013}, Brownian particles with repulsive soft-core interactions \cite{DelfauOLBH2016}, band formation \cite{NunesAT2016}, activity-driven spins \cite{ZakineFvW2018}, and defect morphologies in polyethylene lattices \cite{HouYDZMH2019}. Images of pattern formation allow to extract constitutive relations \cite{ZhaoSBB2020}. The investigation of localized states \cite{Knobloch2015,Knobloch2016,ThieleARGK2013} is a possible direction for future work. A general review of pattern formation is given by Ref.\ \cite{CrossH1993}.

\subsubsection{\label{nucandsol}Nucleation and solidification}
\begin{figure}
\includegraphics[width=\linewidth]{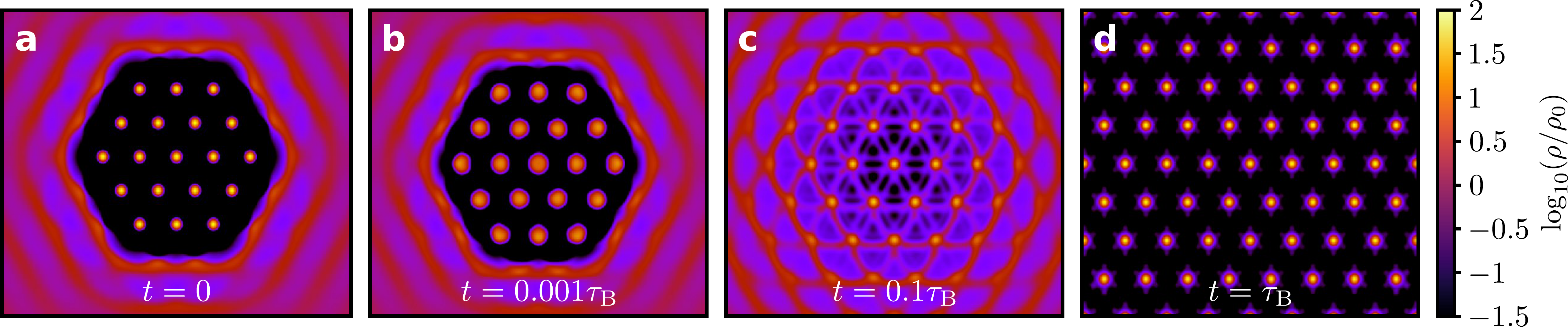}
\caption{\label{fig:nucleation}Time-evolving density field $\rho\rt$ with mean value $\rho_0$ in a DDFT model for solidification. An imposed nucleation cluster grows to a crystal. Adapted from Fig.\ 2 from Ref.\ \protect\cite{vanTeeffelenLL2008b}.}
\end{figure}

A topic that is related to both relaxation and phase separation is the study of nucleation and solidification. Static DFT is a central tool for the study of nucleation and crystallization \cite{SomervilleSAHAB2018} (see Ref.\ \cite{LaaksonenTO1995} for a review). It can be used as a basis for calculating dynamical nucleation rates \cite{TalanquerO1994,TalanquerO1995,LaaksonenTO1995}. DDFT has also been applied to this problem. A brief review can be found in Ref.\ \cite{NeuhausHMSL2014}. In nucleation theory, the critical nucleus can be determined from DFT and is then used as a starting point for a time evolution in DDFT \cite{KahlL2009}. The initial nucleation cluster can also be imposed, DDFT then predicts whether or not a crystal will grow \cite{Loewen2010b}. DDFT is a useful tool for modeling heterogeneous nucleation \cite{LoewenLABvT2007}. Effects that can be analyzed in DDFT include the difference between rapid and slow heating \cite{LoewenLABvT2007}, freezing/melting in interfacial systems \cite{LiuHCXH2011}, crystallization with multiple seeds and grain boundary formation \cite{NeuhausSL2013b}, nucleation and growth of ice crystals \cite{YeNTZM2016}, formation of protein clusters \cite{LutskoN2016,Lutsko2016}, influence of confinement obstacles on nucleation \cite{GonzalezMV2017}, and self-assembly of nanoparticles \cite{HouDZTM2017}. Moreover, crystallization on patterned surfaces has been studied, where the interplay between crystal and substrate structures leads to \ZT{compatibility waves} \cite{NeuhausSL2013}. Stochastic methods are also applied in this field \cite{DuranYKL2018}.

An example is shown in \vref{fig:nucleation}, which is adapted from Ref.\ \cite{vanTeeffelenLL2008b}. Here, crystal growth at externally imposed nucleation clusters was studied. Figure \ref{fig:nucleation} shows snapshots of the time evolution arising from a weakly compressed rhombic nucleation seed consisting of 19 tagged particles. As can be seen, DDFT describes the evolution to the crystalline equilibrium state.

Note that, however, many studies of nucleation processes are performed using PFC models rather than DDFT (see Refs.\ \cite{BackofenRV2007,GranasyTTP2011,TothTPTG2010,TothTPTG2012,TothPTTG2011,GranasyT2014,PodmaniczkyTPG2016,BackofenV2010,KorbulyPTHPG2017} for examples for and Refs.\ \cite{GranasyPTTP2014,PodmaniczkyTTG2015,TothPTG2012,PodmaniczkyTTPG2017,GranasyPD2013,PodmaniczkyTPG2016} for an overview over the use of PFC models in nucleation theory). Although they involve more approximations than DDFT (and are thus quantitatively less accurate \cite{HouYDZM2017}), PFC models are very successful in this area. Comparisons of DDFT and PFC models in solidification can be found in Refs.\ \cite{vanTeeffelenBVL2009,JaatinenA2010,HouYDZM2017,KahlL2009}. Typically, the agreement between PFC models and DDFT is better at longer wavelengths \cite{JaatinenAEA2009}. \citet{ArcherRRS2019} discussed in detail the effect of the PFC approximation on the phase diagram (see \cref{phasefield}). An overview over various theoretical and experimental approaches to nucleation is given by Ref.\ \cite{EmmerichVWS2014}.

For solidification, DDFT allows, in contrast to DFT, to find not only the equilibrium profile, but also effects of the dynamical evolution on the structure. For example, DDFT has been applied to solidification fronts (see \cref{front}). The wavelengths selected by front propagation can differ from that of equilibrium crystals, leading to disorder \cite{ArcherRTK2012,ArcherWTK2014,ArcherWTK2016}. This is particularly interesting for analyzing why some materials form glasses \cite{ArcherWTK2014}. Moreover, phases that are developed during freezing, but which are not the equilibrium state, such as quasicrystals, can also be observed in DDFT \cite{ArcherRK2015,ArcherRK2013}. The freezing transition can also be analyzed in kinetic extensions of DDFT \cite{BaskaranBL2014}. \citet{HouYDZM2017} considered self-healing effects in crystal growth, where, despite initial impurities being present, particles rearrange to the ideal crystal structure. Finally, even if one is only interested in the long-time equilibrium state, DDFT can be used as a minimization procedure and thus as a tool for DFT \cite{vanTeeffelenLL2008}.

In general, nucleation involves that an energy barrier is crossed which has its maximum at the critical nucleus size. A method for the calculation of nucleation pathways is to find the \ZT{minimal free energy path}, with is the most likely route for nucleation \cite{Lutsko2008,Iwamatsu2009}. A particularly interesting strategy was suggested by Lutsko \cite{Lutsko2011,Lutsko2012,Lutsko2012b,Lutsko2019} for obtaining the \ZT{most likely path} for nucleation. The basic idea originates from \citet{OnsagerM1953}. DDFT, being a theory in which the free energy monotonously decreases, can (without noise terms) not describe energy barrier crossing \cite{DuranL2015}. One can, however, start at the top of the barrier and then perform a DDFT time evolution forwards in time to the final state and backwards in time to the initial state (both going downwards the energy barrier), thereby determining the most likely nucleation pathway.

\subsubsection{\label{chemical}Chemical reactions}
Reaction-diffusion equations, which are obtained by adding reaction terms to the diffusion equation \eqref{diffusionequation}, have a long tradition in the description of effects such as pattern formation \cite{Turing1952}. Similar reaction models can also be obtained using the Cahn-Hilliard equation \eqref{cahnhilliardequation} rather than the diffusion equation \cite{GlotzerDMM1995}. Since DDFT is a generalized diffusion equation, one can incorporate chemical reactions by extending DDFT to a reaction-diffusion equation by adding reaction terms \cite{Lutsko2016,LutskoN2016}. This is the idea behind \ZT{reaction-diffusion DFT} (RDDFT), which is applied to NO oxidation in Ref.\ \cite{LiuL2020}. Reactions can only occur if the reacting particles meet, such that simple models are constructed through reaction-diffusion equations in which the number of encounters is limited by diffusion but not by interactions. However, in particular in crowded environments such as cells, the interactions of the particles have effects on the frequency of encounters \cite{DorsazDMPDLRF2010}. This problem can be studied in DDFT, which allows to incorporate such effects \cite{PiazzaDDDF2013}. Similarly, DDFT can be used to analyze the effects of the product species on diffusion-controlled reactions. Products can, e.g., accumulate at catalysts and thereby interact with the reactants which might have very different physical properties \cite{RoaSKD2018}. Diffusion-controlled reactions are moreover discussed in Ref.\ \cite{DzubiellaM2005} (based on a linear theory related to DDFT). Besides these general studies, the formation and stability of protein clusters has been studied using DDFT \cite{LutskoN2016}. A more accurate model takes into account, also within DDFT, that the reaction rates are not fixed but state dependent \cite{Lutsko2016}. Recently, RDDFT was applied to study active switching, the switches then correspond to the reaction terms \cite{MonchoD2020}. It also plays a role in the description of surfactant-covered interfaces \cite{GrawitterS2018}. Nanoreactors \cite{AngiolettiLBD2015} and crowded systems under shear \cite{ZacconeDPDMMF2011} are an additional possible field of application. Nonideal diffusion also plays a role in Brownian aggregation, where FMT and DDFT can be used to predict aggregation rate constants \cite{KelkarFC2014,KelkarFC2015}.

\subsubsection{\label{disease}Disease spreading}
Closely related to RDDFT is the DDFT model for disease spreading proposed by \citet{teVrugtBW2020}. It is an extension of the simple SIR model \cite{KermackM1927} used in theoretical biology, which describes the temporal evolution of the number of susceptible persons $S$, infected persons $I$, and recovered persons $R$ under the assumptions that susceptible persons are infected at a rate $c_{\mathrm{ir}}$ when meeting infected persons and that infected persons recover at a rate $c_{\mathrm{rr}}$ and are immune afterwards. In an extension known as SIRD model, it is additionally assumed that infected persons die at a rate $c_{\mathrm{dr}}$. The SIR-DDFT model is given by 
\begin{align}
\pdif{}{t}S\rt &= \Gamma_S\Nabla\cdot\bigg(S\rt\Nabla\frac{\delta F}{\delta S\rt}\bigg) - c_{\mathrm{ir}}S\rt I\rt \label{sr},\\  
\pdif{}{t}I\rt &= \Gamma_I\Nabla\cdot\bigg(I\rt\Nabla\frac{\delta F}{\delta I\rt}\bigg) + c_{\mathrm{ir}}S\rt I\rt - c_{\mathrm{rr}} I\rt - c_{\mathrm{dr}} I\rt \label{ir},\\ 
\pdif{}{t}R\rt &=\Gamma_R\Nabla\cdot\bigg(R\rt\Nabla\frac{\delta F}{\delta R\rt}\bigg) + c_{\mathrm{rr}} I\rt \label{rr}
\end{align}
and describes persons as interacting particles, where repulsive interactions contained in the functional $F$ incorporate effects of social distancing and quarantine. The mobilities $\Gamma_S$, $\Gamma_I$, and $\Gamma_R$ of susceptible, infected, and recovered persons, respectively, can be different. Numerically, it can be shown that at certain interaction strengths the model undergoes a transition to a phase in which the number of infected persons is significantly reduced as a consequence of isolation and social distancing. The interactions lead to a pattern formation effect in which infected and healthy persons separate. An extension of the SIR-DDFT model incorporating adaptive political interventions was developed in Ref.\ \cite{teVrugtBW2020b}, where the effects of various containment strategies on disease spreading and the conditions under which multiple waves of a pandemic occur are investigated.

\subsubsection{\label{fc}Feedback control}
Another possible reason for pattern formation is time-delayed feedback, as shown by \citet{TaramaEL2019} for traveling bands. DDFT is a useful tool for modeling effects of time-delayed feedback in systems with interactions. In time-delayed feedback control, a (nonlinear) system is controlled using forces that depend on the state of the system at a previous time $t - \tau_d$ where $\tau_d$ is a delay. A current topic of research is the way in which such control mechanisms are affected by particle interactions in the controlled system. This topic is reviewed in Ref.\ \cite{GernertLLK2016}. DDFT allows to incorporate particle interactions \cite{LichtnerPK2012}. As shown by \citet{LichtnerK2010}, a current reversal can be observed in interacting colloidal systems controlled by time-delayed feedback. \citet{GernertK2015} proposed a feedback control scheme for enhancing colloidal motility in interacting systems. \citet{GrawitterS2018} explored a feedback mechanism coupling flow velocities to external light spots in fluids with photoresponsive surfactants (see also Ref.\ \cite{GrawitterS2018b}). 

\subsubsection{\label{sound}Sound waves}
In Ref.\ \cite{Archer2006}, the speed of sound in an atomic fluid is calculated using the DDFT for atomic fluids (see \cref{inertia}). Taylor expanding the excess free energy, decomposing the density field as $\rho\rt = \rho_0 +\Delta\rho\rt$, and using \cref{atomic} gives
\begin{equation}
\pdif{^2}{t^2}\Delta\rho\rt+\nu\pdif{}{t}\Delta\rho\rt =\frac{k_B T}{m}\vec{\nabla}^2\Delta\rho\rt - \frac{k_B T\rho_0}{m}\Nabla^2\INT{}{}{^3r'}\Delta\rho(\vec{r}',t)c^{(2)}(\vec{r}-\vec{r}').
\label{soundequation}
\end{equation}
Making the plane-wave ansatz
\begin{equation}
\Delta\rho\rt = \SmallAmplitude e^{\ii(\vec{k}\cdot\vec{r}-\omega t)}
\end{equation}
with a small amplitude $\SmallAmplitude$ allows to obtain the dispersion relation for the considered fluid. Setting $\vec{k}\cdot\vec{k}=q^2+2\ii q/\AttenuationLength$ with $q, \AttenuationLength \in \mathbb{R}$ (which corresponds to assuming $\AttenuationLength^{-1} \ll q$, i.e., to assuming that propagation and attenuation length scales are well separated) and separating real and imaginary parts in \cref{soundequation} gives the dispersion relation 
\begin{equation}
\omega^2(q)=\frac{q^2k_B T}{mS(q)}
\end{equation}
and the attenuation length
\begin{equation}
\AttenuationLength(q)=\frac{2\omega}{\nu q}.
\end{equation}
In the long-wavelength limit, we obtain the speed of sound
\begin{equation}
\SpeedSound = \frac{\omega}{q}=\frac{1}{\sqrt{\rho_0 k_B T \chi_T}}
\label{speedofsound}
\end{equation}
with the isothermal compressibility $\chi_T$, when using that $S(0)\approx \rho_0 k_B T \chi_T$ \cite{HansenMD2009}. Due to the negligence of temperature fluctuations in \cref{atomic}, the result \eqref{speedofsound} is only valid at low temperatures and high densities. In Ref.\ \cite{DuqueZumajoCdlTCE2019}, sound propagation in confined fluids perpendicular to walls is discussed using a DDFT with momentum density.

While these approaches consider fluids, DFT is also useful for the description of waves in crystals. It allows to calculate the free energy change associated with a small deformation of a crystal, which then gives the phonon dispersion relation \cite{MahatoKR1991,TosiT1994,FerconiT1991b,FerconiT1991}. More recently, a combination of DFT and projection operator methods (see \cref{po}) was employed in Refs.\ \cite{WalzF2010,RasSF2020}. The relevant variables include, in a state with spontaneously broken symmetry, not only the conserved hydrodynamic variables, but also symmetry-restoring variables. For the theory developed in Refs.\ \cite{WalzF2010,RasSF2020}, the relevant variables are therefore given by the density fluctuations near the reciprocal lattice vectors and by the momentum density. If one considers the linear response regime and neglects dissipation, the projection operator formalism gives linear reversible equations of motion where the transport coefficients are given by equilibrium correlation functions. These can be evaluated using DFT.

\subsection{\label{vh}Dynamics of the van Hove function}
Finally, we present a \ZT{theoretical application} of DDFT, namely the investigation of the van Hove function. This is connected to aspects discussed in previous sections, since the van Hove function is important, e.g., for relaxation dynamics (\cref{relax}) and electrochemistry (\cref{electro}). Two further theoretical applications of DDFT, namely the derivation of MCT and PFC models, are discussed in \cref{mct,phasefield}, respectively.

The van Hove function \cite{vanHove1954} is a quantity used in the description of diffusion processes. It can be measured through confocal microscopy and scattering techniques \cite{StopperRH2016}. For a system of $N$ particles, the van Hove function is given by
\begin{equation}
G_{\mathrm{vH}}\rt=\frac{1}{N}\bigg\langle\sum_{i=1}^{N}\sum_{j=1}^{N}\delta(\vec{r} + \vec{r}_j(0)-\vec{r}_i(t))\bigg\rangle
\end{equation}
and measures the probability of finding an arbitrary particle $i$ located at position $\vec{r}$ at time $t$ given that a particle $j$ has been located at the origin at $t=0$ \cite{LiuWDZM2020}. The van Hove function can be splitted into a self-part and a distinct part, distinguishing between the cases $i=j$ and $i\neq j$ \cite{StopperMRH2015}. 

DDFT methods were applied to calculate van Hove functions in ion systems in Refs.\ \cite{ChandraB1999,ChandraB2000,ChandraB2000b,ChandraB2000c,DufrecheJTB2008,DufrecheB2002,ChandraBB1999}. Here, a generalized transport equation including momentum and memory (similar to \cref{pdg0,pdg1}), which is obtained from DDFT, is used. The unknown memory kernel is assumed to be equal to the ion friction (see \cref{electro}). On this basis, a closed equation for the van Hove function of an electrolyte system can be obtained, which is most conveniently dealt with in Fourier-Laplace space \cite{ChandraB1999}.

A method for studying the dynamics of the van Hove function in DDFT via a generalization of Percus' test particle limit \cite{Percus1962} to dynamical correlation functions was derived by \citet{ArcherHS2007}. The idea is that a one-component system can be thought of as a binary mixture of two species $s$ and $d$ with identical interactions and properties. The species $s$ only consists of one particle. This method allows to keep track of a certain \ZT{test particle}. The self-part and the distinct part of the van Hove function can then be identified with the one-body densities. Their dynamics is described by DDFT \cite{ArcherHS2007}. This method has been applied to complex fluids \cite{BiervRDvdS2008}, systems of colloidal hard spheres \cite{HopkinsFAS2010,StopperRH2015,StopperMRH2015}, one-dimensional hard rods \cite{ReinhardtB2012}, colloid-polymer mixtures \cite{StopperRH2016}, Brownian disks \cite{StopperTDR2018}, and polymer nanocomposites \cite{YeNTZM2019}. A comparison to experiments is made in Ref.\ \cite{StopperTDR2018}. An extended dynamical theory for the van Hove function has been obtained by using PFT instead of DDFT \cite{BraderS2015}. The van Hove function can also be determined from the nonequilibrium Ornstein-Zernike relation derived by \citet{BraderS2013}. Superadiabatic van Hove currents are calculated in Ref.\ \cite{SchindlerS2016}.

\section{\label{outlook}Outlook}
After having reviewed the existing literature on DDFT, we here give an outlook over important directions of future work.

An interesting idea would be the application of machine learning methods to DDFT, as it is already successfully done in quantum DFT. Machine learning methods can be used to find approximate free energy functionals \cite{LinO2019} that are required as an input for DDFT.

On top of that, the various lines of DDFT presented in \cref{traditional} have had a relatively independent development. In particular, articles using or extending the DDFT by \citet{MarconiT1999} rarely discuss work based on the DDFT by \citet{Fraaije1993} and vice versa, even though both methods are very widely used. This is a missed opportunity, since results obtained in one \ZT{strain} of DDFT can also be useful for the other one. The method by \citet{MarconiT1999} has undergone a rich theoretical development, while simulation methods based on the work by \citet{FraaijevVMPEHAGW1997} are now used in polymer industry. Combinations would offer further room for both theory and applications. 

Moreover, applications of DDFT have, up to now, been focused on various types of liquids. While this is not surprising given that this is what the theory is constructed for, applications to quantum and astrophysical systems \cite{Schmidt2015,DiawM2016,DiawM2017} show that the general principle can also be applied in other contexts. Further exploring the (possible) connections to theories for many-particle quantum or astrophysical systems could thus inspire a huge number of extensions as well as applications of classical DDFT. On the other hand, methods that are established in quantum mechanics, such as the Car-Parrinello approach \cite{CarP1985}, could have potential for classical DDFT \cite{Loewen2015}.

In soft matter physics, applications in biological contexts have a lot of unexplored potential. DDFT, as a microscopic theory of complex fluids that is applicable to nonspherical particles, confinement, and crowded environments, is well suited to describe intracellular diffusion as well as transport with blood flow. On larger scales, DDFT models for active matter have, of course, a natural application for studying biological microswimmers and active liquid crystals. Applications of classical DFT to crowds \cite{MendezKYSCA2018} as well as descriptions of population dynamics by dynamic equations for the one-body density \cite{MartinezCMOL2013} also indicate the dynamics of crowds or flocks as a possible application of active DDFTs.

Moreover, applications of DDFT to driven and active systems tend to face the problem that DDFT is based on an adiabatic approximation which is not generally possible for such systems. This requires the construction and development of more general methods (such as PFT) as well as the extension of DDFT. A combination of analytical and numerical work can be applied here. For example, in active systems pair-correlation functions \cite{JeggleSW2020,JeggleSW2019b} obtained from simulation results can be used as an input for active DDFTs to replace the equilibrium approximation. Similarly, numerical simulations allow to obtain flow fields \cite{VossW2020} required for hydrodynamic theories. 

Furthermore, given that DDFT is a paradigmatic example of a theory describing the collective dynamics of nonequilibrium systems, studying the foundations of DDFT provides interesting insights into general problems of nonequilibrium statistical mechanics. One such problem is the importance of memory effects. As discussed in \cref{exact}, these are naturally incorporated in theories derived using power functional theory or the projection operator formalism. Standard deterministic DDFT arises as a limiting case after certain simplifications that include neglecting memory effects. Additional topics for which basic research on DDFT can be expected to be fruitful are the description of nonergodic systems \cite{SchindlerWB2019} (see \cref{glass}), the approach to thermodynamic equilibrium \cite{teVrugt2020} (see \cref{thermo}), and origin and importance of fluctuations \cite{ArcherR2004} (see \cref{noise}). A field in which all these aspects are relevant is the study of the glass transition (see \cref{glass}), which can greatly benefit from an improved understanding of the complex relationship between microscopic dynamics, deterministic DDFT, stochastic DDFT, and MCT.

Finally, a lot of unexplored potential can be found in the applications of generalizations of DDFT. Although very general methods have been developed, allowing to include, e.g., nonspherical particles or nonisothermal systems, most applications of DDFT use its simple standard form. Considering more general forms would allow for much more general insights, e.g., into the relaxation and transport dynamics of systems consisting of anisotropic particles, which are a very promising field of research in soft matter physics. This also opens further room for industrial applications of DDFT \cite{DijkstraEtAl2011}.

\section{\label{conclusion}Conclusions}
We have presented a thorough review of classical DDFT, including its relation to other theories and methods as well as its extensions and applications. DDFT is, as should be clear from our presentation, a useful and versatile method as well as an active field of research. It will continue to shape work in statistical mechanics and soft matter physics in the years to come.

\section*{Acknowledgments}
We thank Jens Bickmann, Johannes G.\ E.\ M.\ Fraaije, Kazuhiro Fuchizaki, Ben Goddard, Matthieu Marechal, Roland Roth, Kenneth S.\ Schweizer, Peter Stratmann, Uwe Thiele, and Ren{\'e} Wittmann for helpful discussions.

\section*{Disclosure statement}
No potential conflict of interest was reported by the authors.

\section*{Funding}
H.L.\ and R.W.\ are funded by the Deutsche Forschungsgemeinschaft (DFG, German Research Foundation) -- LO 418/25-1; WI 4170/3-1.

\nocite{apsrev41Control}
\bibliographystyle{apsrev4-1}
\bibliography{control,refs}

\end{document}